\documentclass[fleqn,usenatbib]{mnras}

\usepackage{newtxtext,newtxmath}

\usepackage[T1]{fontenc}
\usepackage{ae,aecompl}
\usepackage{comment}

\usepackage{graphicx}	
\usepackage{amsmath}	
\usepackage{nccmath}
\usepackage{multirow}
\usepackage[dvipsnames]{xcolor}
\usepackage{svg}
\usepackage{academicons}
\newcommand{\orcid}[1]{\href{https://orcid.org/#1}{\textcolor[HTML]{A6CE39}{\aiOrcid}}}
\usepackage{hhline} 

\title[B-D]{Coincidence between morphology and star-formation activity through cosmic time: the impact of the bulge growth}

\author[Dimauro et al.]{
{Paola Dimauro}$^{1,2}$ ,\thanks{E-mail: padim88@gmail.com, paoladimauro@on.br} 
Emanuele Daddi$^{3}$, Francesco Shankar$^{4}$, Andrea Cattaneo$^{2}$, 
\newauthor
Marc Huertas-Company$^{2,5,6}$, Mariangela Bernardi$^{7}$, Fernando Caro$^{1,2}$, Renato Dupke$^{1,8,9}$, 
\newauthor
Boris H\"au\ss ler$^{10}$, 
Johnston, Evelyn$^{11}$, Arianna Cortesi$^{12,15}$, 
Simona Mei$^{13,14}$, Reynier Peletier$^{16}$ \\ 
 \\
$^{1}$ Observatório Nacional, Ministério da Ciencia, Tecnologia, Inovação
e Comunicações, São Cristóvão, 20921-400, Rio de Janeiro,Brazil\\
$^{2}$ Observatoire de Paris, LERMA, PSL University, 61 avenue de l’Observatoire, F-75014 Paris, France\\
$^{3}$ CEA Saclay, Laboratoire AIM-CNRS-Universit\'e Paris Diderot, Irfu/SAp, Orme des Merisiers, 91191, Gif-sur-Yvette, France\\
$^{4}$ School of Physics and Astronomy, University of Southampton, Highfield, SO17 1BJ, UK\\
$^{5}$ Instituto de Astrofısica de Canarias, 38200, La Laguna, Tenerife, Spain\\
$^{6}$ Departamento de Astrofısica, Universidad de La Laguna, 38205, La Laguna, Tenerife, Spain\\
$^{7}$ Department of Physics and Astronomy, University of Pennsylvania, Philadelphia, PA 19104, USA\\
$^{8}$Department of Astronomy, University of Michigan, 311 West Hall,
1085 South University Ave., Ann Arbor, USA\\
$^{9}$Department of Physics and Astronomy, University of Alabama, Box 870324, Tuscaloosa, AL, USA\\
$^{10}$ European Southern Observatory, Alonso de Cordova 3107, Vitacura, Casilla 19001, Santiago, Chile\\
$^{11}$ N\'ucleo de Astronom\'ia de la Facultad de Ingenier\'ia y Ciencias, Universidad Diego Portales, Av Ej\'ercito Libertador 441, Santiago, Chile\\
$^{12}$Observatório do Valongo, Universidade Federal do Rio de Janeiro, Ladeira Pedro Antonio 43, Saude Rio de Janeiro, RJ, 20080-090, Brazil\\
$^{13}$Universit\'e de Paris, CNRS, Astroparticule et Cosmologie, F-75013 Paris, France\\
$^{14}$Jet Propulsion Laboratory and Cahill Center for Astronomy \& Astrophysics, California Institute of Technology, 4800 Oak Grove Drive, Pasadena, California 91011, USA \\
$^{15}$ Centro Brasileiro de Pesquisas Físicas, Rua Dr. Xavier Sigaud 150, CEP 22290-180, Rio de Janeiro, RJ, Brazil \\
$^{16}$ Kapteyn Astronomical Institute, University of Groningen, Postbus 800, NL-9700 Au Groningen, The Netherlands\\
}

\date{Accepted XXX. Received YYY; in original form ZZZ}

\pubyear{2022}

\begin{document}
\label{firstpage}
\pagerange{\pageref{firstpage}--\pageref{lastpage}}
\maketitle

\def \aj {AJ}
\def \mnras {MNRAS}
\def \pasp {PASP}
\def \apj {ApJ}
\def \apjs {ApJS}
\def \apjl {ApJL}
\def \aap {A\&A}
\def \nat {Nature}
\def \araa {ARAA}
\def \iaucirc {IAUC}
\def \aaps {A\&A Suppl.}
\def \qjras {QJRAS}
\def \na {New Astronomy}
\def \aapr {A\&ARv}
\def\lesssim{\mathrel{\hbox{\rlap{\hbox{\lower4pt\hbox{$\sim$}}}\hbox{$<$}}}}
\def\gtrsim{\mathrel{\hbox{\rlap{\hbox{\lower4pt\hbox{$\sim$}}}\hbox{$>$}}}}

\newcommand{\Galfitm}{\textsc{GalfitM}\:}
\newcommand{\GalfitM}{\textsc{GalfitM}\:}
\newcommand{\galfit}{\textsc{Galfit}\:}
\newcommand{\Sersic}{S\'{e}rsic\:}
\newcommand{\BTM}{$B/T_{M_{*}}$}
\newcommand{\BT}{$B/T$}
\newcommand{\SFRM}{$\:log\: SFR-log\: M_{*}$\:}
\newcommand{\SFM}{Main Sequence }
\newcommand{\M}{log $M_{*}$}

\begin{abstract}
The origin of the quenching in galaxies is still highly debated. Different scenarios and processes are proposed. We use multi band ($400-1600$ nm) bulge-disc decompositions of massive galaxies in the redshift range $0<z<2$ to explore the distribution and the evolution of galaxies in the \SFRM plane as a function of the stellar mass weighted bulge-to-total ratio (\BTM) and also for internal galaxy components (bulge/disc) separately. We find evidence of a clear link between the presence of a bulge and the flattening of the Main Sequence in the high-mass end.  All bulgeless galaxies (\BTM<0.2) lie on the main-sequence, and there is little evidence of a quenching channel without bulge growth. Galaxies with a significant bulge component (\BTM>0.2) are equally distributed in number between star forming and passive regions. The vast majority of bulges in the Main Sequence galaxies are quiescent, while star-formation is localized in the disc component. Our current findings underline a strong correlation between the presence of the bulge and the star formation state of the galaxy. A bulge, if present, is often quiescent, indipendently of the morphology or the star formation activity of the host galaxy. Additionally, if a galaxy is quiescent, with a large probability, is hosting a bulge. Conversely, if the galaxy has a disky shape is highly probable to be star forming.

\end{abstract}




\section{Introduction}

The distribution of galaxies in the plane defined by the star-formation rate and the stellar mass (\SFRM) is a powerful diagnostic of how stellar mass is assembled in galaxies. Deep surveys undertaken in the last decade have allowed to extensively investigate how galaxies populate the plane from $z\sim3$. One key result, which has deeply changed our understanding of how galaxies form and evolve, is the presence of the so-called main sequence of star-formation (e.g. \citealp{Brinchmann2004,Elbaz2007,Daddi2007,Whitaker2012}). Star forming galaxies present a remarkable correlation between their stellar mass and the rate at which they form stars. The slope of the main-sequence is close to unity leading to a roughly constant specific star-formation rate for all star forming galaxies. This has been interpreted as evidence for the self-regulation of star formation in galaxies. The rate at which gas is converted into stars is usually interpreted in cosmological dark matter-dominated models as a consequence of smooth mass accretion onto the host dark matter haloes (e.g. \citealp{Dekel2009,Rodriguez2017}). Kinematic studies of the gas in galaxies lying on the main sequence have also shown that for the vast majority the gas is rotating (e.g, \cite{Wisnioski2015}, although this has been challenged by some recent works (e.g, \citealp{Rodriguez2017}). The Main Sequence of star-formation has been observed at least since $z\sim3$ (\citealp{Whitaker2012,Barro2015}), with an evolution with redshift consistent with a decreasing normalization and nearly constant slope, i.e. galaxies at high redshift formed stars at higher rates than low redshift galaxies of the same stellar mass. It is still unclear whether the evolution with redshift on the main sequence stems from more efficient conversion of gas into stars at earlier epochs or from larger reservoirs of cold gas. Several works have pointed out that the evolution of the normalization of the main-sequence closely tracks the increase of the gas fraction with redshift (e.g \citealp{Genzel2012,Magdis2012}). This would suggest the existence of an universal mode of star-formation, essentially driven by the amount of available gas in a reservoir, which is regulated by the galaxy halo which could boost the conversion of gas into stars.
 A 'gas-regulator' approach, generally called 'bathtub' \citep{Bouche2010,Lilly2013,Pipino2014,Dekel2014b,Feldmann2013}, proposes a simple model that links together the mass assembly of the dark matter haloes, the evolution of the gas content, the metal content and the stellar population of the galaxies through cosmic time, and successfully reproduces many of the key galaxy scaling relations \citep{PengM2014}.  
The $SFR-M_*$ relation has also revealed the existence of galaxies, specially at the high mass end, with specific star-formation several orders of magnitude lower than main-sequence galaxies of comparable stellar mass. Quiescent galaxies already exist since at least $z\sim3$, possibly even earlier \citep{Wuyts2007,Whitaker2015}, although their number density increases monotonically with redshift, progressively dominating the high-mass end of the stellar mass function at $z<1$. (e.g \citealp{Ilbert2013,Muzzin2013,Bernardi2013,Bernardi2016,Bernardi2017a,Bernardi2017b}). Understanding how galaxies move from the main-sequence to the \emph{passive cloud}, i.e. how galaxies cease star formation, has become one of the key questions in the field of galaxy evolution. The bimodal distribution of the SFRs/colors was initially seen as evidence that the star formation shutdown must be fast, either through gas removal or heating \citep{Granato2004,Hopkins2006,Shankar2006}. Feedback from AGN/quasar \citep{Silk1998,Harrison2018} or galaxy-galaxy interactions \citep{Toomre1977,Hopkins2010} are commonly invoked processes to explain quenching at the high mass end. Recent works also pointed out that a more gradual transition through strangulation might in fact be a common channel for quenching. This is seen to occur in satellite galaxies (\citealp{Peng2015}), but it might also be relevant for central galaxies. Gas entering massive haloes can indeed be shock heated and prevented from cooling and forming new stars \citep{Dekel2006,Cattaneo2006}. In this context, the flattening of the slope of the main sequence at the high mass end can be interpreted as a decrease in the star-formation activity of massive galaxies. However, it is also well known that galaxies can rebuild a disc through the accretion of gas and eventually move back to the main sequence. The importance of this \emph{rejuvenation} process (\citealp{Mancini2019,Chauke2019,Martin2021}) is still an open issue.

Another key unsolved question is how the galaxy morphology and its stellar populations correlate with the decline of the star-formation. Some models also suggested that the change in galaxy morphology from disky/spirals to spheroidal/ellipticals shape could itself be a driver for quenching \citep{Martig2009}. The advent of large surveys has encouraged the community to develop statistical methods to try and address this open question \citep{Kauffmann2003,Baldry2004,Wuyts2011}. Quantitative measurements of galaxy morphologies are based on the idea that galaxy light
profiles are sufficiently well described by analytic formulae as the Sersic profile \citep{Sersic1968,deVaucouleurs1953}. Different codes have been developed to allow an automatized analysis (some example \citealp{Peng2002,Bertin1996,haeussler2013,Li2021,Akhlaghi2021}). Meanwhile, as the level of detail in observations has increased and the wavelength coverage of data has improved, the focus of the analysis has also expanded to resolve stellar populations. It is today well known that most of the red and passive galaxies tend to present early-type morphologies, while star forming galaxies have a  more disc-dominated structure and a younger stellar population (e.g \citealp{Kauffmann2003,Franx2008,Schawinski2009,Wuyts2011,Whitaker2012,Huertas2015,Huertas2016, Morselli2019}).
A complementar measurement of this correlation is based on the analysis of the stellar mass density. Quiescent galaxies have been shown to have a denser core at all redshifts, which is well captured by the stellar mass density in the central kpc (e.g \citealp{Fang2013,Barro2015,Dimauro2019}). Such phenomena have been seen in numerical and empirical simulations (\citealp{Tacchella2015,Tacchella2016,Rodriguez2017}) and it might be the indication of a compaction that precede the quenching (e.g. \citealp{Zolotov2015,Varma2021}). In \cite{Dimauro2019}, we showed that this increase in the central stellar density is directly correlated with the growth of the bulge component in galaxies. Thus, following this line of thought, the presence of a prominent bulge in galaxies should be correlated with their level of star formation rate.

In order to shed light on how galaxies move in the $SFR-M_*$ plane and how this is related to galaxy morphology, it is however required to spatially resolve the star-formation activity within galaxies. This would allow to track the shutdown of star formation setting additional constraints on how gas is consumed \citep{Johnston2017}. Integral Field Unit (IFU) surveys are the optimal dataset for this analysis. However, today's IFU surveys are limited to the local Universe (e.g. MANGA \citep{Bundy2015}, SAMI \citep{Croom2012}, CALIFA \citep{Sanchez2012}). 

In a series of papers, we are investigating the structural and stellar population properties within galaxies from $z\sim2$ by decomposing the surface brightness profiles in bulges and discs in multiple high-resolution filters. While \cite{Dimauro2019} focused on the analysis of the structural properties, this work aims at unveiling the correlations between the position of a galaxy on the \SFRM plane and its stellar properties, most notably the bulge-over-total stellar mass ratio(\BTM) and its rest-frame colour. The strenght of this work is in the use of mass-weighted quantities (as the \BTM) instead of flux-dependent parameters. Indeed, the computation of the \BTM takes advantange of the entire set of available observations not only single bands. We will dissect where bulges and discs are in the \SFRM plane and show that there is a preference for large \BTM to reside below the Star Forming Main Sequence. We will also investigate the contribution of the bulge and the disc to the star formation activity of the hosting galaxy.\\

This paper is organized as follows. Section \ref{sec:data} provides a description of the data as well as of the external information used in the analysis presented in this work. The distribution of morphologies along the main sequence is discussed in sections \ref{sec:SFR1} and \ref{sec:SFR2}. Section \ref{sec:best_fit} give a quantitative notion of the results  while section \ref{sec:colors} focuses on the rest-frame colors of bulges and discs. Results are discussed in section \ref{sec:discussion} and summarized in \ref{sec:conclusion}.

\section{Data}
\label{sec:data}

The analysis presented in this paper made use of the exquisite high-quality data from the Cosmic Assembly Near-IR Deep Extragalactic Legacy Survey: CANDELS \citep{Grogin2011,Koekemoer2011}. Information about the structure as well as the photometry of CANDELS galaxies are from the multi-wavelength bulge/disc decomposition catalog presented in \cite{Dimauro2018} (hereafter DM18). They were retrieved modeling the surface brightness profile with the Megamorph package (i.e. Galapagos-2, GalfitM,  \citealp{haeussler2013,Barden2012,Vika2014}), to simultaneously apply a 2-component model fit (\Sersic +Exponential) over the $4-7$ filters (depending on the field) covering the spectral range $400-1600$ nm. The structural parameters are well recovered down to the magnitude limit of 23 in the F160W band with a statistical error of $\sim20\%$. In this work, we selected galaxies within this magnitude limit in all the 5 CANDELS fields to build a clean and robust sample of bulges and discs models. 
The final sample is mass complete down to $10^{10.7}M_{\odot}$ at $z\sim2$. The completeness limits are reported in Table \ref{tab:compl}. We provide a brief summary about the modelling and the morphological classification in the next 2 subsections. More details can be found DM18.

\begin{table}
\centering
\begin{tabular}{c|c|c|c}
 \hline
  z &  all & Q & SF \\
 \hline
 \hline
0-0.5   & 9.0  &  9.16 & 8.98 \\
 \hline
0.5-1.0 & 9.75 &  9.91 & 9.79 \\
 \hline
1.0-1.4 & 10.3 & 10.38 & 10.28 \\
 \hline
1.4-2.0 & 10.7 & 10.72 & 10.69 \\
 \hline
 \hline
\end{tabular}
\caption{Stellar mass completeness limits of the sample used in this work. We show the values for all galaxies, quiescent (Q) and star forming (SF). The table is taken from DM18 but reproduced here for clarity.}
\label{tab:compl}
\end{table}

\subsection{B-D catalog and morphological classification}
\label{sec:cat1}

The DM18 catalog contains structural and stellar properties information for $\sim$ 17600 galaxies. Each galaxy profile is fitted using \Galfitm considering two different setups: single \Sersic profile and a 2-component \Sersic + Exponential disc profile. Moreover, since this software allows us to reconstruct galaxy models in a multiwavelength mode, we used several configurations, in which different wavelength dependences of the model, such as constant, linear or higher order polynomial functions, were tested \citep{haeussler2013}.

The new feature of the DM18 catalog is in the 1-2 component selection algorithm. We developed a new decision criteria, based on supervised deep learning, that estimates an a-priori probability for each galaxy to be better described by single or a multiple component model. The probability allows us to chose the optimal solution to reconstruct the surface brightness profile for each galaxy, reducing systematic errors to $\sim10\%$, which otherwise would be of the order of $\sim50\%$. This step is crucial to define a clean sample of discs and bulges that will not introduce obvious systematics in the analysis. Full details of the procedure can be found in \cite{Dimauro2018}. 

The public catalog \footnote{\href{http://paoladimauro.space/morph_cat/}{link} to the catalog} contains the main setup (setups 1 and 4 from Table 1 in DM18) where sizes are modeled with a second-order polynomial function over the wavelength. 
This is also the main catalog used in this work. Other setups, with more restrictive constraints, are used to correct ambiguous cases (according to the deep-learning classification: Setups 4 and 6) and also to estimate random uncertainties on the structural parameters derived from the fitting procedure (following a similar approach to \citealp{vanderWel2014}). The 1-2 component choice is already applied in the final catalog. Consequently, only the best profile is provided in the public release and used in the analysis of this work, in accordance with the selection algorithm.

\subsection{Stellar population analysis and rest-frame colors}
\label{sec:cat2}
Given the availability of the ensemble of multiwavelength photometric informations, we can perform a Spectral Energy Distribution fit (SED fitting with the code FAST (\citealp{Kriek2009}) separately for bulges and discs. We used \cite{Bruzual2003} stellar population models with \cite{Chabrier2003} initial mass function(IMF) and \cite{Calzetti2000} extinction law. As a direct result, we obtain stellar masses of bulges and discs, from which we derived stellar mass bulge-to-total ratio (\BTM) with typical uncertainties < 0.2. In DM1, we have made an extensive effort to test the goodness and the accuracy of the modeling process as well as of the stellar analysis results. In detail, we have tested stellar masses estimation accounting different number of filters and we have compared them with the ones from the official CANDELS catalog. Our estimates are unbiased with a scatter that reach a maximum of $\sim$ 0.3 that increases to $\sim$ 0.4 if less than four filters are used (see Figure 18 from DM18 for more details). 

\begin{figure}

$\begin{array}{c c}
\includegraphics[width=0.2\textwidth]{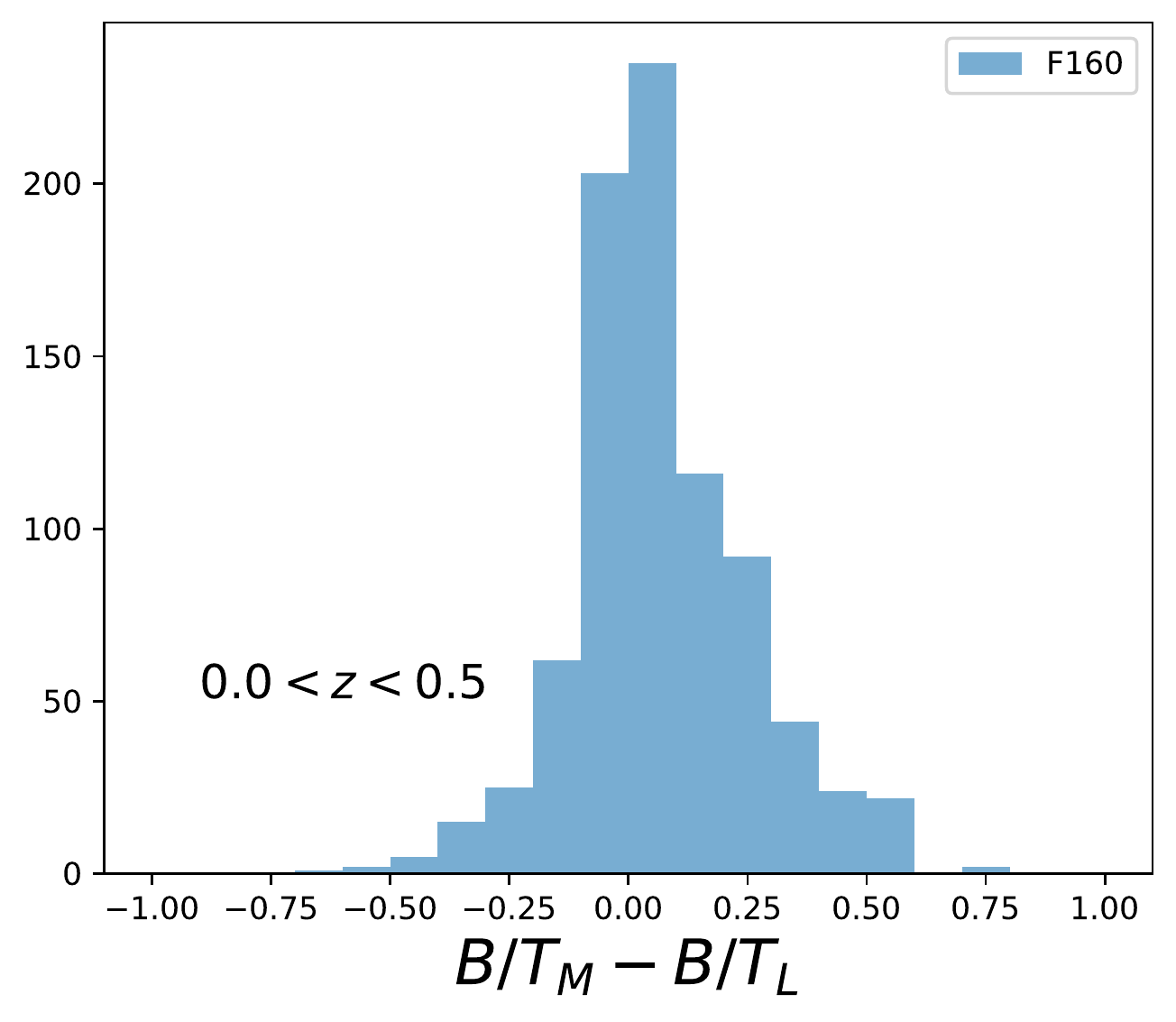}&
\includegraphics[width=0.2\textwidth]{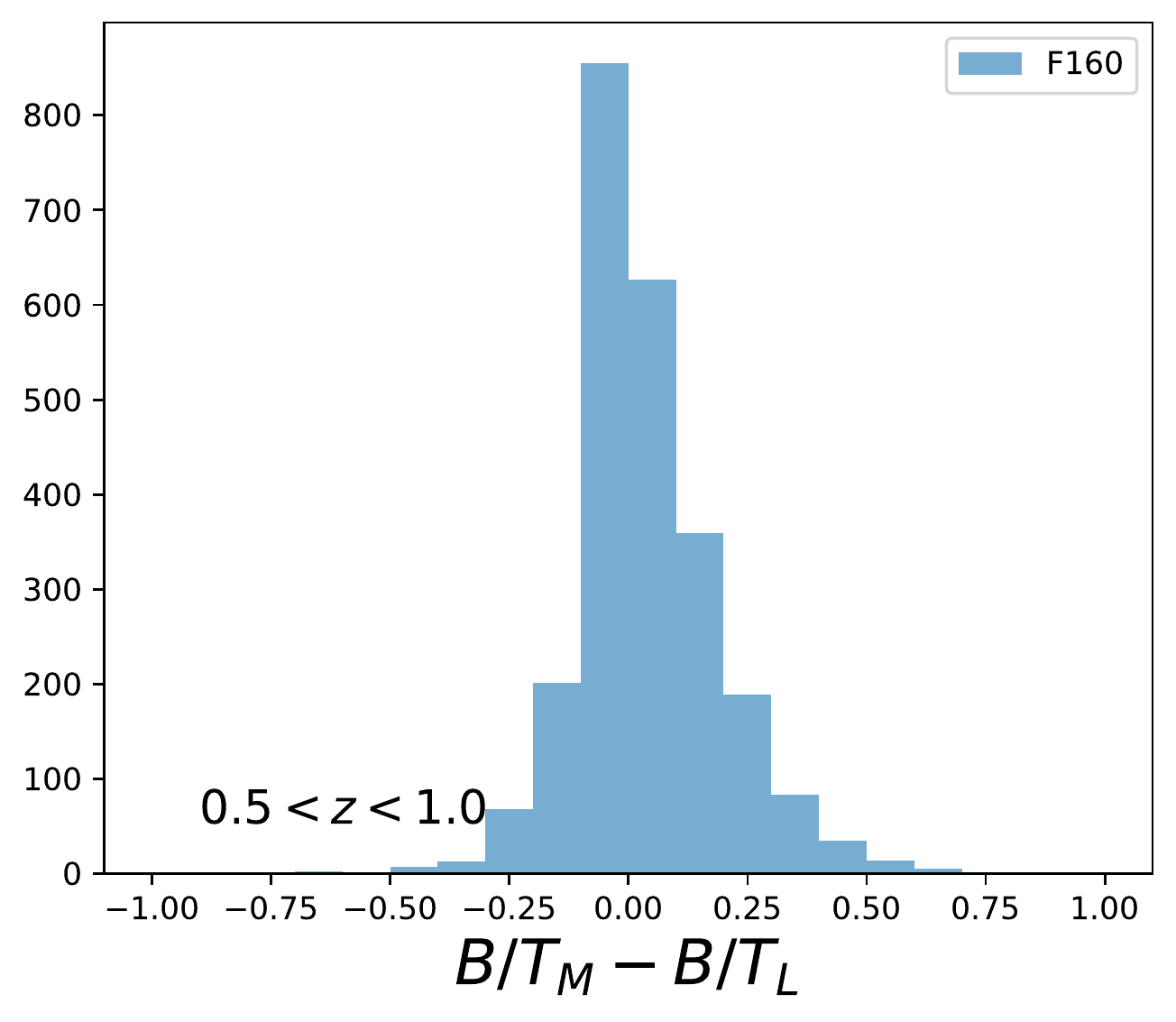}\\
\includegraphics[width=0.2\textwidth]{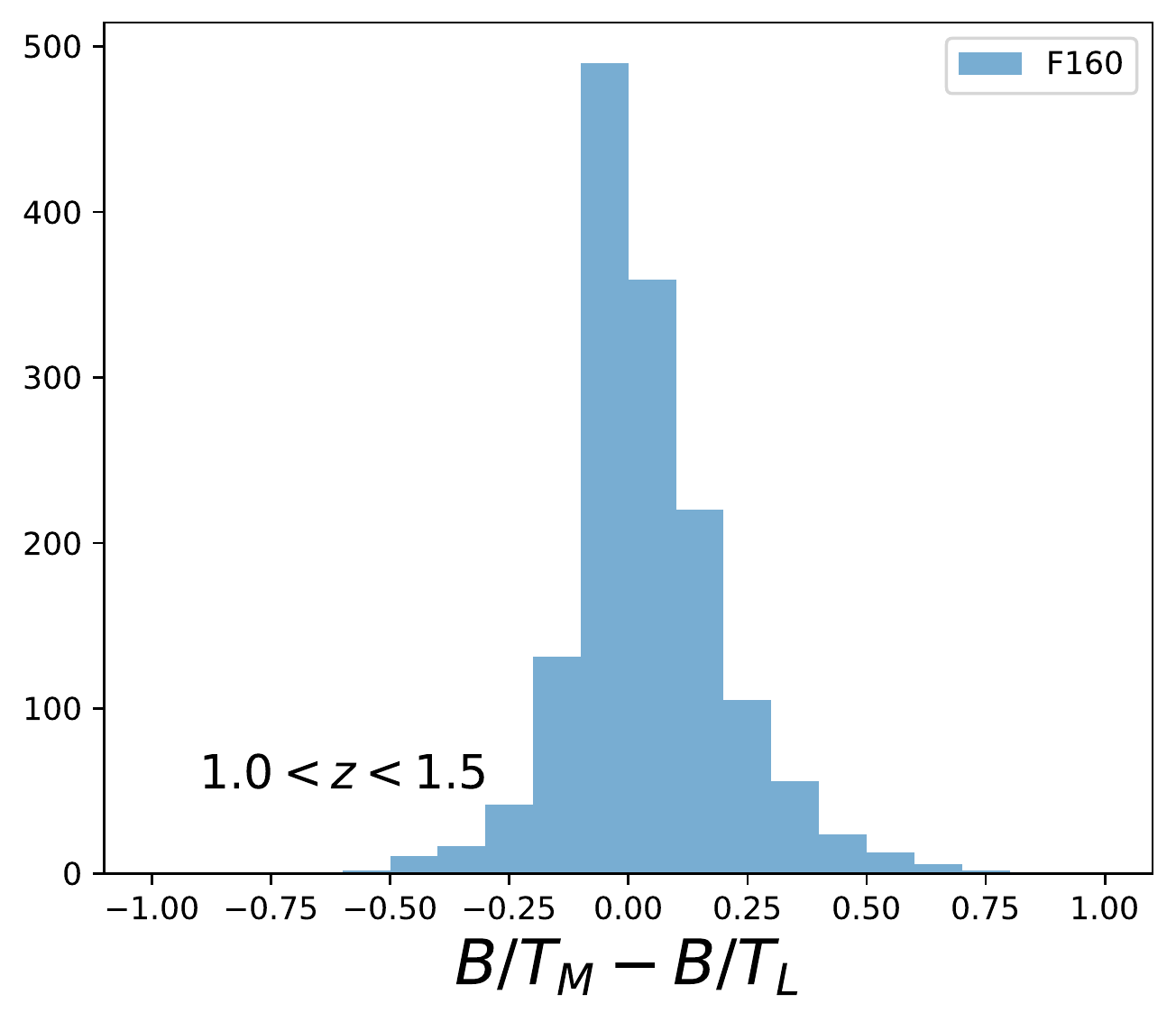}&
\includegraphics[width=0.2\textwidth]{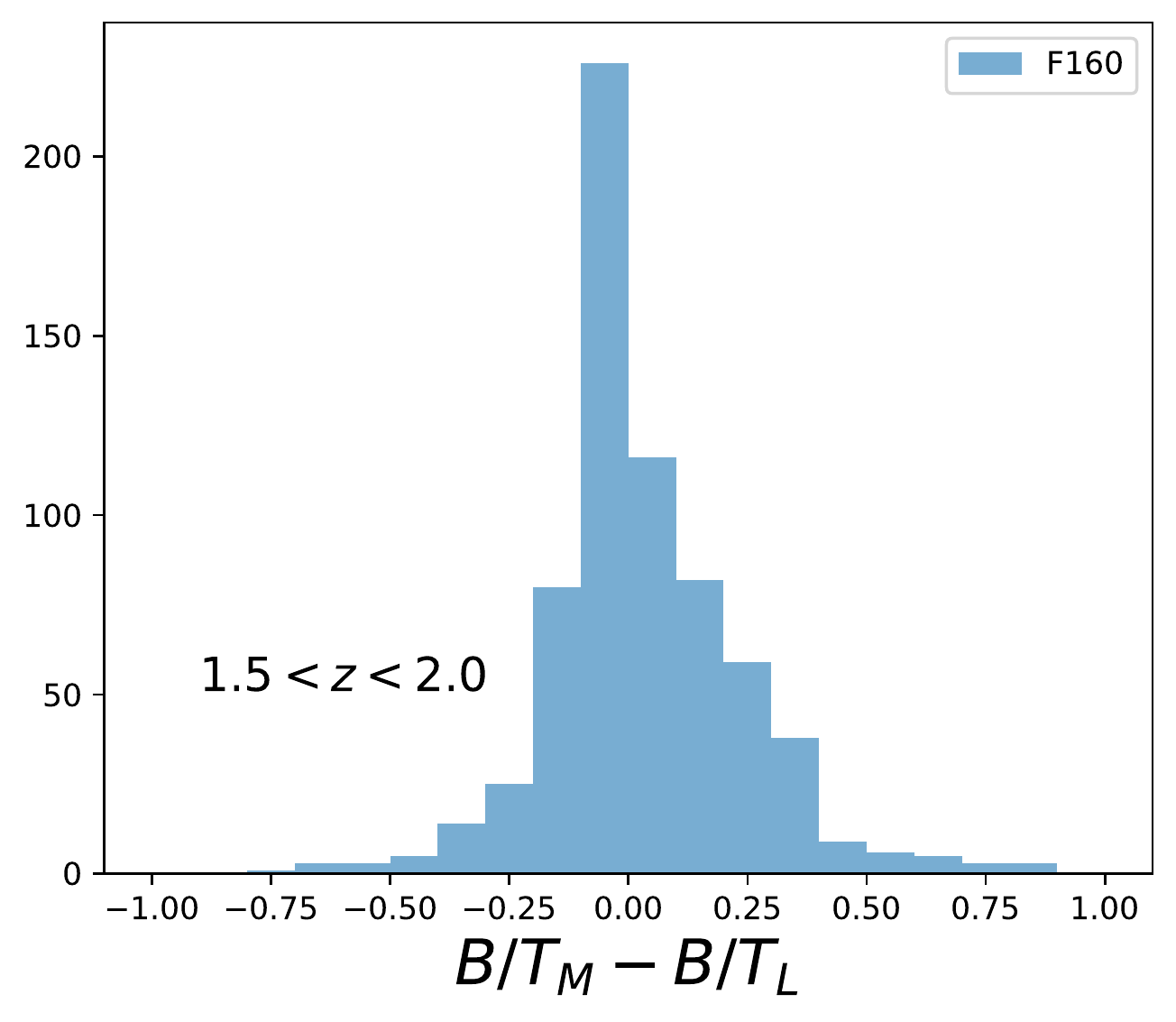}\\
\end{array}$
\caption{Comparison between \BT (F160 filter) and \BTM. The two quantities are in agreement with a mean of the difference of 0.012 $\pm$ 0.2.}
\label{fig:B_T_M}
\end{figure}

Additionally, we estimated U, V, J rest-frame colors using the flux interpolation from the theoretical rest-frame SED. This method avoids the need of applying further corrections. In order to test the accuracy of this method, we compared our measurements with rest-frame colors from the CANDELS catalog and we obtain a statistical error of 10\%. More details are given in the appendix \ref{apx:UVJ}. For each galaxy, i.e. for each set of magnitudes, a distribution of 100 mock SEDs is generated using a MonteCarlo method within uncertanties on each magnitude value. Rest-frame colors are interpolated for all the mock SED. Consequently, uncertanties on the U-V and V-J rest-frame colors are computed taking the median from the U-V,V-J distributions for each galaxy. The choice of the quantity of mock galaxies to be used for each object was taken as a good compromise between computational time and accuracy of the result.
\citep{Bernardi2022} recently showed that a gradient in stellar population can reflect a gradient in the stellar mass-to-light (M/L) ratio within a galaxy, consequently affecting the \BT. However, since we decide to use a constant IMF, the M/L is expected to be constant within a galaxy. Figure \ref{fig:B_T_M} shows a comparison between \BT and \BTM. The 
mean is found to be 0.012 $\pm$ 0.2 and it confirms the absence of strong systematics caused by the approssimation adopted in the SED analysis.

\subsection{Additional information}
\label{sec:CANDELS_cat}
In addition to the DM18 catalog, we used several information from the official published CANDELS catalog  \citep{Galametz2013,Guo2013,Stefanon2017,Barro2019}.
We used the CANDELS spectroscopic redshifts, when available, but also photometric redshifts, derived using a variety of SED fitting codes (for more details see \citealp{Dahlen2013}). The measured errors are of the order of $\Delta_z/(1+z) \sim$3\%. 
Additionally, we used total stellar masses and UVJ rest-frame colors from the CANDELS collaboration catalog, to test the correctness of our results.
Stellar masses are derived through SED fitting using the best available redshift with SYNTHETIZER \citep{Perez2003}. They are computed using a grid of \cite{Bruzual2003} model, \cite{Salpeter1955} IMF, solar metallicity and a \cite{Calzetti2000} extinction law were considered. 
We converted the CANDELS stellar masses to the \cite{Chabrier2003} IMF for consistency with the analysis of the present work but also to directly compare different values, by applying a $0.22$ dex shift \citep{Bernardi2010,Bernardi2013}.

Finally, we also used integrated SFRs from the CANDELS catalog. They are computed by combining IR and UV rest-frame luminosities \citep{Kennicutt1998,Bell2005} and adopting a Chabrier IMF (see \cite{Barro2011} for more details). The following relation was used: $SFR_{UV+IR}=1.09\times10^{-10}(L_{IR}+3.3L_{2800})$. Total IR luminosities are obtained using Chary \& Elbaz (2011) template-fitted MIPS $24\mu m$ fluxes and applying a \emph{Herschel\,based} recalibration. For galaxies undetected in $24\mu m$, SFRs are estimated using rest-frame UV luminosities (\citealp{Wuyts2011}).

\section{Morphology along the main sequence: where are the bulges?}
\label{sec:SFR1}

\begin{figure*}
\centering
$\begin{array}{c c}
\includegraphics[width=0.4\textwidth]{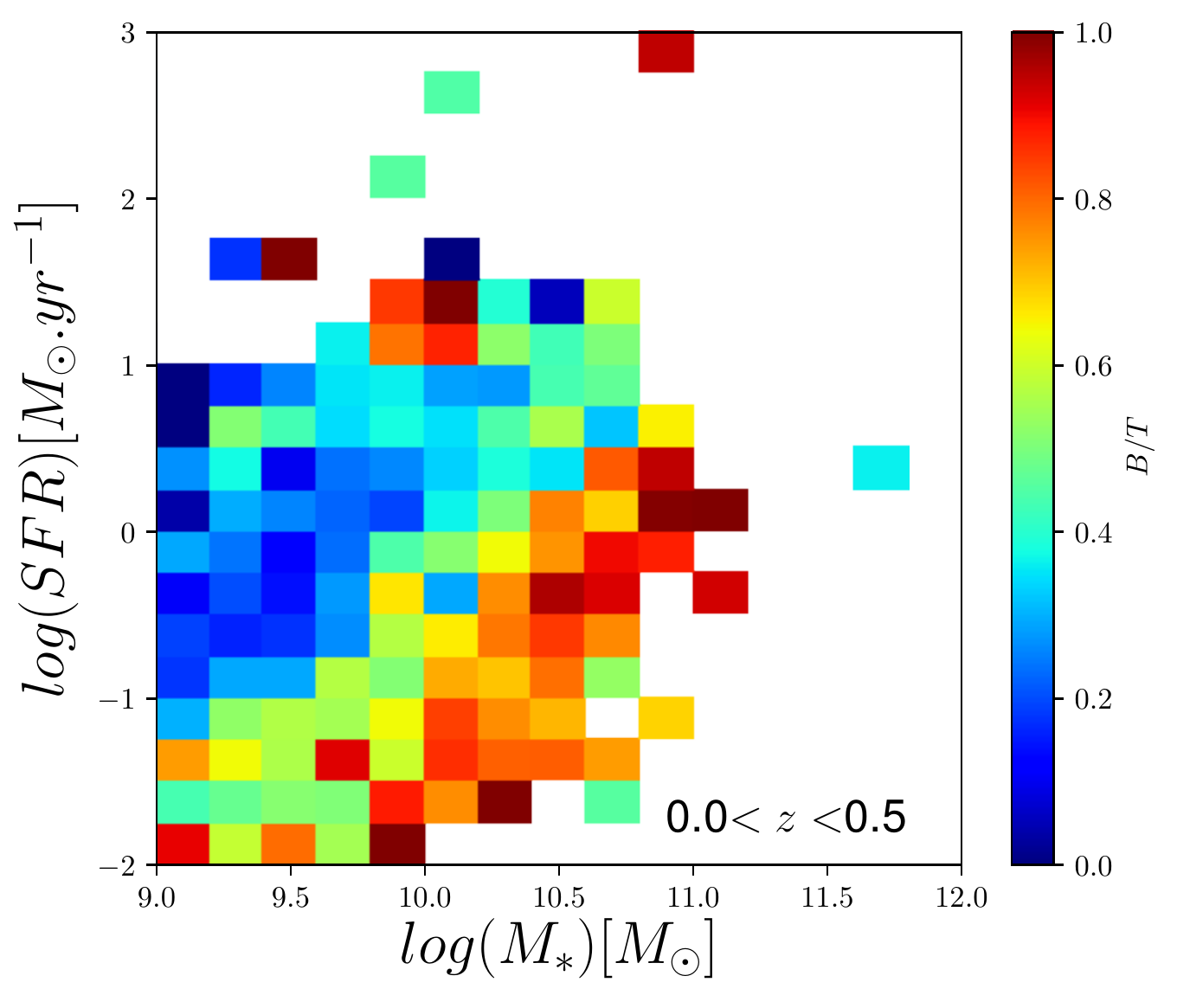}&
\includegraphics[width=0.4\textwidth]{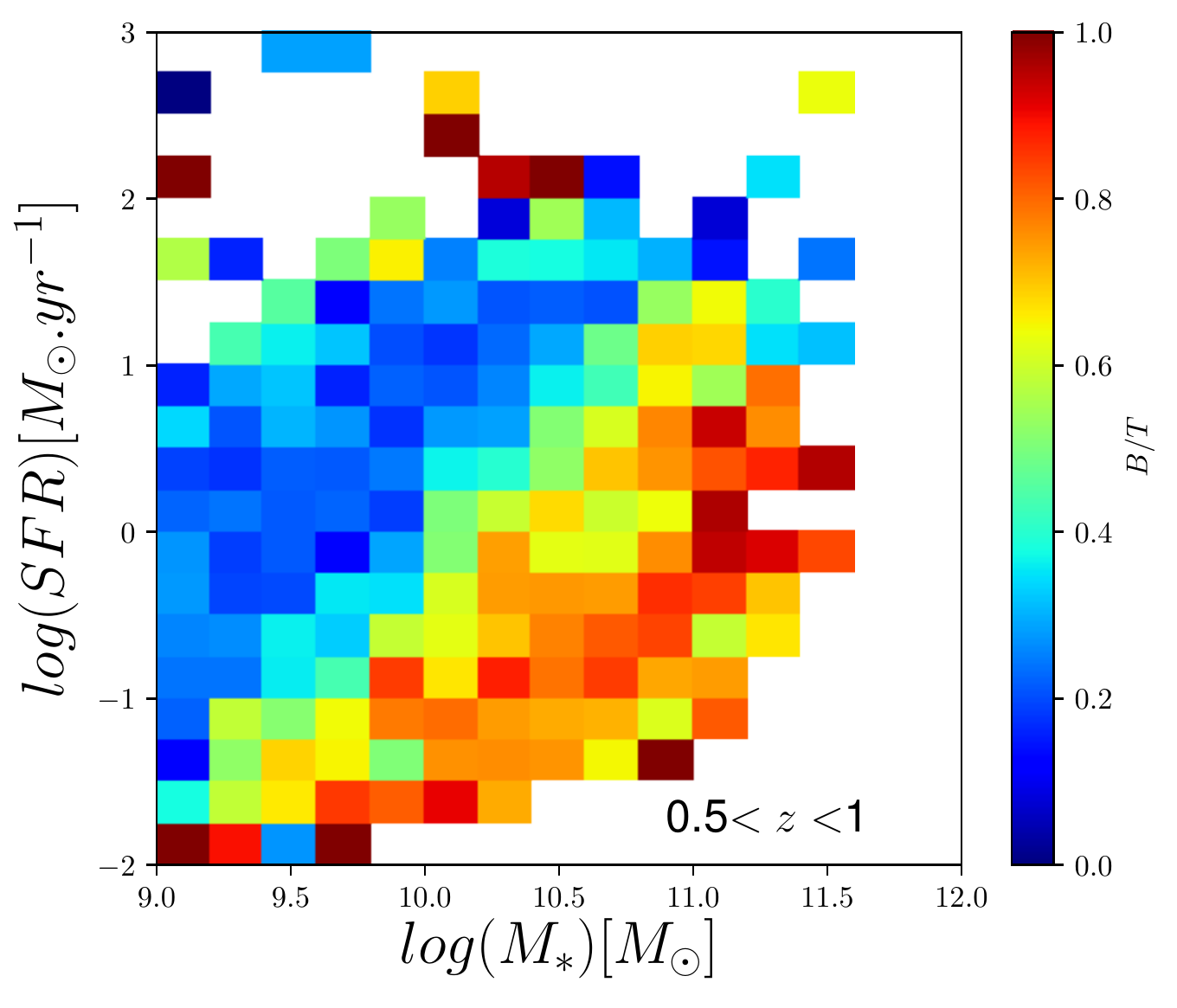}\\
\includegraphics[width=0.4\textwidth]{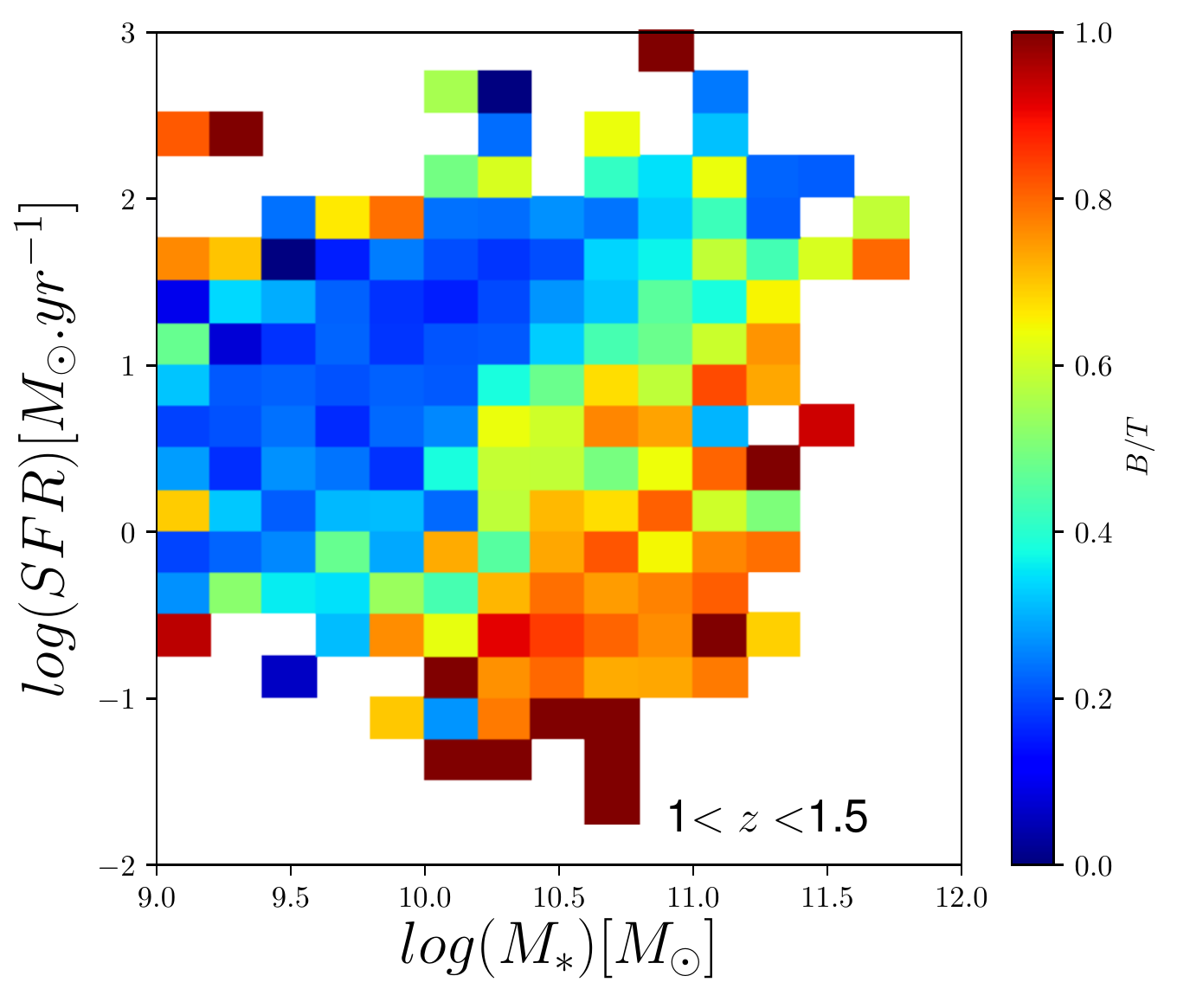}&
\includegraphics[width=0.4\textwidth]{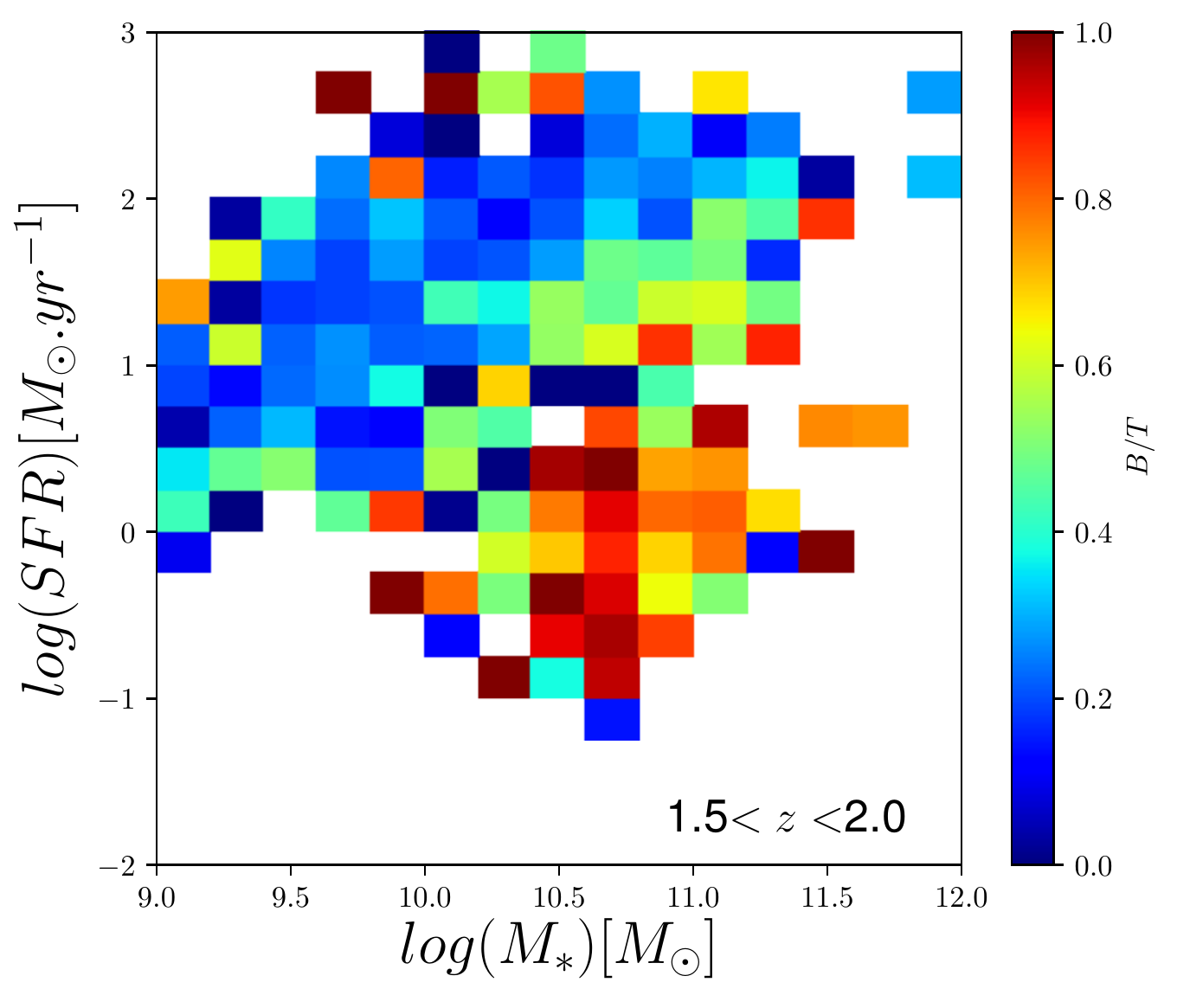}\\
\end{array}$
\caption{Distribution of the morphology along the \SFRM plane. Colors are representing the mean \BTM for each SFR-mass bin. Most of the disky galaxies lie in the MS while galaxies with a more prominent bulge are concentrated in the quenched region.}
\label{fig:SFR_mstar1}
\end{figure*}
 
It is well known today that galaxies in the \SFRM plane, are concentrated in two main regions: the main sequence and the quenched region. The position of a galaxy in the plane is also linked to the shape of its surface brightness profile, i.e. with the \Sersic index and the half-light size (ex: \citealp{Kauffmann2003,Baldry2004,Wuyts2011,Brennan2016}). Passive galaxies are compact and spheroidal, while the star forming ones have a lower stellar density and an extended disky structure \citep{Wuyts2011,Bernardi2014,Whitaker2015,Dimauro2019}. Similarly, it has been shown that there exists a link between the SFR and the bulge-over-total light ratio (e.g.\citealp{vanderWel2014,Lang2014,Whitaker2015,Morselli2019}).
While several works already suggested the existence of a correlation between morphology and star formation activity (\citealp{Wuyts2011,Huertas2015,Whitaker2017}), the link between the presence of stellar bulge and the level of star formation is still not well understood.
For this reason, in this work we explore the cause-consequence relation between the properties of galaxies and the presence of the bulge. 

We start studying the distribution of bulges and discs within the \SFRM plane. The newly added information in this analysis is the use of the bulge-over-total stellar mass ratio (\BTM) as a morphological proxy instead of flux dependent parameters like the bulge-over-total flux ratio. Indeed, the flux relevance of bulges and discs depends on the observed band, and therefore it is not constant over wavelength, introducing possible systematics. Differently, stellar masses are computed taking advantage of the entire set of available observations, making the measurement more robust (statistical uncertainties on \BTM are of the order of$\sim$ 20\%. See DM18 for more informations). Given the above, using stellar mass weighted parameters represents the best solution. 

Figure \ref{fig:SFR_mstar1} shows the distribution of galaxies in the \SFRM plane (SFR values are from the CANDELS catalog and are not estimated within our analysis, see sec: \ref{sec:CANDELS_cat}). The color code represents the mean \BTM in SFR-M bins. As expected, the main sequence is mostly populated by disky galaxies with \BTM<0.2-0.3, while galaxies that are hosting a prominent bulge (\BTM>0.5) tend be dominant in the quiescent region. 
This trend is consistent with the bimodal distribution of morphological types. Massive bulges are more probable to be hosted in passive galaxies, while disc-dominated structures are found more often within star forming galaxies. The mean scatter of the \BTM distribution for each \SFRM bin is also computed and shown in Figure \ref{fig:SFR_mstar5}. No clear features are observed, supporting the reliability of the main result (Appendix \ref{apx:SFR_std}). As an additional test, Figure \ref{fig:SFR_mstar6} shows the same excercise of Figure \ref{fig:SFR_mstar1}, but using the light-weighted $B/T$ (F160 filter). The main trends are still present independently of the different estimations of $B/T$ used.

\begin{figure*}
\centering
$\begin{array}{c c}
\includegraphics[width=0.4\textwidth]{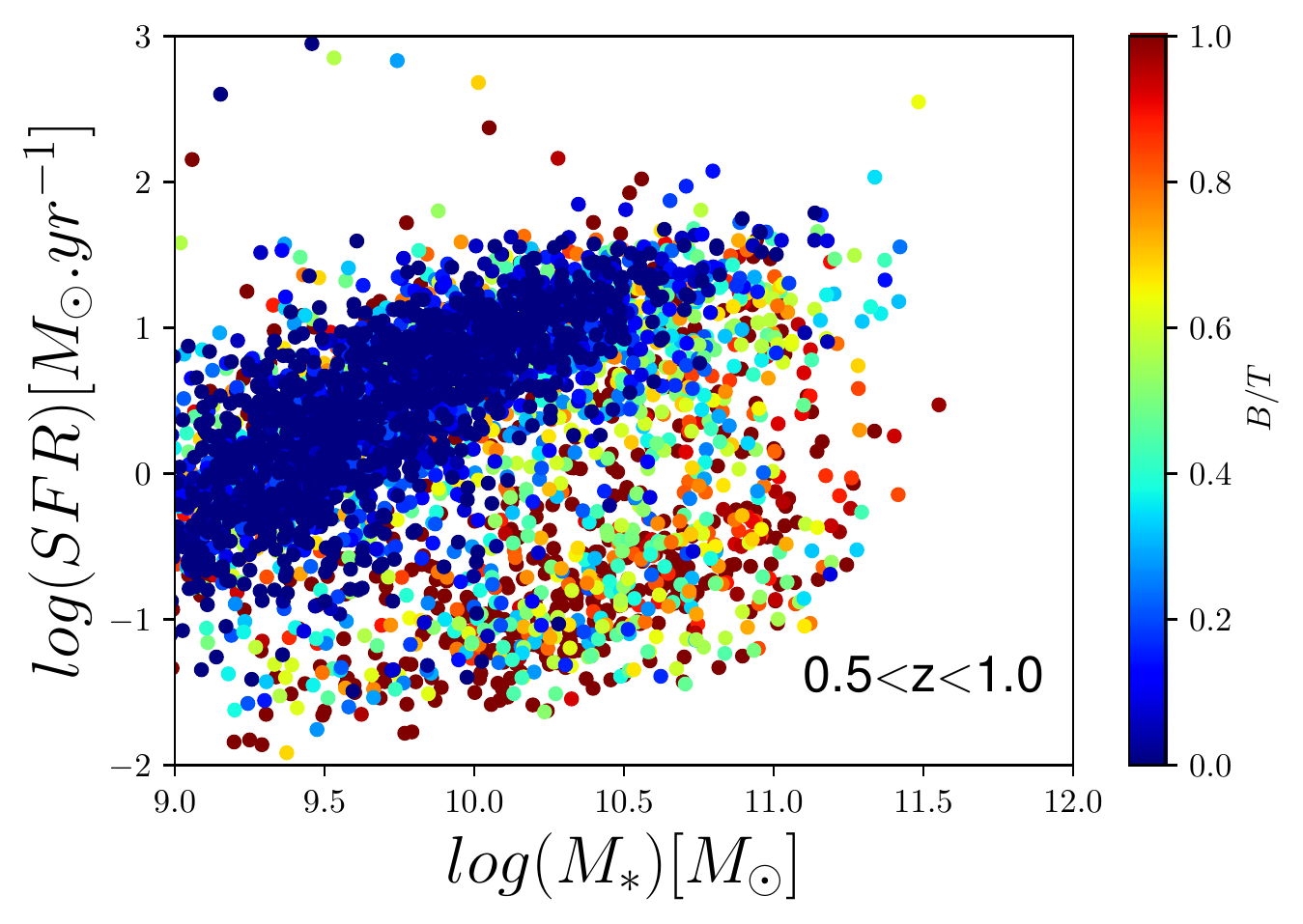}&
\includegraphics[width=0.4\textwidth]{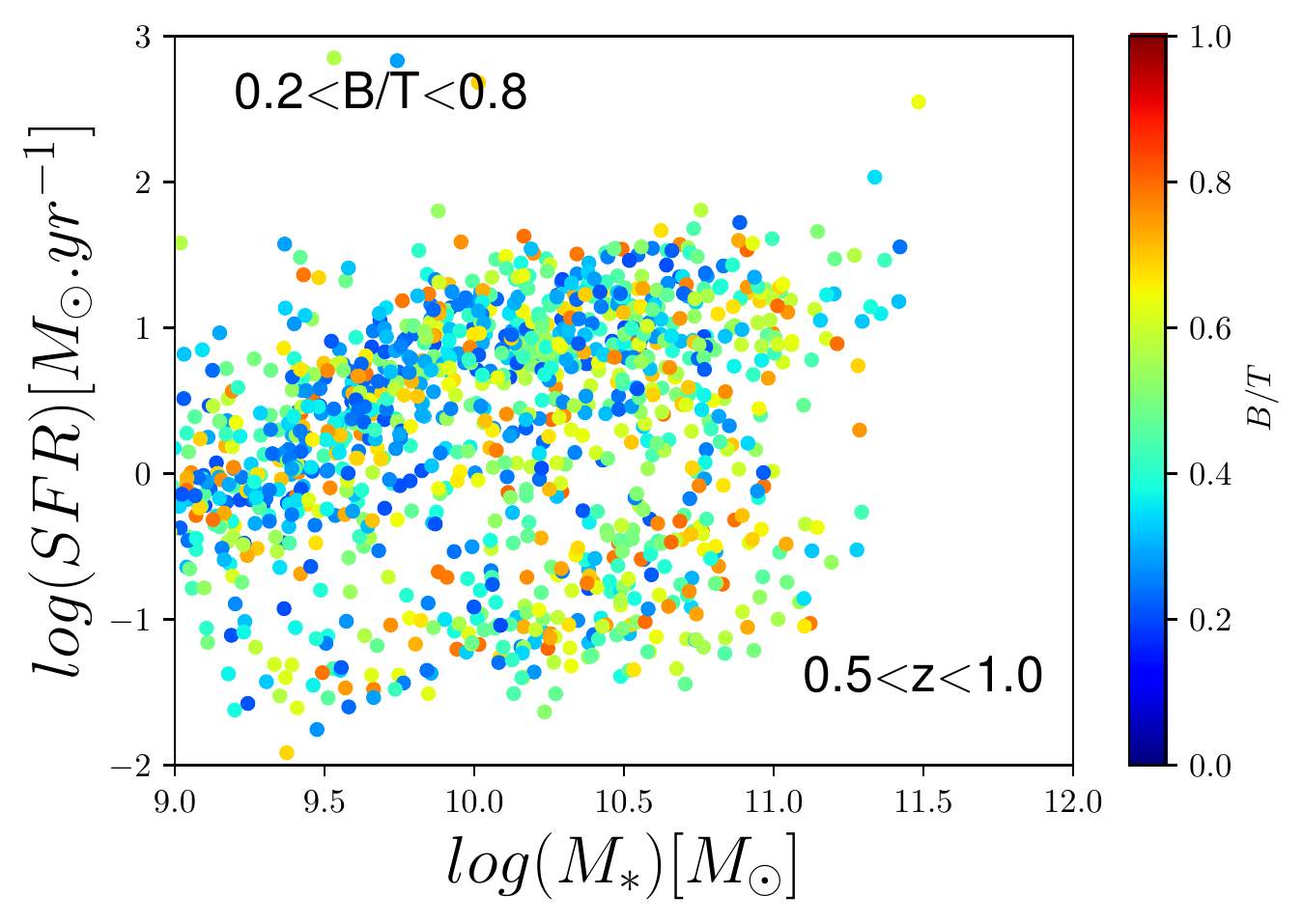}\\
\includegraphics[width=0.4\textwidth]{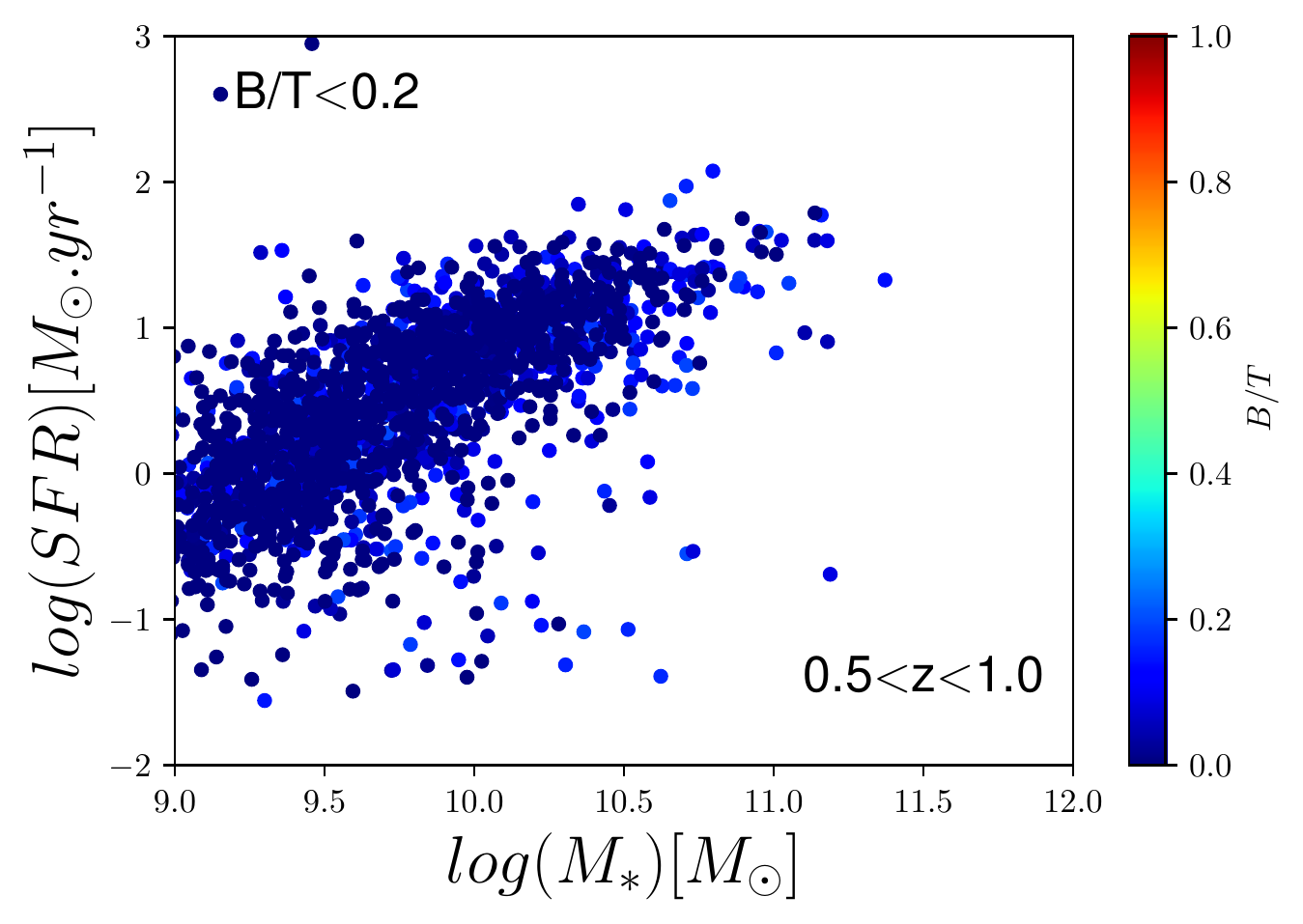}&
\includegraphics[width=0.4\textwidth]{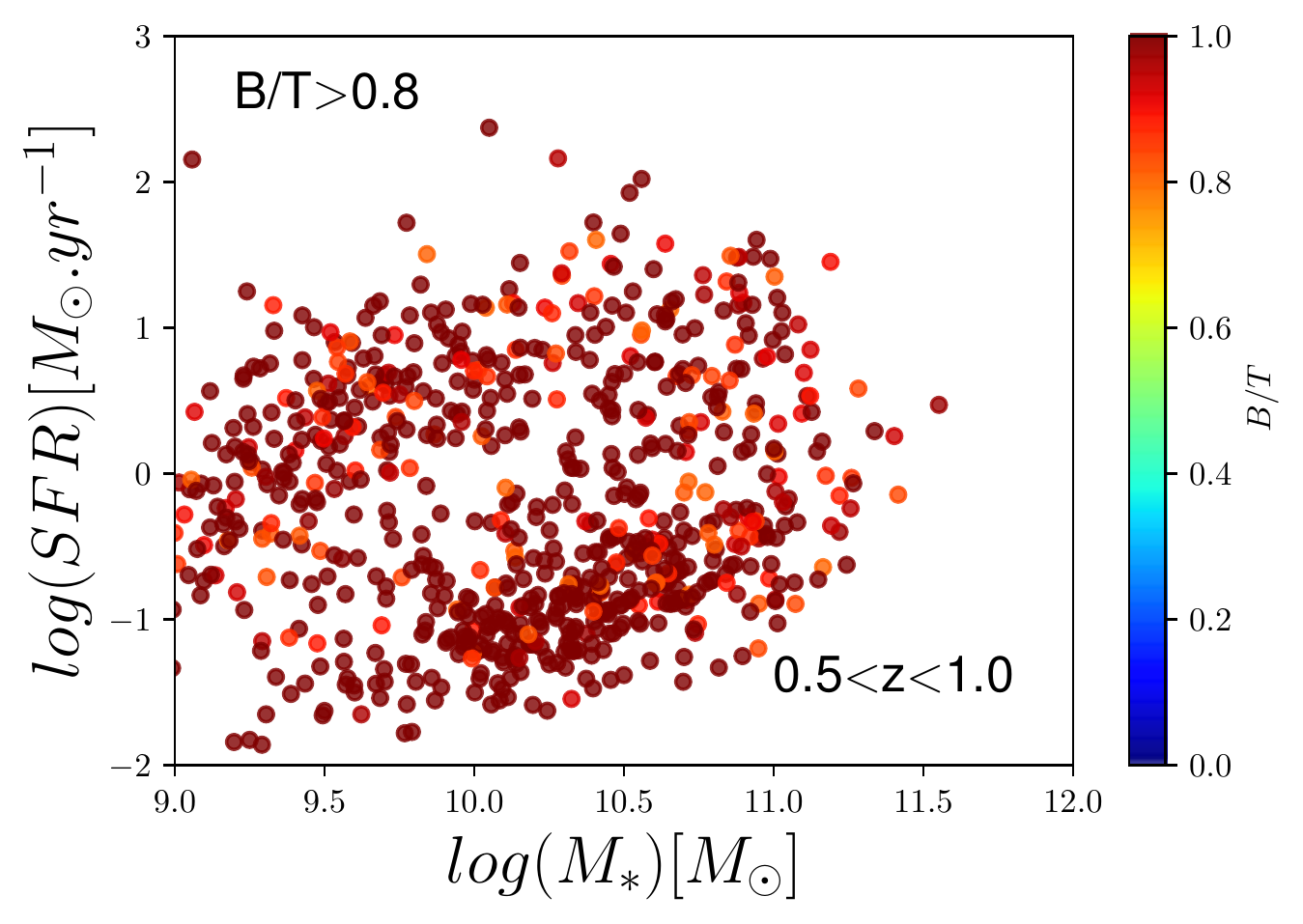}\\

\end{array}$
\caption[SFRM plane]{Distribution of galaxies in the \SFRM plane. From top left to bottom right, the sample divided in bins of \BTM. The color code represent \BTM, while, for the bottom panels, shows the fraction of objects in small regions of the \SFRM compared to the galaxy subsample(\BTM<0.2 or \BTM>0.8). }
\label{fig:SFR_mstar2}
\end{figure*}

In order to better analyse the distribution of galaxy morphologies in the \SFRM plane, the sample was divided in three classes: \BTM <0.2 (disc-dominated or pure disc galaxies, 41\% of the sample), \BTM >0.8 (bulge-dominated or spheroidal galaxies, 16\% of the sample), and  0.2<\BTM<0.8 (double component systems, 42\% of the sample). The three different cases are shown in the sequence of plots in Figure \ref{fig:SFR_mstar2} (the complete sequence of plots covering the entire redshift range can be found in Figures \ref{fig:SFR_mstar4bis}, \ref{fig:SFR_mstar5bis} in the appendix \ref{apx:SFR}). Bottom panels of Figure \ref{fig:SFR_mstar2} show that the majority of the spheroidal systems (\BTM >0.8) are in the quiescent region while almost all the disc-dominated galaxies (\BTM <0.2) lie on the main sequence. This confirms the results from Figure \ref{fig:SFR_mstar1}, and  shows a lack of passive pure disc galaxies (\BTM<0.2). Besides the main trend, in both cases (\BTM <0.2, \BTM >0.8) there are disc galaxies as well as spheroidal ones that scatter towards the quiescent or star forming regions, respectively. A fraction of these objects reflects the non-perfect equivalence between the morphological selection and the star formation activity. 
However, to quantify this effect, a density distribution plot is shown in Figure \ref{fig:SFR_mstar6bis}. While the density peaks of both sub-populations fall in the MS and quenched regions respectively, the sequence of plots also emphasizes the presence of outliers that will be discussed in more detail in Section \ref{sec:pec_pop}. 

The removal of the two extreme classes (\BTM<0.2, \BTM>0.8), as it can be observed in top right panel of Figure \ref{fig:SFR_mstar2}, reveals that the intermediate population is interestingly present in both regions, i.e. galaxies are both star forming and quiescent. This result, first, shows that the \BTM parameter alone cannot be used as a quenching predictor. However, it allows to select the ideal sample for the analysis.
Indeed, having a similar structure but with a wide range of star-formation activities, places this population of galaxies as good candidates representing the intermediate evolutionary step between the two main populations. A detailed analysis of this sub-population, could provide important information to the galaxy evolutionary channels that lead to the quenching. The presence of \BTM >0.2 systems already on the main sequence, suggests that the bulge growth starts while galaxies are still actively forming stars. The latter, combined with the lack of passive pure disc galaxies, stresses the relevance of the central component in the quenching process, discarding possible quenching scenarios that do not account of a bulge growth as main evolutionary path.

\section{The effect of bulges "on" the main sequence slope}
\label{sec:SFR2}
\begin{figure*}
\centering
$\begin{array}{c c}
\includegraphics[width=0.4\textwidth]{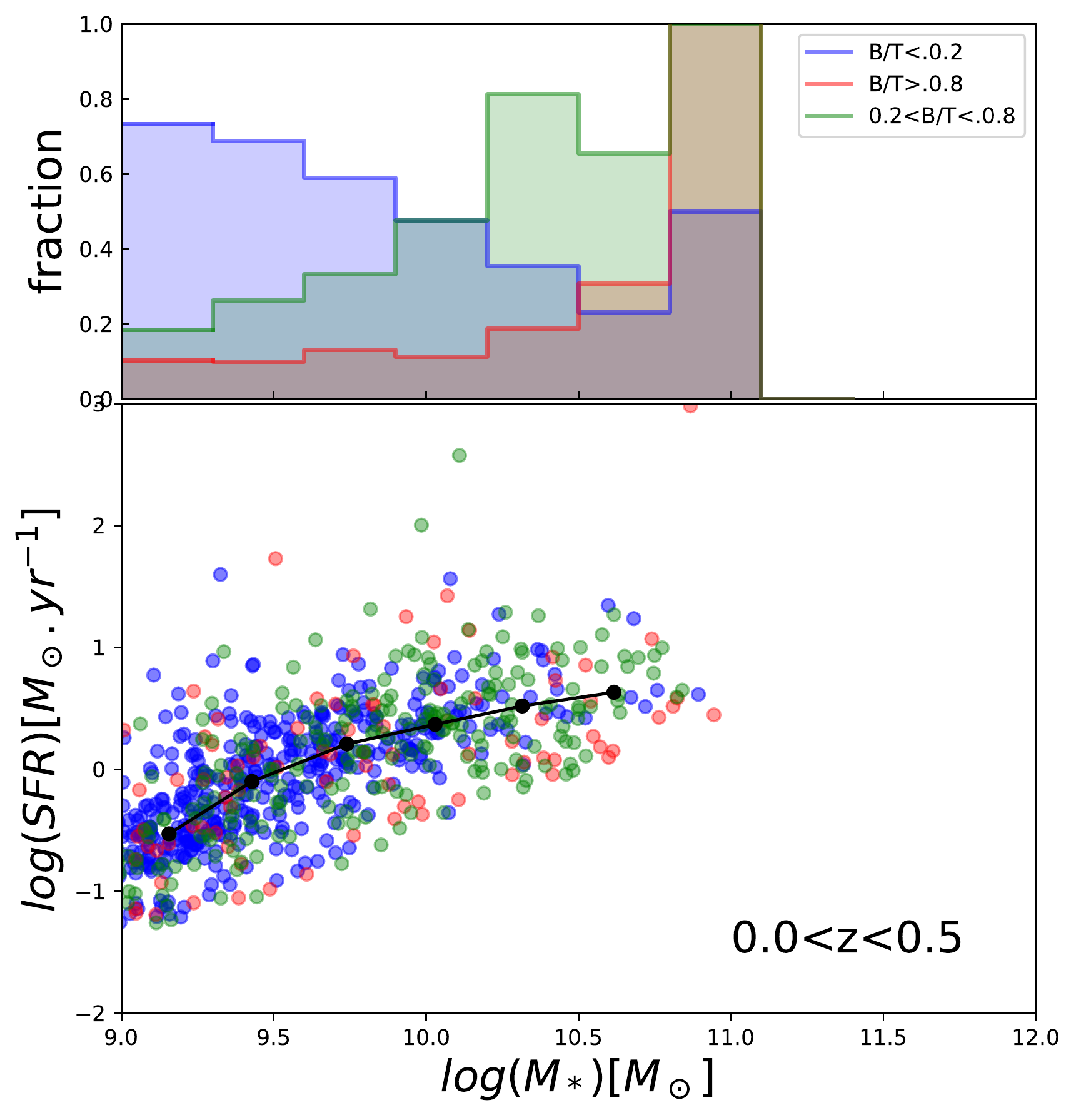}&
\includegraphics[width=0.4\textwidth]{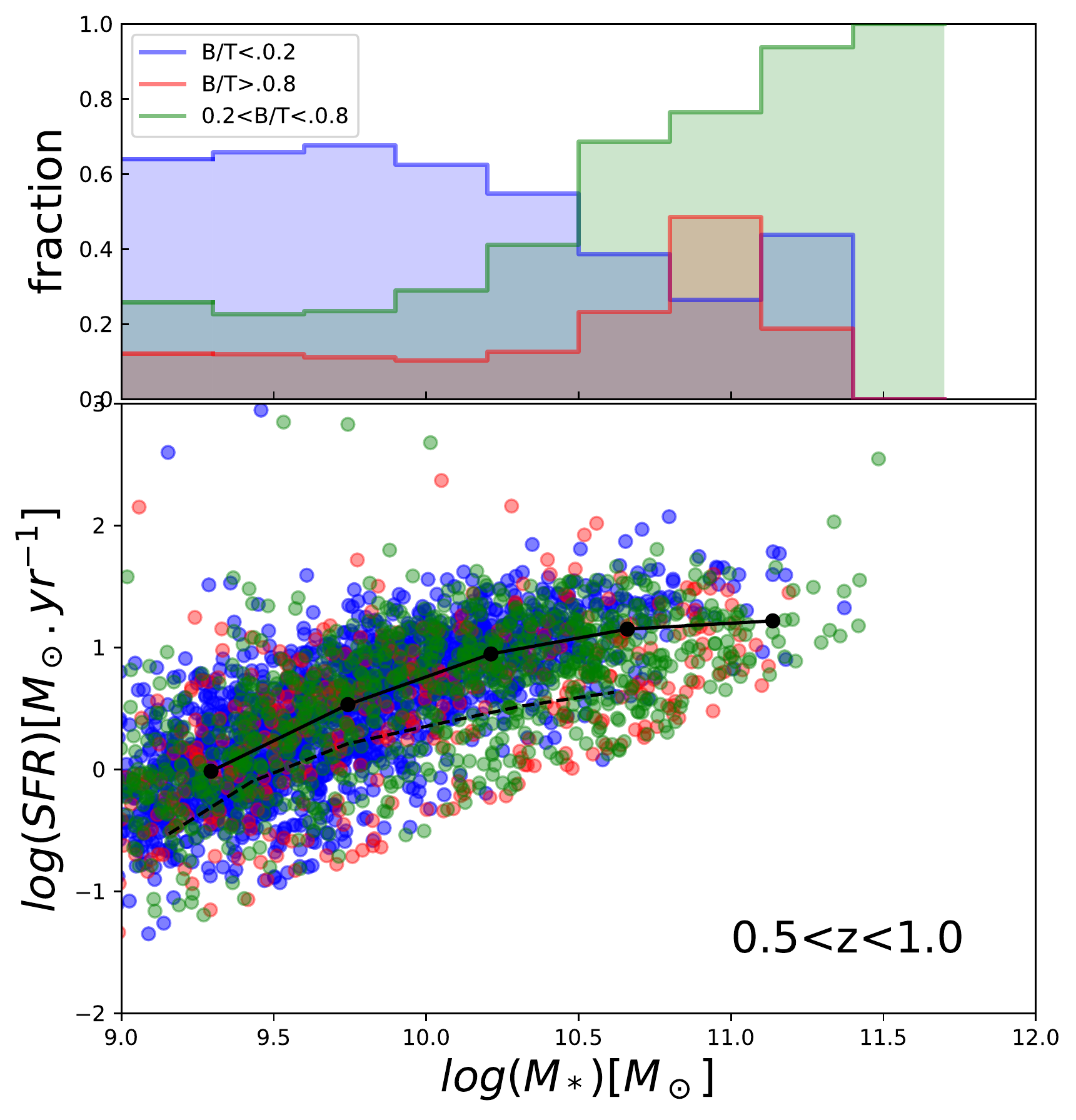}\\
\includegraphics[width=0.4\textwidth]{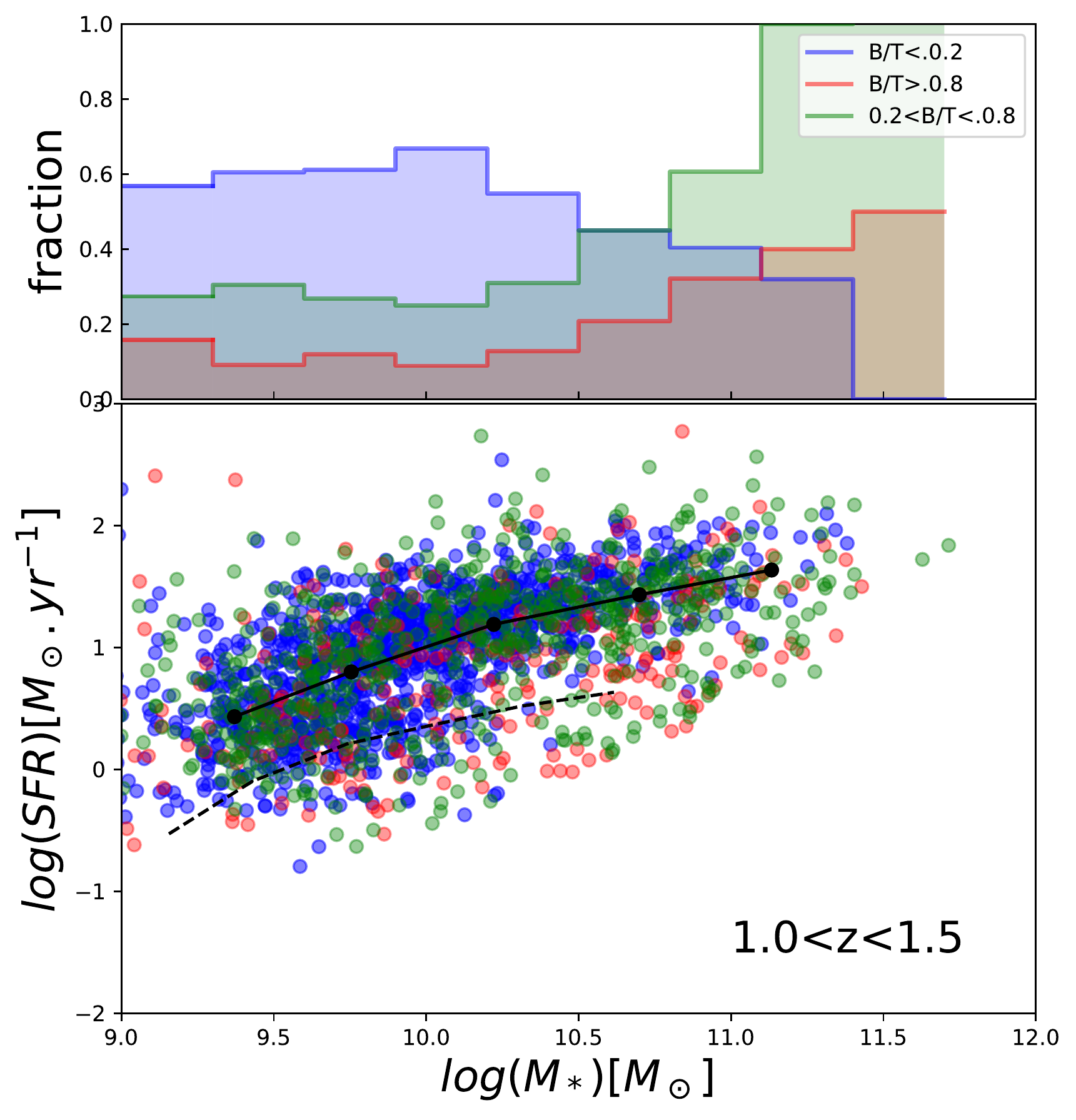}&
\includegraphics[width=0.4\textwidth]{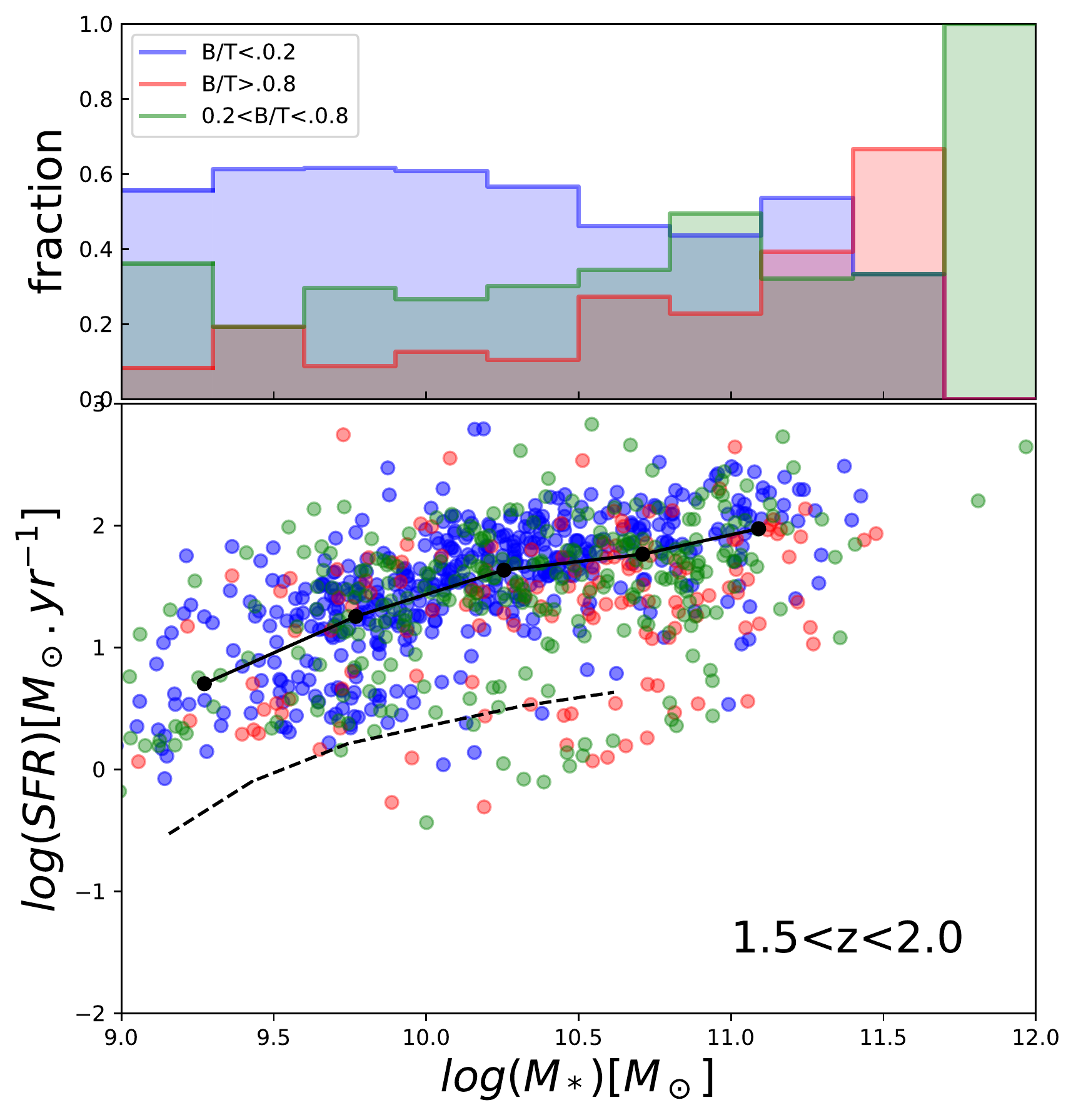}\\
\end{array}$
\caption{Distribution of galaxies along the Main Sequence. Star forming galaxies are selected with log sSFR<-10.5 $M_{\sun} yr^{-1}$. The color code is representative of the \BTM. Black points/lines are the median of the SFR per mass bin. The black dashed line is the median SFR with mass from the lowest redshift plot. Each panel is accompanied by an histogram of the distribution of the three classes of galaxies: \BTM<=0.2, \BTM>=0.8, 0.2<\BTM<0.8 in each mass bin. Galaxies with \BTM>0.2  dominate the massive end of the plane, above $M_*>10.5M_{\odot}$, and they may be driving the bending of the MS. }
\label{fig:SFR_mstar3}
\end{figure*}

Scatter plots from Figure \ref{fig:SFR_mstar3} show the distribution of star forming galaxies along the Main Sequence, color coded by their corresponding \BTM values. They are selected using the sSFR. Black points/lines are the mean of the SFR as a function of the stellar mass, at different redshift. 
The zero-point of the main sequence increases to higher redshift. The main sequence slope slightly changes from low to high masses, a trend that was already noted by some groups but not yet generally confirmed \citep{Whitaker2014, Schreiber2015, Lee2015,Tomczak2016,Sherman2021}. 
Several groups have interpreted the flattening of the \SFRM relation as a consequence of the change in morphology towards earlier type galaxies beyond M$_*\sim 3*10^{10} M_\odot$ \citep{Wuyts2011,Lang2014,Morselli2017,Mancini2019}, possibly related to quenching (eg. \citealp{Tasca2015}), though others do not find any casual correlation among morphology and SFR on the Main Sequence (eg. \citealp{Renzini2015,Whitaker2015,Carollo2016}).

To shed light on the correlation between morphology and location on the $SFR-M_*$ plane, in Figure \ref{fig:SFR_mstar3} we distinguish again between galaxies with distinct \BTM ratios, \BTM<0.2, 0.2<\BTM<0.8, \BTM>0.8. The majority of galaxies that have a bulge start to dominate the main sequence at high stellar masses. Indeed, at \M> 10.5 M$_\odot$ beyond which a flattening in the SFR-$M_\odot$ slope is observed, 60-70\% of main sequence galaxies have \BTM>0.2 . This result confirms the link between the bending of the main sequence and the morphological changes, i.e, the emergence of a bulge. It also suggests that massive galaxies, hosting relevant bulges, have lower star formation activities as a result.

\begin{figure*}
\centering
$\begin{array}{c c c}
\includegraphics[width=0.32\textwidth]{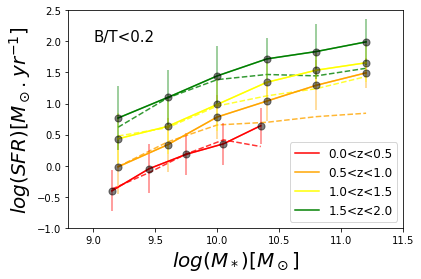}&
\includegraphics[width=0.32\textwidth]{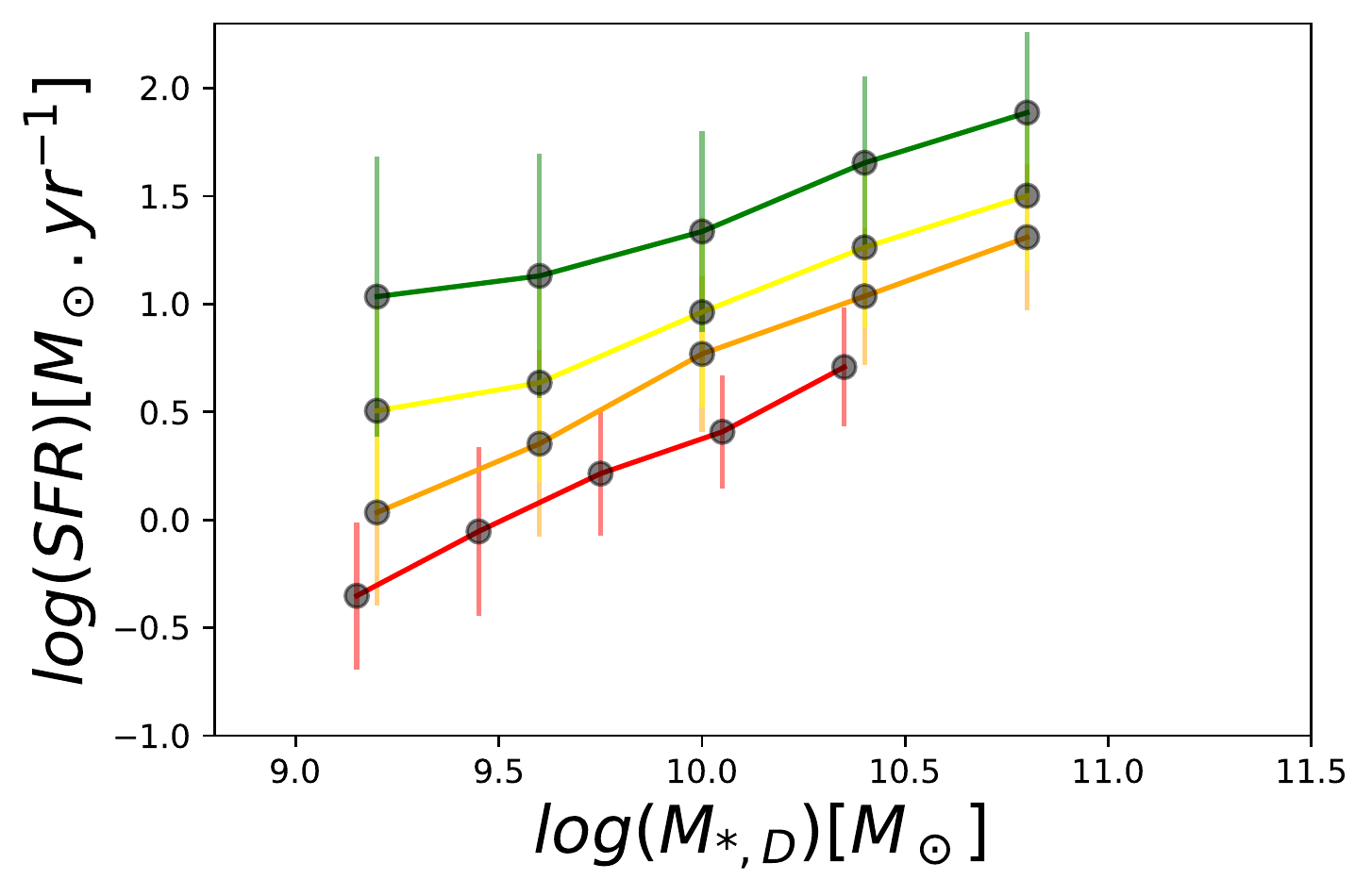}&
\includegraphics[width=0.32\textwidth]{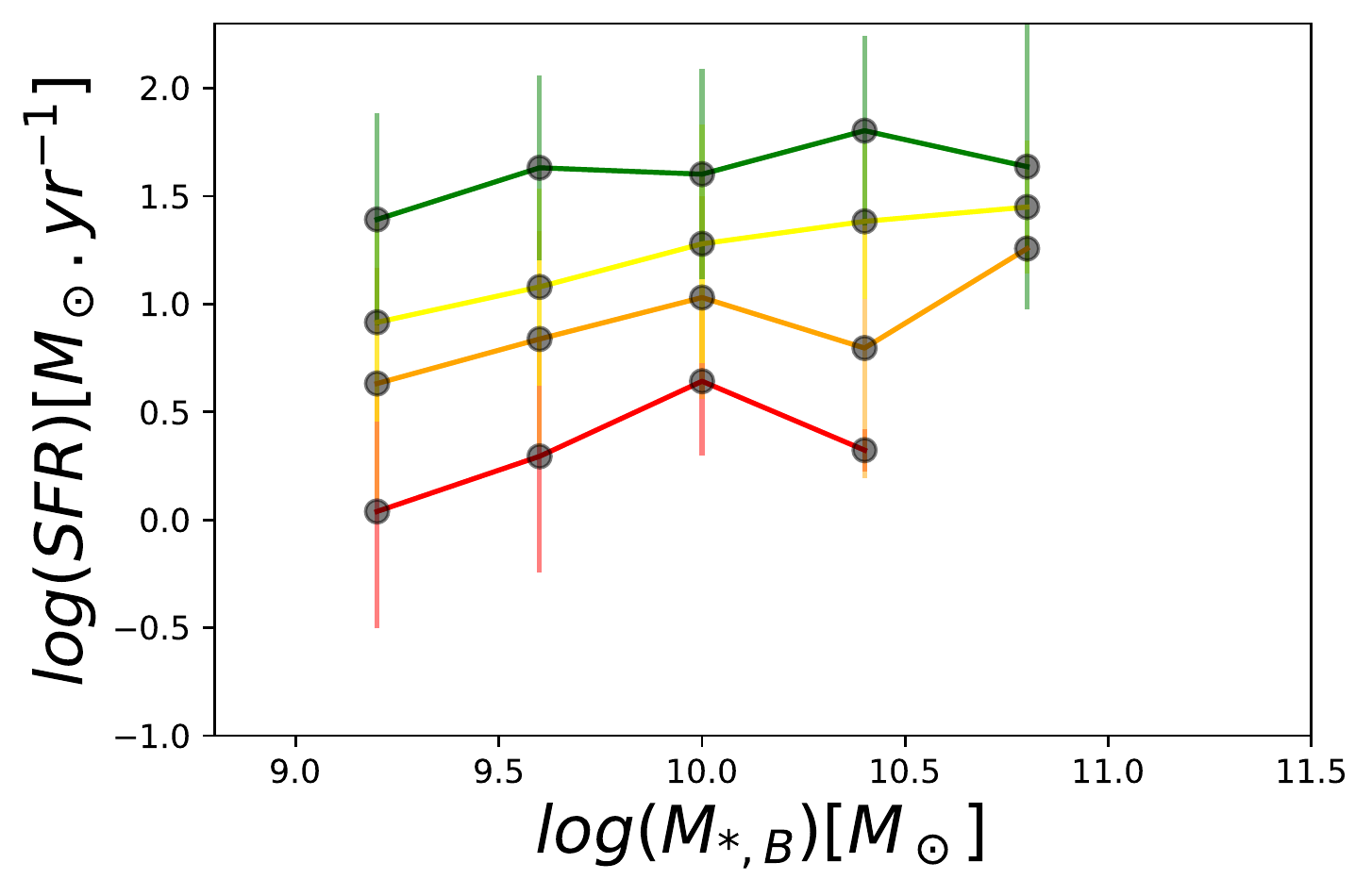}\\
\end{array}$
\caption{Mean sequence of galaxies at different redshift. Left panel shows the evolution of the main sequence for disky galaxies (\BTM<0.2). Dotted lines are main sequences computed considering the entire sample. Middle and right panels report the main sequence at different redshifts, calculated accounting for only the stellar mass of the disc and of the bulge, respectively. The ensemble of the plots shows that the linearity of the main sequence is mostly driven by the star formation activity that resides in the disc component, while the bending is more likely connected to the presence of the central density.}
\label{fig:main_seq}
\end{figure*}

Previous works argued that the bending of the MS is linked to the definition of star forming galaxies. Indeed, the selection of blue star forming galaxies, or galaxies with small \Sersic index, results in a main sequence slope value close to one \citep{Peng2010,Whitaker2012,Whitaker2015}. 
Therefore, the bending is a consequence of the mass infall into the central region of the galaxy \citep{Abramson2014,Schreiber2015,Fraser-McKelvie2021}. An alternative interpretation connects the bending to the presence of rejuvenated galaxies \citep{Mancini2019}. Indeed, galaxies can exhibit a low level of star formation because they are either quenching or starting to form stars again, thanks to new gas accretion (e.g. through minor/major mergers). 

All the interpretations agree on the possible link of the bending with the bulge component. The removal of the stellar mass contribution of the bulge in the \SFRM plane should reduce the bending and suggests that bulges have a little role in the star formation activity.
\cite{Popesso2019} discussed the possibility that the growth of the bulge alone is not enough to explain the MS bending, and therefore additional processes are required to decrease the star formation within the disc. Indeed, \cite{Guo_2015} show that a signature of the bending has also been observed in pure disc galaxies populations. They argue that the decline of star formation in the discs of massive star forming galaxies is a natural consequence of halo quenching, combined with the accretion of central bulges through AGN feedback or morphological quenching.

Our preliminary results in Figure \ref{fig:SFR_mstar2} and \ref{fig:SFR_mstar3} suggest that actively star forming galaxies move along the main sequence and then reduce their growth rate and flatten the MS above \M$ \sim 3* 10^{10} M_\odot$, when they start developing a significant inner bulge. When the bulge becomes prominent, galaxies migrate below the MS and quench. However, this does not imply that bulges are the reason of the bending. To further highlight the importance of the \BTM in the quenching process, Figure \ref{fig:main_seq} plots the MS(i.e. the total SFR) for galaxies with \BTM <0.2 (left), MS in which only the disc (middle) or bulge (left) component stellar mass is accounted for. Dotted lines represent the MS computed accounting of the entire sample. At fixed bulge stellar mass, the mean SFR is equal to or higher than the corresponding SFR at fixed disc stellar mass. This is because bulges are more often present in more massive galaxies with higher SFR than the discs population.

The Main Sequence of disc-dominated galaxies (\BTM <0.2) shows a slope change toward high masses. If only the disc stellar mass is accounted, the mean SFR follows an almost linear relation (in the logarithmic space) over the whole range of mass and redshift, while the mean SFR-$M_{*,BULGE}$ relation is nearly flat.
The bending is significantly reduced if only disc stellar masses are accounted in the analysis. Bulges have a weak contribution to the SF activity of galaxies. The star formation is tightly linked to the stellar mass of the disc, the main driver of the Main Sequence slope. The growth of the central component acts increasing the stellar mass of a galaxy but does not alter its SFR. Consequently, galaxies move from the low to the high mass region of the plane, causing the observed SFR decrease. 

\section{Which is the best quenching predictor parameter?}
\label{sec:best_fit}

\begin{figure*}
\centering
$\begin{array}{c c}
\includegraphics[width=0.45\textwidth]{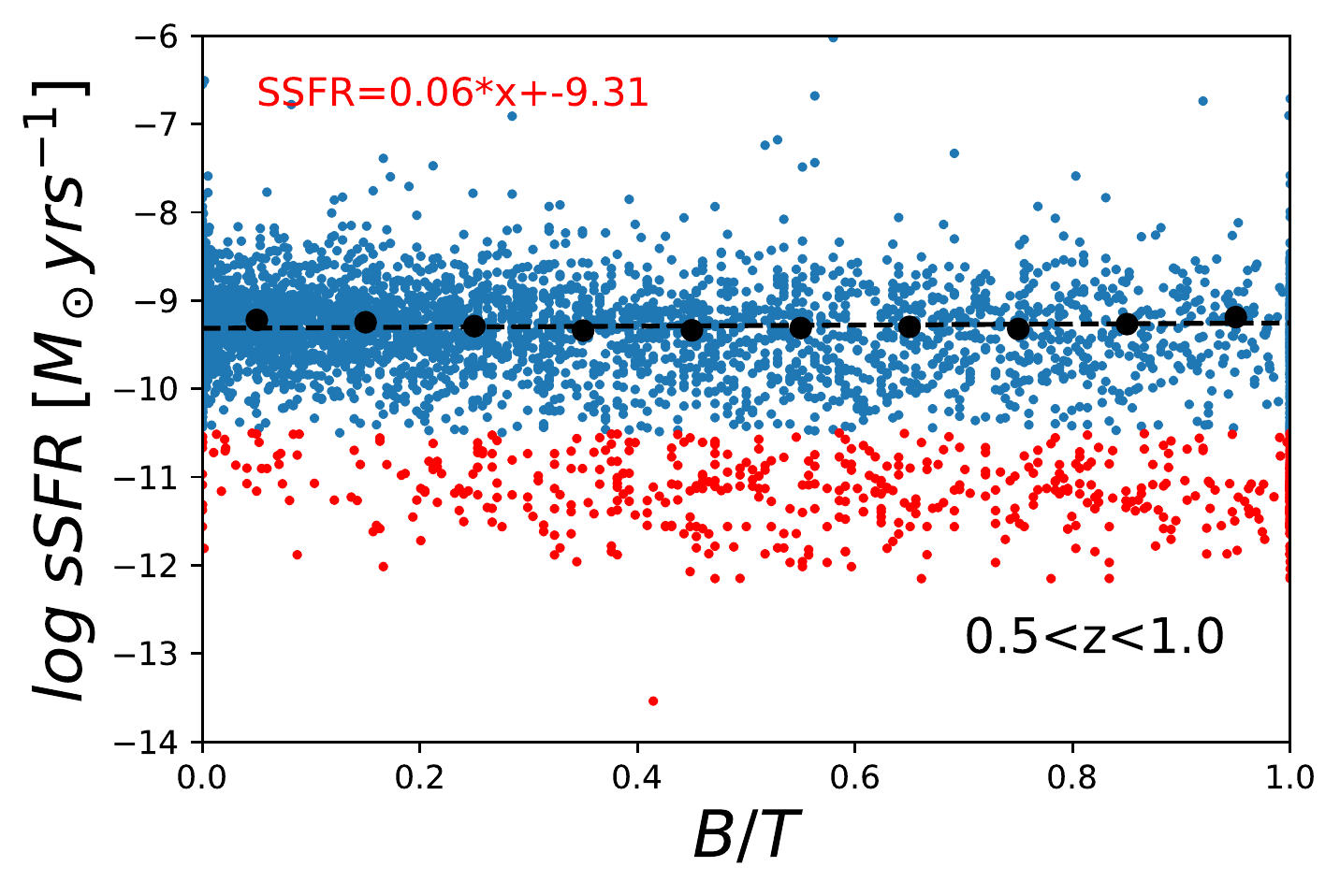}&
\includegraphics[width=0.45\textwidth]{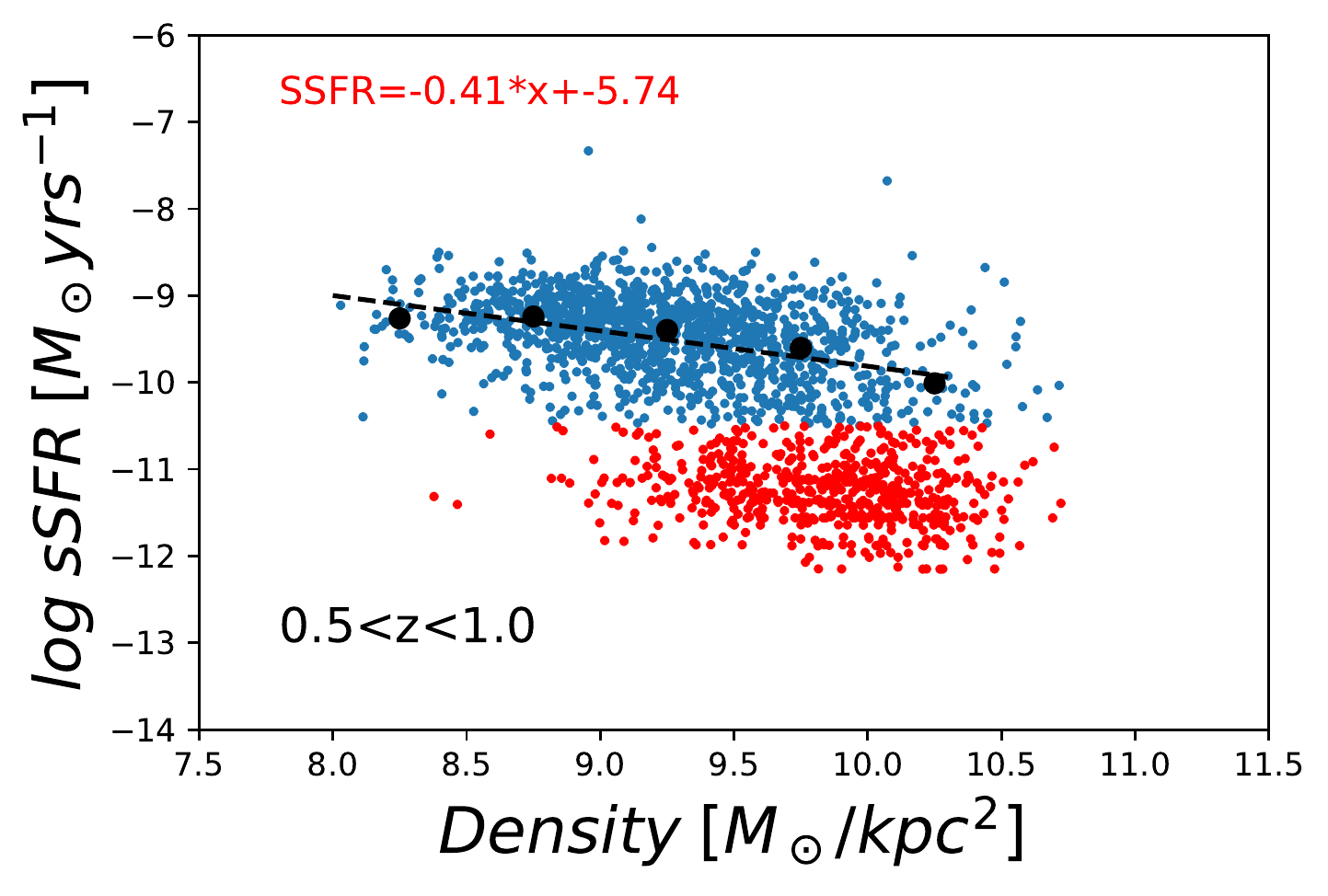}\\
\includegraphics[width=0.45\textwidth]{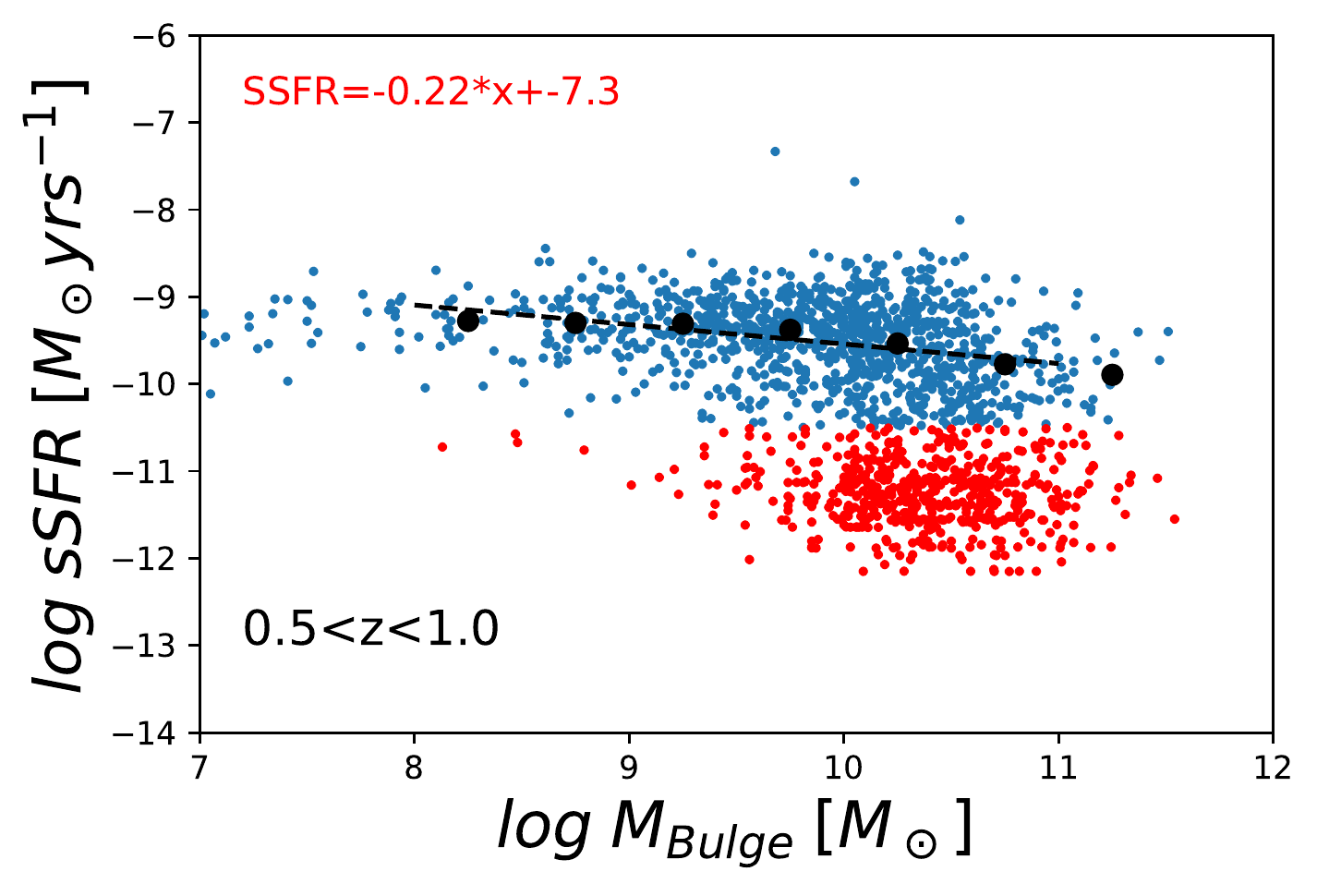}&
\includegraphics[width=0.45\textwidth]{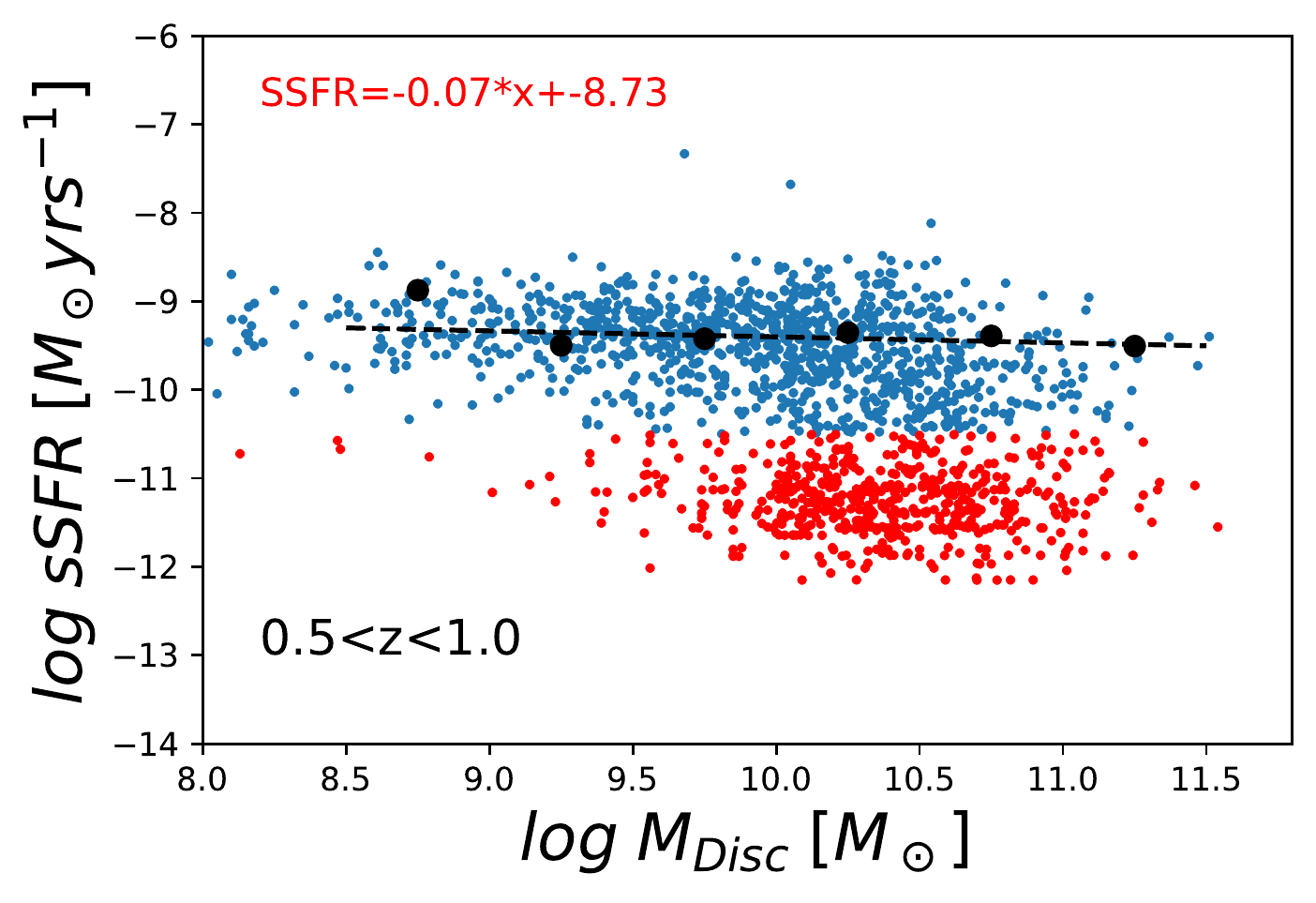}\\
\end{array}$
\caption{Star formation-morphology-quenching. The sequence of the plots aims to investigate the role of the different parameters in decreasing the Star Formation. To do this, each parameter is compared with the sSFR (to remove the dependence on the total stellar mass from the analysis). The upper panels are analyzing the relationship between sSFR and \BTM, central density. The lower panels show the impact of the bulge(left)/disc(right) stellar masses on the Star Formation activity. Star forming(blue) galaxies are differentiate from the quiescent ones(red) using the log sSFR=-10.5 $M_{\sun} yrs^{-1}$ limit. The black points are the median values for which a better linear fit is applied and the final equation is shown at the top of each panel. }
\label{fig:quenching}
\end{figure*}

In the previous sections we found evidence that a (significant) bulge component plays a fundamental role in understanding the quenching. In this section we explore the contribution of other physical parameters. Most notably, we consider the role of galaxy stellar mass density ($D\sim M_*/Re^2$), which was several times put forward in the literature as a possible driver for quenching from theoretical and observational point of view ( e.g. \citealp{Fang2013,Dekel2014,Lang2014,Barro2017,Whitaker2017,Woo2017,Dekel2017}). We will analyse D toghether with the Specific Star Formation rate ($sSFR=SFR/M_*$), stellar mass and the morphology. We here restrict the analysis to only star forming galaxies.  

Figure \ref{fig:quenching} presents a sequence of plots aimed at investigating possible correlations between the sSFR and different parameters: \BTM, D, $M_{Bulge}$, $M_{Disc}$. Only one redshift bin is shown in the Figure, while the entire redshift range can be found in Appendix \ref{apx:quenching_best}.
The top left panel shows the distribution of star forming (blue) and quiescent(red) galaxies in the sSFR-\BTM plane. Even though most of the star forming galaxies are concentrated towards low values of \BTM,  the mean sSFR shows an almost constant distribution (for star forming galaxies) over the entire range of \BTM. Indeed, a linear fit yields to a slope of 0.06 (reported in the top-left the panel).
A linear best-fit is applied to all the parameter-space analysed in Figure \ref{fig:quenching}. The larger values are measured for $D$ and $M_{Bulge}$. Same trend is observed also for the entire sequence plots reported in Figures from the Appendix \ref{apx:quenching_best}. 

To quantify these results with better statistical significance, we computed the Pearson correlation coefficients, that are reported in Table \ref{tab:Pears}. Those values confirm the previous main conclusion. Indeed, the strongest correlation factor is measured for stellar mass density and $M_{Bulge}$. Interestingly, no(or weak) correlation results between the sSFR and the \BTM. 
This result is in agreement with the finding of sections \ref{sec:SFR1} and \ref{sec:SFR2} and underlines that the presence of the bulge is relevant for the quenching but not its relevance compared to the disc component(\BTM). It also confirms the not-direct effect of the bulge growth on the Star Formation activity. Furthermore, the fact that the same main conclusion is obtained using stellar mass density, i.e. a quantity that does not require any morphological information, indirectly supports the robustness of the morphological classification.

\begin{table}
\centering
\begin{tabular}{|c|c|c|c|c|}
 \hline
 z & \BTM & $D[M_* / kpc^2] $ & $M_{Bulge} [M_*]$ & $M_{Disc} [M_*] $\\
 \hline
 \hline
0.0-0.5 &  0,031 & -0,687 & -0,127 & -0,079\\
0.5-1.0 & -0,065 & -0,637 & -0,274 & -0,002\\
1.0-1.5 & -0,066 & -0,746 & -0,293 & -0,076\\
1.5-2.0 & -0,089 & -0,832 & -0,344 & -0,079\\
 \hline
 \hline
\end{tabular}
\caption{Pearson Correlation Coefficient computed between the sSFR and the input parameters of the table. }
\label{tab:Pears}
\end{table}

\begin{figure*}
\centering
$\begin{array}{c c}
\includegraphics[width=0.3\textwidth]{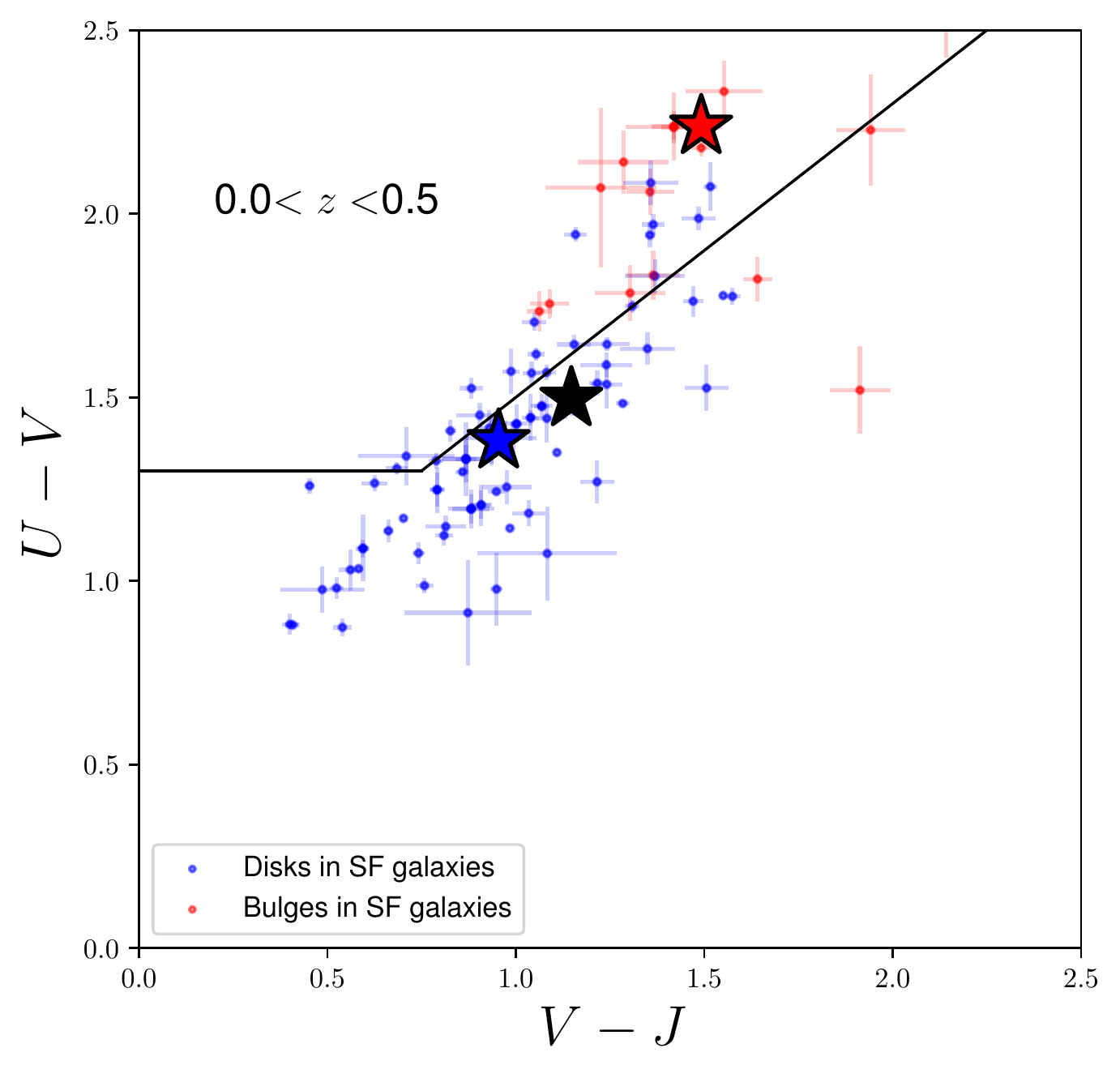}&
\includegraphics[width=0.3\textwidth]{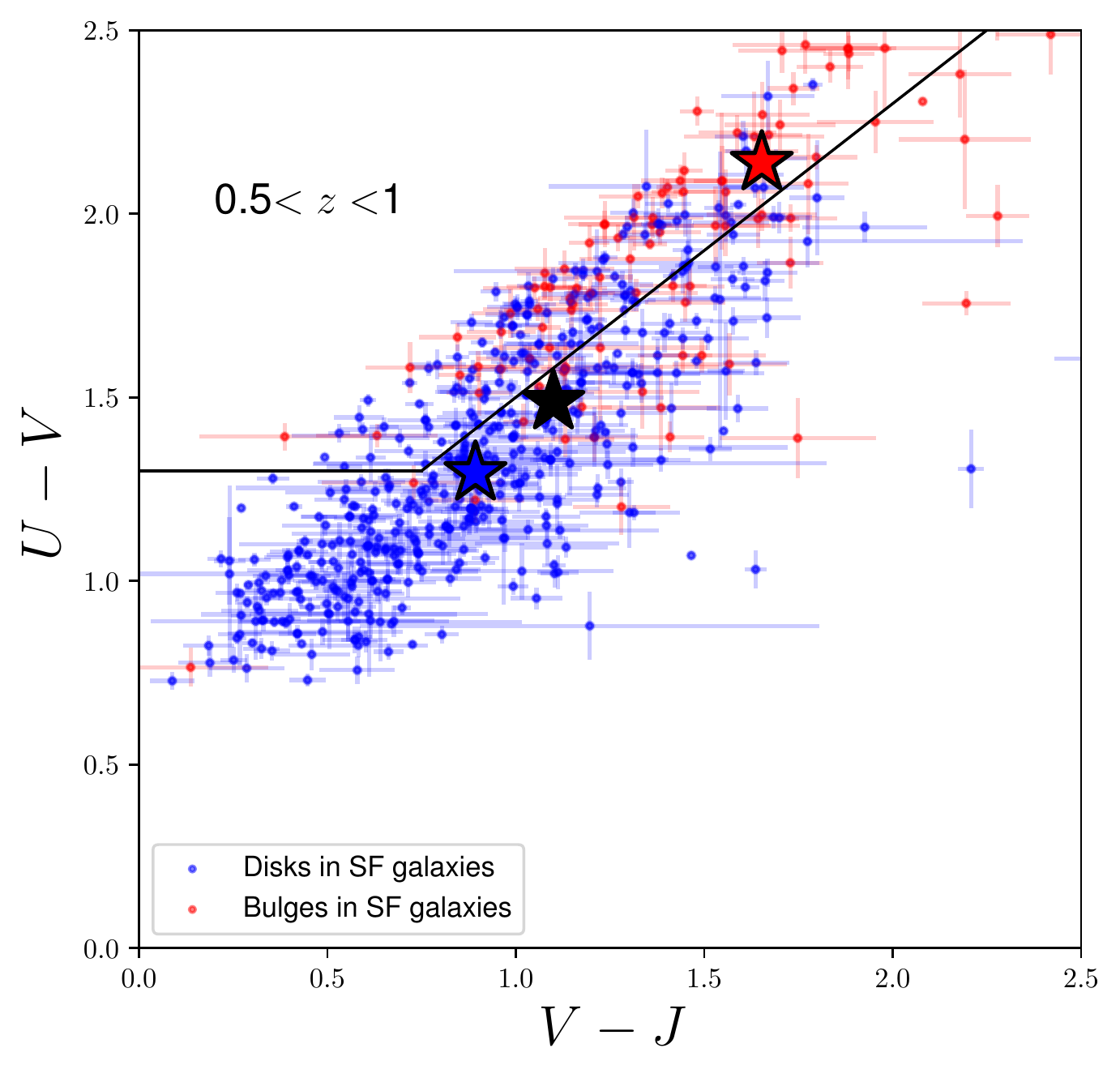}\\
\includegraphics[width=0.3\textwidth]{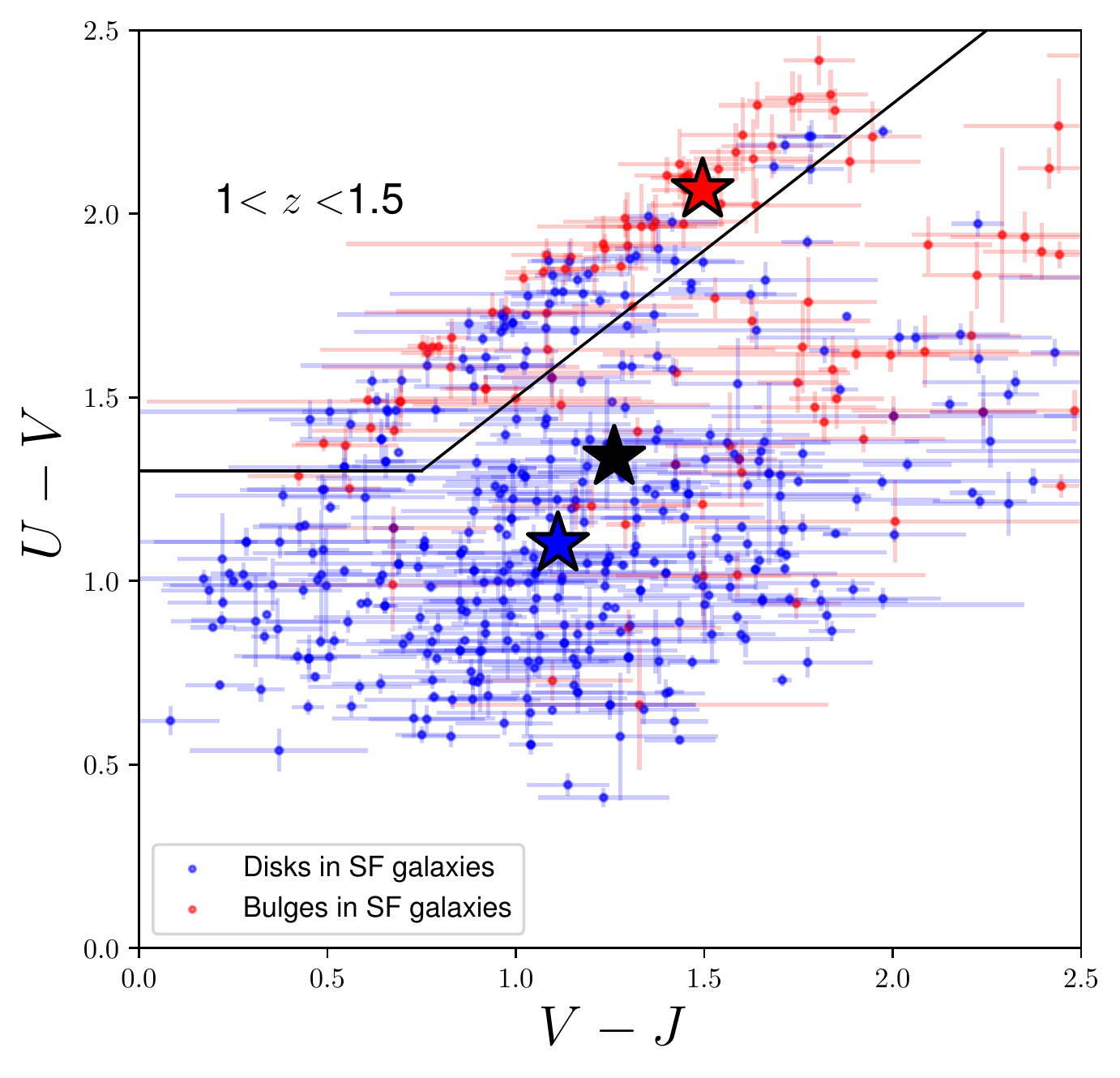}&
\includegraphics[width=0.3\textwidth]{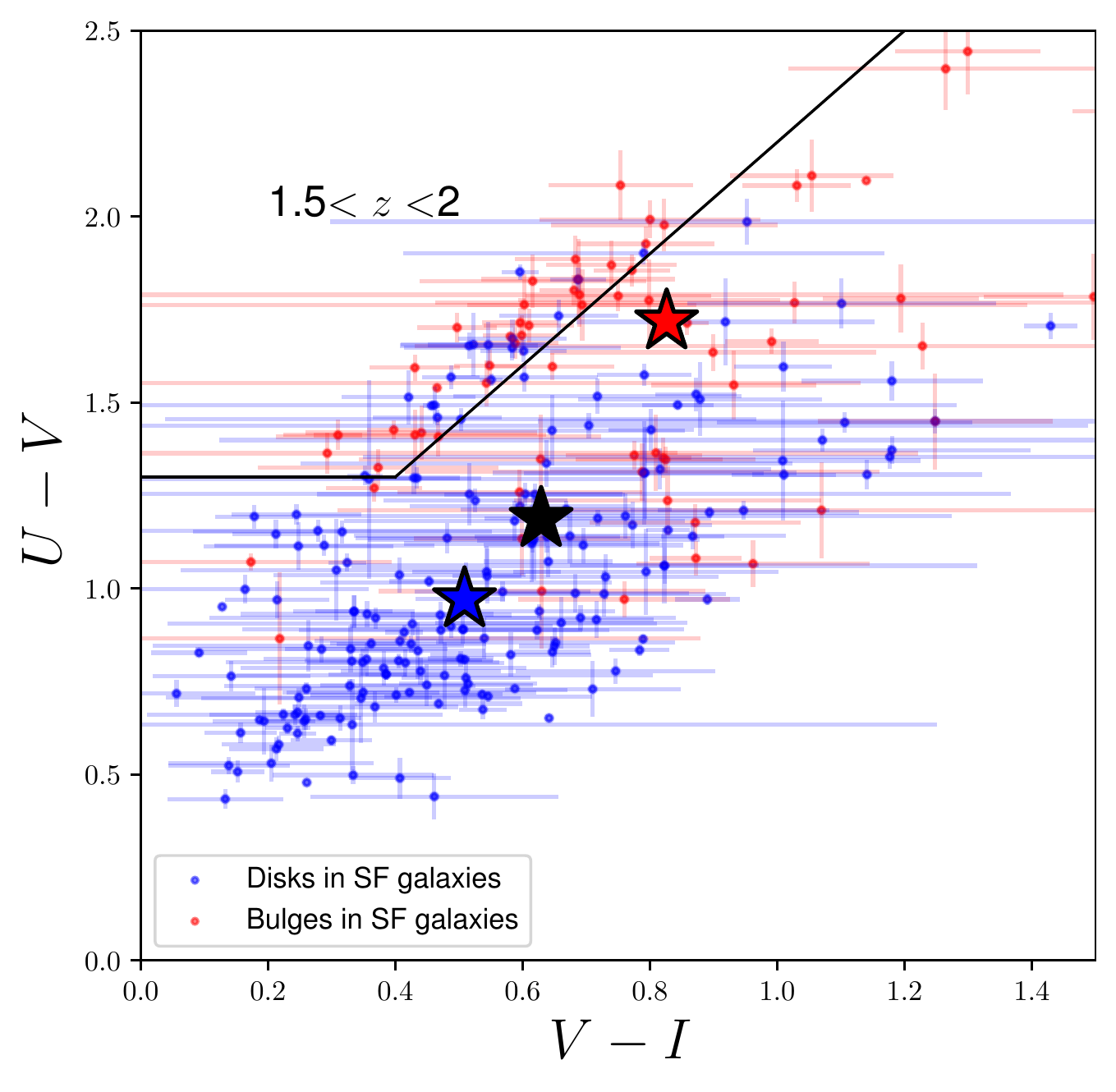}\\
\end{array}$
\caption[UVJ plane ]{Distribution of bulges and discs in the UV,VJ-I rest-frame colors plane hosted by star forming galaxies. Black orizontal and diagonal lines divide the plane in quiescent(top) and star forming regions. The sequence of plots show internal components of galaxies selected as a two-component systems, i.e. bulges and discs for each selected galaxy are shown. Blue and red stars are the color median values. The black one is the median color for the entire galaxy estimated using the CANDELS catalog.}
\label{fig:BDuvj1}
\end{figure*}

\section{Colors of bulges and discs}
\label{sec:colors}

Understanding the possible impact of stellar bulges in modulating galactic star formation requires knowledge of the star formation activity of internal components. However, a reliable SFR estimate requires an optimal wavelength coverage from the FIR to the NUV. To avoid possible systematics induced by the adoption of only optical data, we analized
the UVJ rest-frame colors as a proxy of the SFR activity of the internal components to spatially resolve the origin of the star formation within galaxies (\citealp{Labbe2005,Wuyts2007,Williams2009}). The analysis is restricted to a sub-sample that only includes galaxies covered by 7 bands (i.e, only GOOD-N/S fields), to minimize the increasing uncertainties on the SED fitting that could occur when only few bands are considered.

\begin{figure*}
\centering
$\begin{array}{c c}
\includegraphics[width=0.3\textwidth]{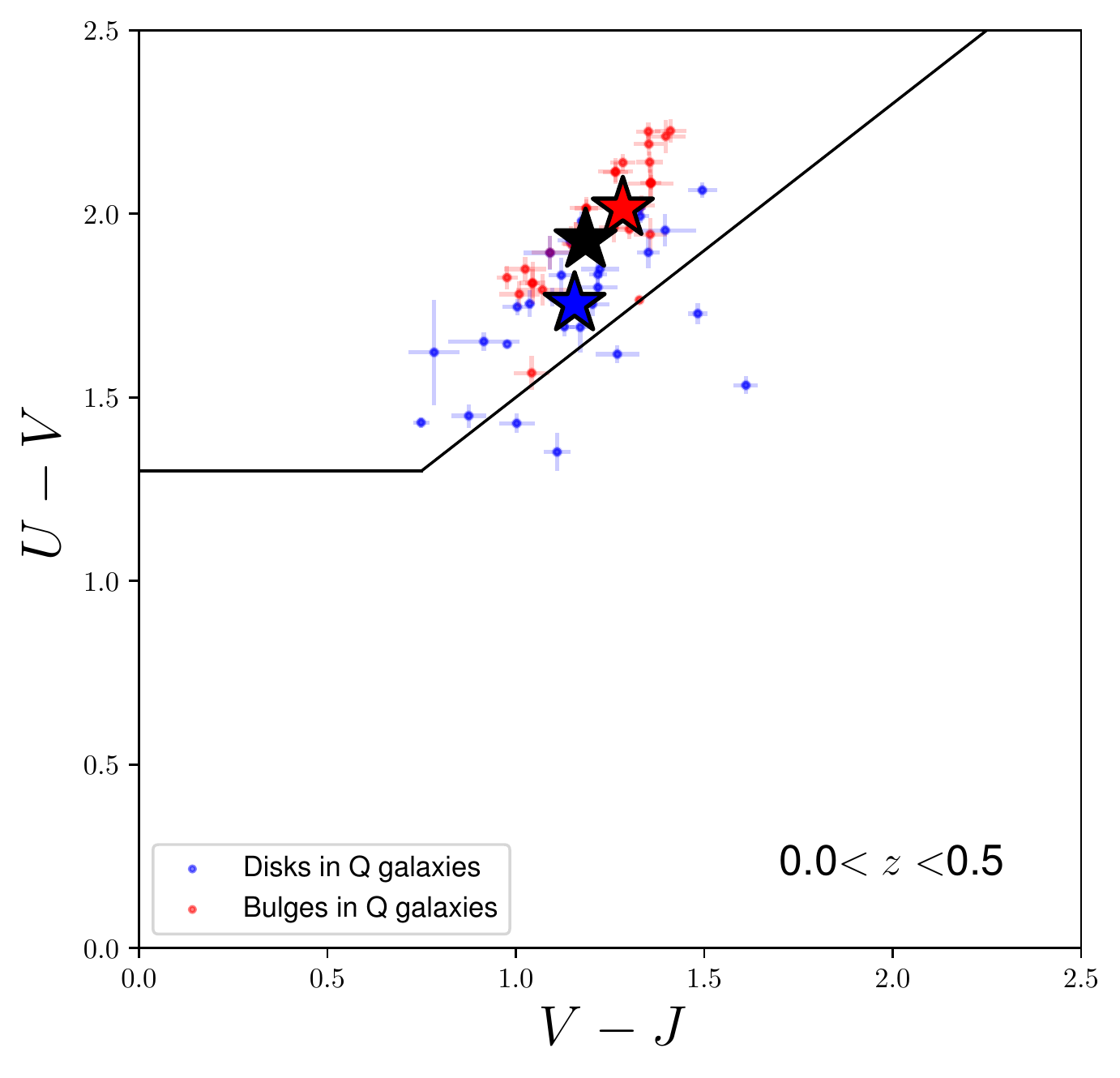}&
\includegraphics[width=0.3\textwidth]{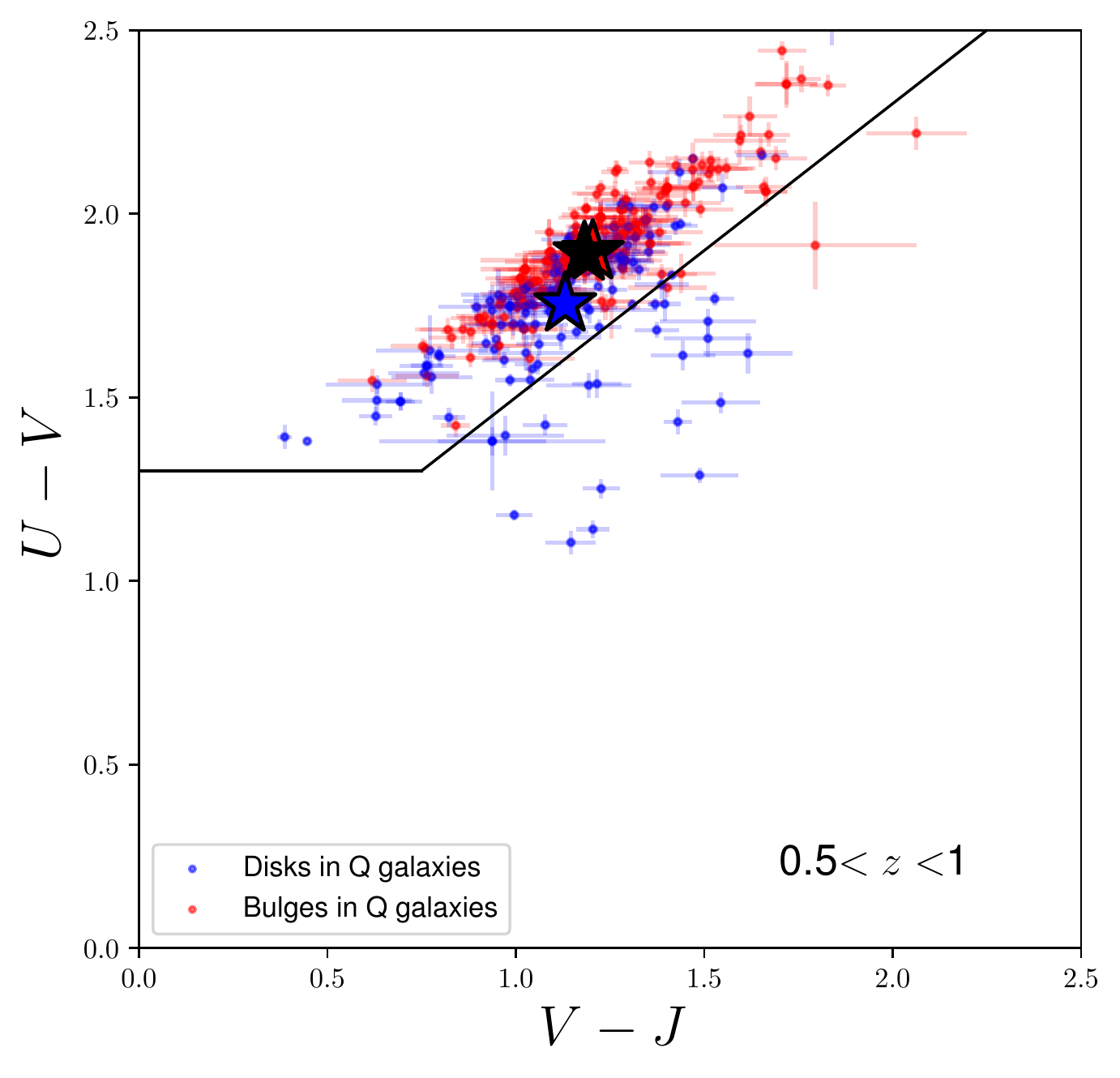}\\
\includegraphics[width=0.3\textwidth]{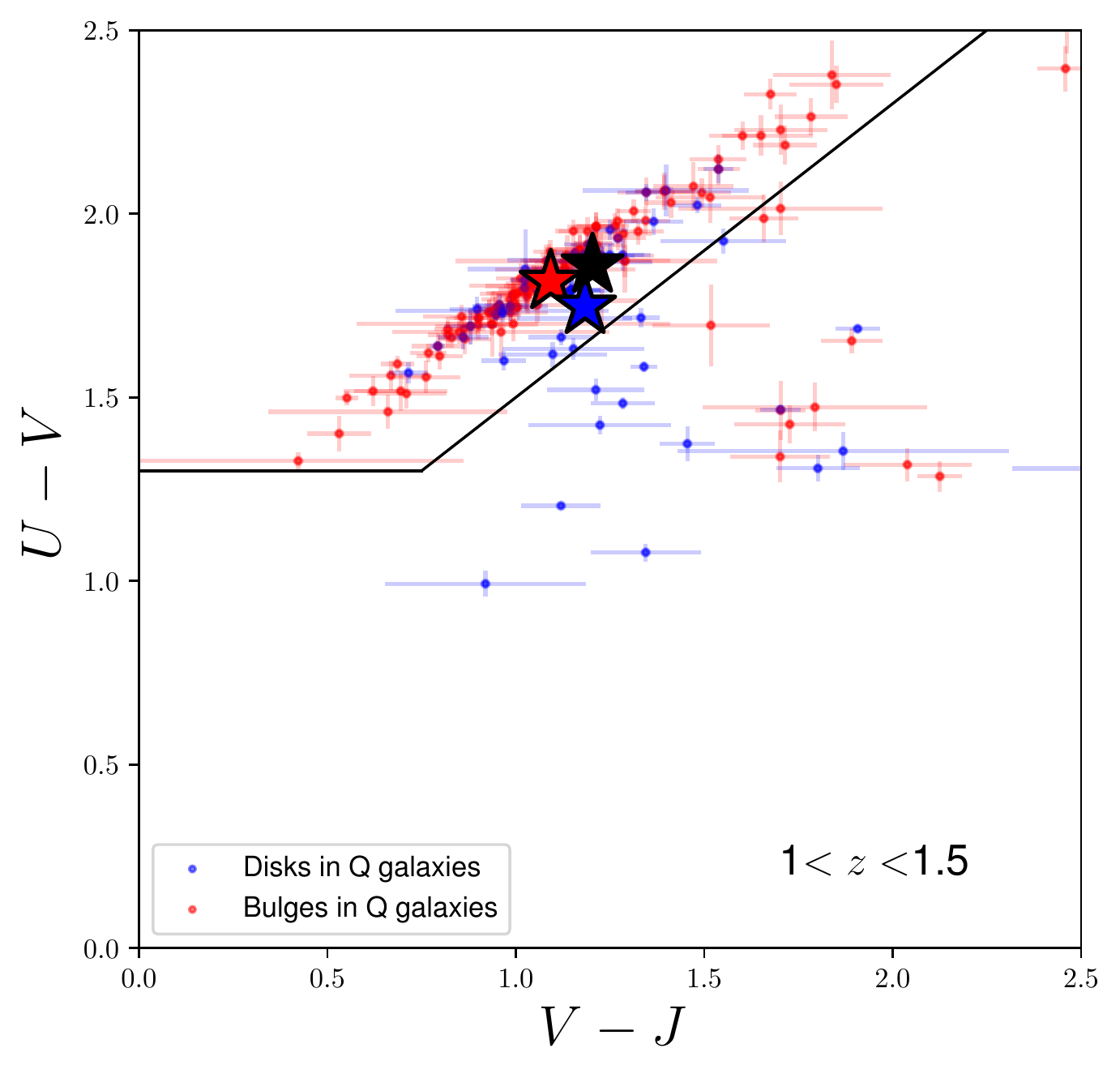}&
\includegraphics[width=0.3\textwidth]{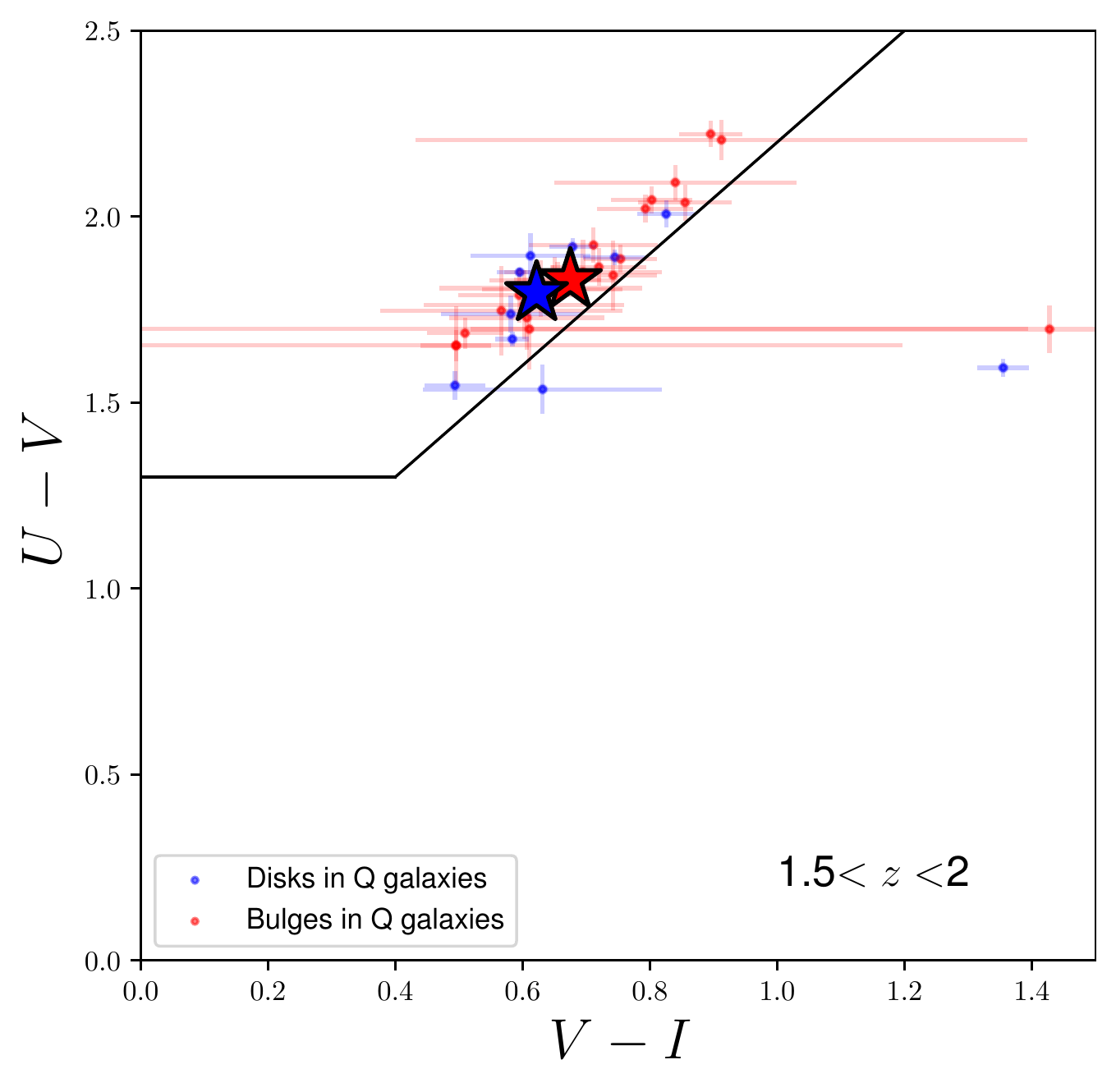}\\
\end{array}$
\caption[UVJ plane ]{Distribution of bulges and discs in the UV,VJ-I rest-frame colors plane hosted by quiescent galaxies.  Black horizontal and diagonal lines divide the plane in quiescent(top) and star forming regions. The sequence of plots show internal components of galaxies selected as a two-component systems, i.e. bulges and discs for each selected galaxy  iare shown. Blue and red stars are the color median values. The black one is the median color for the entire galaxy estimated using the CANDELS catalog.}
\label{fig:BDuvj2}
\end{figure*}

The sample is divided into star forming and quiescent galaxies, using the color selection on the total U-V V-J rest-frame colors and indipendently of the morphology. 
Rest-frame colors of bulges and discs are estimated as explained in Section \ref{sec:cat2} and reported in Figures \ref{fig:BDuvj1} and \ref{fig:BDuvj2}. 
Bulges and discs are represented by, respectively, red and blue points in Figures \ref{fig:BDuvj1} (star forming galaxies) and \ref{fig:BDuvj2} (quenched sample).
In the case of quiescent galaxies both bulges and discs are passive, there is no clear evidence of any star formation activity in either of the two components.
On the countrary, the disc component of star forming galaxies populate the star forming region of the UVJ plane, following a similar distribution as massive star forming galaxies \citep{Fang2018}, while most of bulges are concentrated in the quenched region. 
These trends suggest that the major contribution to the SFR of the entire galaxy comes from the disc component, while bulges are mostly passive. However, as can be seen in Figure \ref{fig:BDuvj1}, the distribution of bulges has a tail of objects that scatters outside of the quiescent region. The relevance of these detections (20\% of bulges within the SF population) and the accuracy of the models for those sub-sample of objects will be addressed in the discussion. However, it is important to note that this fraction is not statistically relevant in terms of the final result.

\begin{figure*}
\centering
$\begin{array}{cc}
\includegraphics[width=0.4\textwidth]{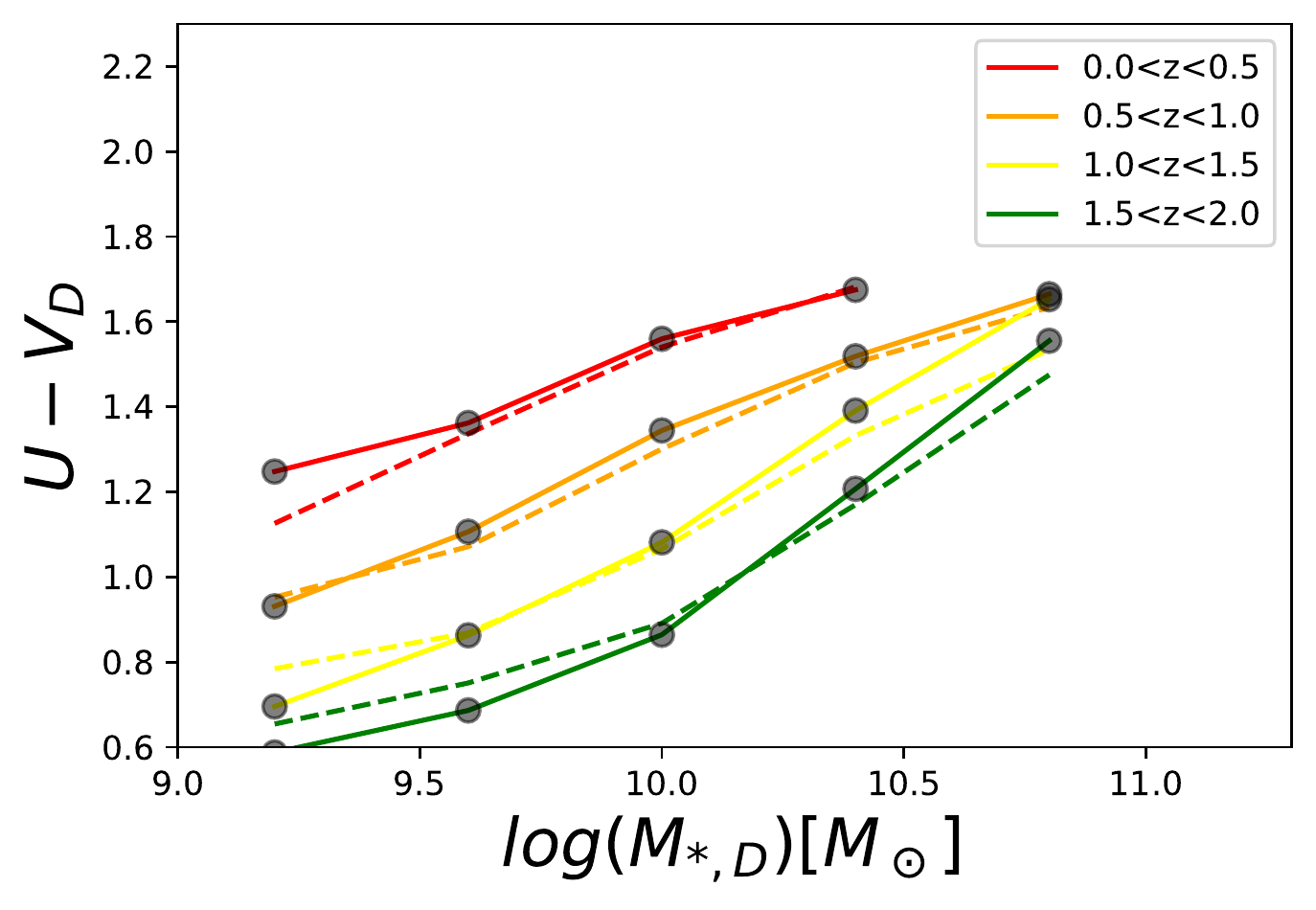}&
\includegraphics[width=0.4\textwidth]{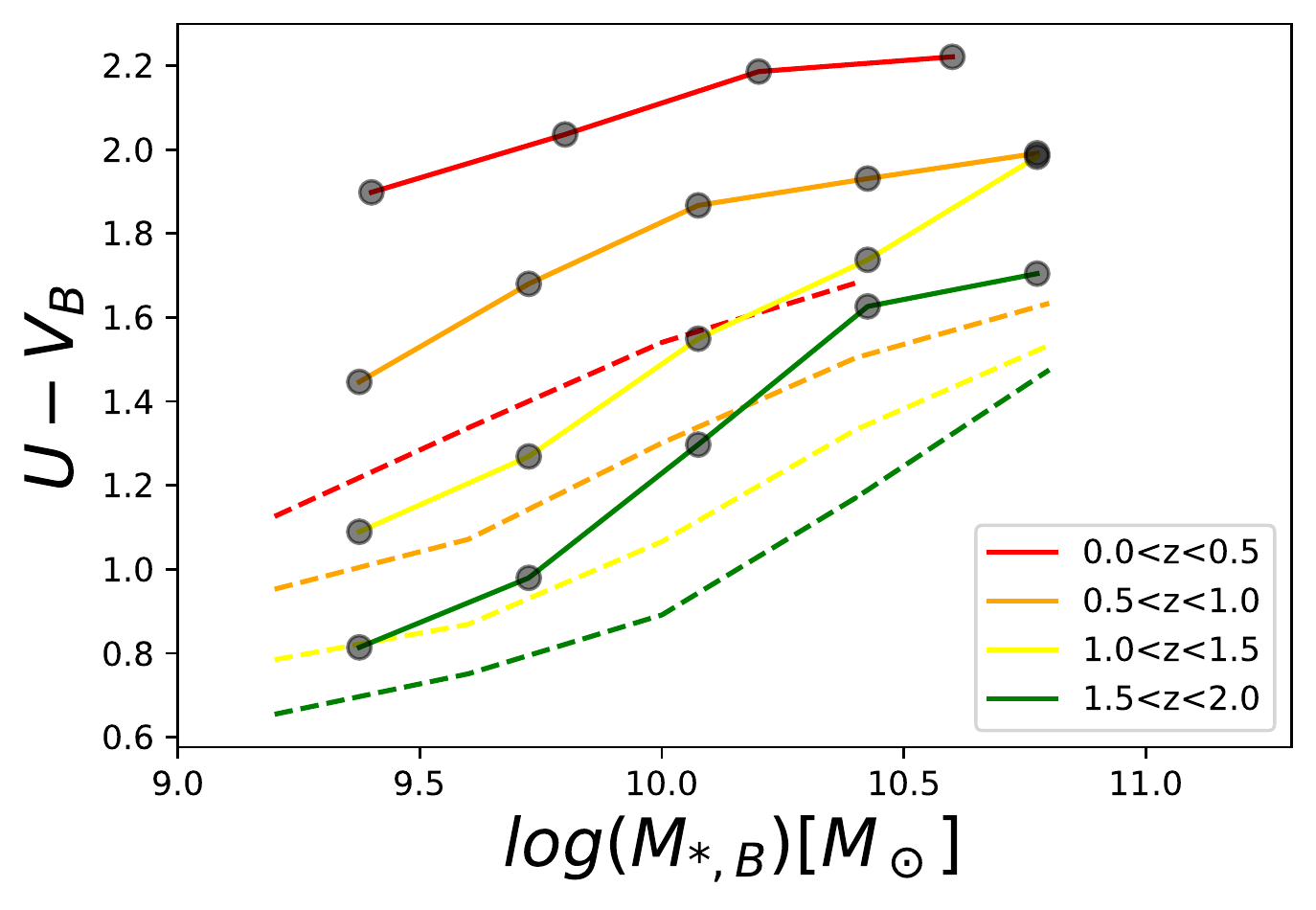}\\
\end{array}$
\caption{Colors of bulges and discs within Main Sequence galaxies. Dotted lines represents the color of the entire galaxies.}
\label{fig:MS_colors}
\end{figure*}

If we consider the colors of the U-V rest frame as a proxy of the SFR, we can calculate the Color Main-Sequence. Figure \ref{fig:MS_colors} shows colors Main Sequence for bulges (right panel) and discs (left panel) at different redshift. Dotted lines in both cases correspond to the colors of galaxies without taking into account the internal decomposition. 
The colors of discs get redder towards high masses with a steeper slope than bulges. Bulges, instead, follow rather flat relationships, showing a redder color than the one of the galaxy. Discs drive the galaxy colors, as it can be seen in the Figure. The contribution of bulges is not relevant. In fact, the slope of the disc colors does not show any sign of flattening at $M_*>10.5 M_{\odot}$ where the Main Sequence is dominated by galaxies hosting a relevant bulge.  
 
\section{Discussion}
\label{sec:discussion}

\subsection{Causes and consequences of the presence of the bulge}

\begin{figure*}
\centering
$\begin{array}{c c}
\includegraphics[width=0.3\textwidth]{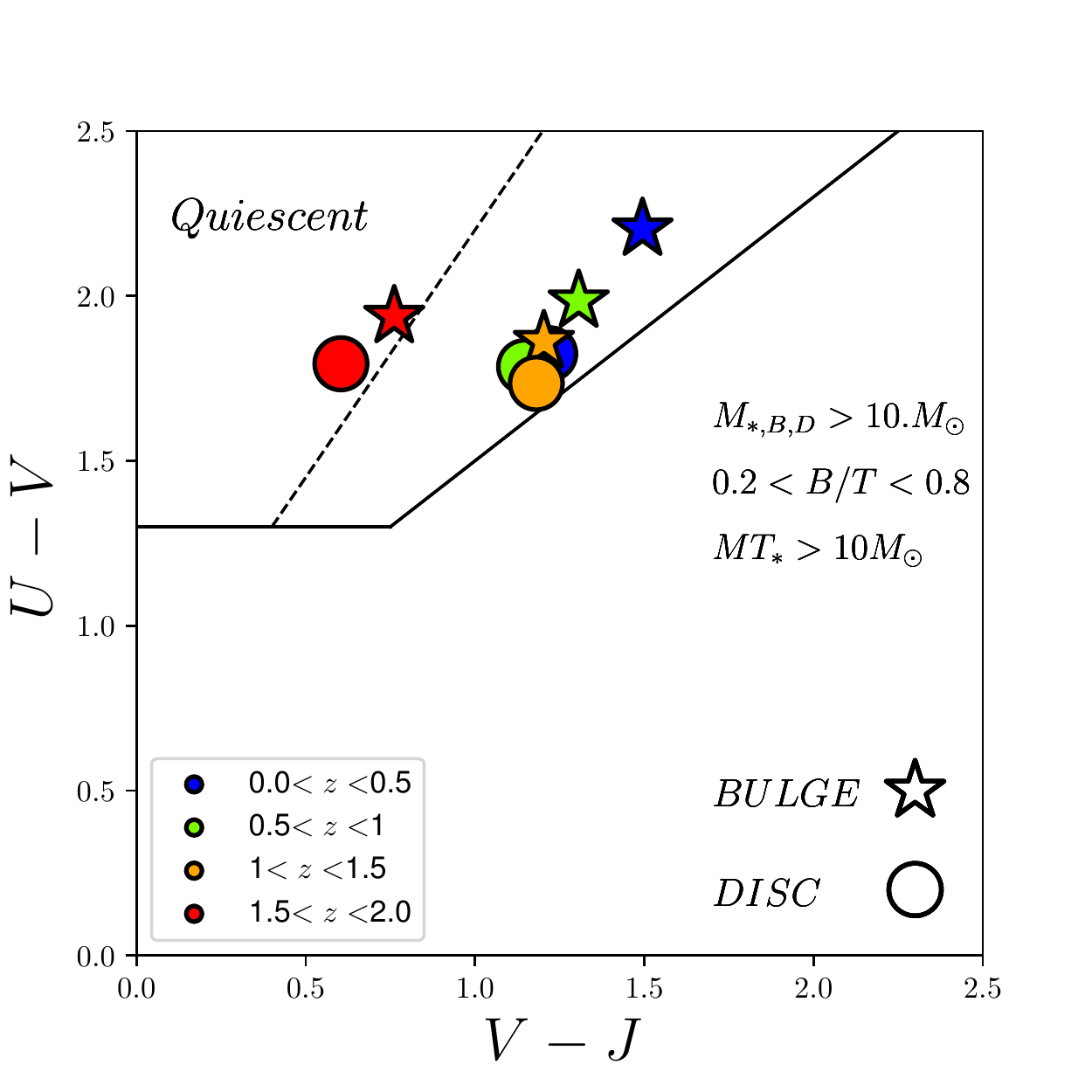}&
\includegraphics[width=0.3\textwidth]{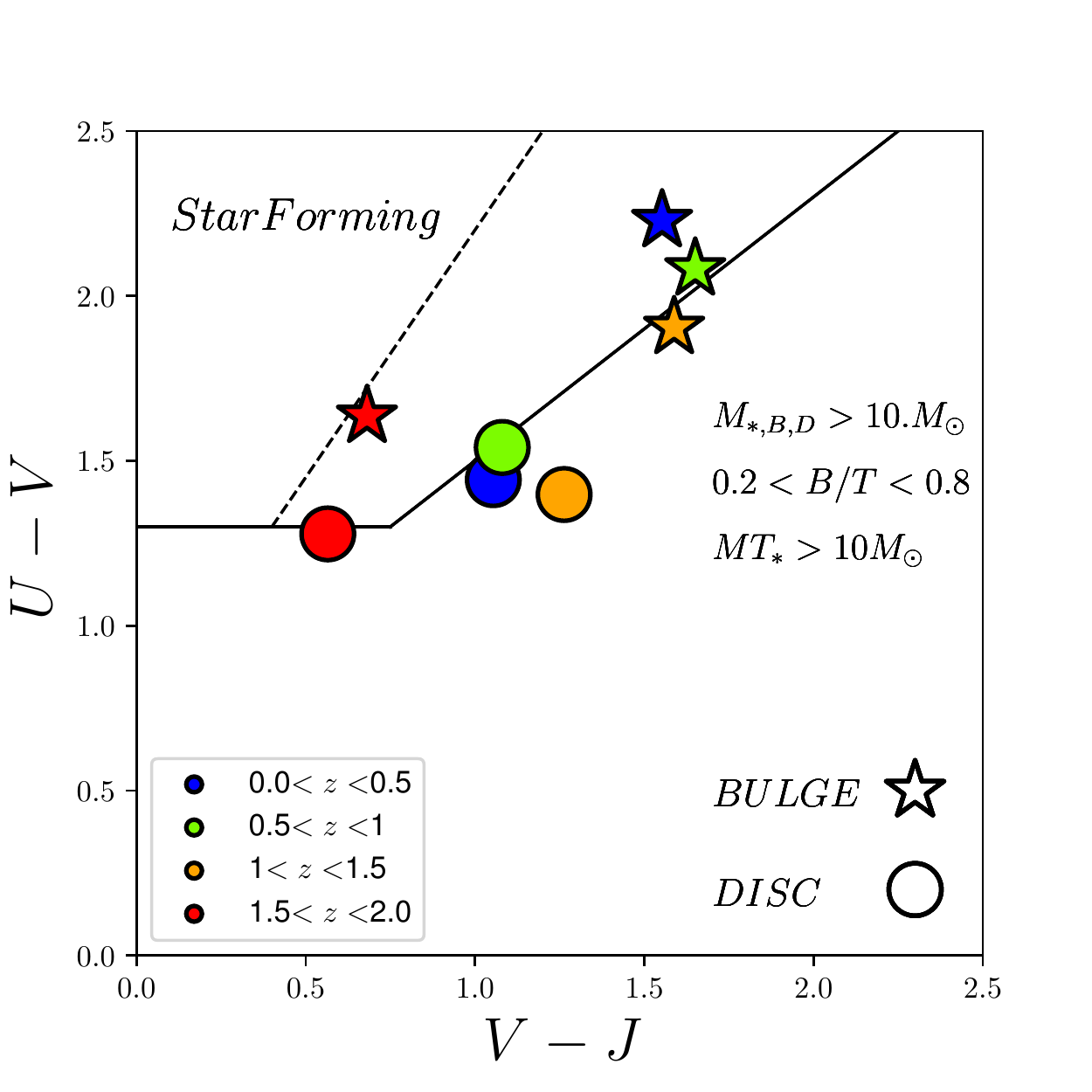}\\
\includegraphics[width=0.3\textwidth]{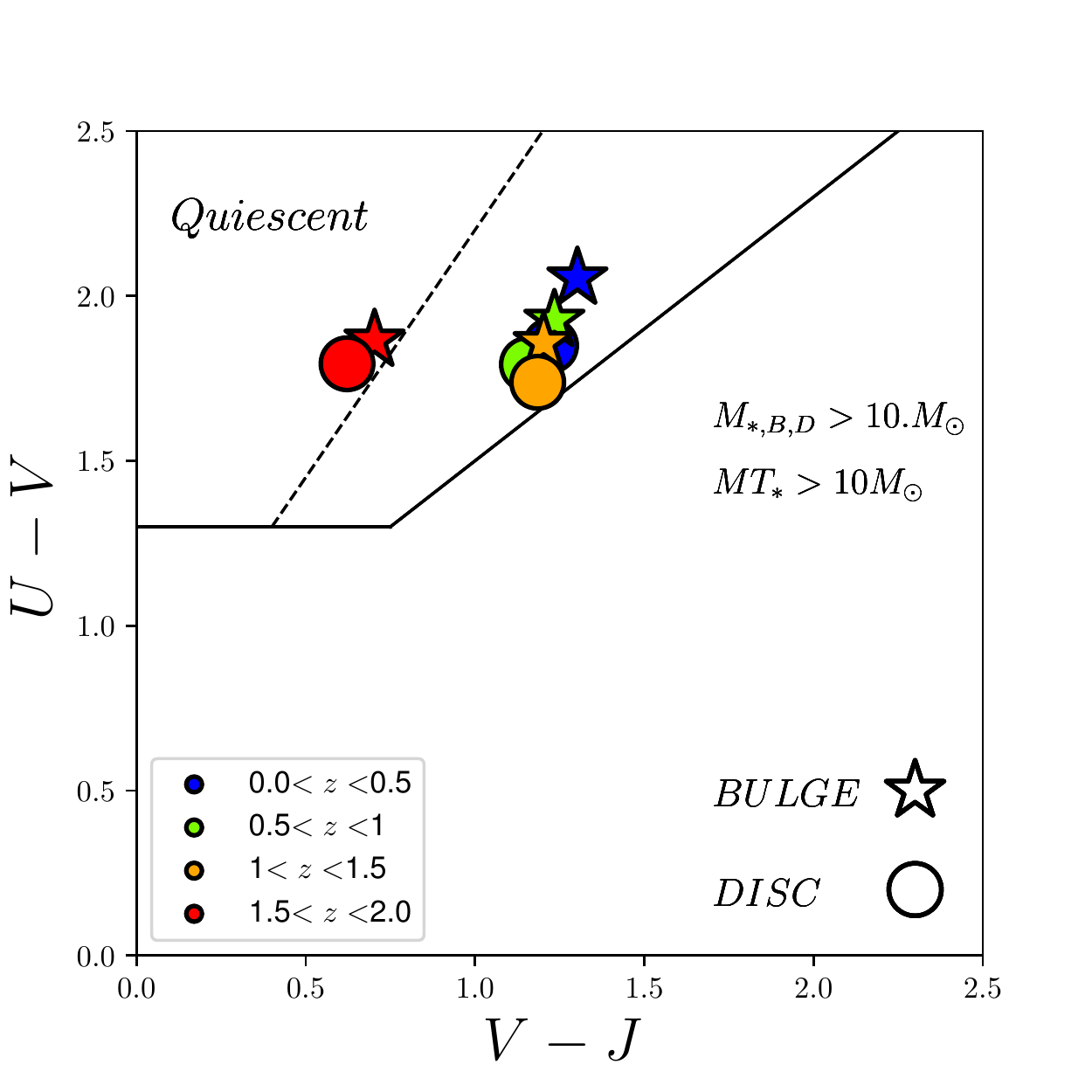}&
\includegraphics[width=0.3\textwidth]{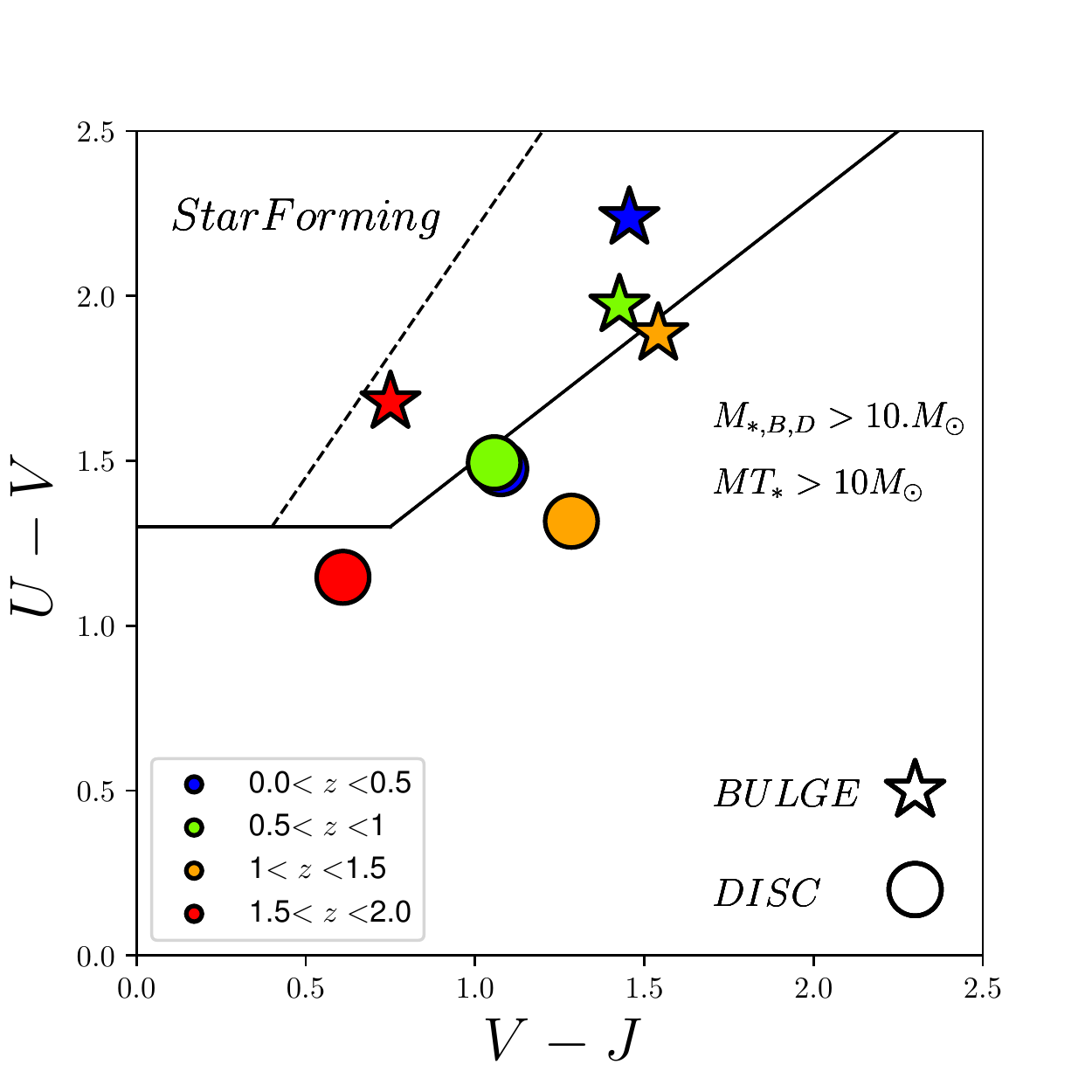}\\
\end{array}$
\caption[UVJ summary]{UVJ rest-frame colors of bulges and discs.
The series of plots show the median position of bulges and discs in the UV-VJ/I plane, for star forming(right) and quiescent(left) galaxies.
To avoid mass selection bias, the same analysis is done using two sample selection: $M_{B,D}>10 M_{\odot}, 0.2 < B/T <0.8, M_T >10M_{\odot}$, bottom: $M_{B,D}>10 M_{\odot},M_T >10M_{\odot}$.
Overall results are in agreement. Galaxy components show a difference in color indipendly of the selection criteria.}
\label{fig:UVJ_all}
\end{figure*}

In \cite{Dimauro2019}, we started analyzing the link between the morphology and the star formation activity with the aim to investigate possible signatures of the quenching process in the internal galaxy structure. 
Although the morphological difference between star forming and quiescent galaxies is confirmed, our analysis shows that the internal components structure weakly depend on the morphology and on the star-formation activity. Quenching signatures, if any, are present only in the bulge structure (we cannot exclude signatures that are beyound our uncertainty limits).
We also found an increase of the mean \Sersic index with cosmic time for bulge-dominated systems (B/T$\geq $0.8), in line with expectation from merger-driven models (e.g. \citealp{Hopkins2009,Nipoti2012,Shankar2018}). Galaxy-galaxy mergers have been often proposed as efficient triggers of star formation, black hole fuelling, and quenching, possibly driven by feedback from the central supermassive black hole itself (e.g, \citealp{Silk1998,Granato2004,Hopkins2006,Lapi2006,Shankar2006}). Additional channels for quenching in (central) galaxies could be linked to the host dark matter halo mass, which could halt the accretion of fresh gas to the galaxy reservoir (\citealp{Hopkins2014,Dekel2009,Cattaneo2009,Shankar2006}).

Furthermore, the stabilization of the gas, due to the growth of a massive central density, may also prevent the formation of new star forming regions within galaxies ('morphological quenching', \citealp{Martig2009}). Rapid halting of the star formation can also occur due to fast gas consumption from a star formation burst or violent disc instability (\citealp{Granato2004,Bournaud2016,Lapi2018}).  
Several works have probed the correlation between morphology and star formation activity (\citealp{Wuyts2011,Lang2014,Bluck2014,Whitaker2017,Barro2015,Barro2017,Peng2020}). However, the causes behind quenching remain debated. \cite{Bluck2019,Bluck2022} showed that central stellar velocity dispersion is the main galactic property linked to the level of sSFR, in the local universe. \cite{Marsden2022} also showed that at fixed stellar mass, the stellar velocity dispersion remains constant for B/T>0.2. These recent results are in line with our findings pointing to a significant role of the bulges in the quenching process.

Figure \ref{fig:SFR_mstar1} shows that bulges are present within the galaxies along the main sequence. The building-up of the central density already started while the galaxy was still forming stars and it keeps growing as the galaxy quenches and reddens.  
Moreover, Figure \ref{fig:SFR_mstar2} also revealed a lack of passive pure-disc galaxies and additionally, it shows that galaxies with 0.2 $<$ \BTM $< $ 0.8 are distributed in both the MS and the quenched region at all redshifts. 

These results show firstly that most of, if not all, galaxies (considering the limits of the galaxy selection used in this paper) without a bulge are star forming. A quenching channel without the growth of a bulge does not seem to be a common path. The absence of the bulge can assure that there is no quenching (\citealp{Lang2014,Barro2017}). Besides that, the observation of main sequence bulges implies that the quenching may not be a direct consequence of the bulge formation, suggesting that the bulge is a required but not sufficient condition to quench.
In this respect, \cite{Dimauro2019} found that the formation of a bulge component does not alter the disc structure nor the level of quenching/sSFR in the galaxy. The two components evolve as independent objects.

Furthermore, Figure \ref{fig:main_seq} shows that if the bulge stellar mass contribution is removed from the analysis, the SF Main Sequence follows a linear relation (in logarithmic space). This suggests that the major contribution to the Star Formation comes from the disc component. The growth of the central density increases the total stellar mass but not its SFR. Consequently, galaxies move from the low to the high mass region of the plane, causing the observed SFR to decrease. 
Complementary analysis of the Pearson coefficient points towards the same conclusion. The highest correlation with the sSFR is observed for the stellar mass density and the stellar mass of the bulge. This result confirms the presence of the bulge as one of the main requirements for the quenching. Indeed, the absence of it can assure that no quenching process are acting.

Many previous works also analysed the link between the quiescent state of a galaxy and its structural properties (e.g, \citealp{Bell2012,Cheung2012,Wake2012,Franx2008})
to investigate which property correlates better with the shutdown of star formation.
They exploited the structural properties of the galaxy as a whole, to infer the presence/absence of bulges, arguing that a prominent bulge is an important condition for quenching star formation \citep{Bell2008,Bell2012}
In this context, this work adds some important pieces of information to the analysis (the \BTM and resolved stellar properties of bulges and discs up to z$\sim$2) which lead to and confirm the presence of prominent stellar bulges in quenched galaxies.

The natural step forward is to look for any residual star formation in the core of Main Sequence galaxies. So far the discussion has focused on the total SFR of a galaxy. However, a detailed analysis requires to resolve stellar properties within galaxies and consequently, to distinguish between the total and the local SF. In a similar manner, quenching can be discussed considering the galaxy as whole or distinguishing internal components.
 In the local universe, it has be shown that galaxies present radial colour gradients (e.g. \citealp{Kennedy2016,Vika2013,Perez2013,Perez2010}) that reflect the distribution of the stellar populations from the centre to the outskirts. Young stellar population is the signature of an ongoing formation of stars which yields, as a result, to a blue rest-frame color. If a redder color is observed, it means that the stellar population is dominated by the emission of old stars, revealing a weak/absent SF. More detailed analysis, which aims to resolve the internal components, showed that the bulges exhibit redder colors than the disks when hosted in late-type galaxies, while both components exhibit a similar red color in the early-type ones (e.g. \citealp{Lange2016,Ghosh2020}). However, the topic is still debated at high-redshift. Several studies present different results. Few works underlined the presence of a sample of blue bulges and red discs \citep{Pan2014, Dominguez2009}.
The blue bulge color can be related to the presence of a central bar component \citep{Gadotti2008} or it can be the signature of an on-going rejuvenation process (\citealp{Mancini2015,Fang2013}). The central blue color can also be related to the presence of not-resolved young central components as star forming rings, bars etc,. On the contrary, the red color in the central region may be also explained by star formation that is obscured by dust, leading to a mis-classificationas as quenched object \citep{Tadaki2020,Whitaker2017b,vanDokkum2015}.
Additionally, internal color gradients link to ages gradient that also connect to the mass assembly of galaxies. This is reflected on the colors of the components. Bulges can be younger (bluer) or older (redder) than discs, depending on whether they are the result of an outside-in or inside-out assembly process (eg. \citealp{Perez2013,Pan2015,Tuttle2020,Costantin2021}). 

The analysis from this work shows that for most of the galaxies, the internal components (bulge and disc) show a clear color difference in the UVJ rest-frame colors, at all redshifts. A summary of median values at each redshift bin is shown in Figure \ref{fig:UVJ_all}. The analysis is focused on a subsample of galaxies with intermediate \BTM values (0.2<\BTM<0.8). These galaxies have a similar morphology and are either star forming or quiescent. Therefore, they are the ideal sample to investigate the effect of the quenching process. We find that discs are always bluer than bulges in star forming galaxies, while both components lie in the quiescent region when hosted in passive galaxies.

\begin{figure}
\includegraphics[width=0.4\textwidth]{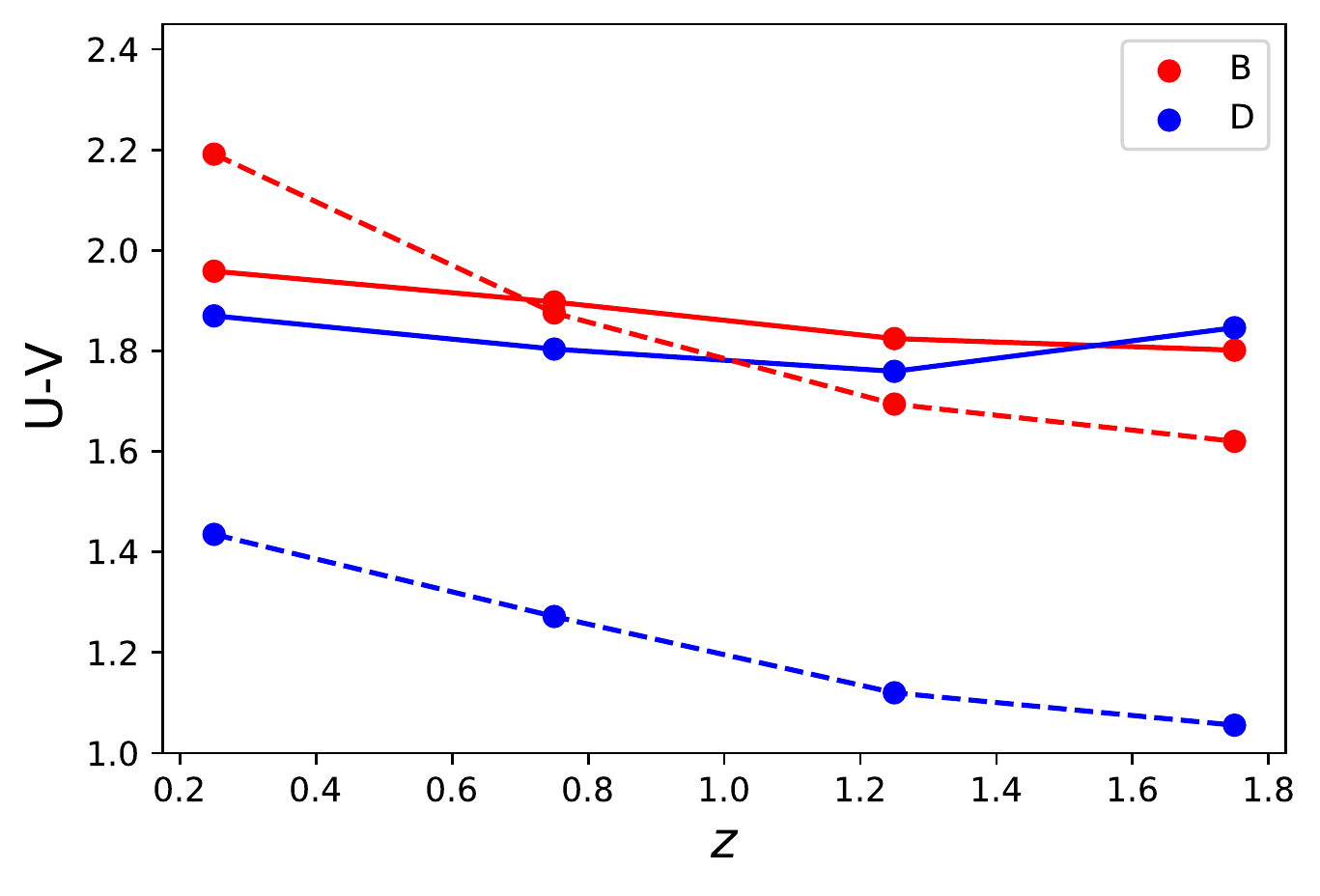}
\caption{Color of bulges and discs in SF(dotted lines) and quiescent galaxies at different redshift bins}
\label{fig:UV_z}
\end{figure}

To verify the impact of mass selection on these results, the same analysis is performed on the entire sample of galaxies considered in this work (bottom plot in Figure \ref{fig:UVJ_all}). The two sequences of plots present results that are in agreement with each other. Mean colors of bulges and ellipticals are concentrated in the quiescent regions of the UVJ plane while discs and disky galaxies have a blue or red mean color when hosted by star forming or quiescent galaxies. 
A complementary analysis, shown in Figure \ref{fig:UV_z}, further supports this result. The average color of bulges is constantly red over the entire redshift range, while discs exhibit different behaviors depending on whether they are hosted by star forming or passive galaxies. It is interesting to note that in the latter case the U-V color is almost constant, while, in star forming galaxies, it shows an increase (i.e, a decrease of SF) from high to low redshift, in agreement with the statistical decrease of the SFR with redshift.

Considering the U-V rest-frame color as a proxy of the SFR, a color Main Sequence can be built and analyzed (see Figure \ref{fig:MS_colors}). Colors of discs get redder towards high masses with a steeper slope than bulges. Bulges, instead, follow rather flat relationships, showing a redder color than the one of the galaxy. Discs are statistically driving the  galaxy colors, as it can be seen in Figure \ref{fig:MS_colors}. Moreover, the slope of the disc colors do not show any changes at log $M_*>10.5 \, M_{\sun}$ where the main sequence is dominated by galaxies hosting a relevant bulge. This result supports the weak contribution of bulges to the star formation activity of galaxies, but also that their presence does not affect the star formation activity of the disc component. 

The ensemble of the results listed above confirms that most of the star-formation activity is taking place within the disc component. The bulge component does not show signatures of residual star formation (within the limit of the present analysis). Such result is in agreement with Figure \ref{fig:main_seq} but also with the main findings of \cite{Dimauro2019}. 
Bulges and discs evolve as independent components. The presence of the bulge does not strongly affect the structure or the SF of the disc. 
Despite that, the lack of bulgeless passive galaxies suggests that the morphological transformation is a common evolutionary channel that may precede the quenching.  

Recent high-resolution cosmological simulations have shown that galaxies, during their time on the main sequence, pass through multiple gas-compaction events that create an overdensity of stars and provoke a loss of angular momentum \citep{Dekel2014,Zolotov2015}. These compaction phases can lead to quenching if the halo reaches a critical mass that prevents accretion from the outer regions. Observations that support this scenario \citep{Tacchella2017,Barro2015} propose an evolutionary path, typically referred as inside-out quenching, in which galaxies start to quench first in the central core and then later in the outskirts. Such process results in a gradient color signature. In the present work, we moved a step forward, estimating the colors (i.e, the star formation activity) of the internal component individually.  

From the data we can picture the SFR state of the galaxy in order to understand where the majority of SFR is localised. However this does not give us notions about the timeline of the events. In this context, it can only be stated that bulges are formed before the host galaxy quenches in most of the cases. Discs may have survived the quenching or re-accreated at a later time. Both cases would be compatible with the results observed. Consequently, inference of potential inside-out quenching have to be carefully verified. Furthermore, since bulges are almost all passive, they cannot be quenched further. Therefore, the quenching process affects the discs. The presence of the bulge is a required condition for the quenching but it is not the direct cause of the SF decrease. Consequently, there is room for additional processes, acting in decreasing the star formation rate within discs. 

Moreover, the analysis of this work does not completely rule out the progenitor bias effect. In fact, galaxies, now quenched, were formed in the past, when the universe was denser, i.e, they are more compact and dense than MS galaxies. The latter ones are consequently less dense and disky. This does not necessarly imply a morphological transformation. The presence of bulges and discs in MS galaxies can be explained by a later disc accretion (e.g. \citealp{Costantin2021,Mendez2021}).

\subsection{Coincidence between morphology and star-formation activity}
 In this work, we showed that bulges are already present in star forming galaxies, suggesting that they are formed during the life-time of a galaxy along the Main Sequence. This finding could suggest a late formation of this component compared to the galaxy as a whole, and consequently points towards the idea of a young central component. However, we do not observe bulge in formation. Indeed, section \ref{sec:colors} shows that bulges are more passive (and presumably older) than the discs, i.e. bulges are already formed and almost all already quenched at the time of the observation. This may involve a mass assembly which is growing inside-out. 
 These two interpretations are, at first glance, contradictory. Discs may only look younger because they keep forming stars, whereas younger bulges do not. This scenario could partially explain the results and solve the contradiction. A detailed investigation requires a spectroscopical analysis in order to resolve the SFH. An example of work done in this direction is presented by \cite{Mancini2019}. Through the study of the SFH of green valley galaxies, they conlude that the bulge mass is fully assembled already at z>2. Thereafter, the galaxy/bulge, already passive, passes through a process, called rejuvenation, that will accrete a star forming disc.
However, while this explation is a possible reconciliation between the presence of an old-looking bulge, in almost all galaxies, with the picture of a bulge emergence in pre-existing MS galaxies, it does not completely answer the main question. Indeed, if a component of a galaxy, such as a bulge or a disc, appears with an evolved stellar population, it does not directly imply that it is old. Conversely, a young stellar population does not strictly reflect its late formation time.

The present work underlines a strong correlation (even considering the presence of outliers) between the presence of the bulge and the star formation state of the galaxy. If there is a bulge, indipendently of the morphology or the star formation activity of the host galaxy, it is quiescent. Or, if a galaxy is quiescent, with a large probability, is hosts a significant bulge. Conversely, if the galaxy has a disky shape is highly probable to be star forming. This suggests that the quenching process is bounded to the morphological transformations \citep{DEugenio2021}, highlighting a strong evidence of the presence of the bulge being a necessary and suffucient condition for quiescence. 
This result puts constraints on our understanding of galaxy evolution. 
Often in the literature, when discussing about quiescence phase and quenching process, one usually refers to the galaxy as a whole while a peculiar analysis should be done on components indipendently. How do bulges quench? There are no direct observations of bulges in the transition phase. This may be explained by a fast evolution and quenching due to quasar feedback or mergers \citep{Toft2014}. They may be formed in a galaxy that already lost most of its gas content, i.e. already quenching. As an alternative explanation, the star formation signature may be obscured by an high dust content in the central compact region of massive main-sequence galaxies. Consequently, bulge formation might be dust-enshrouded, and only once the galaxy clean up of thick dust, an already quenched structure emerges ( e.g., for SMG galaxies \citealp{Cimatti2008} and ALMA compact galaxies \citealp{Puglisi2019,Puglisi2021}). \cite{Tadaki2020} showed that in massive galaxies (M$_*$>11M$\odot$) at z>2, most of the star formation is concentrated in the central region, but obscured by dust, suggesting the presence of blue bulges for this sample of galaxies. The finding of extended massive discs with a very dense and high star formation in the central region ($\sim 1 Kpc$), may be the first step to understand bulge formation (Kalita et al 2022, in prep.). Their existence can be explained by the occurrence of many processes, as mergers, thanks to which the disc can be destroyed and re-accreted. All of this combined with the present work, it might suggest an evolutionary path for bulges from beeing blue at high-z and quench later on. However, the present work is not in the position to either confirm or deny this conclusion since the sample analyzed contains few very massive galaxies at high-z, observed in the optical. 
 Recent analysis of the TNG Simulation link the AGN activity to the decrease of SF, proposing the AGN feedback as the main quenching driver. The AGN warms up the intergalactic medium on a long time-scale altering the accretion of cold gas (strangulation). The galaxy will actively forms stars  consuming its gas reservoire until exaustion. This would lead to a slowly decrease of SFR without altering the morphology \citep{Pillepich2017,Weinberger2017,Donnari2021}. While this work does not discard this mechanism as one of the processes affecting the SF, it does not support it as the main quenching path.

\subsection{Peculiar populations of red discs and blue bulges}
\label{sec:pec_pop}
The bimodal distribution of galaxy properties points towards an exemplified galaxy classification in two main categories: red ellipticals and blue spirals (from the local Universe to high-z, \citealp{Baldry2004,Whitaker2014}). As a consequence, bulges and discs are expected to be red and blue respectively. While from a statistical perspective both assumptions are confirmed by the results presented in this work and also previous works from the literature, the existence of blue ellipticals and red spirals is not excluded (eg. \citealp{Wuyts2011}).
 Similarly, but at a smaller scale, i.e, internal structure of galaxies, the morphological mix is also present. Indeed, from Figures \ref{fig:SFR_mstar2} and \ref{fig:UVJ_all}, two peculiar sub-populations of blue bulges and red discs are observed.
Blue spheroids and red discs have already been observed in the local Universe (eg. \citealp{Schawinski2009,Kannappan2009,Rowlands2012}). 
From a theoretical perspective, numerical simulations introduced the existence of blue compact galaxies, called 'blue nuggets' (i.e. blue spheroid or galaxies with a compact blue central density, \citealp{Dekel2013,Zolotov2015}), as probable progenitors of red, compact galaxies observed at high-redshift \citep{Zolotov2015,Barro2017}. 
Moreover, they may be the final result of alternative evolutionary paths as rejuvenation or disc re-accretion (eg. \citealp{Kannappan2009,Fang2013,Rowlands2018,Mancini2019}). 
Consequently, the analysis of their properties is important to improve our understanding of the galaxy evolution scenario.
The next sub-sections are focused on the investigations of the nature of those outliers. However, it is worth noticing that, quantitatively, blue bulges and red discs represent a small fraction compared to the entire sample as it can also be seen in Figure \ref{fig:SFR_mstar1}, and thus they are not statistically relevant for the analysis and do not affect the main results.

\subsection*{Red discs}
Figure \ref{fig:SFR_mstar2} shows the presence of a tail of objects with \BTM<0.2 extending below the main sequence. Those galaxies represent the $2-3\%$ of the entire galaxy sample with \BTM<0.2 . After proceeding to a visual inspection, we classify half of the sample as double component systems, with a bulge fraction < 20\%. Most of them have a star forming disc and a red bulge. Consequently, they may be galaxies passing through the green valley before entering in the quenching region.
  The remaining 30\% of the sub-sample, (i.e, 1\% of pure disc galaxy sample) are red disky galaxies. The machine learning classification discards the presence of a bulge for them. As an additional test, we verified that the single model fit has actually a disky shape. Properties of the selected sub-sample of red discs are shown in Figure \ref{fig:sph_disk}. The top right panel compares the U-V rest-frame color distribution of star forming  and quenched galaxies with red discs. It shows that the distribution of the latter ones peaks towards red colors in a similar manner than the one of red galaxies. The bottom panel shows the distribution of the \Sersic index values. The whole sub-sample has a range of values <2, showing that total best-fits for this population are in agreement with the definition of a disc profile. An example of a red disc is reported in Figure \ref{fig:example}. The main plot shows the Spectral Energy distribution, while the right corner panel shows the position of the galaxy in the UVJ rest-frame color plane. Both the UVJ rest-frame colors and the overall shape of the SED, confirm that this galaxy, and the selected sub-sample, are passive but with a disky surface brightness profile. The remaining galaxies of the outlier sample are mostly irregular objects that lead to mainly wrong fits. 

The existence of red discs was already proposed by previous works \citep{Toft2017}. 
Their observations suggest an alternative quenching path where galaxies, during the quenching process, must pass through major changes, not only on their structure, but also on their kinematics \citep{Peng2015,Toft2017} with the result of halting/stopping the gas accretion and therefore reducing the star formation activity without altering their structure \citep{Dekel2006}. High-density environments, as galaxy groups or clusters, can also alter the structure and the stellar populations of galaxies \citep{Masters2010,Vulcani2014,Lopes2016,Barsanti2021}

\subsection*{Blue bulges}

Figure \ref{fig:BDuvj1} shows a sub-population of bulges scattered towards the blue main sequence. The fraction (20\% of bulges hosted by SF galaxies) of those bulges increases towards high redshift. 
In a similar manner as for red discs, we proceeded to a visual inspection to verify the morphology and the goodness of the fit but also to quantify how many within them are blue bulges or blue ellipticals.
A fraction of these outliers is composed by irregular or clumpy galaxies that cause a wrong detection of the two components that result in wrong models. This is more likely to happen at high-redshift since the amount of clumpy galaxies is higher \citep{Guo2015, Huertas2020}. Besides that, a large fraction of those outliers are bulge+disc galaxies. Many of them are hosting a blue bulge and a red disc. The visual inspection of their morphology as well as their \Sersic profile ($n_B,n_T<2$) suggest for a fraction of them, the presence of a bar that could explain the blue color observed in the center \citep{Coelho2011, Gadotti2009}. These cases are at the limit of our method. Indeed, due to their peculiar structure, they may need a specific fitting configuration. Due to the variety of galaxy morphologies, we had to make a choice that suits most of the sample. 
The remaining fraction is composed by blue spheroidal galaxies ($n_B>2$). Structural properties are shown in Figure \ref{fig:sph_disk}. While the top left panel shows a typical \Sersic index distribution of a spheroidal population (they peak at n=2), the rest-frame color histogram shows a population that is in the middle between star forming and quiescent galaxies.
An example of a blue spheroid is shown in Figure \ref{fig:example}. 

\section{Summary and conclusions}
\label{sec:conclusion}
Using the multi-wavelength bulge-disc decomposition models from \citealp{Dimauro2018}, we have analyzed the location of galaxies in the $SFR-M_*$ plane as a function of the bulge-to-total ratio and how the internal components of galaxies (bulge and disc) populate such plane. In addition to that, we have analysed the rest-frame color distribution of the components. The sample of galaxies covers the redshift range [$0,2$]\\

The main results of this work are as follows:
\begin{itemize}
\item The bending of the main sequence at the high-mass end is directly linked to the population of galaxies with bulges (\BTM>0.2)

\item All the galaxies with no bulge (\BTM<0.2) are on the main sequence of star formation at all redshifts, which suggests that the growth of a bulge is a necessary condition for quenching. A quenching channel without bulge growth seems very rare, at least in the general field environment probed by our sample.

\item Galaxies with a significant bulge component (\BTM>0.2) populate both the main sequence and the quenched region with similar abundances. Growing a bulge is a required but not sufficient condition to quench or the bulge grows before the quenching occurs.

\item Disc survival or re-accretion is a common process due to the number of galaxies($\simeq 40\%$ of the sample) we observe with 0.2< \BTM<0.8.

\item Most of the galaxies present a difference in color. The vast majority of bulges in star forming galaxies lies in the quiescent region of the UVJ plane, thus do not form stars, while discs reside in the star forming region. This allow us to conclude that most of the star formation is localized in the disc component.

\item In the case of passive galaxies neither the disc nor the bulge form stars. This demonstrates that passive discs do exist (as a galaxy component) whereas star forming bulges do not. 

\item  The morphological transformation plays a relevant role in the quenching process. Most of the bulges do not show clear signs of star formation activity, suggesting a scenario in which the quenching process mostly affects the disc. 

All in all, our data point to the presence of a bulge as a necessary step towards quenching galaxies, which may be linked to the presence of a central black hole, and/or simply to a sort of morphological quenching.

\end{itemize}

\section*{DATA AVAILABILITY}
The analysis of the paper is based on the morphological catalog publicly released in \cite{Dimauro2018} and can  be downloaded from the web page:  \href{http://paoladimauro.space/morph_cat/}{link}

\section*{Acknowledgments}
We thank the anonymous referee whose comments and suggestions have improved the quality of this paper.\\
\noindent
The present work has been funded by a French national PhD scholarship and by the Brazilian PCI CNPq post-doctoral fellowship.\\
E.J.J. acknowledges support from FONDECYT Iniciacion en investigacion Project 11200263\\
R.A.D. acknowledges partial support support from CNPq grant 308105/2018-4\\

\bibliographystyle{mnras}
\bibliography{biblio}  

\begin{thebibliography}{}
\makeatletter
\relax
\def\mn@urlcharsother{\let\do\@makeother \do\$\do\&\do\#\do\^\do\_\do\%\do\~}
\def\mn@doi{\begingroup\mn@urlcharsother \@ifnextchar [ {\mn@doi@}
  {\mn@doi@[]}}
\def\mn@doi@[#1]#2{\def\@tempa{#1}\ifx\@tempa\@empty \href
  {http://dx.doi.org/#2} {doi:#2}\else \href {http://dx.doi.org/#2} {#1}\fi
  \endgroup}
\def\mn@eprint#1#2{\mn@eprint@#1:#2::\@nil}
\def\mn@eprint@arXiv#1{\href {http://arxiv.org/abs/#1} {{\tt arXiv:#1}}}
\def\mn@eprint@dblp#1{\href {http://dblp.uni-trier.de/rec/bibtex/#1.xml}
  {dblp:#1}}
\def\mn@eprint@#1:#2:#3:#4\@nil{\def\@tempa {#1}\def\@tempb {#2}\def\@tempc
  {#3}\ifx \@tempc \@empty \let \@tempc \@tempb \let \@tempb \@tempa \fi \ifx
  \@tempb \@empty \def\@tempb {arXiv}\fi \@ifundefined
  {mn@eprint@\@tempb}{\@tempb:\@tempc}{\expandafter \expandafter \csname
  mn@eprint@\@tempb\endcsname \expandafter{\@tempc}}}

\bibitem[\protect\citeauthoryear{{Abramson}, {Kelson}, {Dressler}, {Poggianti},
  {Gladders}, {Oemler}  \& {Vulcani}}{{Abramson} et~al.}{2014}]{Abramson2014}
{Abramson} L.~E.,  {Kelson} D.~D.,  {Dressler} A.,  {Poggianti} B.,  {Gladders}
  M.~D.,  {Oemler} Augustus J.,   {Vulcani} B.,  2014, \mn@doi [\apjl]
  {10.1088/2041-8205/785/2/L36}, \href
  {https://ui.adsabs.harvard.edu/abs/2014ApJ...785L..36A} {785, L36}

\bibitem[\protect\citeauthoryear{{Akhlaghi}, {Infante-Sainz}, {Roukema},
  {Khellat}, {Valls-Gabaud}  \& {Baena-Gall{\'e}}}{{Akhlaghi}
  et~al.}{2021}]{Akhlaghi2021}
{Akhlaghi} M.,  {Infante-Sainz} R.,  {Roukema} B.,  {Khellat} M.,
  {Valls-Gabaud} D.,   {Baena-Gall{\'e}} R.,  2021, {Maneage: Managing data
  lineage} (\mn@eprint {ascl} {2106.010})

\bibitem[\protect\citeauthoryear{{Baldry}, {Glazebrook}, {Brinkmann},
  {Ivezi{\'c}}, {Lupton}, {Nichol}  \& {Szalay}}{{Baldry}
  et~al.}{2004}]{Baldry2004}
{Baldry} I.~K.,  {Glazebrook} K.,  {Brinkmann} J.,  {Ivezi{\'c}} {\v Z}.,
  {Lupton} R.~H.,  {Nichol} R.~C.,   {Szalay} A.~S.,  2004, \mn@doi [\apj]
  {10.1086/380092}, \href {http://adsabs.harvard.edu/abs/2004ApJ...600..681B}
  {600, 681}

\bibitem[\protect\citeauthoryear{{Barden}, {H{\"a}u{\ss}ler}, {Peng},
  {McIntosh}  \& {Guo}}{{Barden} et~al.}{2012}]{Barden2012}
{Barden} M.,  {H{\"a}u{\ss}ler} B.,  {Peng} C.~Y.,  {McIntosh} D.~H.,   {Guo}
  Y.,  2012, \mn@doi [\mnras] {10.1111/j.1365-2966.2012.20619.x}, \href
  {http://adsabs.harvard.edu/abs/2012MNRAS.422..449B} {422, 449}

\bibitem[\protect\citeauthoryear{{Barro} et~al.,}{{Barro}
  et~al.}{2011}]{Barro2011}
{Barro} G.,  et~al., 2011, VizieR Online Data Catalog, \href
  {https://ui.adsabs.harvard.edu/abs/2011yCat..21930013B} {p. J/ApJS/193/13}

\bibitem[\protect\citeauthoryear{{Barro} et~al.,}{{Barro}
  et~al.}{2015}]{Barro2015}
{Barro} G.,  et~al., 2015, preprint, \href
  {http://adsabs.harvard.edu/abs/2015arXiv150900469B} {} (\mn@eprint {arXiv}
  {1509.00469})

\bibitem[\protect\citeauthoryear{{Barro} et~al.,}{{Barro}
  et~al.}{2017}]{Barro2017}
{Barro} G.,  et~al., 2017, \mn@doi [\apj] {10.3847/1538-4357/aa6b05}, \href
  {http://adsabs.harvard.edu/abs/2017ApJ...840...47B} {840, 47}

\bibitem[\protect\citeauthoryear{{Barro} et~al.,}{{Barro}
  et~al.}{2019}]{Barro2019}
{Barro} G.,  et~al., 2019, \mn@doi [\apjs] {10.3847/1538-4365/ab23f2}, \href
  {https://ui.adsabs.harvard.edu/abs/2019ApJS..243...22B} {243, 22}

\bibitem[\protect\citeauthoryear{Barsanti et~al.,}{Barsanti
  et~al.}{2021}]{Barsanti2021}
Barsanti S.,  et~al., 2021, \mn@doi [The Astrophysical Journal]
  {10.3847/1538-4357/abe5ac}, 911, 21

\bibitem[\protect\citeauthoryear{{Bell}}{{Bell}}{2008}]{Bell2008}
{Bell} E.~F.,  2008, \mn@doi [\apj] {10.1086/589551}, \href
  {https://ui.adsabs.harvard.edu/abs/2008ApJ...682..355B} {682, 355}

\bibitem[\protect\citeauthoryear{{Bell} et~al.,}{{Bell}
  et~al.}{2005}]{Bell2005}
{Bell} E.~F.,  et~al., 2005, \mn@doi [\apj] {10.1086/429552}, \href
  {https://ui.adsabs.harvard.edu/abs/2005ApJ...625...23B} {625, 23}

\bibitem[\protect\citeauthoryear{{Bell} et~al.,}{{Bell}
  et~al.}{2012}]{Bell2012}
{Bell} E.~F.,  et~al., 2012, \mn@doi [\apj] {10.1088/0004-637X/753/2/167},
  \href {https://ui.adsabs.harvard.edu/abs/2012ApJ...753..167B} {753, 167}

\bibitem[\protect\citeauthoryear{{Bernardi}, {Shankar}, {Hyde}, {Mei},
  {Marulli}  \& {Sheth}}{{Bernardi} et~al.}{2010}]{Bernardi2010}
{Bernardi} M.,  {Shankar} F.,  {Hyde} J.~B.,  {Mei} S.,  {Marulli} F.,
  {Sheth} R.~K.,  2010, \mn@doi [\mnras] {10.1111/j.1365-2966.2010.16425.x},
  \href {https://ui.adsabs.harvard.edu/abs/2010MNRAS.404.2087B} {404, 2087}

\bibitem[\protect\citeauthoryear{{Bernardi}, {Meert}, {Sheth}, {Vikram},
  {Huertas-Company}, {Mei}  \& {Shankar}}{{Bernardi}
  et~al.}{2013}]{Bernardi2013}
{Bernardi} M.,  {Meert} A.,  {Sheth} R.~K.,  {Vikram} V.,  {Huertas-Company}
  M.,  {Mei} S.,   {Shankar} F.,  2013, \mn@doi [\mnras]
  {10.1093/mnras/stt1607}, \href
  {https://ui.adsabs.harvard.edu/abs/2013MNRAS.436..697B} {436, 697}

\bibitem[\protect\citeauthoryear{{Bernardi}, {Meert}, {Vikram},
  {Huertas-Company}, {Mei}, {Shankar}  \& {Sheth}}{{Bernardi}
  et~al.}{2014}]{Bernardi2014}
{Bernardi} M.,  {Meert} A.,  {Vikram} V.,  {Huertas-Company} M.,  {Mei} S.,
  {Shankar} F.,   {Sheth} R.~K.,  2014, \mn@doi [\mnras]
  {10.1093/mnras/stu1106}, \href
  {http://adsabs.harvard.edu/abs/2014MNRAS.443..874B} {443, 874}

\bibitem[\protect\citeauthoryear{{Bernardi}, {Meert}, {Sheth},
  {Huertas-Company}, {Maraston}, {Shankar}  \& {Vikram}}{{Bernardi}
  et~al.}{2016}]{Bernardi2016}
{Bernardi} M.,  {Meert} A.,  {Sheth} R.~K.,  {Huertas-Company} M.,  {Maraston}
  C.,  {Shankar} F.,   {Vikram} V.,  2016, \mn@doi [\mnras]
  {10.1093/mnras/stv2487}, \href
  {https://ui.adsabs.harvard.edu/abs/2016MNRAS.455.4122B} {455, 4122}

\bibitem[\protect\citeauthoryear{{Bernardi}, {Meert}, {Sheth}, {Fischer},
  {Huertas-Company}, {Maraston}, {Shankar}  \& {Vikram}}{{Bernardi}
  et~al.}{2017a}]{Bernardi2017a}
{Bernardi} M.,  {Meert} A.,  {Sheth} R.~K.,  {Fischer} J.~L.,
  {Huertas-Company} M.,  {Maraston} C.,  {Shankar} F.,   {Vikram} V.,  2017a,
  \mn@doi [\mnras] {10.1093/mnras/stx176}, \href
  {https://ui.adsabs.harvard.edu/abs/2017MNRAS.467.2217B} {467, 2217}

\bibitem[\protect\citeauthoryear{{Bernardi}, {Fischer}, {Sheth}, {Meert},
  {Huertas-Company}, {Shankar}  \& {Vikram}}{{Bernardi}
  et~al.}{2017b}]{Bernardi2017b}
{Bernardi} M.,  {Fischer} J.~L.,  {Sheth} R.~K.,  {Meert} A.,
  {Huertas-Company} M.,  {Shankar} F.,   {Vikram} V.,  2017b, \mn@doi [\mnras]
  {10.1093/mnras/stx677}, \href
  {https://ui.adsabs.harvard.edu/abs/2017MNRAS.468.2569B} {468, 2569}

\bibitem[\protect\citeauthoryear{{Bernardi}, {Sheth}, {Dominguez Sanchez},
  {Margalef-Bentabol}, {Bizyaev}  \& {Lane}}{{Bernardi}
  et~al.}{2022}]{Bernardi2022}
{Bernardi} M.,  {Sheth} R.~K.,  {Dominguez Sanchez} H.,  {Margalef-Bentabol}
  B.,  {Bizyaev} D.,   {Lane} R.~R.,  2022, arXiv e-prints, \href
  {https://ui.adsabs.harvard.edu/abs/2022arXiv220107810B} {p. arXiv:2201.07810}

\bibitem[\protect\citeauthoryear{{Bertin} \& {Arnouts}}{{Bertin} \&
  {Arnouts}}{1996}]{Bertin1996}
{Bertin} E.,  {Arnouts} S.,  1996, \mn@doi [\aaps] {10.1051/aas:1996164}, \href
  {https://ui.adsabs.harvard.edu/abs/1996A%26AS..117..393B} {117, 393}

\bibitem[\protect\citeauthoryear{{Bluck}, {Mendel}, {Ellison}, {Moreno},
  {Simard}, {Patton}  \& {Starkenburg}}{{Bluck} et~al.}{2014}]{Bluck2014}
{Bluck} A.~F.~L.,  {Mendel} J.~T.,  {Ellison} S.~L.,  {Moreno} J.,  {Simard}
  L.,  {Patton} D.~R.,   {Starkenburg} E.,  2014, \mn@doi [\mnras]
  {10.1093/mnras/stu594}, \href
  {http://adsabs.harvard.edu/abs/2014MNRAS.441..599B} {441, 599}

\bibitem[\protect\citeauthoryear{{Bluck} et~al.,}{{Bluck}
  et~al.}{2019}]{Bluck2019}
{Bluck} A. F.~L.,  et~al., 2019, \mn@doi [\mnras] {10.1093/mnras/stz363}, \href
  {https://ui.adsabs.harvard.edu/abs/2019MNRAS.485..666B} {485, 666}

\bibitem[\protect\citeauthoryear{{Bluck}, {Maiolino}, {Brownson}, {Conselice},
  {Ellison}, {Piotrowska}  \& {Thorp}}{{Bluck} et~al.}{2022}]{Bluck2022}
{Bluck} A. F.~L.,  {Maiolino} R.,  {Brownson} S.,  {Conselice} C.~J.,
  {Ellison} S.~L.,  {Piotrowska} J.~M.,   {Thorp} M.~D.,  2022, arXiv e-prints,
  \href {https://ui.adsabs.harvard.edu/abs/2022arXiv220107814B} {p.
  arXiv:2201.07814}

\bibitem[\protect\citeauthoryear{{Bouch{\'e}} et~al.,}{{Bouch{\'e}}
  et~al.}{2010}]{Bouche2010}
{Bouch{\'e}} N.,  et~al., 2010, \mn@doi [\apj] {10.1088/0004-637X/718/2/1001},
  \href {https://ui.adsabs.harvard.edu/abs/2010ApJ...718.1001B} {718, 1001}

\bibitem[\protect\citeauthoryear{{Bournaud}}{{Bournaud}}{2016}]{Bournaud2016}
{Bournaud} F.,  2016, in {Laurikainen} E.,  {Peletier} R.,   {Gadotti} D.,
  eds,  Astrophysics and Space Science Library Vol. 418, Galactic Bulges.
  p.~355 (\mn@eprint {arXiv} {1503.07660}),
  \mn@doi{10.1007/978-3-319-19378-6_13}

\bibitem[\protect\citeauthoryear{Brennan et~al.,}{Brennan
  et~al.}{2016}]{Brennan2016}
Brennan R.,  et~al., 2016, \mn@doi [Monthly Notices of the Royal Astronomical
  Society] {10.1093/mnras/stw2690}, 465, 619

\bibitem[\protect\citeauthoryear{{Brinchmann}, {Charlot}, {White}, {Tremonti},
  {Kauffmann}, {Heckman}  \& {Brinkmann}}{{Brinchmann}
  et~al.}{2004}]{Brinchmann2004}
{Brinchmann} J.,  {Charlot} S.,  {White} S.~D.~M.,  {Tremonti} C.,  {Kauffmann}
  G.,  {Heckman} T.,   {Brinkmann} J.,  2004, \mn@doi [\mnras]
  {10.1111/j.1365-2966.2004.07881.x}, \href
  {http://adsabs.harvard.edu/abs/2004MNRAS.351.1151B} {351, 1151}

\bibitem[\protect\citeauthoryear{{Bruzual} \& {Charlot}}{{Bruzual} \&
  {Charlot}}{2003}]{Bruzual2003}
{Bruzual} G.,  {Charlot} S.,  2003, \mn@doi [\mnras]
  {10.1046/j.1365-8711.2003.06897.x}, \href
  {http://adsabs.harvard.edu/abs/2003MNRAS.344.1000B} {344, 1000}

\bibitem[\protect\citeauthoryear{{Bundy} et~al.,}{{Bundy}
  et~al.}{2015}]{Bundy2015}
{Bundy} K.,  et~al., 2015, \mn@doi [\apj] {10.1088/0004-637X/798/1/7}, \href
  {https://ui.adsabs.harvard.edu/abs/2015ApJ...798....7B} {798, 7}

\bibitem[\protect\citeauthoryear{{Calzetti}, {Armus}, {Bohlin}, {Kinney},
  {Koornneef}  \& {Storchi-Bergmann}}{{Calzetti} et~al.}{2000}]{Calzetti2000}
{Calzetti} D.,  {Armus} L.,  {Bohlin} R.~C.,  {Kinney} A.~L.,  {Koornneef} J.,
   {Storchi-Bergmann} T.,  2000, \mn@doi [\apj] {10.1086/308692}, \href
  {http://adsabs.harvard.edu/abs/2000ApJ...533..682C} {533, 682}

\bibitem[\protect\citeauthoryear{{Carollo} et~al.,}{{Carollo}
  et~al.}{2016}]{Carollo2016}
{Carollo} C.~M.,  et~al., 2016, \mn@doi [\apj] {10.3847/0004-637X/818/2/180},
  \href {http://adsabs.harvard.edu/abs/2016ApJ...818..180C} {818, 180}

\bibitem[\protect\citeauthoryear{{Cattaneo}, {Dekel}, {Devriendt}, {Guiderdoni}
   \& {Blaizot}}{{Cattaneo} et~al.}{2006}]{Cattaneo2006}
{Cattaneo} A.,  {Dekel} A.,  {Devriendt} J.,  {Guiderdoni} B.,   {Blaizot} J.,
  2006, \mn@doi [\mnras] {10.1111/j.1365-2966.2006.10608.x}, \href
  {http://adsabs.harvard.edu/abs/2006MNRAS.370.1651C} {370, 1651}

\bibitem[\protect\citeauthoryear{{Cattaneo} et~al.,}{{Cattaneo}
  et~al.}{2009}]{Cattaneo2009}
{Cattaneo} A.,  et~al., 2009, \mn@doi [\nat] {10.1038/nature08135}, \href
  {http://adsabs.harvard.edu/abs/2009Natur.460..213C} {460, 213}

\bibitem[\protect\citeauthoryear{{Chabrier}}{{Chabrier}}{2003}]{Chabrier2003}
{Chabrier} G.,  2003, \mn@doi [\pasp] {10.1086/376392}, \href
  {http://adsabs.harvard.edu/abs/2003PASP..115..763C} {115, 763}

\bibitem[\protect\citeauthoryear{Chauke et~al.,}{Chauke
  et~al.}{2019}]{Chauke2019}
Chauke P.,  et~al., 2019, \mn@doi [The Astrophysical Journal]
  {10.3847/1538-4357/ab164d}, 877, 48

\bibitem[\protect\citeauthoryear{{Cheung} et~al.,}{{Cheung}
  et~al.}{2012}]{Cheung2012}
{Cheung} E.,  et~al., 2012, \mn@doi [\apj] {10.1088/0004-637X/760/2/131}, \href
  {https://ui.adsabs.harvard.edu/abs/2012ApJ...760..131C} {760, 131}

\bibitem[\protect\citeauthoryear{{Cimatti} et~al.,}{{Cimatti}
  et~al.}{2008}]{Cimatti2008}
{Cimatti} A.,  et~al., 2008, \mn@doi [\aap] {10.1051/0004-6361:20078739}, \href
  {https://ui.adsabs.harvard.edu/abs/2008A&A...482...21C} {482, 21}

\bibitem[\protect\citeauthoryear{{Coelho} \& {Gadotti}}{{Coelho} \&
  {Gadotti}}{2011}]{Coelho2011}
{Coelho} P.,  {Gadotti} D.~A.,  2011, \mn@doi [\apjl]
  {10.1088/2041-8205/743/1/L13}, \href
  {https://ui.adsabs.harvard.edu/abs/2011ApJ...743L..13C} {743, L13}

\bibitem[\protect\citeauthoryear{{Costantin} et~al.,}{{Costantin}
  et~al.}{2021}]{Costantin2021}
{Costantin} L.,  et~al., 2021, \mn@doi [apj] {10.3847/1538-4357/abef72}, \href
  {https://ui.adsabs.harvard.edu/abs/2021ApJ...913..125C} {913, 125}

\bibitem[\protect\citeauthoryear{{Croom} et~al.,}{{Croom}
  et~al.}{2012}]{Croom2012}
{Croom} S.~M.,  et~al., 2012, \mn@doi [\mnras]
  {10.1111/j.1365-2966.2011.20365.x}, \href
  {https://ui.adsabs.harvard.edu/abs/2012MNRAS.421..872C} {421, 872}

\bibitem[\protect\citeauthoryear{{D'Eugenio} et~al.,}{{D'Eugenio}
  et~al.}{2021}]{DEugenio2021}
{D'Eugenio} C.,  et~al., 2021, \mn@doi [\aap] {10.1051/0004-6361/202040067},
  \href {https://ui.adsabs.harvard.edu/abs/2021A&A...653A..32D} {653, A32}

\bibitem[\protect\citeauthoryear{{Daddi} et~al.,}{{Daddi}
  et~al.}{2007}]{Daddi2007}
{Daddi} E.,  et~al., 2007, \mn@doi [\apj] {10.1086/521818}, \href
  {http://adsabs.harvard.edu/abs/2007ApJ...670..156D} {670, 156}

\bibitem[\protect\citeauthoryear{{Dahlen} et~al.,}{{Dahlen}
  et~al.}{2013}]{Dahlen2013}
{Dahlen} T.,  et~al., 2013, \mn@doi [\apj] {10.1088/0004-637X/775/2/93}, \href
  {http://adsabs.harvard.edu/abs/2013ApJ...775...93D} {775, 93}

\bibitem[\protect\citeauthoryear{{Dekel} \& {Birnboim}}{{Dekel} \&
  {Birnboim}}{2006}]{Dekel2006}
{Dekel} A.,  {Birnboim} Y.,  2006, \mn@doi [\mnras]
  {10.1111/j.1365-2966.2006.10145.x}, \href
  {http://adsabs.harvard.edu/abs/2006MNRAS.368....2D} {368, 2}

\bibitem[\protect\citeauthoryear{Dekel \& Burkert}{Dekel \&
  Burkert}{2013}]{Dekel2013}
Dekel A.,  Burkert A.,  2013, \mn@doi [Monthly Notices of the Royal
  Astronomical Society] {10.1093/mnras/stt2331}, 438, 1870

\bibitem[\protect\citeauthoryear{{Dekel} \& {Burkert}}{{Dekel} \&
  {Burkert}}{2014}]{Dekel2014}
{Dekel} A.,  {Burkert} A.,  2014, \mn@doi [\mnras] {10.1093/mnras/stt2331},
  \href {http://adsabs.harvard.edu/abs/2014MNRAS.438.1870D} {438, 1870}

\bibitem[\protect\citeauthoryear{{Dekel} \& {Mandelker}}{{Dekel} \&
  {Mandelker}}{2014}]{Dekel2014b}
{Dekel} A.,  {Mandelker} N.,  2014, \mn@doi [\mnras] {10.1093/mnras/stu1427},
  \href {https://ui.adsabs.harvard.edu/abs/2014MNRAS.444.2071D} {444, 2071}

\bibitem[\protect\citeauthoryear{{Dekel} et~al.,}{{Dekel}
  et~al.}{2009}]{Dekel2009}
{Dekel} A.,  et~al., 2009, \mn@doi [\nat] {10.1038/nature07648}, \href
  {http://adsabs.harvard.edu/abs/2009Natur.457..451D} {457, 451}

\bibitem[\protect\citeauthoryear{{Dekel}, {Ishai}, {Dutton}  \&
  {Maccio}}{{Dekel} et~al.}{2017}]{Dekel2017}
{Dekel} A.,  {Ishai} G.,  {Dutton} A.~A.,   {Maccio} A.~V.,  2017, \mn@doi
  [\mnras] {10.1093/mnras/stx486}, \href
  {https://ui.adsabs.harvard.edu/abs/2017MNRAS.468.1005D} {468, 1005}

\bibitem[\protect\citeauthoryear{{Dimauro} et~al.,}{{Dimauro}
  et~al.}{2018}]{Dimauro2018}
{Dimauro} P.,  et~al., 2018, \mn@doi [MNRA] {10.1093/mnras/sty1379}, \href
  {http://adsabs.harvard.edu/abs/2018MNRAS.478.5410D} {478, 5410}

\bibitem[\protect\citeauthoryear{{Dimauro} et~al.,}{{Dimauro}
  et~al.}{2019}]{Dimauro2019}
{Dimauro} P.,  et~al., 2019, \mn@doi [MNRA] {10.1093/mnras/stz2421}, \href
  {https://ui.adsabs.harvard.edu/abs/2019MNRAS.489.4135D} {489, 4135}

\bibitem[\protect\citeauthoryear{{Dom{\'{\i}}nguez-Palmero} \&
  {Balcells}}{{Dom{\'{\i}}nguez-Palmero} \& {Balcells}}{2009}]{Dominguez2009}
{Dom{\'{\i}}nguez-Palmero} L.,  {Balcells} M.,  2009, \mn@doi [\apjl]
  {10.1088/0004-637X/694/1/L69}, \href
  {http://adsabs.harvard.edu/abs/2009ApJ...694L..69D} {694, L69}

\bibitem[\protect\citeauthoryear{{Donnari} et~al.,}{{Donnari}
  et~al.}{2021}]{Donnari2021}
{Donnari} M.,  et~al., 2021, \mn@doi [\mnras] {10.1093/mnras/staa3006}, \href
  {https://ui.adsabs.harvard.edu/abs/2021MNRAS.500.4004D} {500, 4004}

\bibitem[\protect\citeauthoryear{{Elbaz} et~al.,}{{Elbaz}
  et~al.}{2007}]{Elbaz2007}
{Elbaz} D.,  et~al., 2007, \mn@doi [\aap] {10.1051/0004-6361:20077525}, \href
  {http://adsabs.harvard.edu/abs/2007A%26A...468...33E} {468, 33}

\bibitem[\protect\citeauthoryear{{Fang}, {Faber}, {Koo}  \& {Dekel}}{{Fang}
  et~al.}{2013}]{Fang2013}
{Fang} J.~J.,  {Faber} S.~M.,  {Koo} D.~C.,   {Dekel} A.,  2013, \mn@doi [\apj]
  {10.1088/0004-637X/776/1/63}, \href
  {http://adsabs.harvard.edu/abs/2013ApJ...776...63F} {776, 63}

\bibitem[\protect\citeauthoryear{Fang et~al.,}{Fang et~al.}{2018}]{Fang2018}
Fang J.~J.,  et~al., 2018, \mn@doi [The Astrophysical Journal]
  {10.3847/1538-4357/aabcba}, 858, 100

\bibitem[\protect\citeauthoryear{{Feldmann}}{{Feldmann}}{2013}]{Feldmann2013}
{Feldmann} R.,  2013, \mn@doi [\mnras] {10.1093/mnras/stt851}, \href
  {https://ui.adsabs.harvard.edu/abs/2013MNRAS.433.1910F} {433, 1910}

\bibitem[\protect\citeauthoryear{{Franx}, {van Dokkum}, {F{\"o}rster
  Schreiber}, {Wuyts}, {Labb{\'e}}  \& {Toft}}{{Franx}
  et~al.}{2008}]{Franx2008}
{Franx} M.,  {van Dokkum} P.~G.,  {F{\"o}rster Schreiber} N.~M.,  {Wuyts} S.,
  {Labb{\'e}} I.,   {Toft} S.,  2008, \mn@doi [\apj] {10.1086/592431}, \href
  {https://ui.adsabs.harvard.edu/abs/2008ApJ...688..770F} {688, 770}

\bibitem[\protect\citeauthoryear{{Fraser-McKelvie} et~al.,}{{Fraser-McKelvie}
  et~al.}{2021}]{Fraser-McKelvie2021}
{Fraser-McKelvie} A.,  et~al., 2021, \mn@doi [\mnras] {10.1093/mnras/stab573},
  \href {https://ui.adsabs.harvard.edu/abs/2021MNRAS.503.4992F} {503, 4992}

\bibitem[\protect\citeauthoryear{{Gadotti}}{{Gadotti}}{2008}]{Gadotti2008}
{Gadotti} D.~A.,  2008, in {Bureau} M.,  {Athanassoula} E.,   {Barbuy} B.,
  eds,  IAU Symposium Vol. 245, Formation and Evolution of Galaxy Bulges. pp
  117--120 (\mn@eprint {arXiv} {0708.2842}), \mn@doi{10.1017/S1743921308017420}

\bibitem[\protect\citeauthoryear{{Gadotti}}{{Gadotti}}{2009}]{Gadotti2009}
{Gadotti} D.~A.,  2009, \mn@doi [\mnras] {10.1111/j.1365-2966.2008.14257.x},
  \href {http://adsabs.harvard.edu/abs/2009MNRAS.393.1531G} {393, 1531}

\bibitem[\protect\citeauthoryear{{Galametz} et~al.,}{{Galametz}
  et~al.}{2013}]{Galametz2013}
{Galametz} A.,  et~al., 2013, \mn@doi [\apjs] {10.1088/0067-0049/206/2/10},
  \href {http://adsabs.harvard.edu/abs/2013ApJS..206...10G} {206, 10}

\bibitem[\protect\citeauthoryear{{Genzel} et~al.,}{{Genzel}
  et~al.}{2012}]{Genzel2012}
{Genzel} R.,  et~al., 2012, \mn@doi [\apj] {10.1088/0004-637X/746/1/69}, \href
  {https://ui.adsabs.harvard.edu/abs/2012ApJ...746...69G} {746, 69}

\bibitem[\protect\citeauthoryear{Ghosh, Urry, Wang, Schawinski, Turp  \&
  Powell}{Ghosh et~al.}{2020}]{Ghosh2020}
Ghosh A.,  Urry C.~M.,  Wang Z.,  Schawinski K.,  Turp D.,   Powell M.~C.,
  2020, \mn@doi [The Astrophysical Journal] {10.3847/1538-4357/ab8a47}, 895,
  112

\bibitem[\protect\citeauthoryear{Gonzalez-Perez, Castander  \&
  Kauffmann}{Gonzalez-Perez et~al.}{2010}]{Perez2010}
Gonzalez-Perez V.,  Castander F.~J.,   Kauffmann G.,  2010, \mn@doi [Monthly
  Notices of the Royal Astronomical Society]
  {10.1111/j.1365-2966.2010.17744.x}, 411, 1151–1166

\bibitem[\protect\citeauthoryear{{Granato}, {De Zotti}, {Silva}, {Bressan}  \&
  {Danese}}{{Granato} et~al.}{2004}]{Granato2004}
{Granato} G.~L.,  {De Zotti} G.,  {Silva} L.,  {Bressan} A.,   {Danese} L.,
  2004, \mn@doi [\apj] {10.1086/379875}, \href
  {https://ui.adsabs.harvard.edu/abs/2004ApJ...600..580G} {600, 580}

\bibitem[\protect\citeauthoryear{{Grogin} et~al.,}{{Grogin}
  et~al.}{2011}]{Grogin2011}
{Grogin} N.~A.,  et~al., 2011, \mn@doi [\apjs] {10.1088/0067-0049/197/2/35},
  \href {http://adsabs.harvard.edu/abs/2011ApJS..197...35G} {197, 35}

\bibitem[\protect\citeauthoryear{{Guo} et~al.,}{{Guo} et~al.}{2013}]{Guo2013}
{Guo} Y.,  et~al., 2013, \mn@doi [\apjs] {10.1088/0067-0049/207/2/24}, \href
  {http://adsabs.harvard.edu/abs/2013ApJS..207...24G} {207, 24}

\bibitem[\protect\citeauthoryear{{Guo} et~al.,}{{Guo} et~al.}{2015a}]{Guo2015}
{Guo} Y.,  et~al., 2015a, \mn@doi [\apj] {10.1088/0004-637X/800/1/39}, \href
  {http://adsabs.harvard.edu/abs/2015ApJ...800...39G} {800, 39}

\bibitem[\protect\citeauthoryear{Guo, Zheng, Wang  \& Fu}{Guo
  et~al.}{2015b}]{Guo_2015}
Guo K.,  Zheng X.~Z.,  Wang T.,   Fu H.,  2015b, \mn@doi [The Astrophysical
  Journal] {10.1088/2041-8205/808/2/l49}, 808, L49

\bibitem[\protect\citeauthoryear{{Harrison}, {Costa}, {Tadhunter},
  {Fl{\"u}tsch}, {Kakkad}, {Perna}  \& {Vietri}}{{Harrison}
  et~al.}{2018}]{Harrison2018}
{Harrison} C.~M.,  {Costa} T.,  {Tadhunter} C.~N.,  {Fl{\"u}tsch} A.,  {Kakkad}
  D.,  {Perna} M.,   {Vietri} G.,  2018, \mn@doi [Nature Astronomy]
  {10.1038/s41550-018-0403-6}, \href
  {https://ui.adsabs.harvard.edu/abs/2018NatAs...2..198H} {2, 198}

\bibitem[\protect\citeauthoryear{{H{\"a}u{\ss}ler} et~al.,}{{H{\"a}u{\ss}ler}
  et~al.}{2013}]{haeussler2013}
{H{\"a}u{\ss}ler} B.,  et~al., 2013, \mn@doi [\mnras] {10.1093/mnras/sts633},
  \href {http://adsabs.harvard.edu/abs/2013MNRAS.430..330H} {430, 330}

\bibitem[\protect\citeauthoryear{{Hopkins}, {Somerville}, {Hernquist}, {Cox},
  {Robertson}  \& {Li}}{{Hopkins} et~al.}{2006}]{Hopkins2006}
{Hopkins} P.~F.,  {Somerville} R.~S.,  {Hernquist} L.,  {Cox} T.~J.,
  {Robertson} B.,   {Li} Y.,  2006, \mn@doi [\apj] {10.1086/508503}, \href
  {https://ui.adsabs.harvard.edu/abs/2006ApJ...652..864H} {652, 864}

\bibitem[\protect\citeauthoryear{{Hopkins}, {Bundy}, {Murray}, {Quataert},
  {Lauer}  \& {Ma}}{{Hopkins} et~al.}{2009}]{Hopkins2009}
{Hopkins} P.~F.,  {Bundy} K.,  {Murray} N.,  {Quataert} E.,  {Lauer} T.~R.,
  {Ma} C.-P.,  2009, \mn@doi [\mnras] {10.1111/j.1365-2966.2009.15062.x}, \href
  {https://ui.adsabs.harvard.edu/abs/2009MNRAS.398..898H} {398, 898}

\bibitem[\protect\citeauthoryear{{Hopkins} et~al.,}{{Hopkins}
  et~al.}{2010}]{Hopkins2010}
{Hopkins} P.~F.,  et~al., 2010, \mn@doi [\apj] {10.1088/0004-637X/715/1/202},
  \href {http://adsabs.harvard.edu/abs/2010ApJ...715..202H} {715, 202}

\bibitem[\protect\citeauthoryear{{Hopkins}, {Kocevski}  \& {Bundy}}{{Hopkins}
  et~al.}{2014}]{Hopkins2014}
{Hopkins} P.~F.,  {Kocevski} D.~D.,   {Bundy} K.,  2014, \mn@doi [\mnras]
  {10.1093/mnras/stu1736}, \href
  {https://ui.adsabs.harvard.edu/abs/2014MNRAS.445..823H} {445, 823}

\bibitem[\protect\citeauthoryear{{Huertas-Company} et~al.,}{{Huertas-Company}
  et~al.}{2015}]{Huertas2015}
{Huertas-Company} M.,  et~al., 2015, \mn@doi [\apjs]
  {10.1088/0067-0049/221/1/8}, \href
  {http://adsabs.harvard.edu/abs/2015ApJS..221....8H} {221, 8}

\bibitem[\protect\citeauthoryear{{Huertas-Company} et~al.,}{{Huertas-Company}
  et~al.}{2016}]{Huertas2016}
{Huertas-Company} M.,  et~al., 2016, \mn@doi [\mnras] {10.1093/mnras/stw1866},
  \href {http://adsabs.harvard.edu/abs/2016MNRAS.462.4495H} {462, 4495}

\bibitem[\protect\citeauthoryear{{Huertas-Company} et~al.,}{{Huertas-Company}
  et~al.}{2020}]{Huertas2020}
{Huertas-Company} M.,  et~al., 2020, \mn@doi [\mnras] {10.1093/mnras/staa2777},
  \href {https://ui.adsabs.harvard.edu/abs/2020MNRAS.499..814H} {499, 814}

\bibitem[\protect\citeauthoryear{{Ilbert} et~al.,}{{Ilbert}
  et~al.}{2013}]{Ilbert2013}
{Ilbert} O.,  et~al., 2013, \mn@doi [\aap] {10.1051/0004-6361/201321100}, \href
  {http://adsabs.harvard.edu/abs/2013A%26A...556A..55I} {556, A55}

\bibitem[\protect\citeauthoryear{{Johnston} et~al.,}{{Johnston}
  et~al.}{2017}]{Johnston2017}
{Johnston} E.~J.,  et~al., 2017, \mn@doi [\mnras] {10.1093/mnras/stw2823},
  \href {https://ui.adsabs.harvard.edu/abs/2017MNRAS.465.2317J} {465, 2317}

\bibitem[\protect\citeauthoryear{{Kannappan}, {Guie}  \& {Baker}}{{Kannappan}
  et~al.}{2009}]{Kannappan2009}
{Kannappan} S.~J.,  {Guie} J.~M.,   {Baker} A.~J.,  2009, \mn@doi [\aj]
  {10.1088/0004-6256/138/2/579}, \href
  {https://ui.adsabs.harvard.edu/abs/2009AJ....138..579K} {138, 579}

\bibitem[\protect\citeauthoryear{{Kauffmann} et~al.,}{{Kauffmann}
  et~al.}{2003}]{Kauffmann2003}
{Kauffmann} G.,  et~al., 2003, \mn@doi [\mnras]
  {10.1046/j.1365-8711.2003.06292.x}, \href
  {https://ui.adsabs.harvard.edu/abs/2003MNRAS.341...54K} {341, 54}

\bibitem[\protect\citeauthoryear{Kennedy, Bamford, Häußler, Brough, Holwerda,
  Hopkins, Vika  \& Vulcani}{Kennedy et~al.}{2016}]{Kennedy2016}
Kennedy R.,  Bamford S.~P.,  Häußler B.,  Brough S.,  Holwerda B.,  Hopkins
  A.~M.,  Vika M.,   Vulcani B.,  2016, \mn@doi [Astronomy & Astrophysics]
  {10.1051/0004-6361/201628715}, 593, A84

\bibitem[\protect\citeauthoryear{{Kennicutt}}{{Kennicutt}}{1998}]{Kennicutt1998}
{Kennicutt} Robert~C. J.,  1998, \mn@doi [\araa]
  {10.1146/annurev.astro.36.1.189}, \href
  {https://ui.adsabs.harvard.edu/abs/1998ARA&A..36..189K} {36, 189}

\bibitem[\protect\citeauthoryear{{Koekemoer} et~al.,}{{Koekemoer}
  et~al.}{2011}]{Koekemoer2011}
{Koekemoer} A.~M.,  et~al., 2011, \mn@doi [\apjs] {10.1088/0067-0049/197/2/36},
  \href {http://adsabs.harvard.edu/abs/2011ApJS..197...36K} {197, 36}

\bibitem[\protect\citeauthoryear{{Kriek}, {van Dokkum}, {Labb{\'e}}, {Franx},
  {Illingworth}, {Marchesini}  \& {Quadri}}{{Kriek} et~al.}{2009}]{Kriek2009}
{Kriek} M.,  {van Dokkum} P.~G.,  {Labb{\'e}} I.,  {Franx} M.,  {Illingworth}
  G.~D.,  {Marchesini} D.,   {Quadri} R.~F.,  2009, \mn@doi [\apj]
  {10.1088/0004-637X/700/1/221}, \href
  {http://adsabs.harvard.edu/abs/2009ApJ...700..221K} {700, 221}

\bibitem[\protect\citeauthoryear{{Labb{\'e}} et~al.,}{{Labb{\'e}}
  et~al.}{2005}]{Labbe2005}
{Labb{\'e}} I.,  et~al., 2005, \mn@doi [\apjl] {10.1086/430700}, \href
  {https://ui.adsabs.harvard.edu/abs/2005ApJ...624L..81L} {624, L81}

\bibitem[\protect\citeauthoryear{{Lang} et~al.,}{{Lang}
  et~al.}{2014}]{Lang2014}
{Lang} P.,  et~al., 2014, \mn@doi [\apj] {10.1088/0004-637X/788/1/11}, \href
  {http://adsabs.harvard.edu/abs/2014ApJ...788...11L} {788, 11}

\bibitem[\protect\citeauthoryear{{Lange} et~al.,}{{Lange}
  et~al.}{2016}]{Lange2016}
{Lange} R.,  et~al., 2016, \mn@doi [\mnras] {10.1093/mnras/stw1495}, \href
  {http://adsabs.harvard.edu/abs/2016MNRAS.462.1470L} {462, 1470}

\bibitem[\protect\citeauthoryear{{Lapi}, {Shankar}, {Mao}, {Granato}, {Silva},
  {De Zotti}  \& {Danese}}{{Lapi} et~al.}{2006}]{Lapi2006}
{Lapi} A.,  {Shankar} F.,  {Mao} J.,  {Granato} G.~L.,  {Silva} L.,  {De Zotti}
  G.,   {Danese} L.,  2006, \mn@doi [\apj] {10.1086/507122}, \href
  {https://ui.adsabs.harvard.edu/abs/2006ApJ...650...42L} {650, 42}

\bibitem[\protect\citeauthoryear{{Lapi} et~al.,}{{Lapi}
  et~al.}{2018}]{Lapi2018}
{Lapi} A.,  et~al., 2018, \mn@doi [\apj] {10.3847/1538-4357/aab6af}, \href
  {https://ui.adsabs.harvard.edu/abs/2018ApJ...857...22L} {857, 22}

\bibitem[\protect\citeauthoryear{{Lee} et~al.,}{{Lee} et~al.}{2015}]{Lee2015}
{Lee} N.,  et~al., 2015, \mn@doi [\apj] {10.1088/0004-637X/801/2/80}, \href
  {https://ui.adsabs.harvard.edu/abs/2015ApJ...801...80L} {801, 80}

\bibitem[\protect\citeauthoryear{Li, Napolitano, Roy, Tortora, Barbera,
  Sonnenfeld, Qiu  \& Liu}{Li et~al.}{2021}]{Li2021}
Li R.,  Napolitano N.~R.,  Roy N.,  Tortora C.,  Barbera F.~L.,  Sonnenfeld A.,
   Qiu C.,   Liu S.,  2021, GAlaxy Light profile convolutional neural NETworks
  (GaLNets). I. fast and accurate structural parameters for billion galaxy
  samples (\mn@eprint {arXiv} {2111.05434})

\bibitem[\protect\citeauthoryear{{Lilly}, {Carollo}, {Pipino}, {Renzini}  \&
  {Peng}}{{Lilly} et~al.}{2013}]{Lilly2013}
{Lilly} S.~J.,  {Carollo} C.~M.,  {Pipino} A.,  {Renzini} A.,   {Peng} Y.,
  2013, \mn@doi [\apj] {10.1088/0004-637X/772/2/119}, \href
  {http://adsabs.harvard.edu/abs/2013ApJ...772..119L} {772, 119}

\bibitem[\protect\citeauthoryear{{Lopes}, {Rembold}, {Ribeiro}, {Nascimento}
  \& {Vajgel}}{{Lopes} et~al.}{2016}]{Lopes2016}
{Lopes} P.~A.~A.,  {Rembold} S.~B.,  {Ribeiro} A.~L.~B.,  {Nascimento} R.~S.,
  {Vajgel} B.,  2016, \mn@doi [\mnras] {10.1093/mnras/stw1497}, \href
  {http://adsabs.harvard.edu/abs/2016MNRAS.461.2559L} {461, 2559}

\bibitem[\protect\citeauthoryear{{Magdis} et~al.,}{{Magdis}
  et~al.}{2012}]{Magdis2012}
{Magdis} G.~E.,  et~al., 2012, \mn@doi [\apjl] {10.1088/2041-8205/758/1/L9},
  \href {https://ui.adsabs.harvard.edu/abs/2012ApJ...758L...9M} {758, L9}

\bibitem[\protect\citeauthoryear{{Mancini}, {Renzini}, {Daddi}, {Rodighiero},
  {Berta}, {Grogin}, {Kocevski}  \& {Koekemoer}}{{Mancini}
  et~al.}{2015}]{Mancini2015}
{Mancini} C.,  {Renzini} A.,  {Daddi} E.,  {Rodighiero} G.,  {Berta} S.,
  {Grogin} N.,  {Kocevski} D.,   {Koekemoer} A.,  2015, \mn@doi [\mnras]
  {10.1093/mnras/stv608}, \href
  {http://adsabs.harvard.edu/abs/2015MNRAS.450..763M} {450, 763}

\bibitem[\protect\citeauthoryear{{Mancini} et~al.,}{{Mancini}
  et~al.}{2019}]{Mancini2019}
{Mancini} C.,  et~al., 2019, \mn@doi [\mnras] {10.1093/mnras/stz2130}, \href
  {https://ui.adsabs.harvard.edu/abs/2019MNRAS.489.1265M} {489, 1265}

\bibitem[\protect\citeauthoryear{{Marsden}, {Shankar}, {Bernardi}, {Sheth},
  {Fu}  \& {Lapi}}{{Marsden} et~al.}{2022}]{Marsden2022}
{Marsden} C.,  {Shankar} F.,  {Bernardi} M.,  {Sheth} R.~K.,  {Fu} H.,   {Lapi}
  A.,  2022, \mn@doi [\mnras] {10.1093/mnras/stab3705}, \href
  {https://ui.adsabs.harvard.edu/abs/2022MNRAS.510.5639M} {510, 5639}

\bibitem[\protect\citeauthoryear{{Martig}, {Bournaud}, {Teyssier}  \&
  {Dekel}}{{Martig} et~al.}{2009}]{Martig2009}
{Martig} M.,  {Bournaud} F.,  {Teyssier} R.,   {Dekel} A.,  2009, \mn@doi
  [\apj] {10.1088/0004-637X/707/1/250}, \href
  {http://adsabs.harvard.edu/abs/2009ApJ...707..250M} {707, 250}

\bibitem[\protect\citeauthoryear{{Mart{\'\i}n-Navarro}, {Shankar}  \&
  {Mezcua}}{{Mart{\'\i}n-Navarro} et~al.}{2021}]{Martin2021}
{Mart{\'\i}n-Navarro} I.,  {Shankar} F.,   {Mezcua} M.,  2021, \mn@doi [\mnras]
  {10.1093/mnrasl/slab112}, \href
  {https://ui.adsabs.harvard.edu/abs/2021MNRAS.tmpL..99M} {}

\bibitem[\protect\citeauthoryear{{Masters} et~al.,}{{Masters}
  et~al.}{2010}]{Masters2010}
{Masters} K.~L.,  et~al., 2010, \mn@doi [\mnras]
  {10.1111/j.1365-2966.2010.16503.x}, \href
  {https://ui.adsabs.harvard.edu/abs/2010MNRAS.405..783M} {405, 783}

\bibitem[\protect\citeauthoryear{{M{\'e}ndez-Abreu}, {de Lorenzo-C{\'a}ceres}
  \& {S{\'a}nchez}}{{M{\'e}ndez-Abreu} et~al.}{2021}]{Mendez2021}
{M{\'e}ndez-Abreu} J.,  {de Lorenzo-C{\'a}ceres} A.,   {S{\'a}nchez} S.~F.,
  2021, \mn@doi [\mnras] {10.1093/mnras/stab1064}, \href
  {https://ui.adsabs.harvard.edu/abs/2021MNRAS.504.3058M} {504, 3058}

\bibitem[\protect\citeauthoryear{{Morselli}, {Popesso}, {Erfanianfar}  \&
  {Concas}}{{Morselli} et~al.}{2017}]{Morselli2017}
{Morselli} L.,  {Popesso} P.,  {Erfanianfar} G.,   {Concas} A.,  2017, \mn@doi
  [\aap] {10.1051/0004-6361/201629409}, \href
  {http://adsabs.harvard.edu/abs/2017A%26A...597A..97M} {597, A97}

\bibitem[\protect\citeauthoryear{{Morselli}, {Popesso}, {Cibinel}, {Oesch},
  {Montes}, {Atek}, {Illingworth}  \& {Holden}}{{Morselli}
  et~al.}{2019}]{Morselli2019}
{Morselli} L.,  {Popesso} P.,  {Cibinel} A.,  {Oesch} P.~A.,  {Montes} M.,
  {Atek} H.,  {Illingworth} G.~D.,   {Holden} B.,  2019, \mn@doi [\aap]
  {10.1051/0004-6361/201834559}, \href
  {https://ui.adsabs.harvard.edu/abs/2019A&A...626A..61M} {626, A61}

\bibitem[\protect\citeauthoryear{{Muzzin} et~al.,}{{Muzzin}
  et~al.}{2013}]{Muzzin2013}
{Muzzin} A.,  et~al., 2013, \mn@doi [\apj] {10.1088/0004-637X/777/1/18}, \href
  {http://adsabs.harvard.edu/abs/2013ApJ...777...18M} {777, 18}

\bibitem[\protect\citeauthoryear{{Nipoti}}{{Nipoti}}{2012}]{Nipoti2012}
{Nipoti} C.,  2012, in {Capuzzo-Dolcetta} R.,  {Limongi} M.,   {Tornamb{\`e}}
  A.,  eds,  Astronomical Society of the Pacific Conference Series Vol. 453,
  Advances in Computational Astrophysics: Methods, Tools, and Outcome. p.~233
  (\mn@eprint {arXiv} {1109.1669})

\bibitem[\protect\citeauthoryear{Pan, Li, Lin, Wang  \& Kong}{Pan
  et~al.}{2014}]{Pan2014}
Pan Z.,  Li J.,  Lin W.,  Wang J.,   Kong X.,  2014, \mn@doi [The Astrophysical
  Journal] {10.1088/2041-8205/792/1/l4}, 792, L4

\bibitem[\protect\citeauthoryear{Pan, Li, Lin, Wang, Fan  \& Kong}{Pan
  et~al.}{2015}]{Pan2015}
Pan Z.,  Li J.,  Lin W.,  Wang J.,  Fan L.,   Kong X.,  2015, \mn@doi [The
  Astrophysical Journal] {10.1088/2041-8205/804/2/l42}, 804, L42

\bibitem[\protect\citeauthoryear{{Peng} \& {Maiolino}}{{Peng} \&
  {Maiolino}}{2014}]{PengM2014}
{Peng} Y.-j.,  {Maiolino} R.,  2014, \mn@doi [\mnras] {10.1093/mnras/stu1288},
  \href {https://ui.adsabs.harvard.edu/abs/2014MNRAS.443.3643P} {443, 3643}

\bibitem[\protect\citeauthoryear{{Peng} \& {Renzini}}{{Peng} \&
  {Renzini}}{2020}]{Peng2020}
{Peng} Y.-j.,  {Renzini} A.,  2020, \mn@doi [\mnras] {10.1093/mnrasl/slz163},
  \href {https://ui.adsabs.harvard.edu/abs/2020MNRAS.491L..51P} {491, L51}

\bibitem[\protect\citeauthoryear{{Peng}, {Ho}, {Impey}  \& {Rix}}{{Peng}
  et~al.}{2002}]{Peng2002}
{Peng} C.~Y.,  {Ho} L.~C.,  {Impey} C.~D.,   {Rix} H.-W.,  2002, \mn@doi [\aj]
  {10.1086/340952}, \href {http://adsabs.harvard.edu/abs/2002AJ....124..266P}
  {124, 266}

\bibitem[\protect\citeauthoryear{{Peng} et~al.,}{{Peng}
  et~al.}{2010}]{Peng2010}
{Peng} Y.-j.,  et~al., 2010, \mn@doi [\apj] {10.1088/0004-637X/721/1/193},
  \href {http://adsabs.harvard.edu/abs/2010ApJ...721..193P} {721, 193}

\bibitem[\protect\citeauthoryear{{Peng}, {Maiolino}  \& {Cochrane}}{{Peng}
  et~al.}{2015}]{Peng2015}
{Peng} Y.,  {Maiolino} R.,   {Cochrane} R.,  2015, \mn@doi [\nat]
  {10.1038/nature14439}, \href
  {http://adsabs.harvard.edu/abs/2015Natur.521..192P} {521, 192}

\bibitem[\protect\citeauthoryear{{P{\'e}rez} et~al.,}{{P{\'e}rez}
  et~al.}{2013}]{Perez2013}
{P{\'e}rez} E.,  et~al., 2013, \mn@doi [\apjl] {10.1088/2041-8205/764/1/L1},
  \href {https://ui.adsabs.harvard.edu/abs/2013ApJ...764L...1P} {764}

\bibitem[\protect\citeauthoryear{Pillepich et~al.,}{Pillepich
  et~al.}{2017}]{Pillepich2017}
Pillepich A.,  et~al., 2017, \mn@doi [Monthly Notices of the Royal Astronomical
  Society] {10.1093/mnras/stx2656}, 473, 4077–4106

\bibitem[\protect\citeauthoryear{{Pipino}, {Lilly}  \& {Carollo}}{{Pipino}
  et~al.}{2014}]{Pipino2014}
{Pipino} A.,  {Lilly} S.~J.,   {Carollo} C.~M.,  2014, \mn@doi [\mnras]
  {10.1093/mnras/stu579}, \href
  {https://ui.adsabs.harvard.edu/abs/2014MNRAS.441.1444P} {441, 1444}

\bibitem[\protect\citeauthoryear{{Popesso} et~al.,}{{Popesso}
  et~al.}{2019}]{Popesso2019}
{Popesso} P.,  et~al., 2019, \mn@doi [\mnras] {10.1093/mnras/sty3210}, \href
  {https://ui.adsabs.harvard.edu/abs/2019MNRAS.483.3213P} {483, 3213}

\bibitem[\protect\citeauthoryear{{Puglisi} et~al.,}{{Puglisi}
  et~al.}{2019}]{Puglisi2019}
{Puglisi} A.,  et~al., 2019, \mn@doi [\apjl] {10.3847/2041-8213/ab1f92}, \href
  {https://ui.adsabs.harvard.edu/abs/2019ApJ...877L..23P} {877, L23}

\bibitem[\protect\citeauthoryear{{Puglisi} et~al.,}{{Puglisi}
  et~al.}{2021}]{Puglisi2021}
{Puglisi} A.,  et~al., 2021, \mn@doi [\mnras] {10.1093/mnras/stab2914}, \href
  {https://ui.adsabs.harvard.edu/abs/2021MNRAS.508.5217P} {508, 5217}

\bibitem[\protect\citeauthoryear{Pérez-González, Gil~de Paz, Zamorano,
  Gallego, Alonso-Herrero  \& Aragón-Salamanca}{Pérez-González
  et~al.}{2003}]{Perez2003}
Pérez-González P.~G.,  Gil~de Paz A.,  Zamorano J.,  Gallego J.,
  Alonso-Herrero A.,   Aragón-Salamanca A.,  2003, \mn@doi [Monthly Notices of
  the Royal Astronomical Society] {10.1046/j.1365-8711.2003.06077.x}, 338, 508

\bibitem[\protect\citeauthoryear{{Renzini} \& {Peng}}{{Renzini} \&
  {Peng}}{2015}]{Renzini2015}
{Renzini} A.,  {Peng} Y.-j.,  2015, \mn@doi [\apjl]
  {10.1088/2041-8205/801/2/L29}, \href
  {https://ui.adsabs.harvard.edu/abs/2015ApJ...801L..29R} {801, L29}

\bibitem[\protect\citeauthoryear{{Rodr{\'{\i}}guez-Puebla}, {Primack},
  {Avila-Reese}  \& {Faber}}{{Rodr{\'{\i}}guez-Puebla}
  et~al.}{2017}]{Rodriguez2017}
{Rodr{\'{\i}}guez-Puebla} A.,  {Primack} J.~R.,  {Avila-Reese} V.,   {Faber}
  S.~M.,  2017, \mn@doi [\mnras] {10.1093/mnras/stx1172}, \href
  {http://adsabs.harvard.edu/abs/2017MNRAS.470..651R} {470, 651}

\bibitem[\protect\citeauthoryear{{Rowlands} et~al.,}{{Rowlands}
  et~al.}{2012}]{Rowlands2012}
{Rowlands} K.,  et~al., 2012, \mn@doi [\mnras]
  {10.1111/j.1365-2966.2011.19905.x}, \href
  {https://ui.adsabs.harvard.edu/abs/2012MNRAS.419.2545R} {419, 2545}

\bibitem[\protect\citeauthoryear{{Rowlands} et~al.,}{{Rowlands}
  et~al.}{2018}]{Rowlands2018}
{Rowlands} K.,  et~al., 2018, \mn@doi [\mnras] {10.1093/mnras/stx1903}, \href
  {https://ui.adsabs.harvard.edu/abs/2018MNRAS.473.1168R} {473, 1168}

\bibitem[\protect\citeauthoryear{{Salpeter}}{{Salpeter}}{1955}]{Salpeter1955}
{Salpeter} E.~E.,  1955, \mn@doi [\apj] {10.1086/145971}, \href
  {http://adsabs.harvard.edu/abs/1955ApJ...121..161S} {121, 161}

\bibitem[\protect\citeauthoryear{{S{\'a}nchez} et~al.,}{{S{\'a}nchez}
  et~al.}{2012}]{Sanchez2012}
{S{\'a}nchez} S.~F.,  et~al., 2012, \mn@doi [\aap]
  {10.1051/0004-6361/201117353}, \href
  {https://ui.adsabs.harvard.edu/abs/2012A&A...538A...8S} {538, A8}

\bibitem[\protect\citeauthoryear{{Schawinski} et~al.,}{{Schawinski}
  et~al.}{2009}]{Schawinski2009}
{Schawinski} K.,  et~al., 2009, \mn@doi [\mnras]
  {10.1111/j.1365-2966.2009.14793.x}, \href
  {https://ui.adsabs.harvard.edu/abs/2009MNRAS.396..818S} {396, 818}

\bibitem[\protect\citeauthoryear{{Schreiber} et~al.,}{{Schreiber}
  et~al.}{2015}]{Schreiber2015}
{Schreiber} C.,  et~al., 2015, \mn@doi [\aap] {10.1051/0004-6361/201425017},
  \href {https://ui.adsabs.harvard.edu/abs/2015A&A...575A..74S} {575, A74}

\bibitem[\protect\citeauthoryear{{Sersic}}{{Sersic}}{1968}]{Sersic1968}
{Sersic} J.~L.,  1968, {Atlas de Galaxias Australes}

\bibitem[\protect\citeauthoryear{{Shankar}, {Lapi}, {Salucci}, {De Zotti}  \&
  {Danese}}{{Shankar} et~al.}{2006}]{Shankar2006}
{Shankar} F.,  {Lapi} A.,  {Salucci} P.,  {De Zotti} G.,   {Danese} L.,  2006,
  \mn@doi [\apj] {10.1086/502794}, \href
  {https://ui.adsabs.harvard.edu/abs/2006ApJ...643...14S} {643, 14}

\bibitem[\protect\citeauthoryear{{Shankar} et~al.,}{{Shankar}
  et~al.}{2018}]{Shankar2018}
{Shankar} F.,  et~al., 2018, \mn@doi [\mnras] {10.1093/mnras/stx3086}, \href
  {http://adsabs.harvard.edu/abs/2018MNRAS.475.2878S} {475, 2878}

\bibitem[\protect\citeauthoryear{{Sherman} et~al.,}{{Sherman}
  et~al.}{2021}]{Sherman2021}
{Sherman} S.,  et~al., 2021, \mn@doi [\mnras] {10.1093/mnras/stab1350}, \href
  {https://ui.adsabs.harvard.edu/abs/2021MNRAS.505..947S} {505, 947}

\bibitem[\protect\citeauthoryear{{Silk} \& {Rees}}{{Silk} \&
  {Rees}}{1998}]{Silk1998}
{Silk} J.,  {Rees} M.~J.,  1998, \aap, \href
  {https://ui.adsabs.harvard.edu/abs/1998A&A...331L...1S} {331, L1}

\bibitem[\protect\citeauthoryear{{Stefanon} et~al.,}{{Stefanon}
  et~al.}{2017}]{Stefanon2017}
{Stefanon} M.,  et~al., 2017, \mn@doi [\apjs] {10.3847/1538-4365/aa66cb}, \href
  {http://adsabs.harvard.edu/abs/2017ApJS..229...32S} {229, 32}

\bibitem[\protect\citeauthoryear{{Tacchella} et~al.,}{{Tacchella}
  et~al.}{2015}]{Tacchella2015}
{Tacchella} S.,  et~al., 2015, \mn@doi [Science] {10.1126/science.1261094},
  \href {http://adsabs.harvard.edu/abs/2015Sci...348..314T} {348, 314}

\bibitem[\protect\citeauthoryear{{Tacchella}, {Dekel}, {Carollo}, {Ceverino},
  {DeGraf}, {Lapiner}, {Mandelker}  \& {Primack}}{{Tacchella}
  et~al.}{2016}]{Tacchella2016}
{Tacchella} S.,  {Dekel} A.,  {Carollo} C.~M.,  {Ceverino} D.,  {DeGraf} C.,
  {Lapiner} S.,  {Mandelker} N.,   {Primack} J.~R.,  2016, \mn@doi [\mnras]
  {10.1093/mnras/stw303}, \href
  {http://adsabs.harvard.edu/abs/2016MNRAS.458..242T} {458, 242}

\bibitem[\protect\citeauthoryear{{Tacchella} et~al.,}{{Tacchella}
  et~al.}{2018}]{Tacchella2017}
{Tacchella} S.,  et~al., 2018, \mn@doi [\apj] {10.3847/1538-4357/aabf8b}, \href
  {http://adsabs.harvard.edu/abs/2018ApJ...859...56T} {859, 56}

\bibitem[\protect\citeauthoryear{{Tadaki} et~al.,}{{Tadaki}
  et~al.}{2020}]{Tadaki2020}
{Tadaki} K.-i.,  et~al., 2020, \mn@doi [\apj] {10.3847/1538-4357/abaf4a}, \href
  {https://ui.adsabs.harvard.edu/abs/2020ApJ...901...74T} {901, 74}

\bibitem[\protect\citeauthoryear{{Tasca} et~al.,}{{Tasca}
  et~al.}{2015}]{Tasca2015}
{Tasca} L.~A.~M.,  et~al., 2015, \mn@doi [\aap] {10.1051/0004-6361/201425379},
  \href {https://ui.adsabs.harvard.edu/abs/2015A&A...581A..54T} {581, A54}

\bibitem[\protect\citeauthoryear{{Toft} et~al.,}{{Toft}
  et~al.}{2014}]{Toft2014}
{Toft} S.,  et~al., 2014, \mn@doi [\apj] {10.1088/0004-637X/782/2/68}, \href
  {https://ui.adsabs.harvard.edu/abs/2014ApJ...782...68T} {782, 68}

\bibitem[\protect\citeauthoryear{{Toft} et~al.,}{{Toft}
  et~al.}{2017}]{Toft2017}
{Toft} S.,  et~al., 2017, \mn@doi [\nat] {10.1038/nature22388}, \href
  {http://adsabs.harvard.edu/abs/2017Natur.546..510T} {546, 510}

\bibitem[\protect\citeauthoryear{{Tomczak} et~al.,}{{Tomczak}
  et~al.}{2016}]{Tomczak2016}
{Tomczak} A.~R.,  et~al., 2016, \mn@doi [\apj] {10.3847/0004-637X/817/2/118},
  \href {https://ui.adsabs.harvard.edu/abs/2016ApJ...817..118T} {817, 118}

\bibitem[\protect\citeauthoryear{{Toomre}}{{Toomre}}{1977}]{Toomre1977}
{Toomre} A.,  1977, in {Tinsley} B.~M.,  {Larson} D.~Campbell R.~B.~G.,  eds,
  Evolution of Galaxies and Stellar Populations. p.~401

\bibitem[\protect\citeauthoryear{Tuttle \& Tonnesen}{Tuttle \&
  Tonnesen}{2020}]{Tuttle2020}
Tuttle S.~E.,  Tonnesen S.,  2020, \mn@doi [The Astrophysical Journal]
  {10.3847/1538-4357/ab5dbb}, 889, 188

\bibitem[\protect\citeauthoryear{{Varma} et~al.,}{{Varma}
  et~al.}{2021}]{Varma2021}
{Varma} S.,  et~al., 2021, arXiv e-prints, \href
  {https://ui.adsabs.harvard.edu/abs/2021arXiv211011989V} {p. arXiv:2110.11989}

\bibitem[\protect\citeauthoryear{{Vika}, {Bamford}, {H{\"a}u{\ss}ler}, {Rojas},
  {Borch}  \& {Nichol}}{{Vika} et~al.}{2013}]{Vika2013}
{Vika} M.,  {Bamford} S.~P.,  {H{\"a}u{\ss}ler} B.,  {Rojas} A.~L.,  {Borch}
  A.,   {Nichol} R.~C.,  2013, \mn@doi [\mnras] {10.1093/mnras/stt1320}, \href
  {http://adsabs.harvard.edu/abs/2013MNRAS.435..623V} {435, 623}

\bibitem[\protect\citeauthoryear{{Vika}, {Bamford}, {H{\"a}u{\ss}ler}  \&
  {Rojas}}{{Vika} et~al.}{2014}]{Vika2014}
{Vika} M.,  {Bamford} S.~P.,  {H{\"a}u{\ss}ler} B.,   {Rojas} A.~L.,  2014,
  \mn@doi [\mnras] {10.1093/mnras/stu1696}, \href
  {http://adsabs.harvard.edu/abs/2014MNRAS.444.3603V} {444, 3603}

\bibitem[\protect\citeauthoryear{Vulcani, Poggianti, Fritz, Fasano, Moretti,
  Calvi  \& Paccagnella}{Vulcani et~al.}{2014}]{Vulcani2014}
Vulcani B.,  Poggianti B.~M.,  Fritz J.,  Fasano G.,  Moretti A.,  Calvi R.,
  Paccagnella A.,  2014, \mn@doi [The Astrophysical Journal]
  {10.1088/0004-637x/798/1/52}, 798, 52

\bibitem[\protect\citeauthoryear{{Wake}, {van Dokkum}  \& {Franx}}{{Wake}
  et~al.}{2012}]{Wake2012}
{Wake} D.~A.,  {van Dokkum} P.~G.,   {Franx} M.,  2012, \mn@doi [\apjl]
  {10.1088/2041-8205/751/2/L44}, \href
  {https://ui.adsabs.harvard.edu/abs/2012ApJ...751L..44W} {751, L44}

\bibitem[\protect\citeauthoryear{Wang et~al.,}{Wang et~al.}{2017}]{Wang2017}
Wang W.,  et~al., 2017, \mn@doi [Monthly Notices of the Royal Astronomical
  Society] {10.1093/mnras/stx1148}, 469, 4063

\bibitem[\protect\citeauthoryear{{Weinberger} et~al.,}{{Weinberger}
  et~al.}{2017}]{Weinberger2017}
{Weinberger} R.,  et~al., 2017, \mn@doi [\mnras] {10.1093/mnras/stw2944}, \href
  {https://ui.adsabs.harvard.edu/abs/2017MNRAS.465.3291W} {465, 3291}

\bibitem[\protect\citeauthoryear{{Whitaker}, {van Dokkum}, {Brammer}  \&
  {Franx}}{{Whitaker} et~al.}{2012}]{Whitaker2012}
{Whitaker} K.~E.,  {van Dokkum} P.~G.,  {Brammer} G.,   {Franx} M.,  2012,
  \mn@doi [\apjl] {10.1088/2041-8205/754/2/L29}, \href
  {http://adsabs.harvard.edu/abs/2012ApJ...754L..29W} {754, L29}

\bibitem[\protect\citeauthoryear{{Whitaker} et~al.,}{{Whitaker}
  et~al.}{2014}]{Whitaker2014}
{Whitaker} K.~E.,  et~al., 2014, \mn@doi [apj] {10.1088/0004-637X/795/2/104},
  \href {http://adsabs.harvard.edu/abs/2014ApJ...795..104W} {795, 104}

\bibitem[\protect\citeauthoryear{{Whitaker} et~al.,}{{Whitaker}
  et~al.}{2015}]{Whitaker2015}
{Whitaker} K.~E.,  et~al., 2015, \mn@doi [\apjl] {10.1088/2041-8205/811/1/L12},
  \href {http://adsabs.harvard.edu/abs/2015ApJ...811L..12W} {811, L12}

\bibitem[\protect\citeauthoryear{{Whitaker} et~al.,}{{Whitaker}
  et~al.}{2017a}]{Whitaker2017}
{Whitaker} K.~E.,  et~al., 2017a, \mn@doi [\apj] {10.3847/1538-4357/aa6258},
  \href {http://adsabs.harvard.edu/abs/2017ApJ...838...19W} {838, 19}

\bibitem[\protect\citeauthoryear{{Whitaker}, {Pope}, {Cybulski}, {Casey},
  {Popping}  \& {Yun}}{{Whitaker} et~al.}{2017b}]{Whitaker2017b}
{Whitaker} K.~E.,  {Pope} A.,  {Cybulski} R.,  {Casey} C.~M.,  {Popping} G.,
  {Yun} M.~S.,  2017b, \mn@doi [\apj] {10.3847/1538-4357/aa94ce}, \href
  {https://ui.adsabs.harvard.edu/abs/2017ApJ...850..208W} {850, 208}

\bibitem[\protect\citeauthoryear{{Williams}, {Quadri}, {Franx}, {van Dokkum}
  \& {Labb{\'e}}}{{Williams} et~al.}{2009}]{Williams2009}
{Williams} R.~J.,  {Quadri} R.~F.,  {Franx} M.,  {van Dokkum} P.,   {Labb{\'e}}
  I.,  2009, \mn@doi [\apj] {10.1088/0004-637X/691/2/1879}, \href
  {https://ui.adsabs.harvard.edu/abs/2009ApJ...691.1879W} {691, 1879}

\bibitem[\protect\citeauthoryear{{Wisnioski} et~al.,}{{Wisnioski}
  et~al.}{2015}]{Wisnioski2015}
{Wisnioski} E.,  et~al., 2015, \mn@doi [\apj] {10.1088/0004-637X/799/2/209},
  \href {http://adsabs.harvard.edu/abs/2015ApJ...799..209W} {799, 209}

\bibitem[\protect\citeauthoryear{{Woo}, {Carollo}, {Faber}, {Dekel}  \&
  {Tacchella}}{{Woo} et~al.}{2017}]{Woo2017}
{Woo} J.,  {Carollo} C.~M.,  {Faber} S.~M.,  {Dekel} A.,   {Tacchella} S.,
  2017, \mn@doi [\mnras] {10.1093/mnras/stw2403}, \href
  {https://ui.adsabs.harvard.edu/abs/2017MNRAS.464.1077W} {464, 1077}

\bibitem[\protect\citeauthoryear{{Wuyts} et~al.,}{{Wuyts}
  et~al.}{2007}]{Wuyts2007}
{Wuyts} S.,  et~al., 2007, \mn@doi [\apj] {10.1086/509708}, \href
  {https://ui.adsabs.harvard.edu/abs/2007ApJ...655...51W} {655, 51}

\bibitem[\protect\citeauthoryear{{Wuyts} et~al.,}{{Wuyts}
  et~al.}{2011}]{Wuyts2011}
{Wuyts} S.,  et~al., 2011, \mn@doi [apj] {10.1088/0004-637X/742/2/96}, \href
  {http://adsabs.harvard.edu/abs/2011ApJ...742...96W} {742, 96}

\bibitem[\protect\citeauthoryear{{Zolotov} et~al.,}{{Zolotov}
  et~al.}{2015}]{Zolotov2015}
{Zolotov} A.,  et~al., 2015, \mn@doi [\mnras] {10.1093/mnras/stv740}, \href
  {http://adsabs.harvard.edu/abs/2015MNRAS.450.2327Z} {450, 2327}

\bibitem[\protect\citeauthoryear{{de Vaucouleurs}}{{de
  Vaucouleurs}}{1953}]{deVaucouleurs1953}
{de Vaucouleurs} G.,  1953, \mn@doi [\mnras] {10.1093/mnras/113.2.134}, \href
  {https://ui.adsabs.harvard.edu/abs/1953MNRAS.113..134D} {113, 134}

\bibitem[\protect\citeauthoryear{{van Dokkum} et~al.,}{{van Dokkum}
  et~al.}{2015}]{vanDokkum2015}
{van Dokkum} P.~G.,  et~al., 2015, \mn@doi [\apj] {10.1088/0004-637X/813/1/23},
  \href {http://adsabs.harvard.edu/abs/2015ApJ...813...23V} {813, 23}

\bibitem[\protect\citeauthoryear{{van der Wel} et~al.,}{{van der Wel}
  et~al.}{2014}]{vanderWel2014}
{van der Wel} A.,  et~al., 2014, \mn@doi [\apj] {10.1088/0004-637X/788/1/28},
  \href {http://adsabs.harvard.edu/abs/2014ApJ...788...28V} {788, 28}

\makeatother
\end{thebibliography}





\appendix

\section{Standard deviation of \BTM on the main sequence}
\label{apx:SFR_std}
\begin{figure*}
$\begin{array}{cc}
\includegraphics[width=0.3\textwidth]{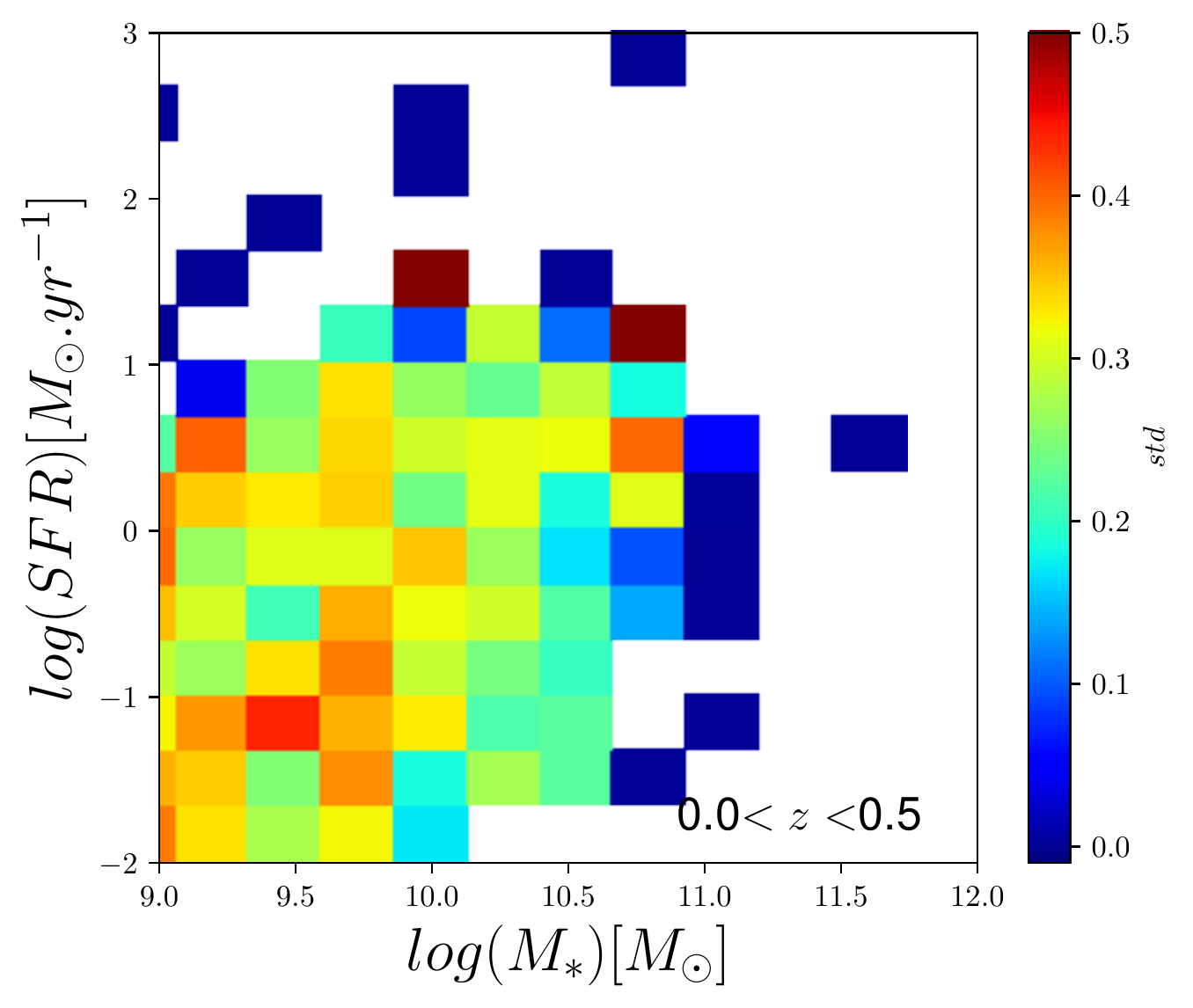}&
\includegraphics[width=0.3\textwidth]{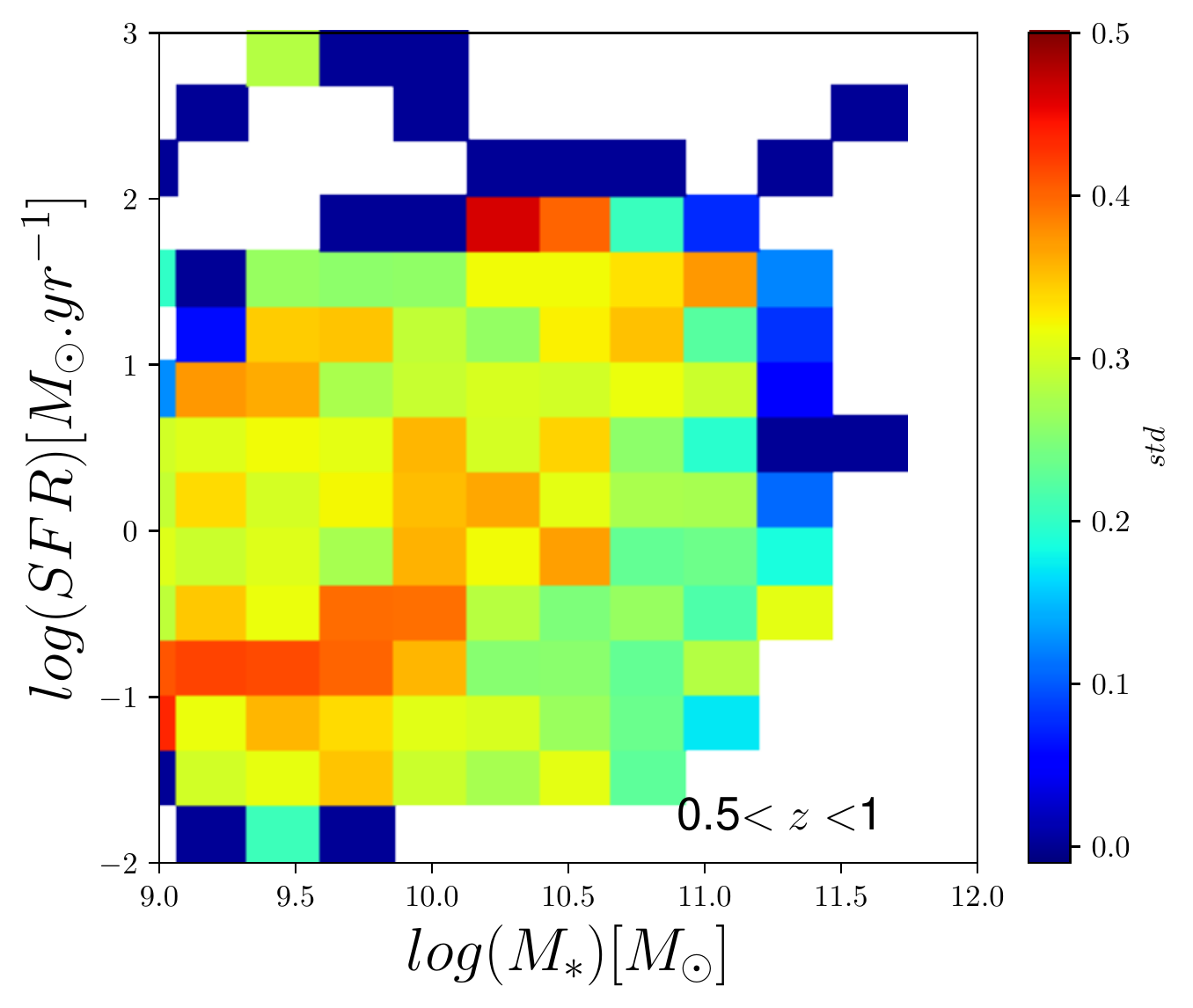}\\
\includegraphics[width=0.3\textwidth]{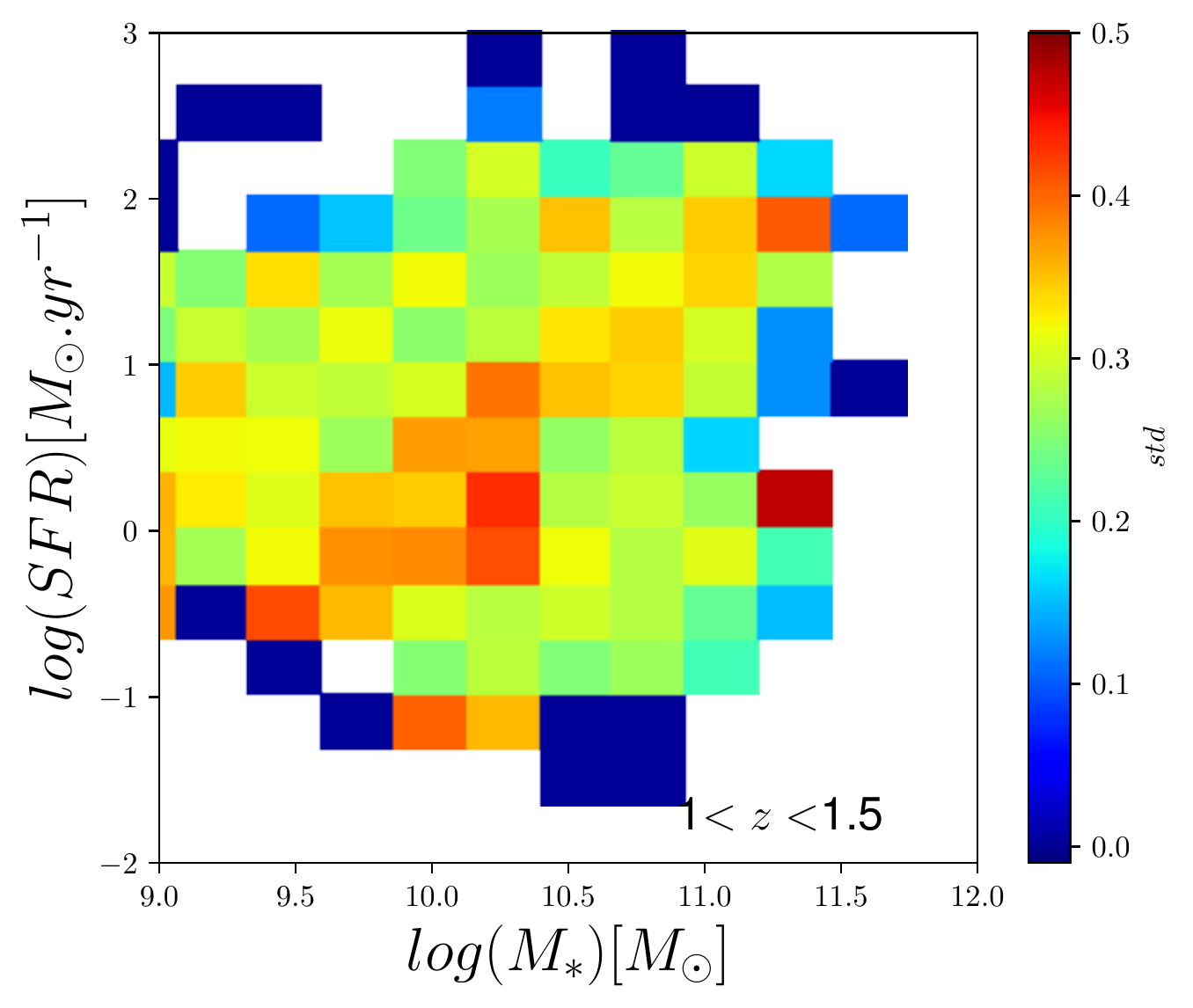}&
\includegraphics[width=0.3\textwidth]{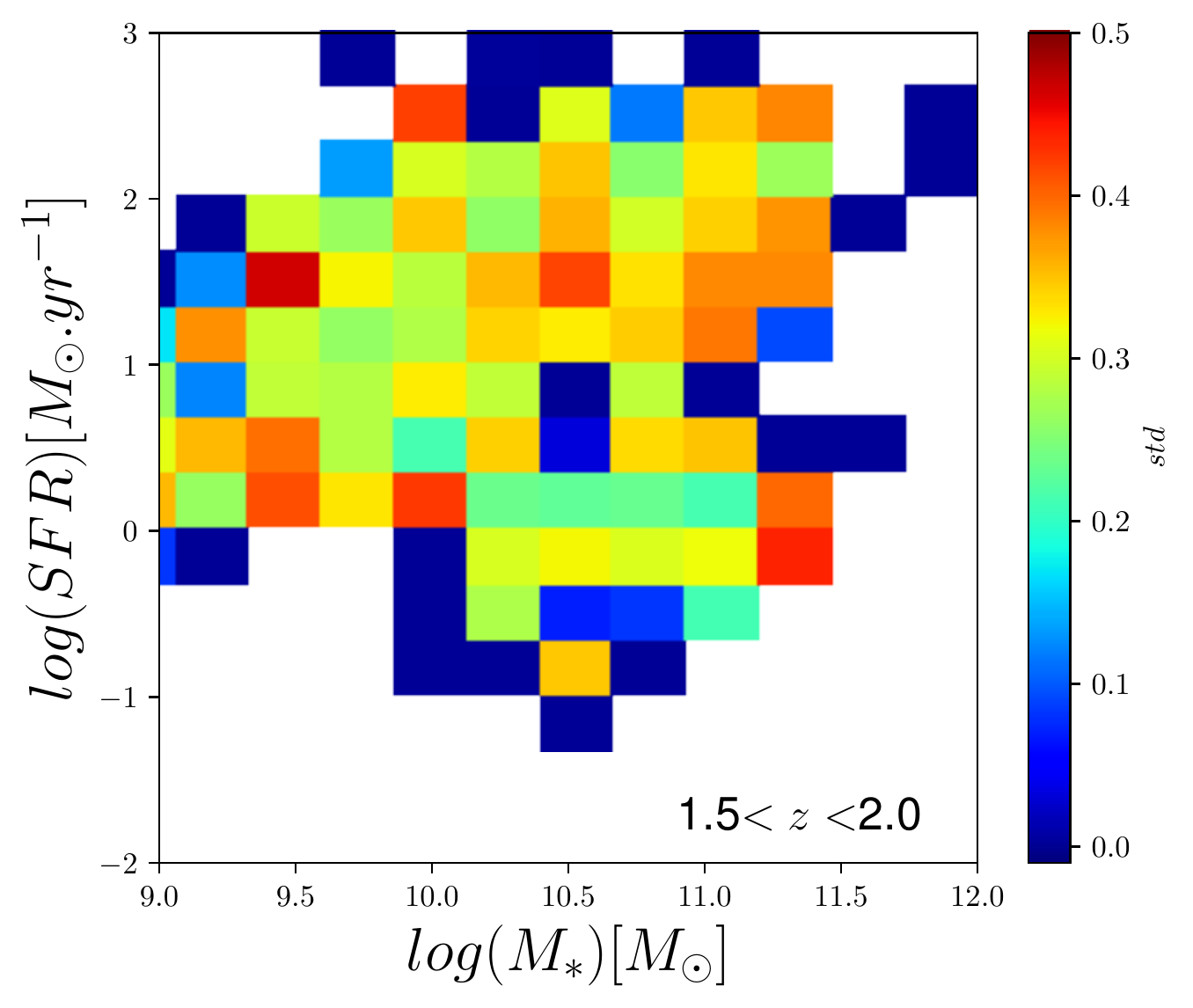}\\
\end{array}$
\caption{Distribution of the standard deviation on median \BTM for SFR-$M_*$ bins.}
\label{fig:SFR_mstar5}
\end{figure*}

\begin{figure*}
$\begin{array}{cc}
\includegraphics[width=0.3\textwidth]{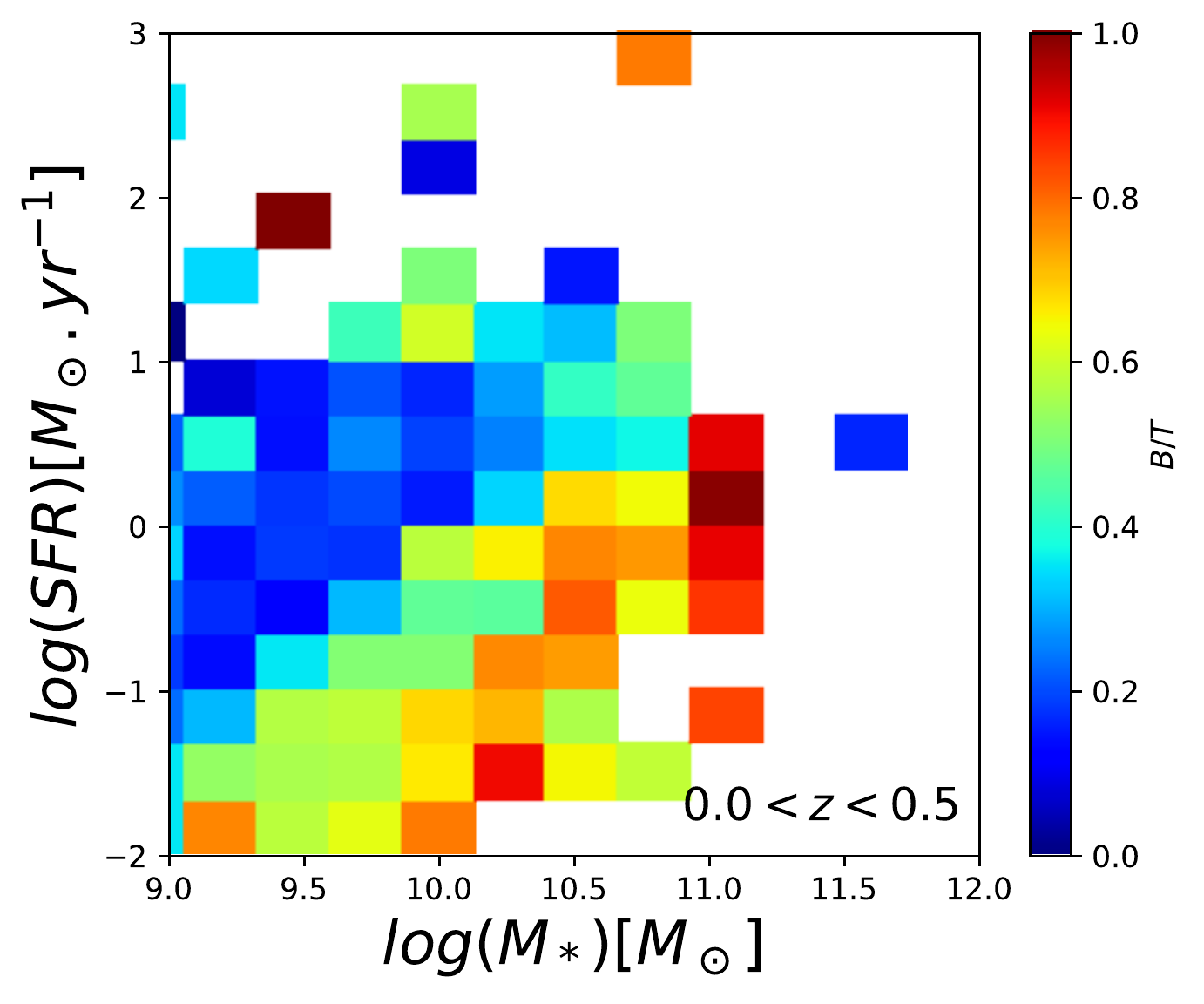}&
\includegraphics[width=0.3\textwidth]{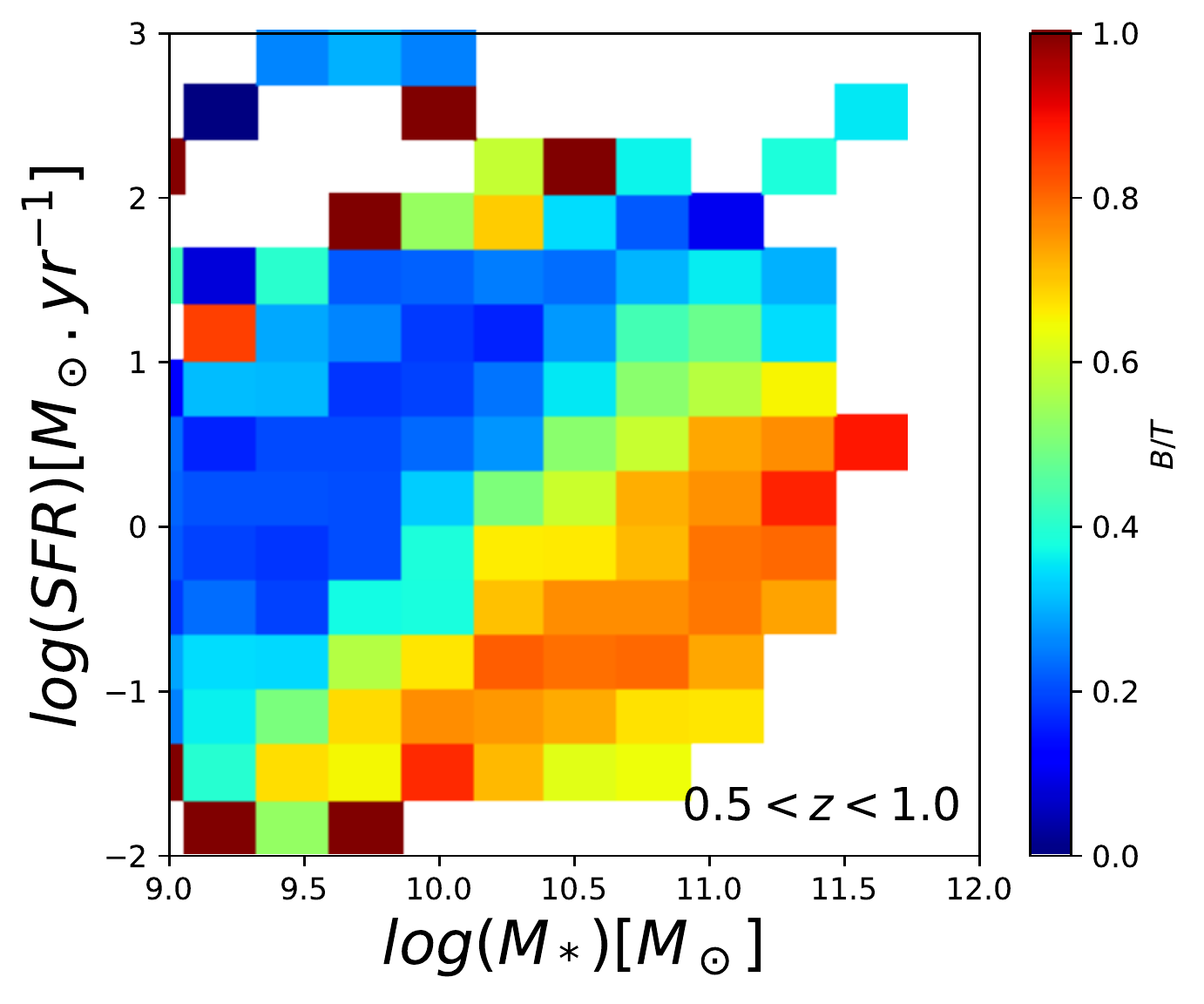}\\
\includegraphics[width=0.3\textwidth]{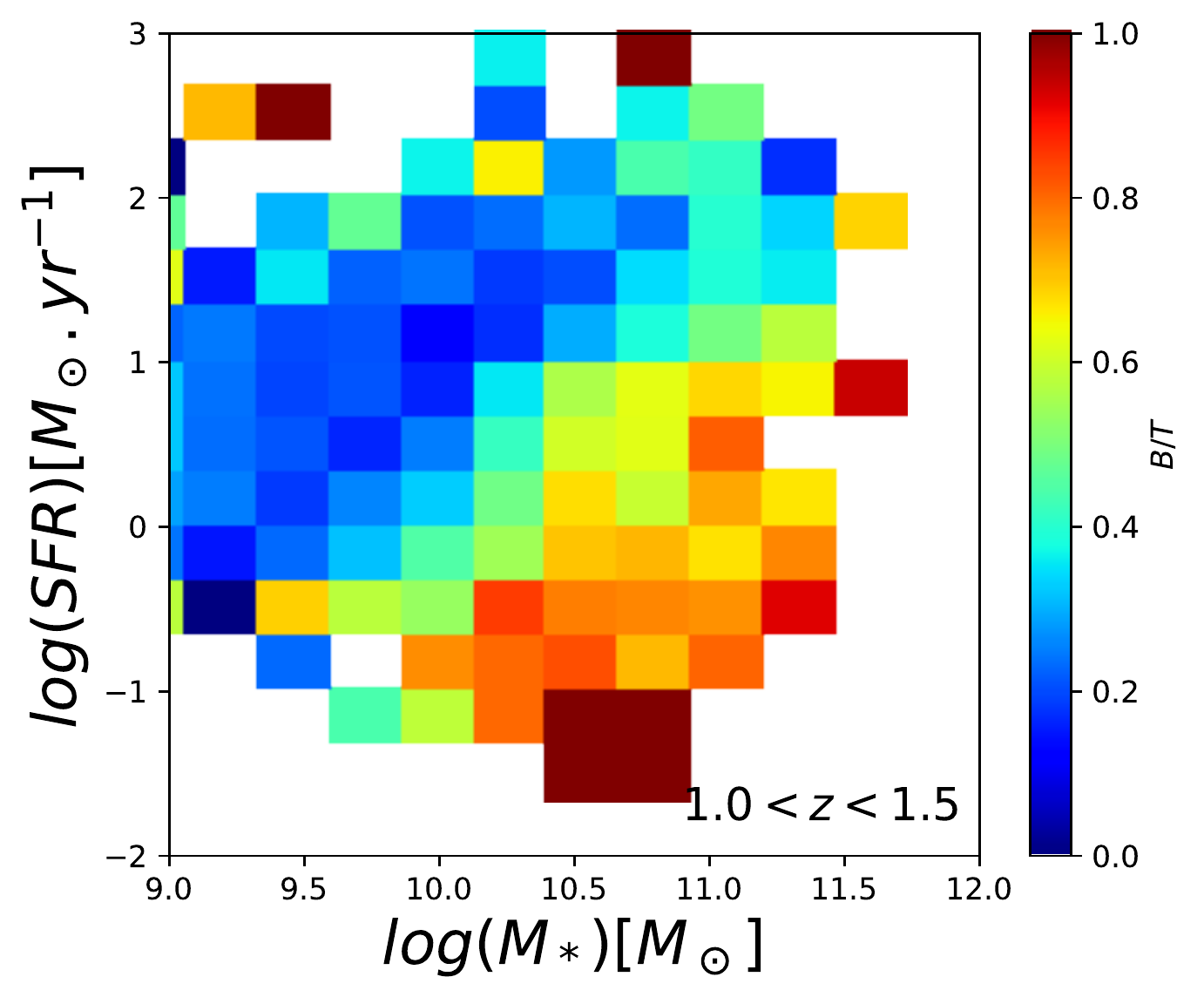}&
\includegraphics[width=0.3\textwidth]{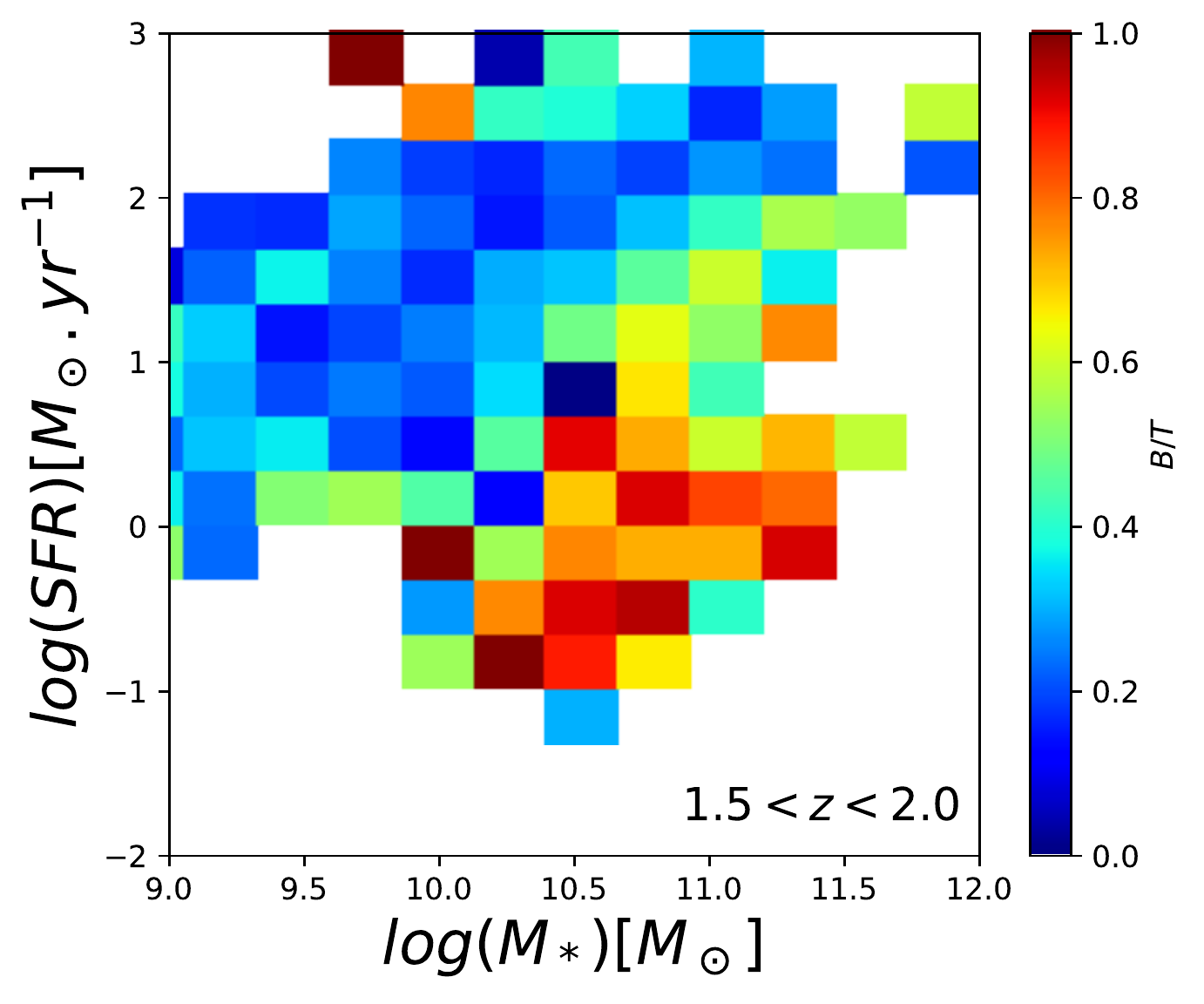}\\
\end{array}$
\caption{Equivalent of Figure 1 but color coded with B/T in the filter F160.}
\label{fig:SFR_mstar6}
\end{figure*}

Figure \ref{fig:SFR_mstar5} show the distribution of \BTM standard deviation per SFR-stellar mass  SFR-M bin. No clear features are present, supporting the robustness of the result. In addition to that, and for completeness of the analysis, Figure \ref{fig:SFR_mstar6} report the equivalent excercise of Figure \ref{fig:SFR_mstar1} but using the bulge-total flux ratio in the F160 filter. Main sequence galaxies have median B/T larger then 0.3 while in the quiescent region is mostly populated by larger values. That result confirms the main trend observed in Figure \ref{fig:SFR_mstar1}.

\section{SFR-$M_*$}
\label{apx:SFR}
\begin{figure*}
\centering
$\begin{array}{c c}
\includegraphics[width=0.3\textwidth]{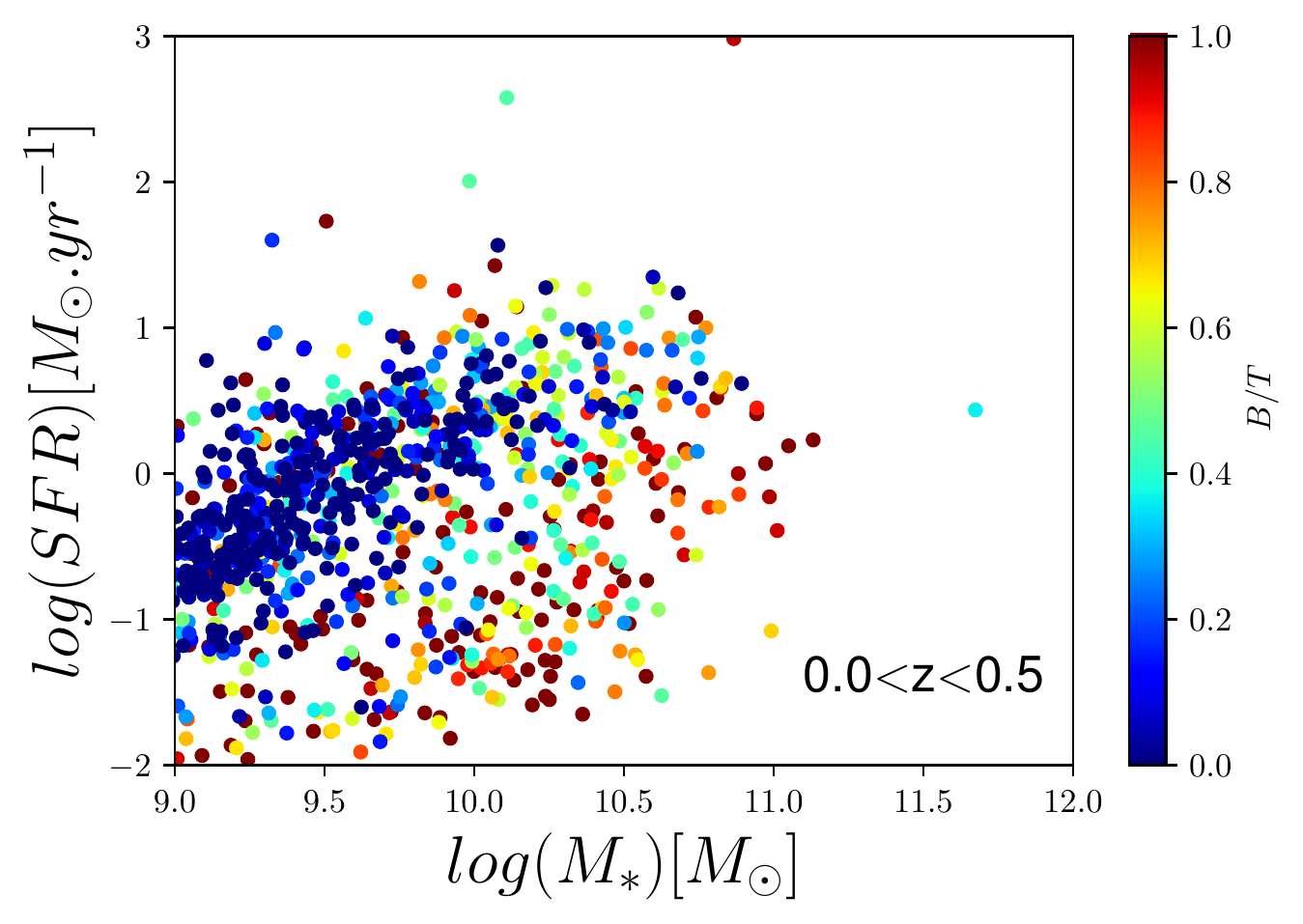}&
\includegraphics[width=0.3\textwidth]{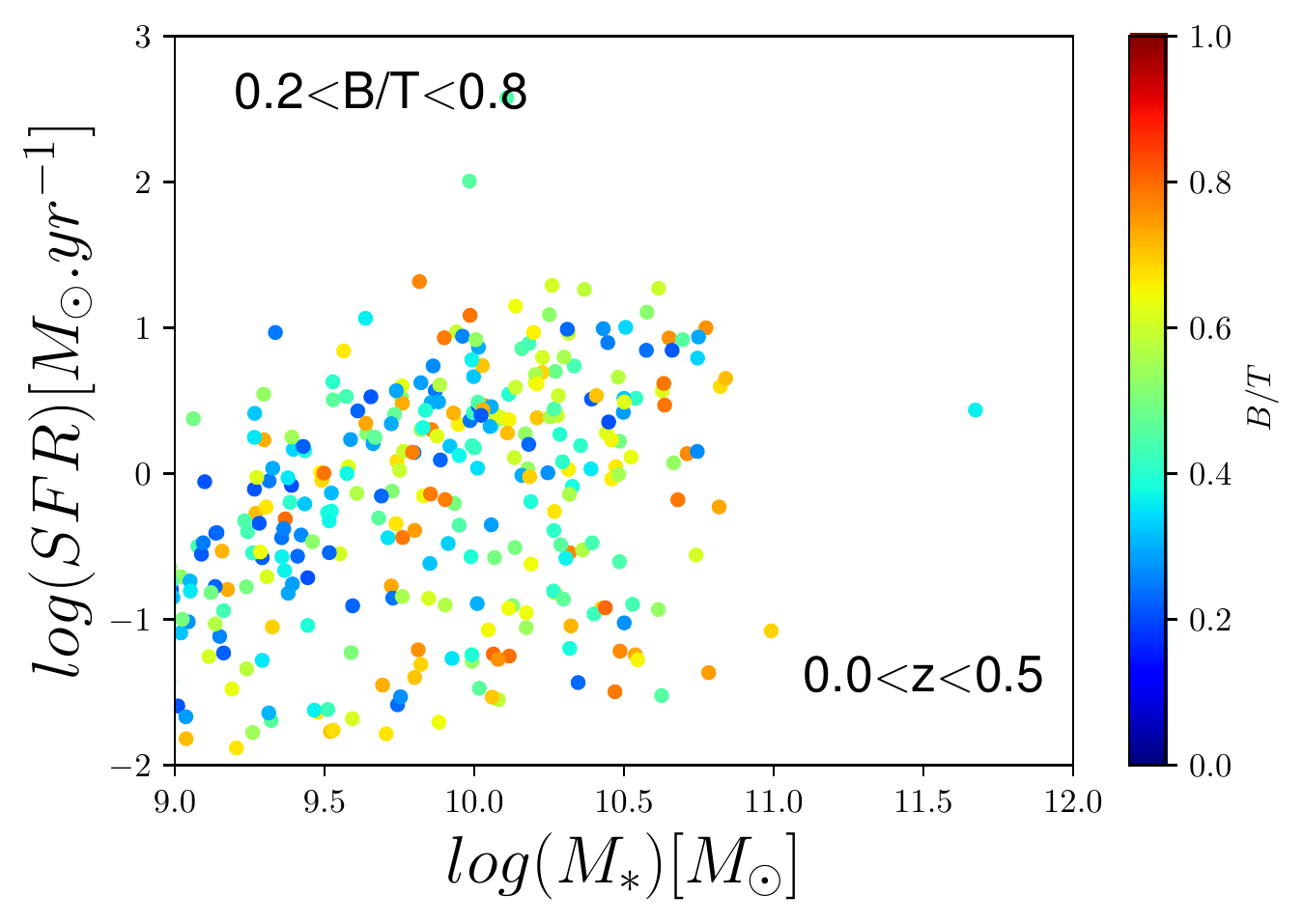}\\
\includegraphics[width=0.3\textwidth]{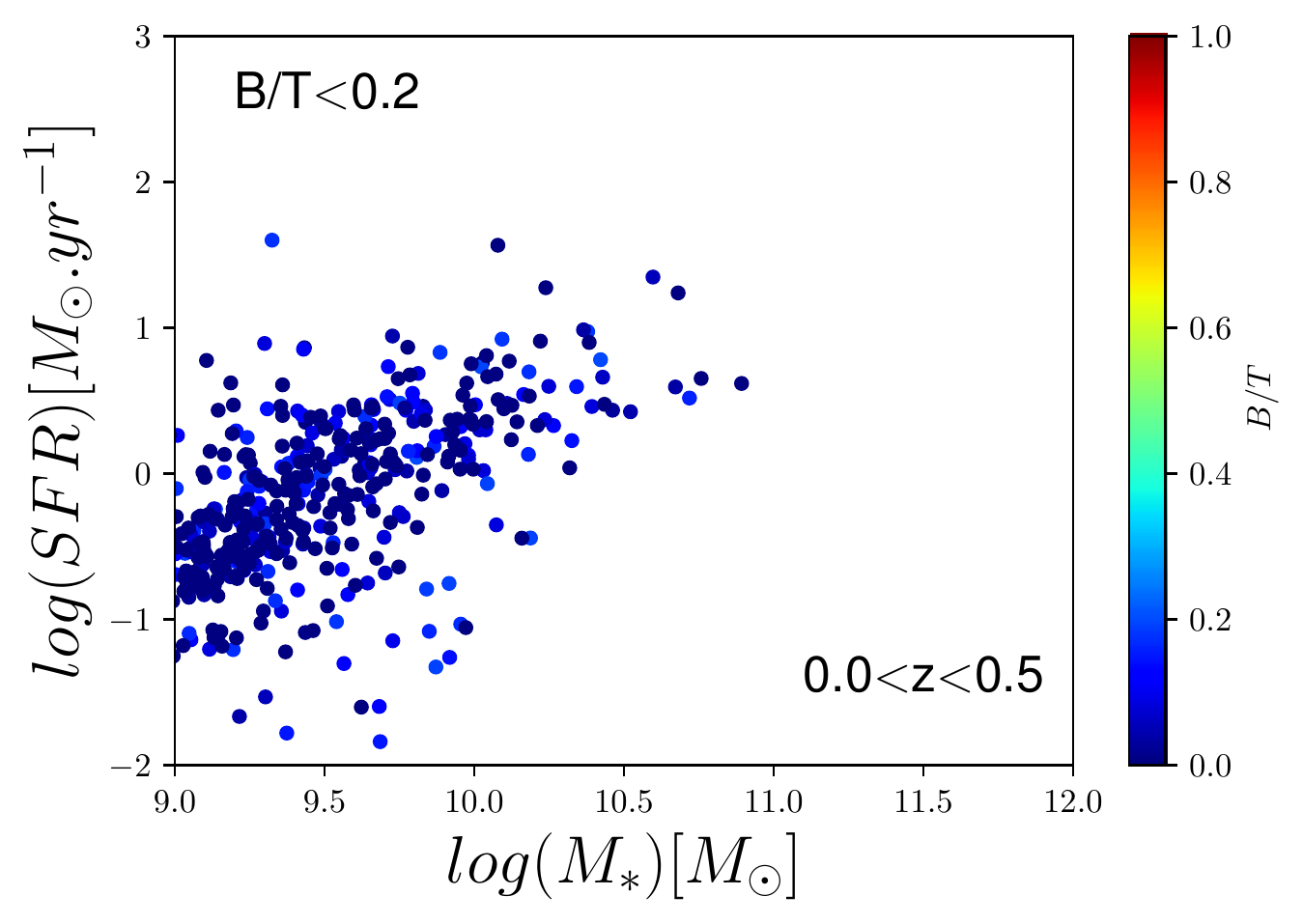}&
\includegraphics[width=0.3\textwidth]{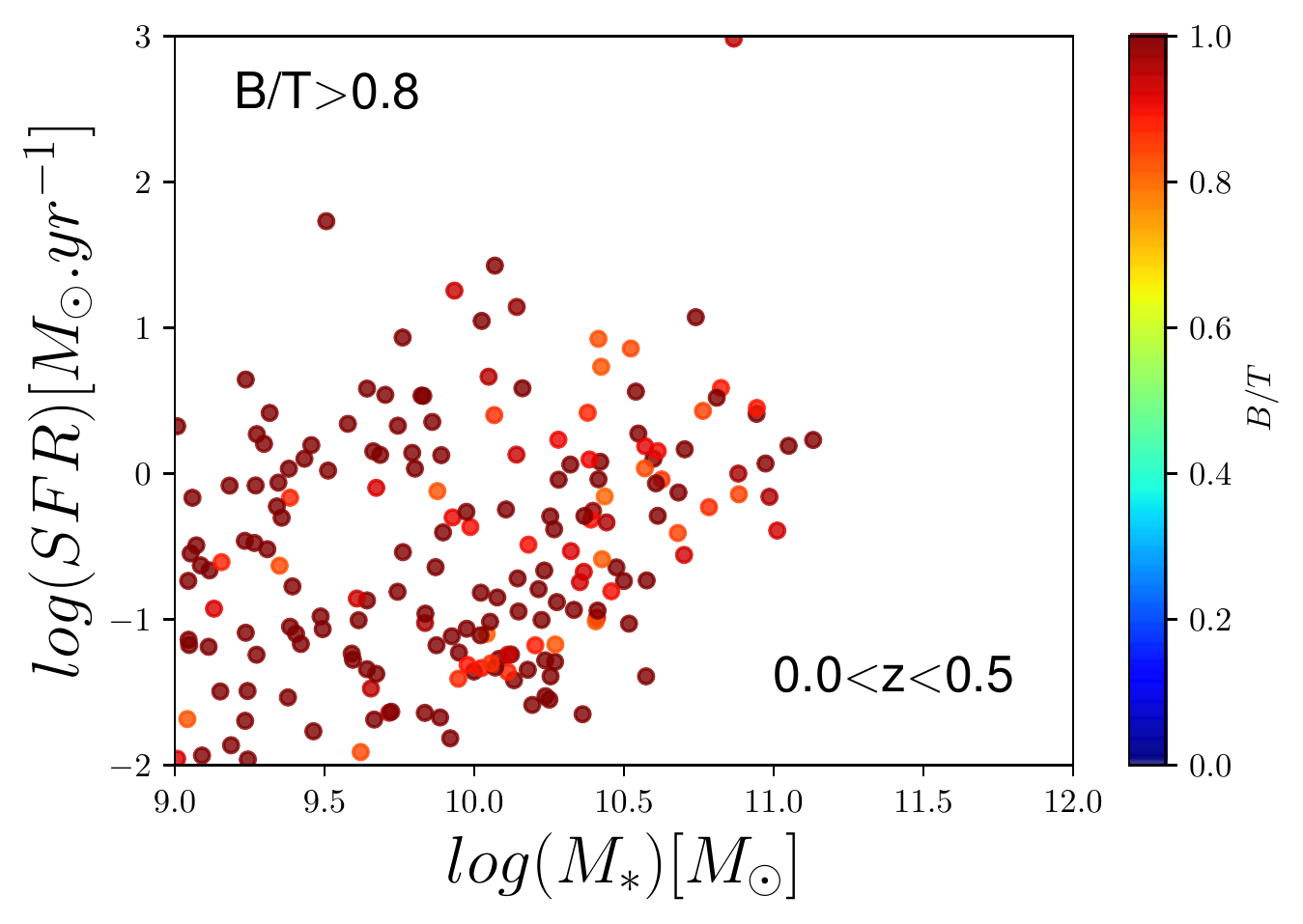}\\

\includegraphics[width=0.3\textwidth]{plots/mstar_SFR_BT2.pdf}&
\includegraphics[width=0.3\textwidth]{plots/mstar_SFR_BT_02_08_bin2.pdf}\\
\includegraphics[width=0.3\textwidth]{plots/mstar_SFR_BT_bin2_bis2.pdf}&
\includegraphics[width=0.3\textwidth]{plots/sfr_mass_dens_BT08_bis_2.pdf}\\
\end{array}$
\caption{Distribution of morphologies in the \SFRM plane. Color code, and \BTM bins are the same as \ref{fig:SFR_mstar2} }
\label{fig:SFR_mstar4bis}
\end{figure*}

\begin{figure*}
\centering
$\begin{array}{c c}
\includegraphics[width=0.3\textwidth]{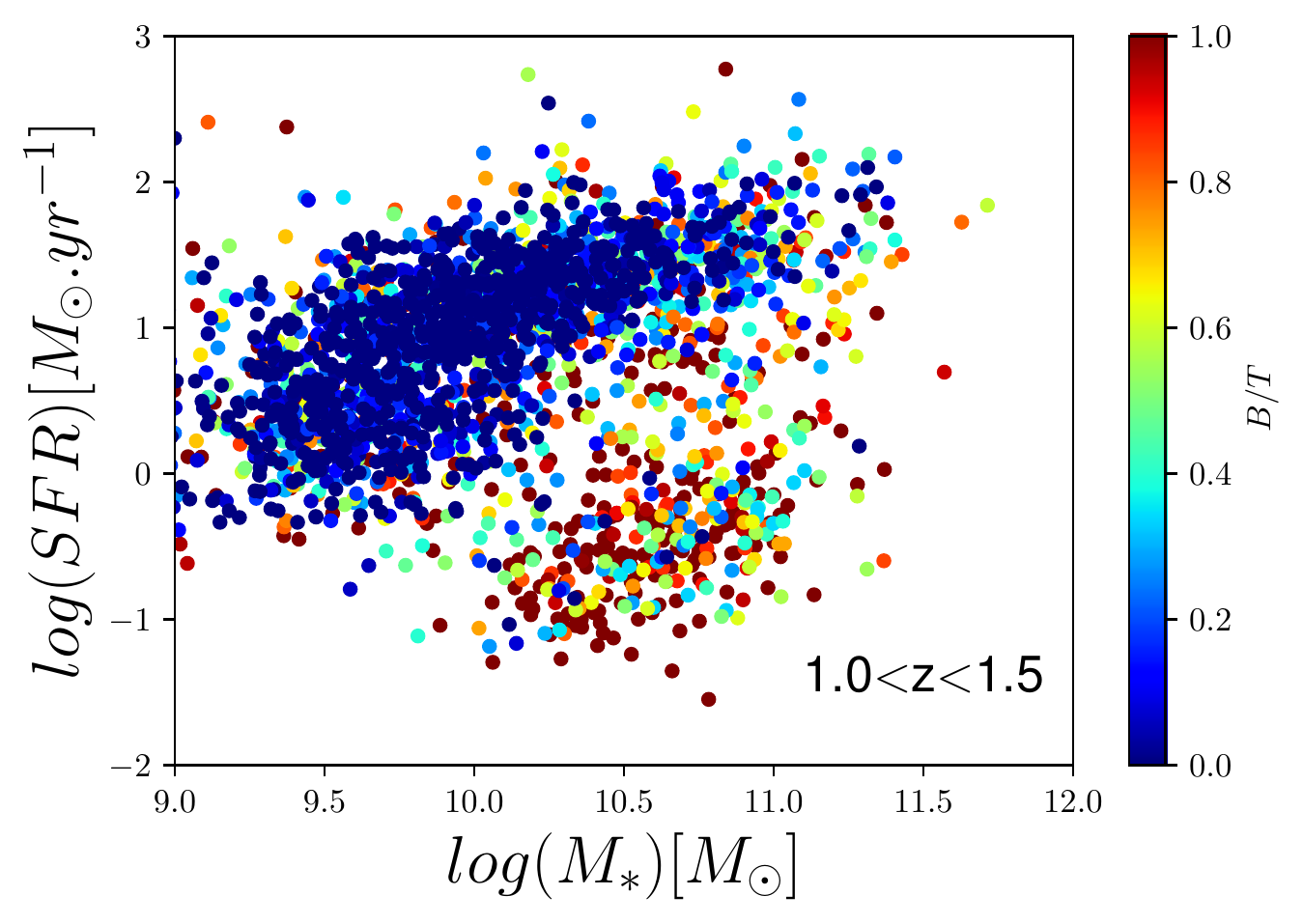}&
\includegraphics[width=0.3\textwidth]{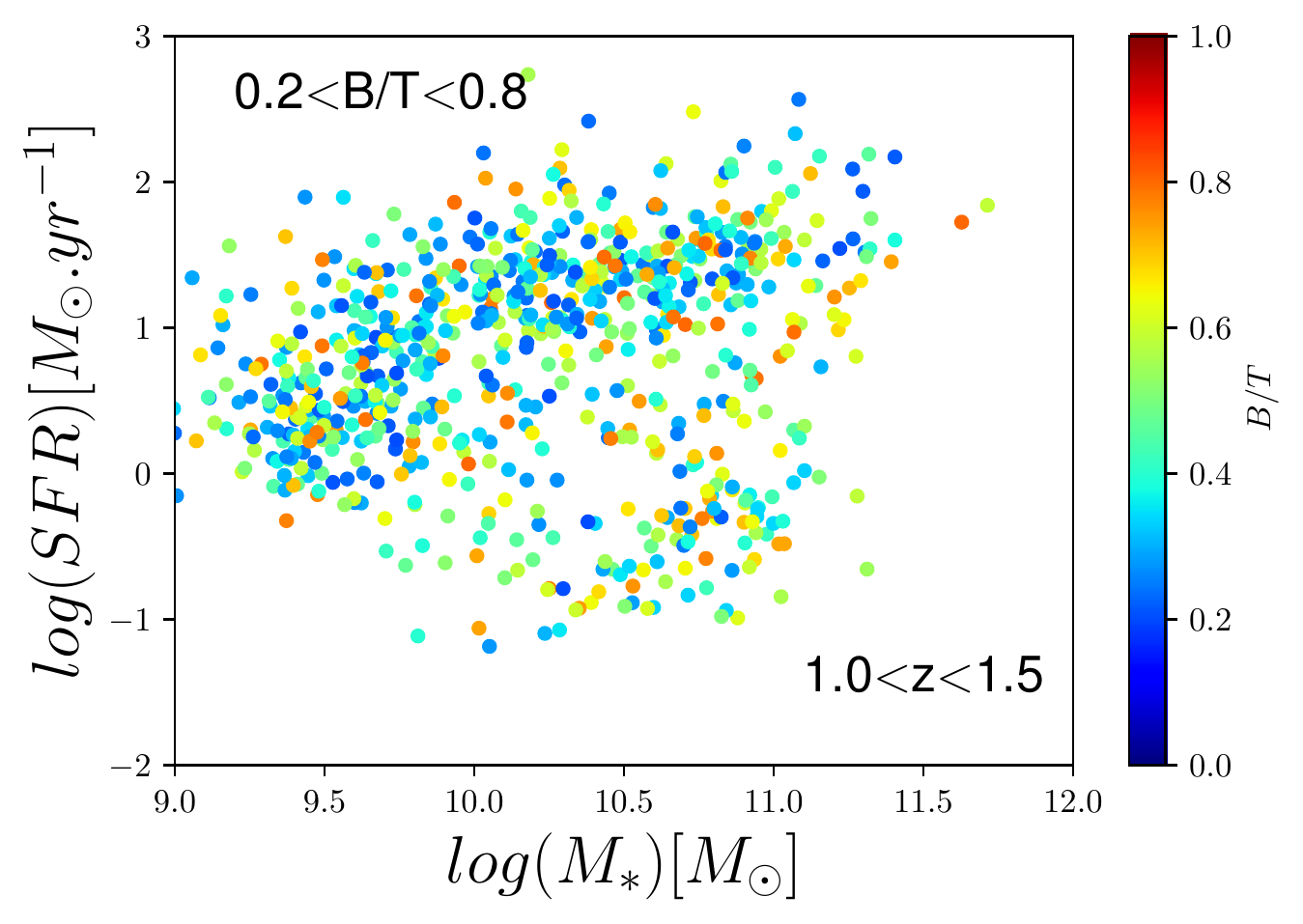}\\
\includegraphics[width=0.3\textwidth]{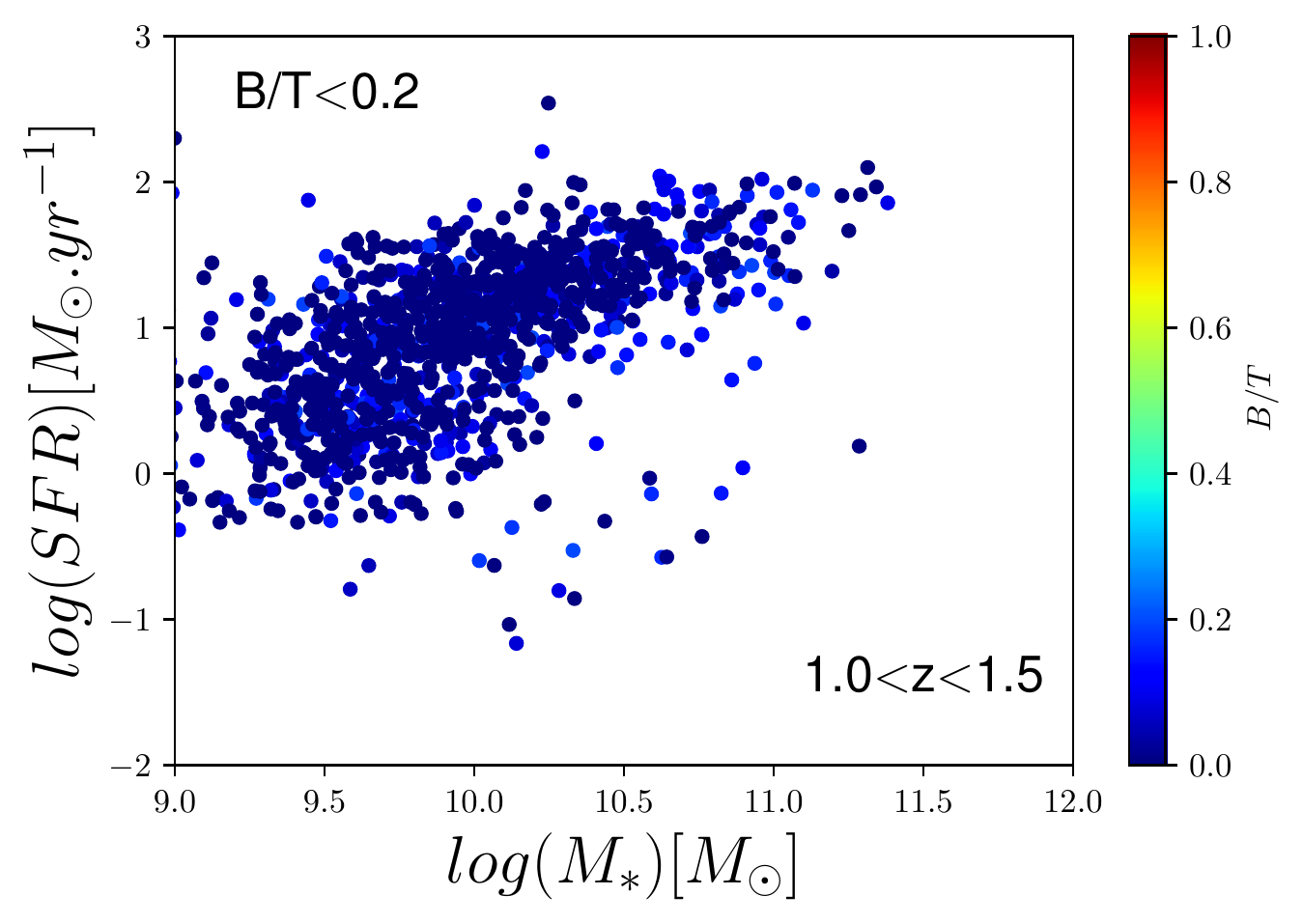}&
\includegraphics[width=0.3\textwidth]{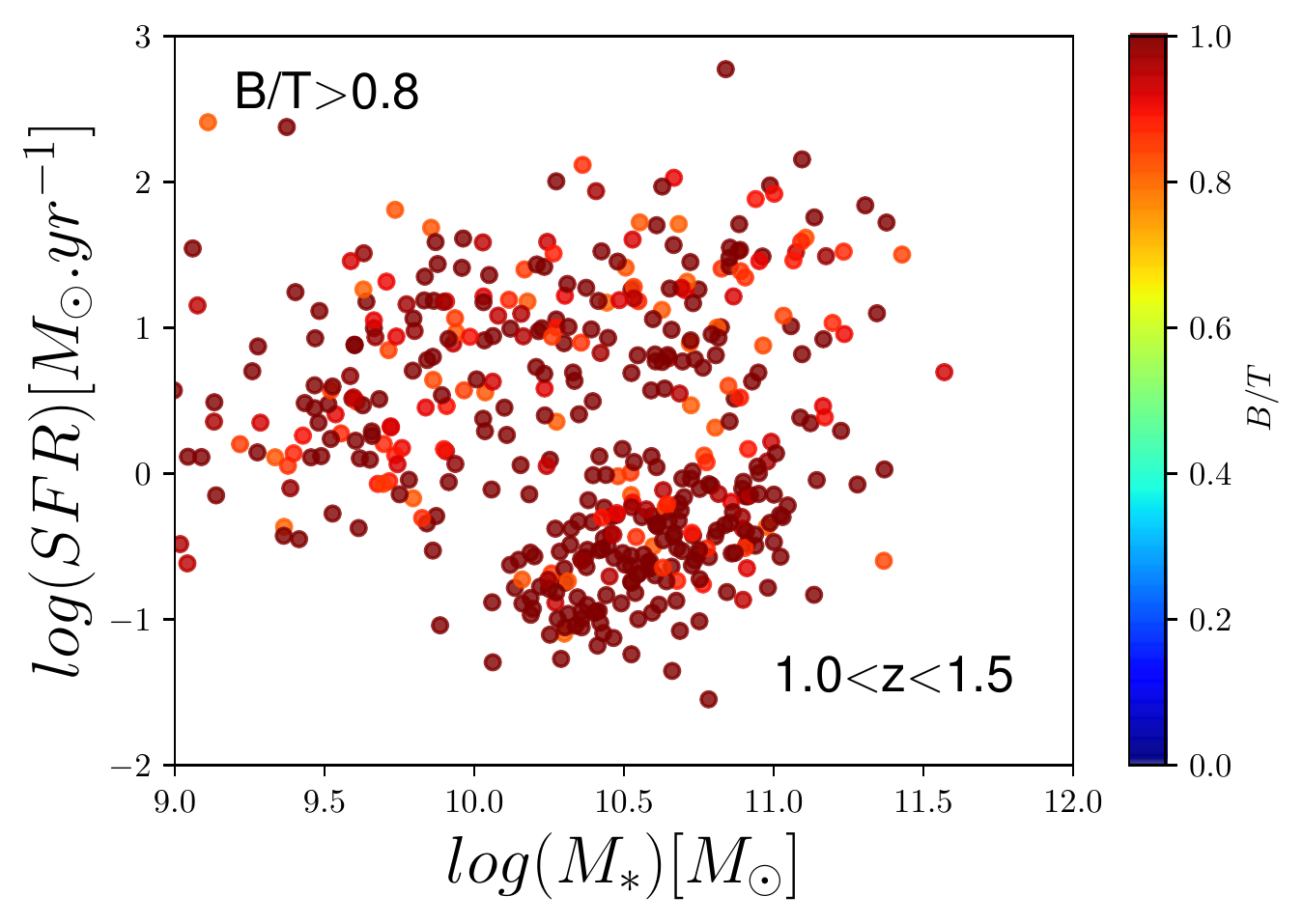}\\

\includegraphics[width=0.3\textwidth]{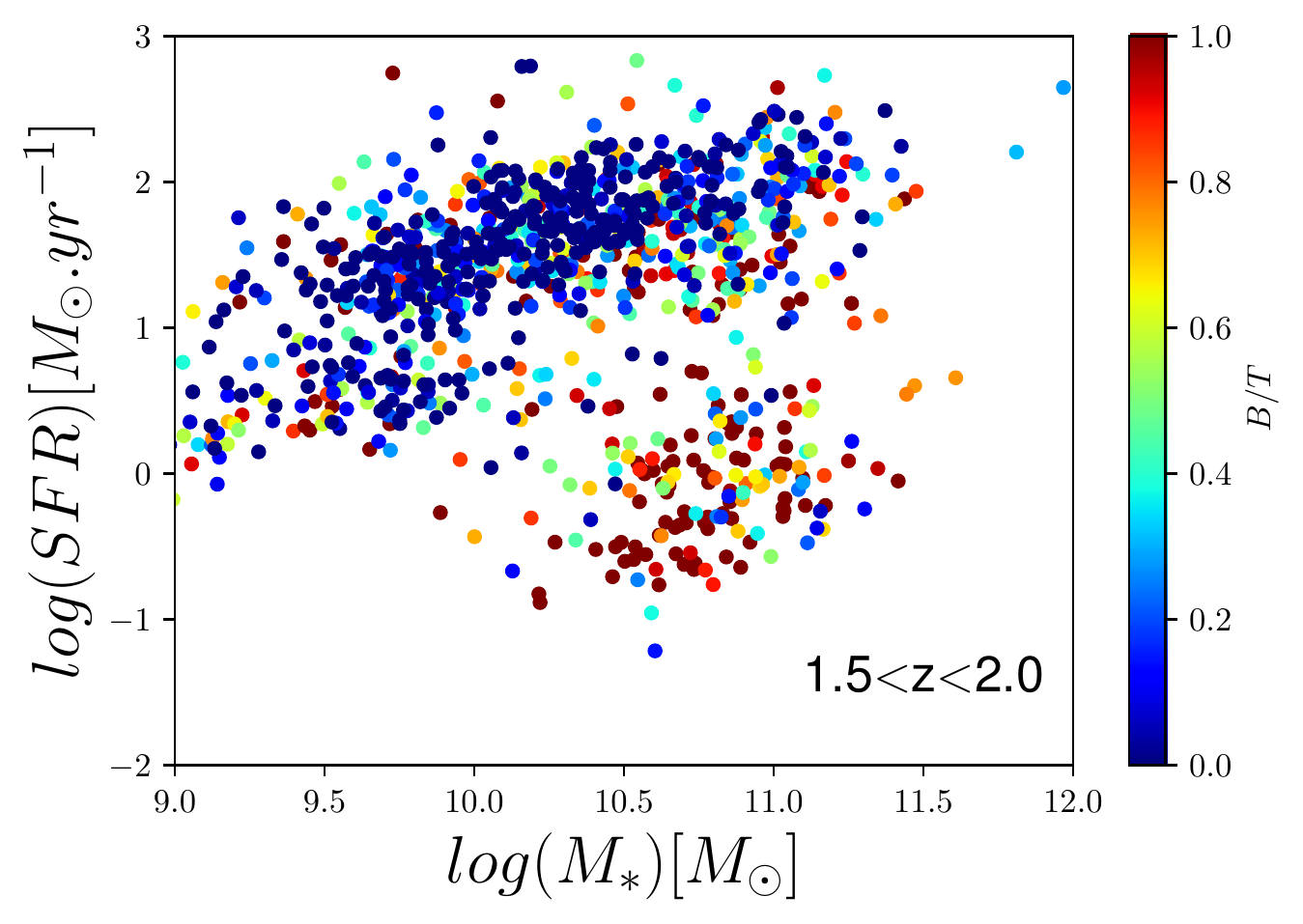}&
\includegraphics[width=0.3\textwidth]{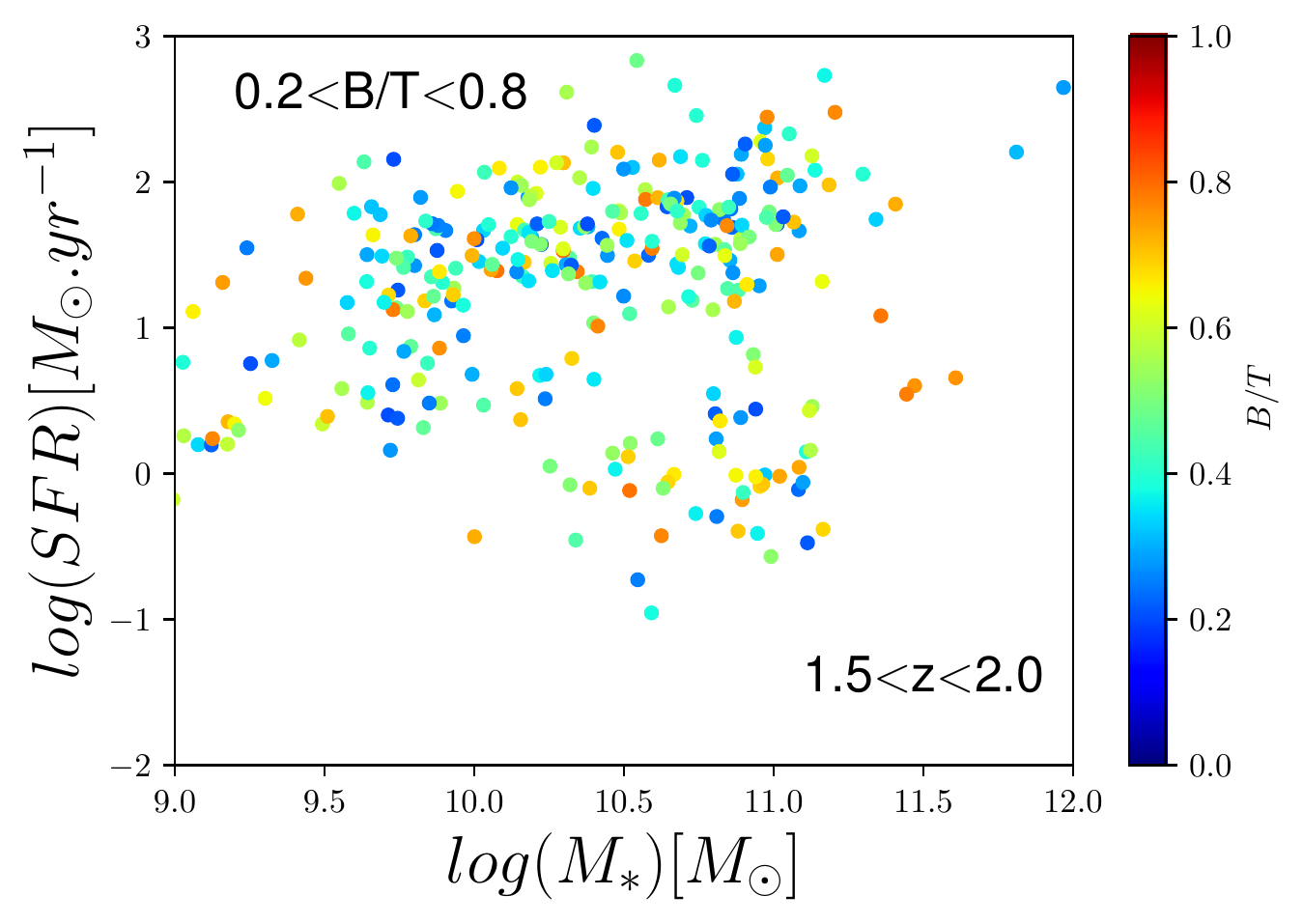}\\
\includegraphics[width=0.3\textwidth]{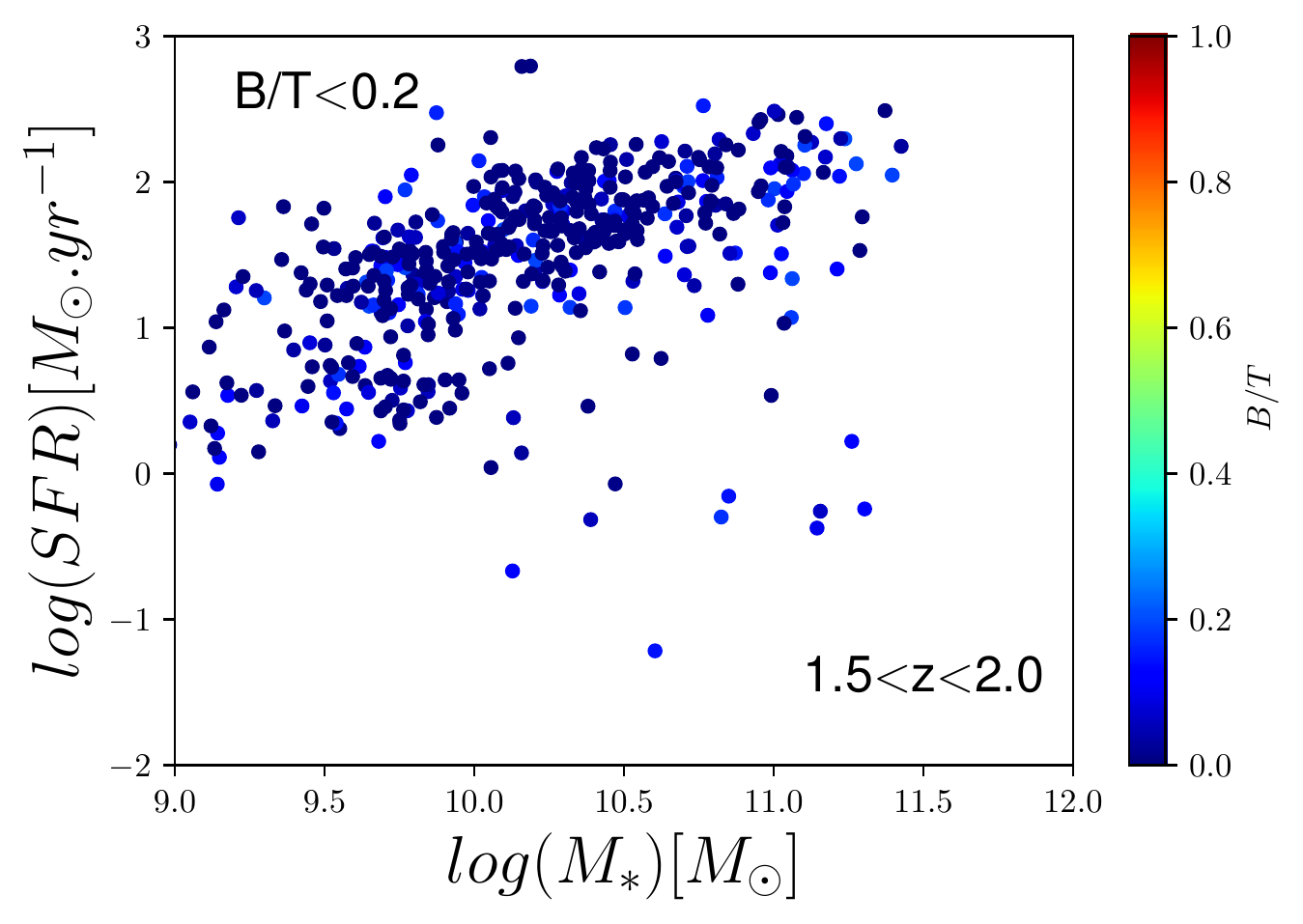}&
\includegraphics[width=0.3\textwidth]{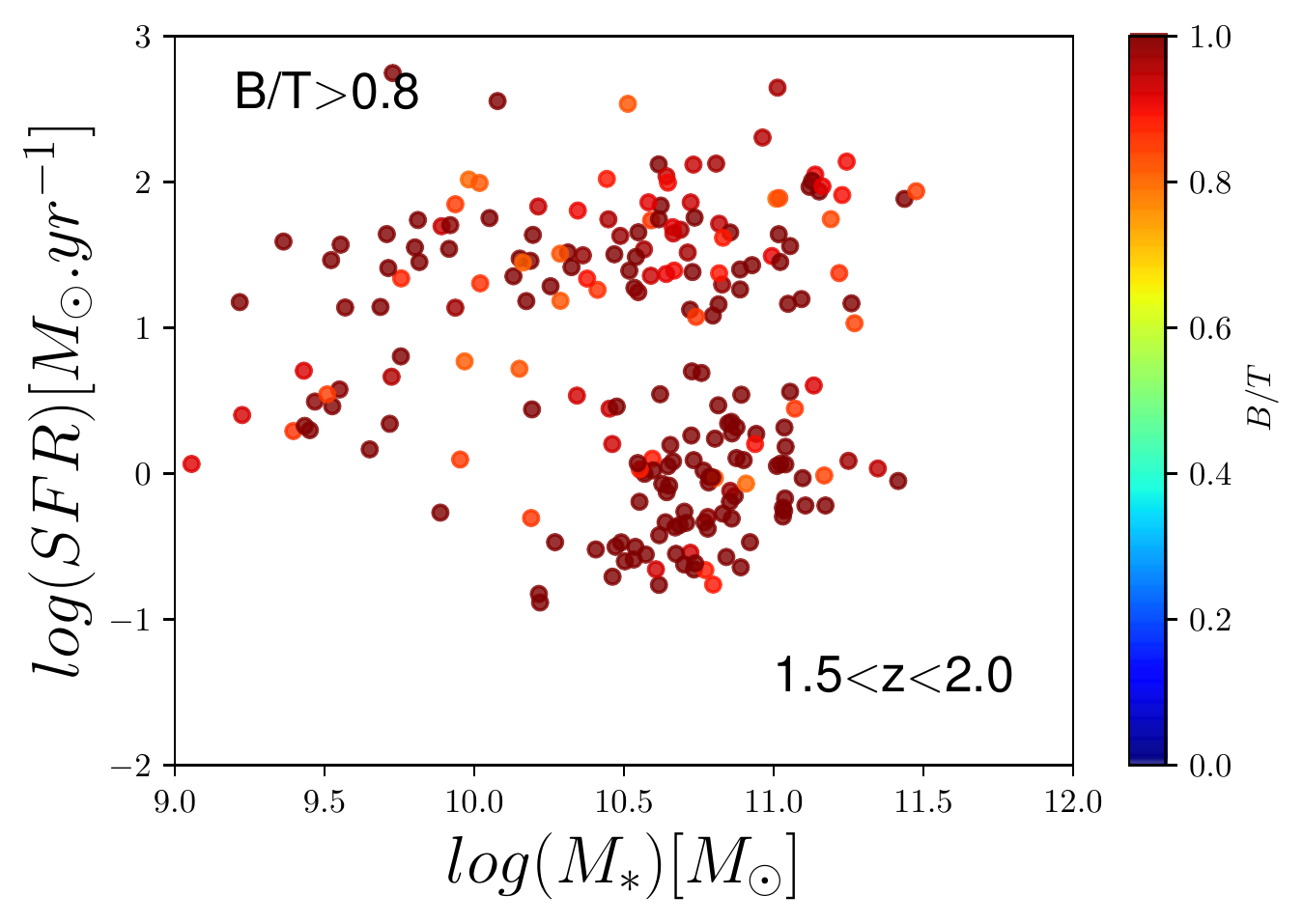}\\
\end{array}$
\caption{Distribution of morphologies in the \SFRM plane. Color code, and \BTM bins are the same as \ref{fig:SFR_mstar2} }
\label{fig:SFR_mstar5bis}
\end{figure*}

The position of bulges and discs along the main sequence gives clues on how galaxy evolve and quench. Figures \ref{fig:SFR_mstar4bis} and \ref{fig:SFR_mstar5bis} show the distribution of galaxies in the \M -SFR plane in the redshift range of (0,2), color coded with \BTM. As discussed in the main text, some general trends are observed. Galaxies wih \BTM<0.2 are mostly concentrated in the main sequence, while objects with intermediate values (0.2<z<0.8) are distributed between the Star forming and the quenched region. 

\begin{figure*}
\centering
$\begin{array}{cccc}
\includegraphics[width=0.25\textwidth]{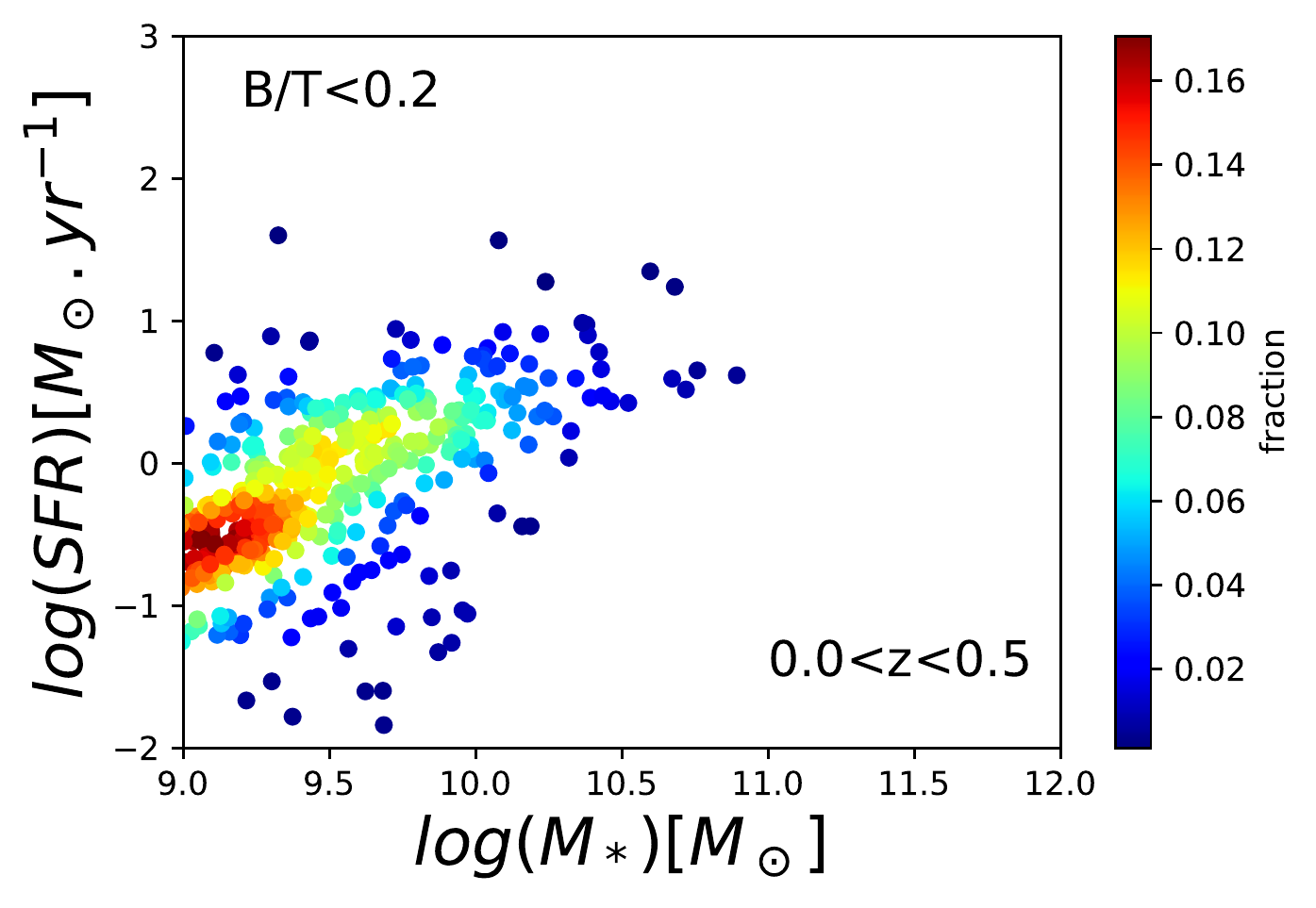}&
\includegraphics[width=0.25\textwidth]{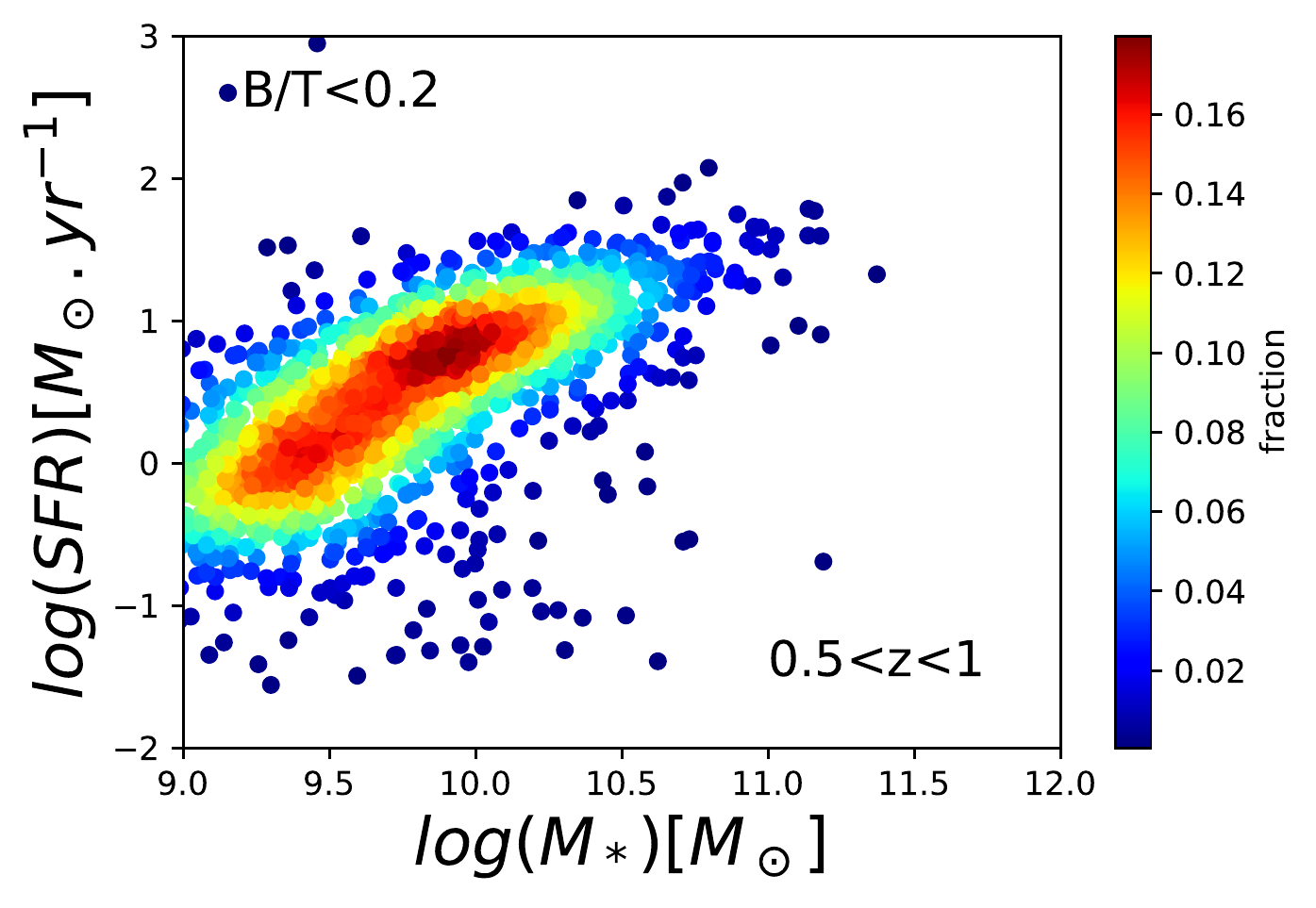}&
\includegraphics[width=0.25\textwidth]{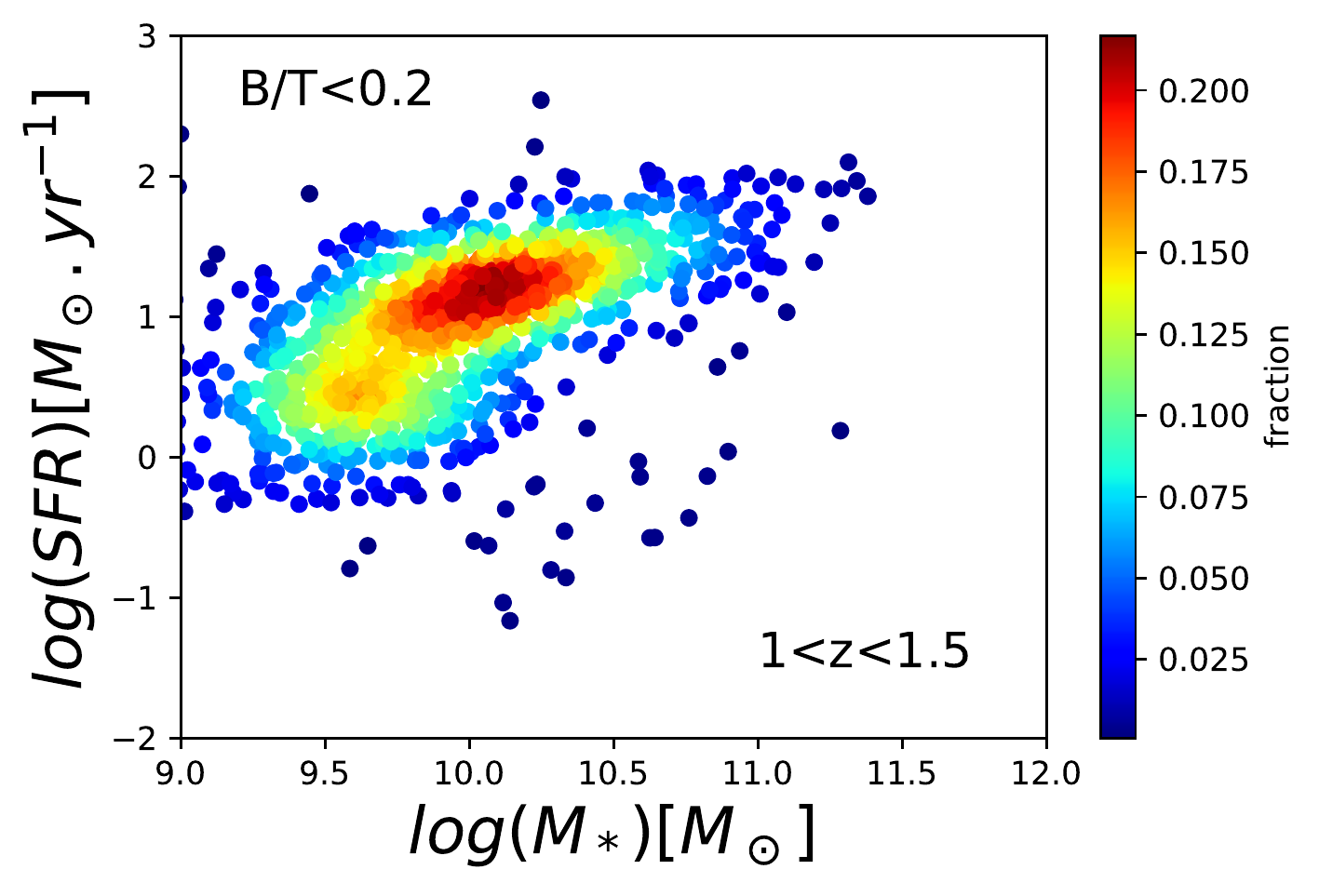}&
\includegraphics[width=0.25\textwidth]{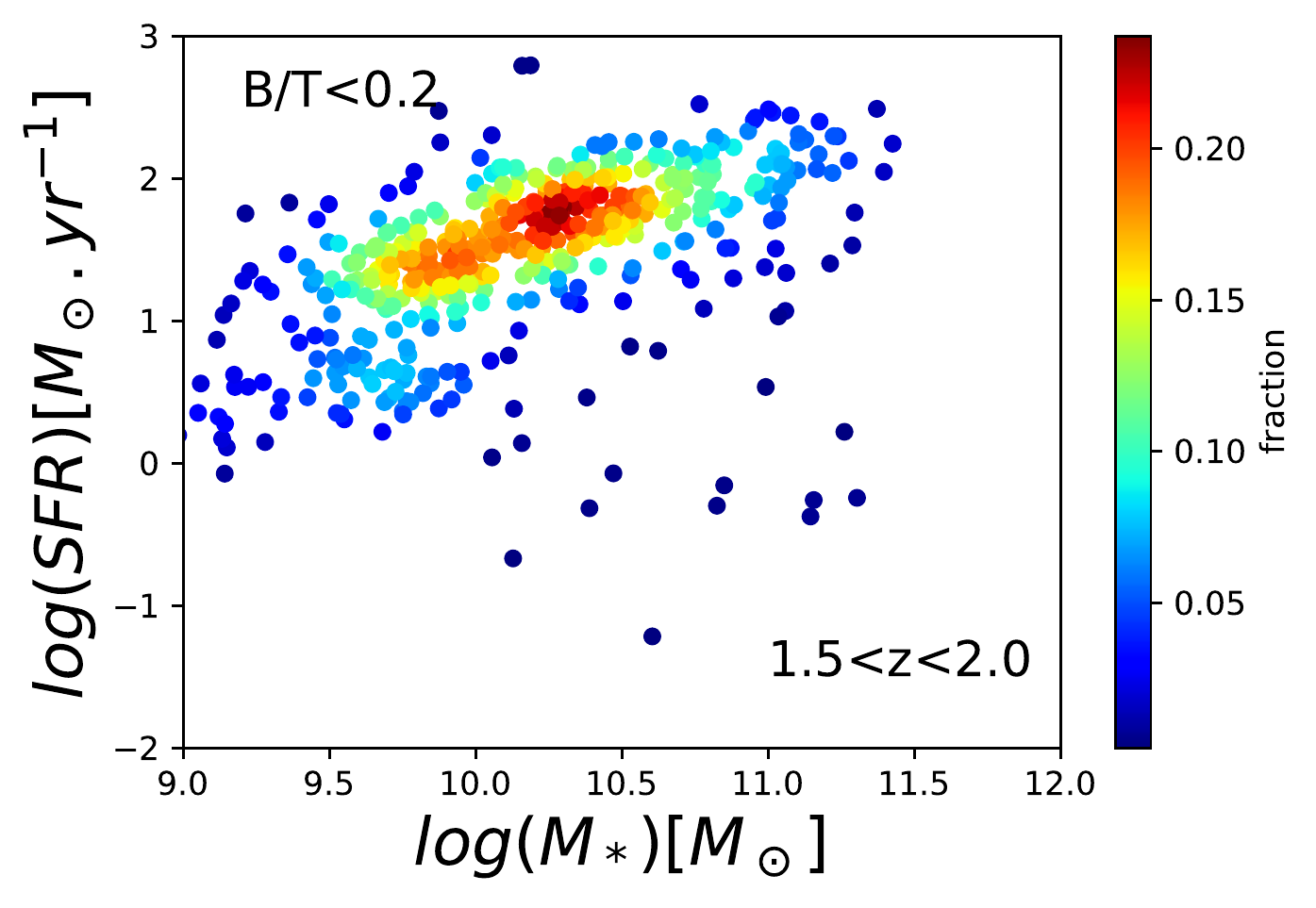}\\

\includegraphics[width=0.25\textwidth]{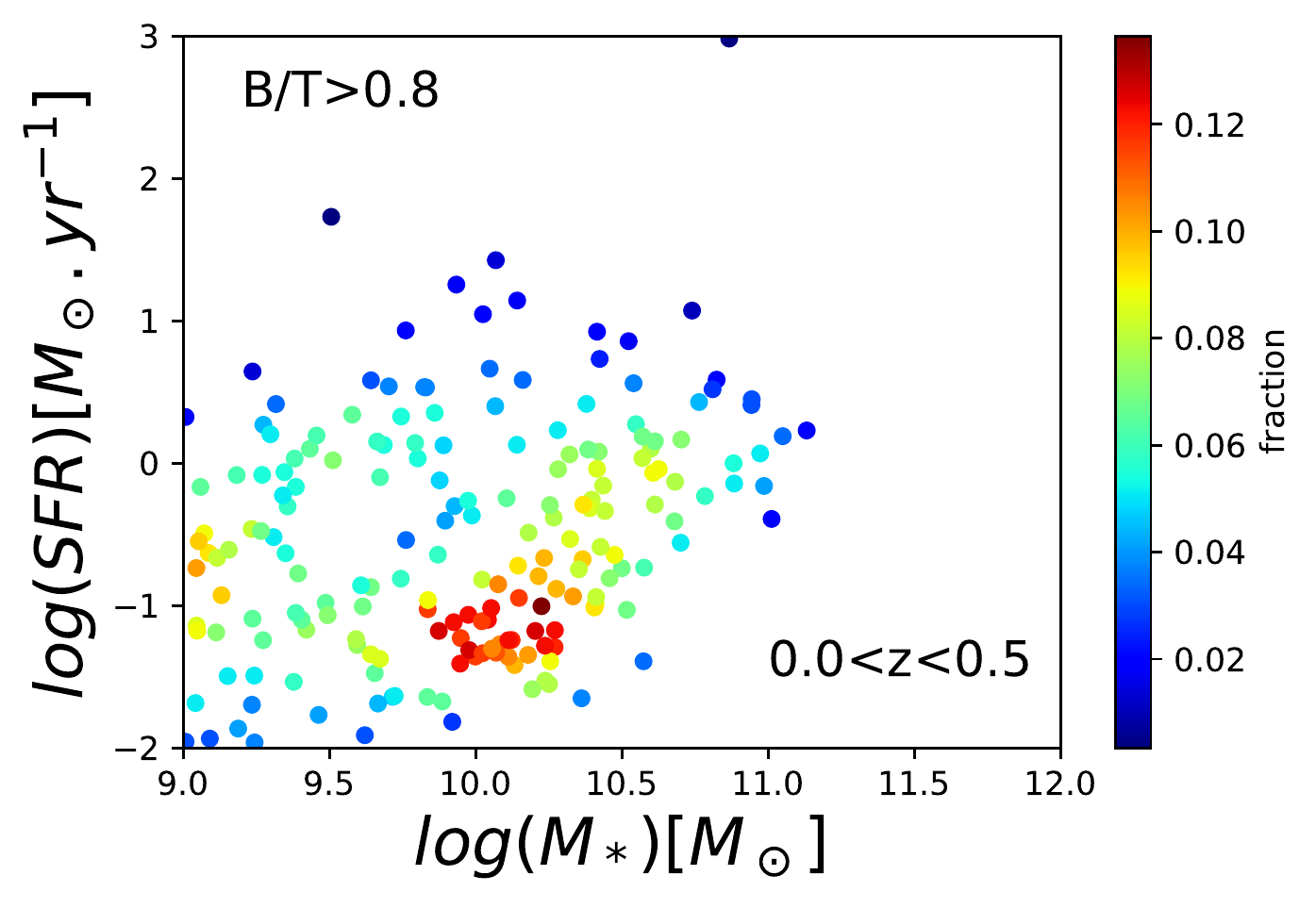}&
\includegraphics[width=0.25\textwidth]{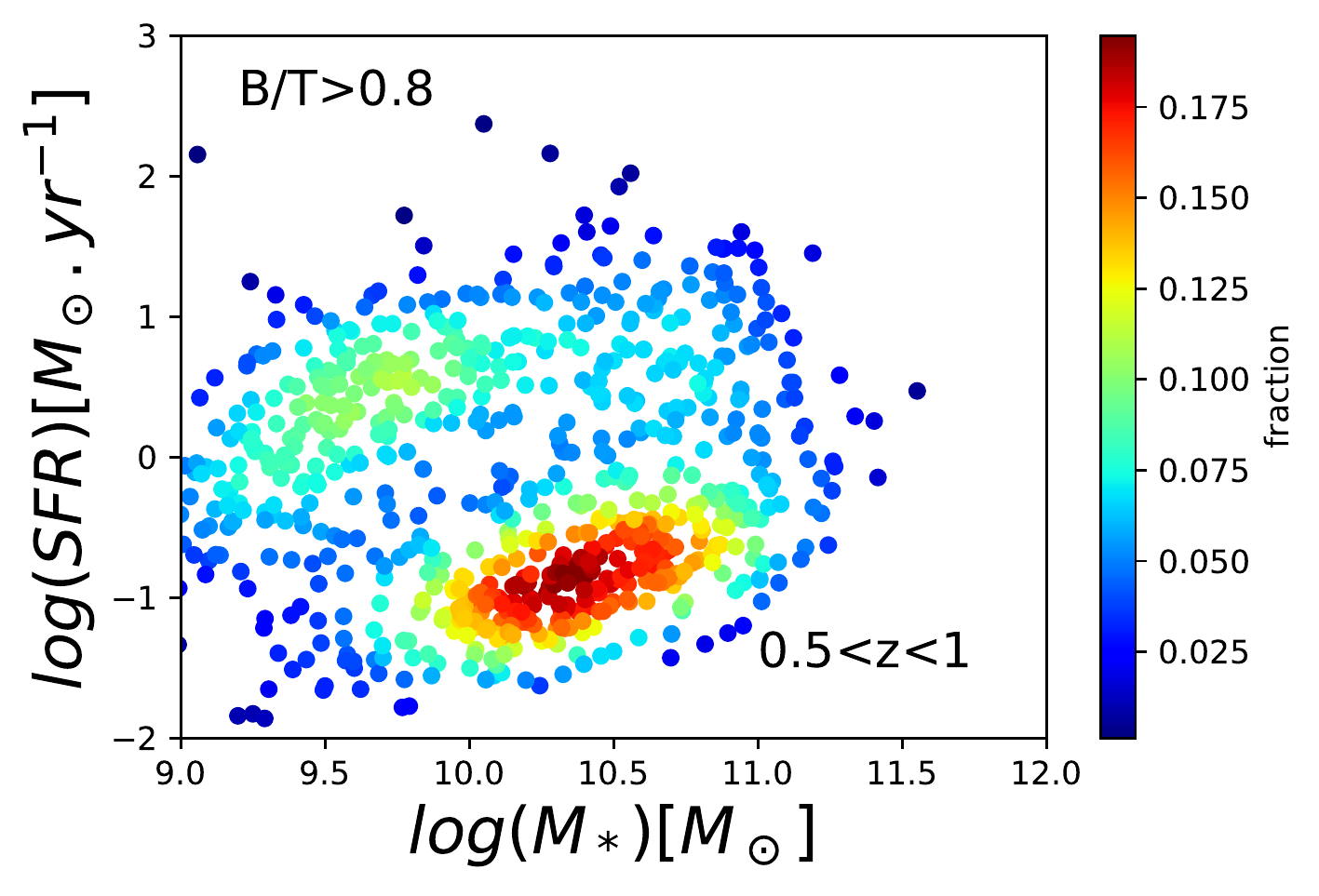}&
\includegraphics[width=0.25\textwidth]{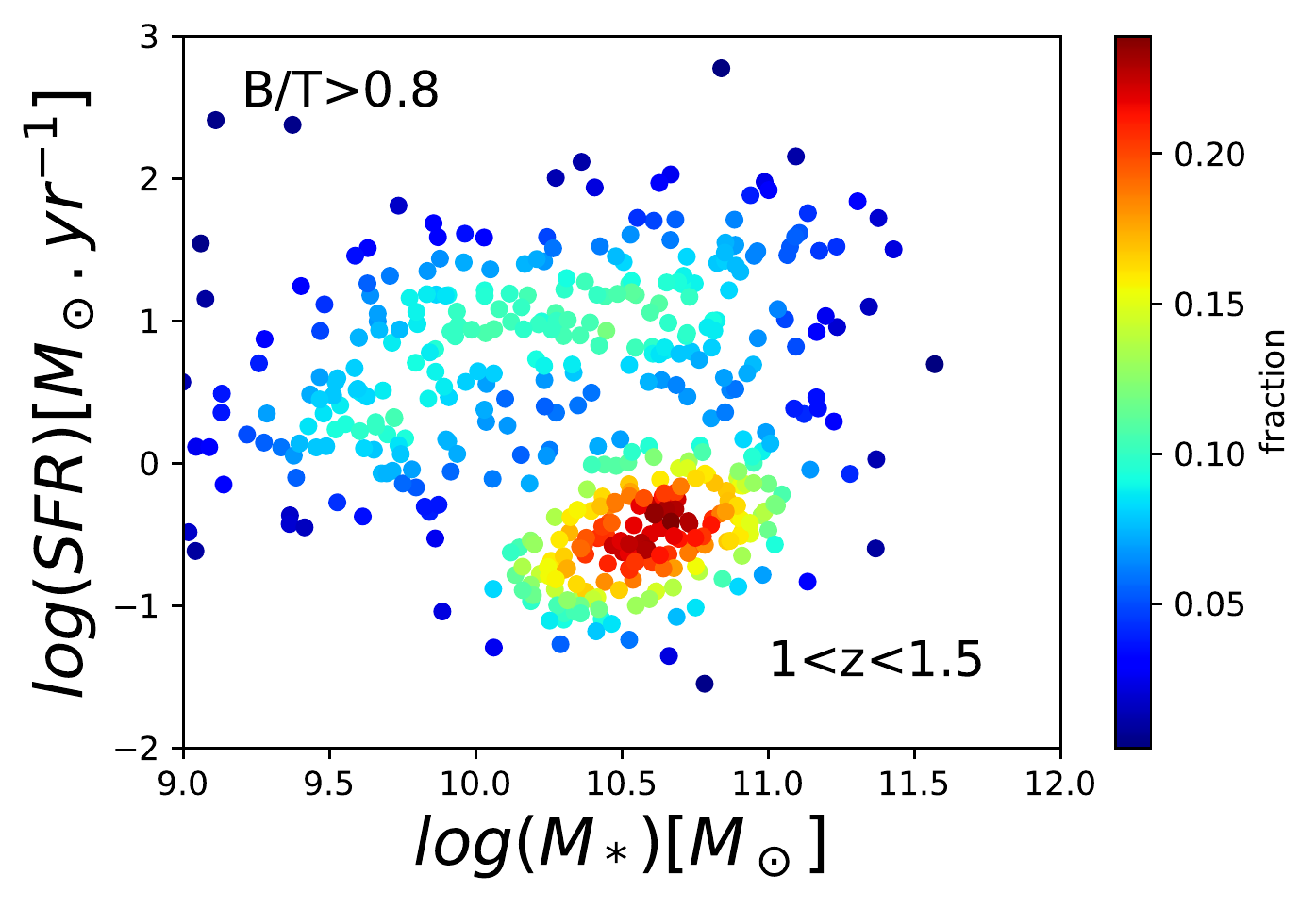}&
\includegraphics[width=0.25\textwidth]{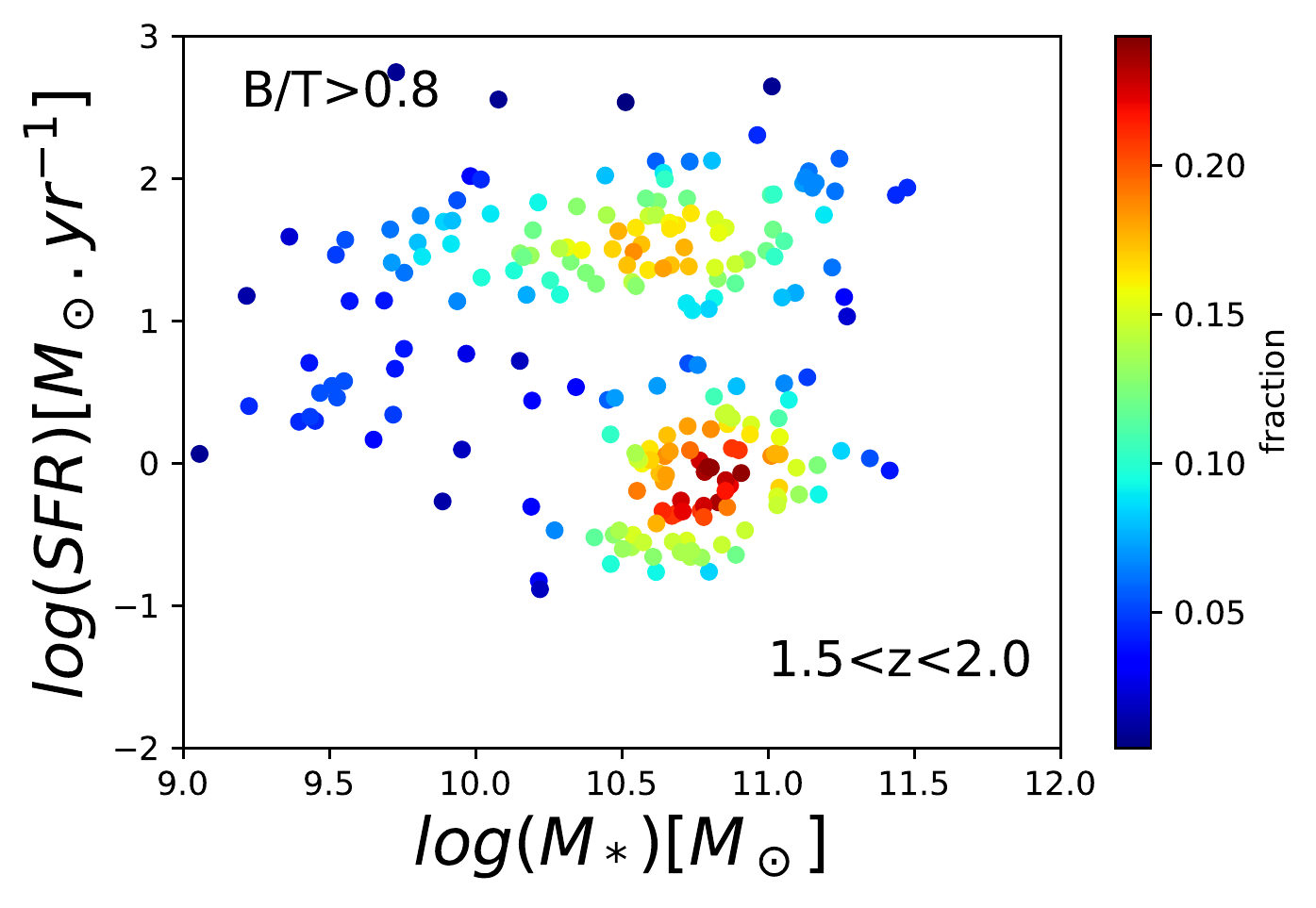}\\
\end{array}$
\caption{Density distribution of galaxies in the \SFRM plane. Top and bottom series of plots are showing the two extreme sub-sample: disky(\BTM<0.2) and bulge-dominated(\BTM>0.8) galaxies. Color code represents the number fraction of objects in the specific region of the plane. It can be pointed out that most of the \BTM<0.2 are concentrated in the MS with few outliers, while the density distribution of galaxies with \BTM>0.8 has a pick in the quenched region with a larger scatter. }
\label{fig:SFR_mstar6bis}
\end{figure*}

\section{Star-formation,morphology and quenching}
\label{apx:quenching_best}
\begin{figure*}
\centering
$\begin{array}{c c}
\includegraphics[width=0.45\textwidth]{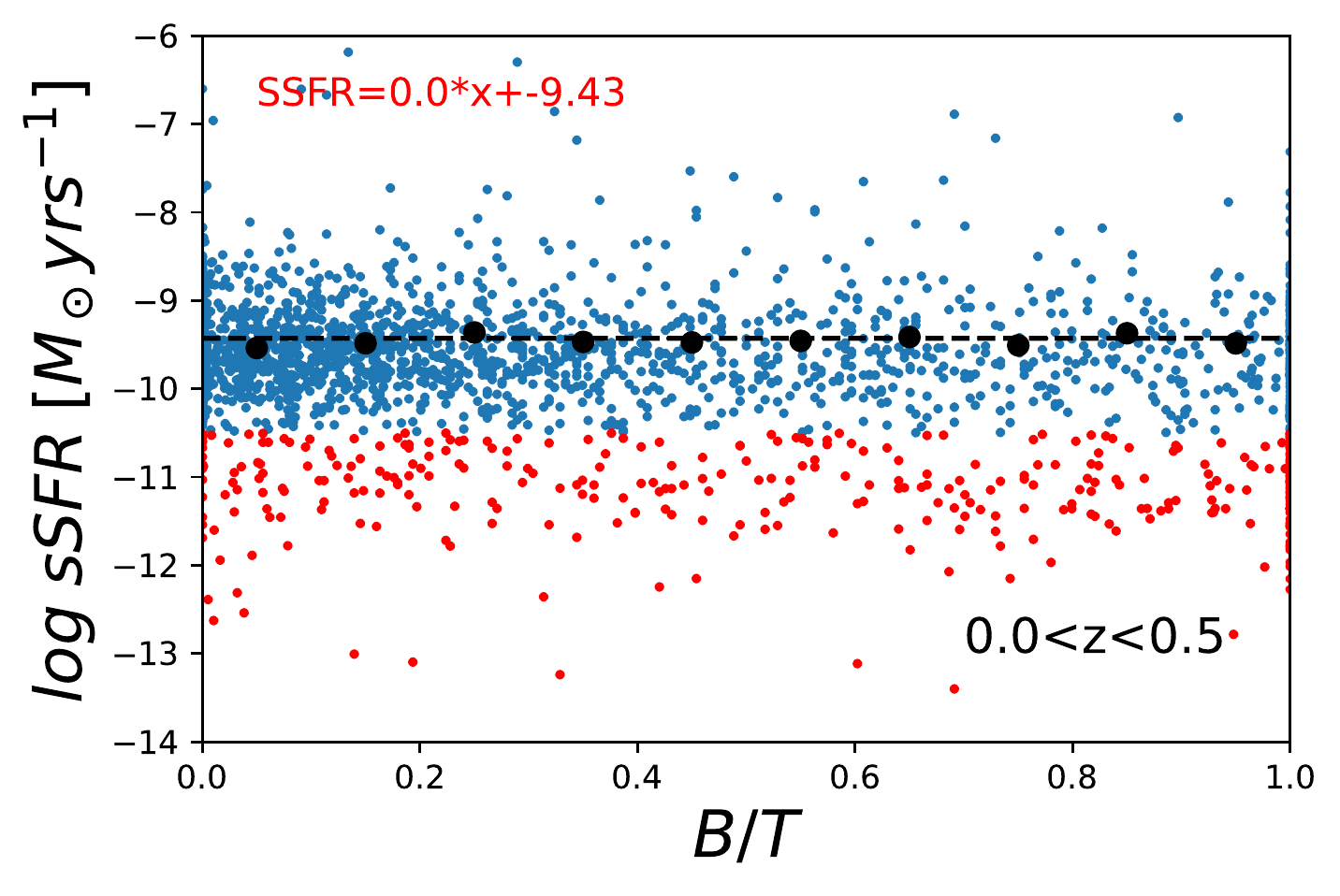}&
\includegraphics[width=0.45\textwidth]{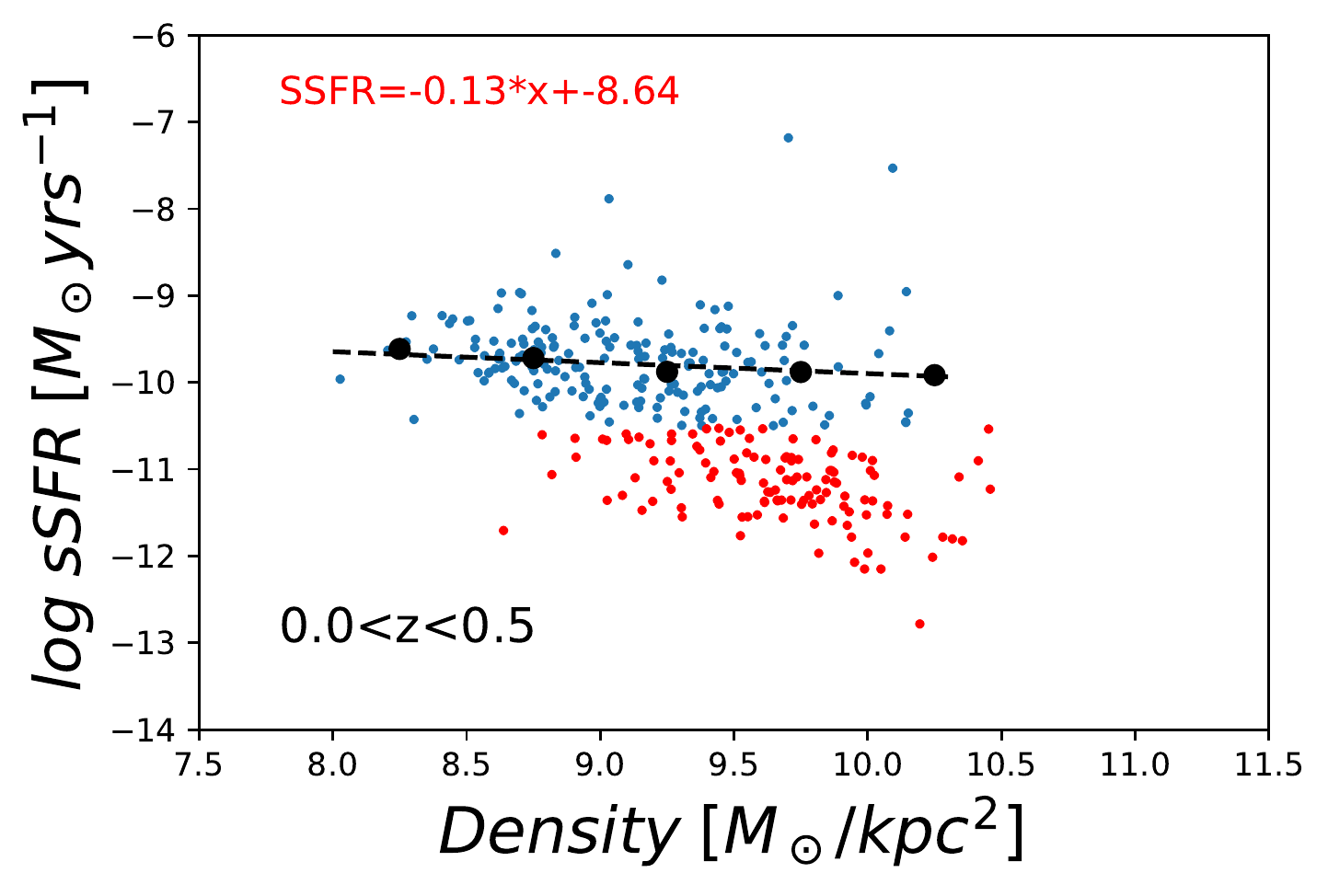}\\
\includegraphics[width=0.45\textwidth]{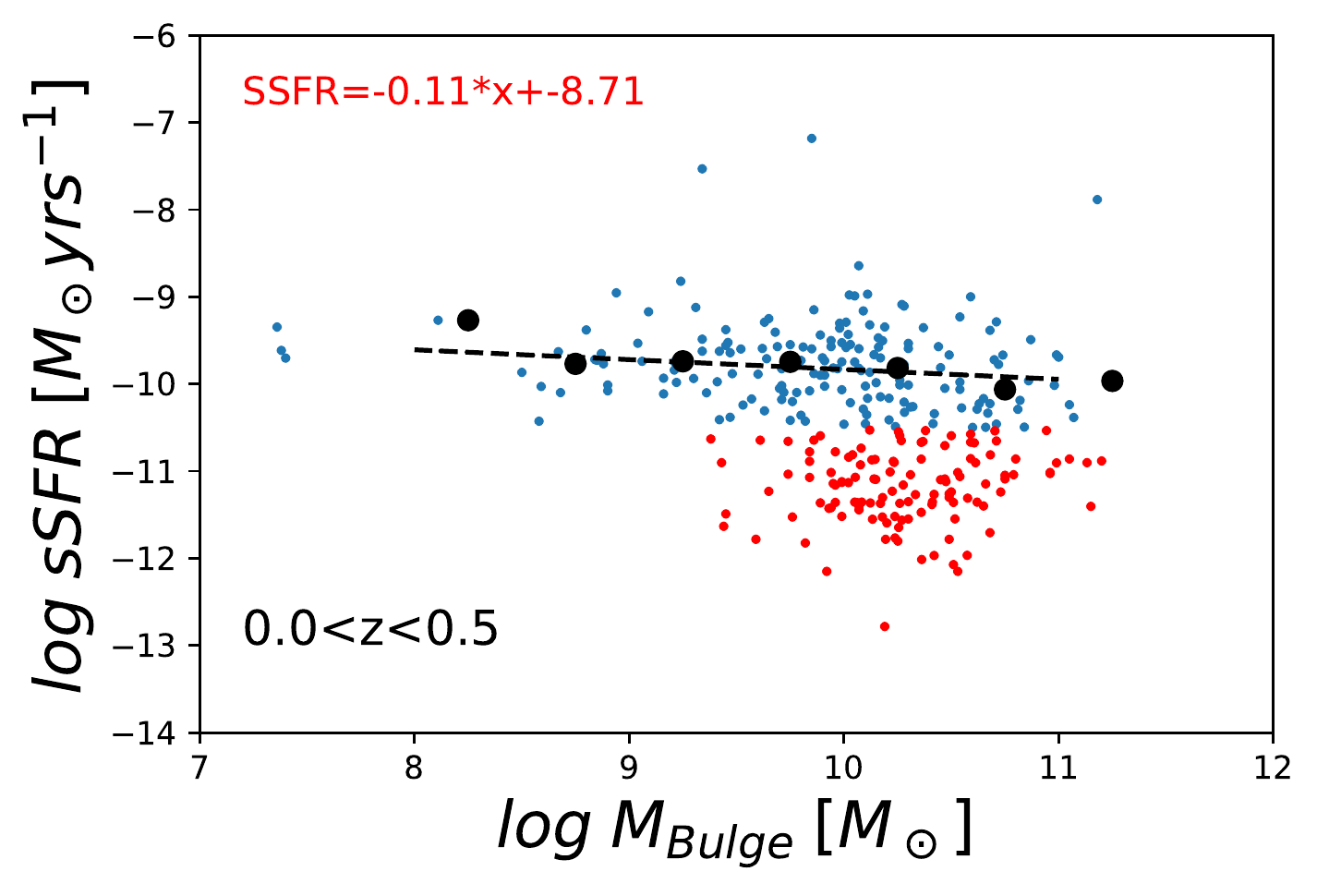}&
\includegraphics[width=0.45\textwidth]{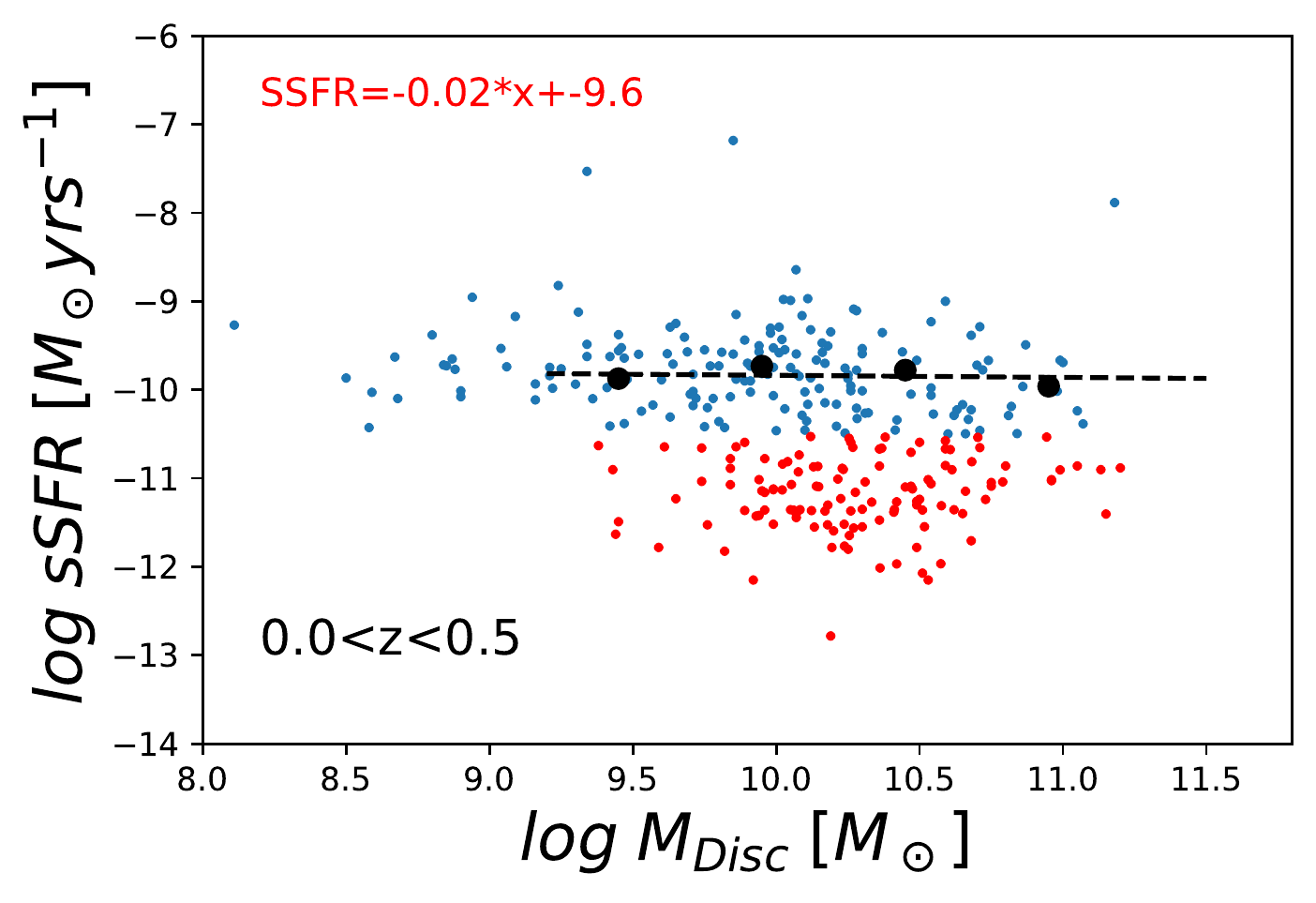}\\
\end{array}$
\caption{Star formation-morphology-quenching. The sequence of the plots aims to investigate the role of the different parameters in decreasing the star formation. To do this, each parameter is compared with the specific star formation rate (to remove the dependence on the total stellar mass from the analysis). The upper panels are analyzing the relationship between sSFR and \BTM, central density. The lower panels show the impact of the bulge(left)/disc(right) stella mass on the star formation activity. The black points are the median values for which a better linear fit is applied and the final equation is shown at the top of each panel. }
\label{fig:quenching2}
\end{figure*}

\begin{figure*}
\centering
$\begin{array}{c c}
\includegraphics[width=0.45\textwidth]{plots/SSFR_BT_1_v3.pdf}&
\includegraphics[width=0.45\textwidth]{plots/SSFR_Den_v31.pdf}\\
\includegraphics[width=0.45\textwidth]{plots/SSFR_massB_v31.pdf}&
\includegraphics[width=0.45\textwidth]{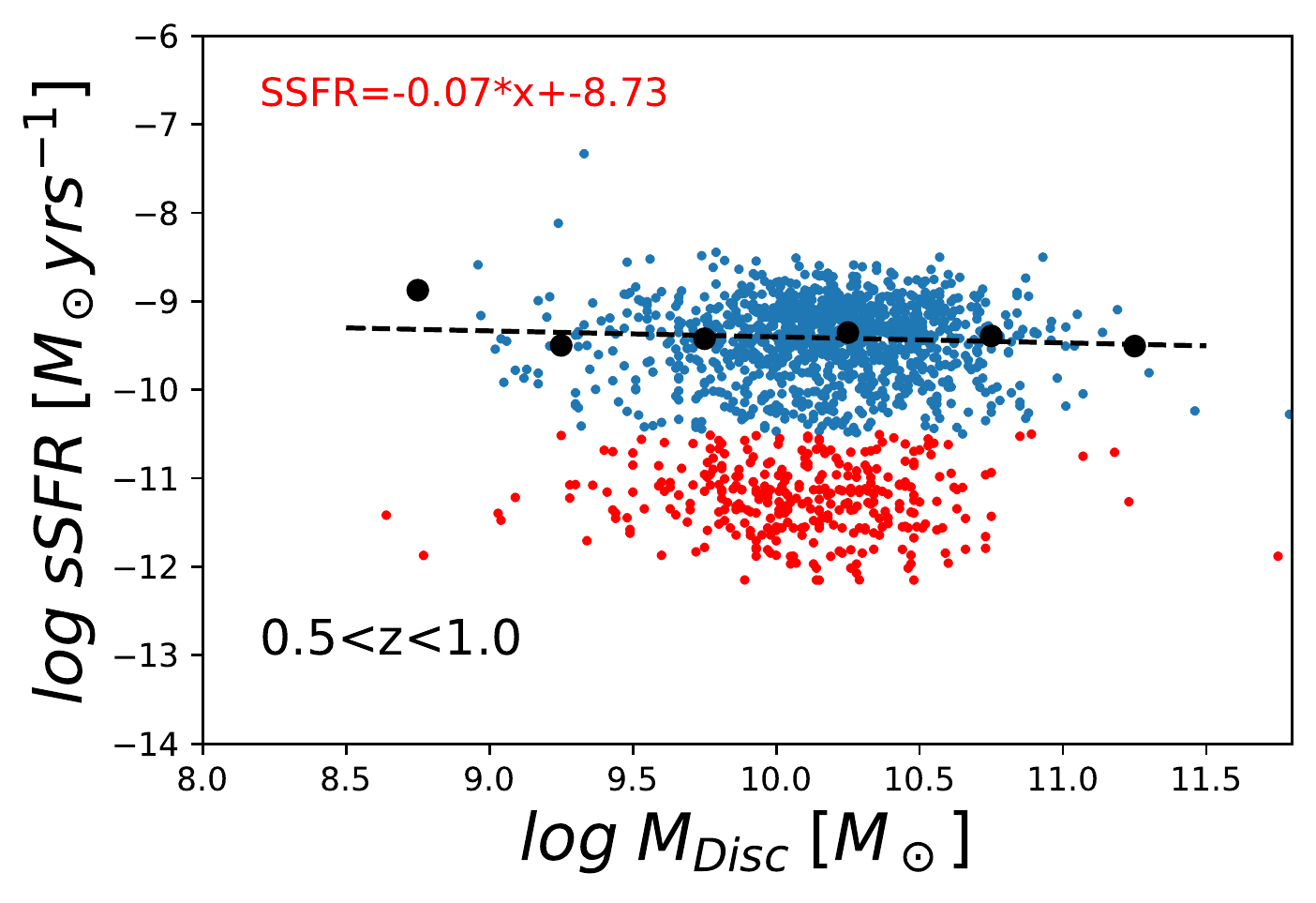}\\
\end{array}$
\caption{Star formation-morphology-quenching. Same as the previous Figure.}
\label{fig:quenching3}
\end{figure*}

\begin{figure*}
\centering
$\begin{array}{c c}
\includegraphics[width=0.45\textwidth]{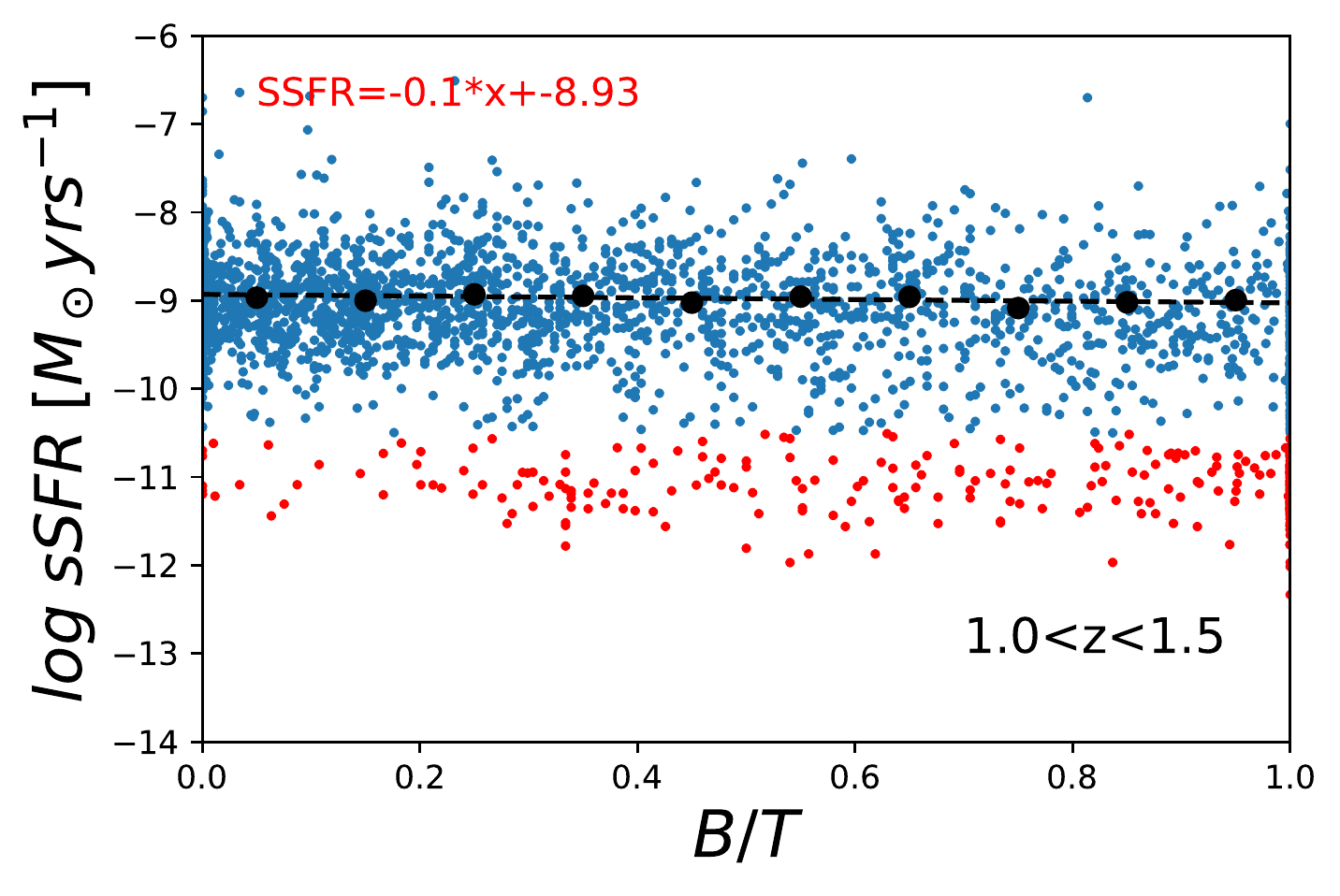}&
\includegraphics[width=0.45\textwidth]{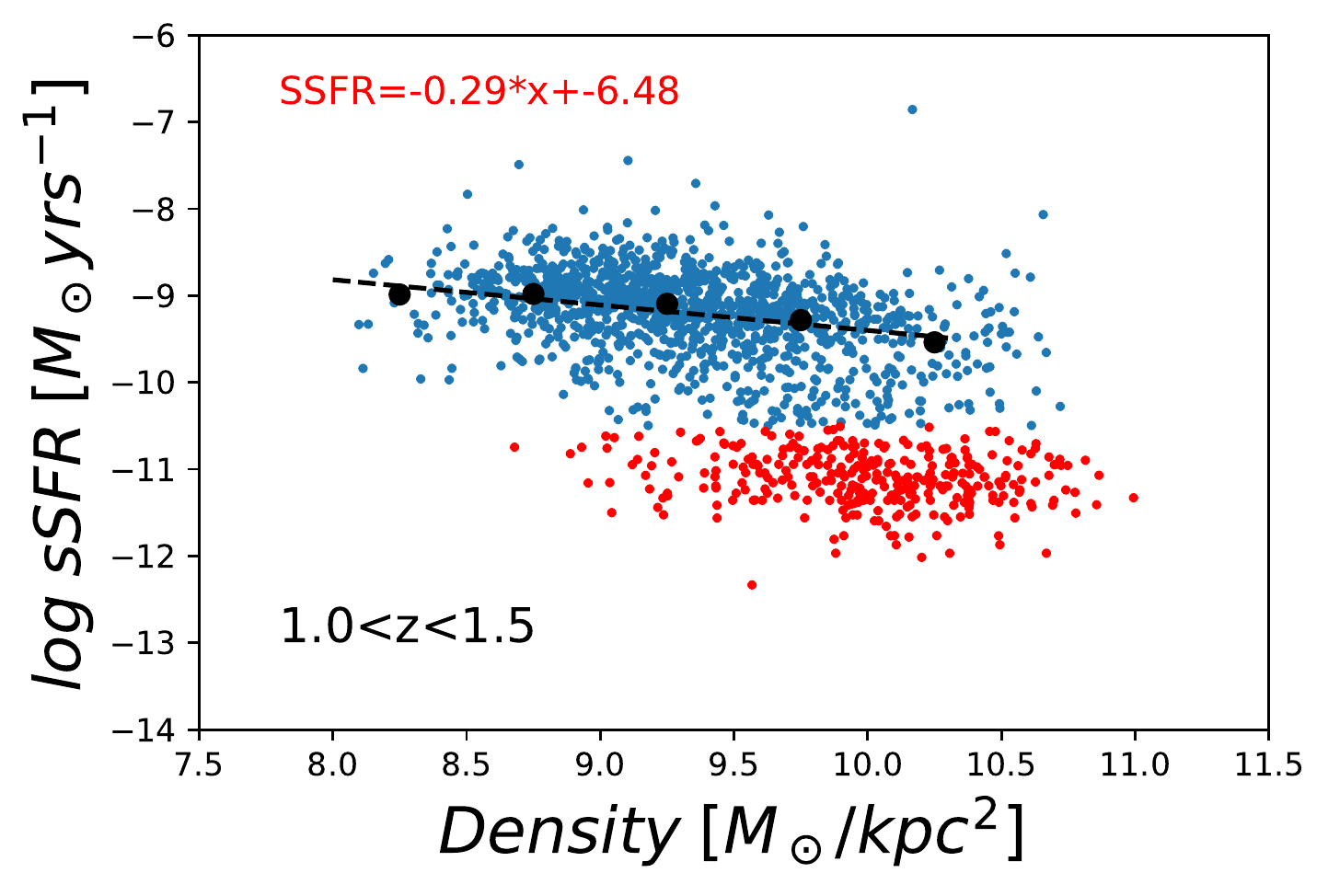}\\
\includegraphics[width=0.45\textwidth]{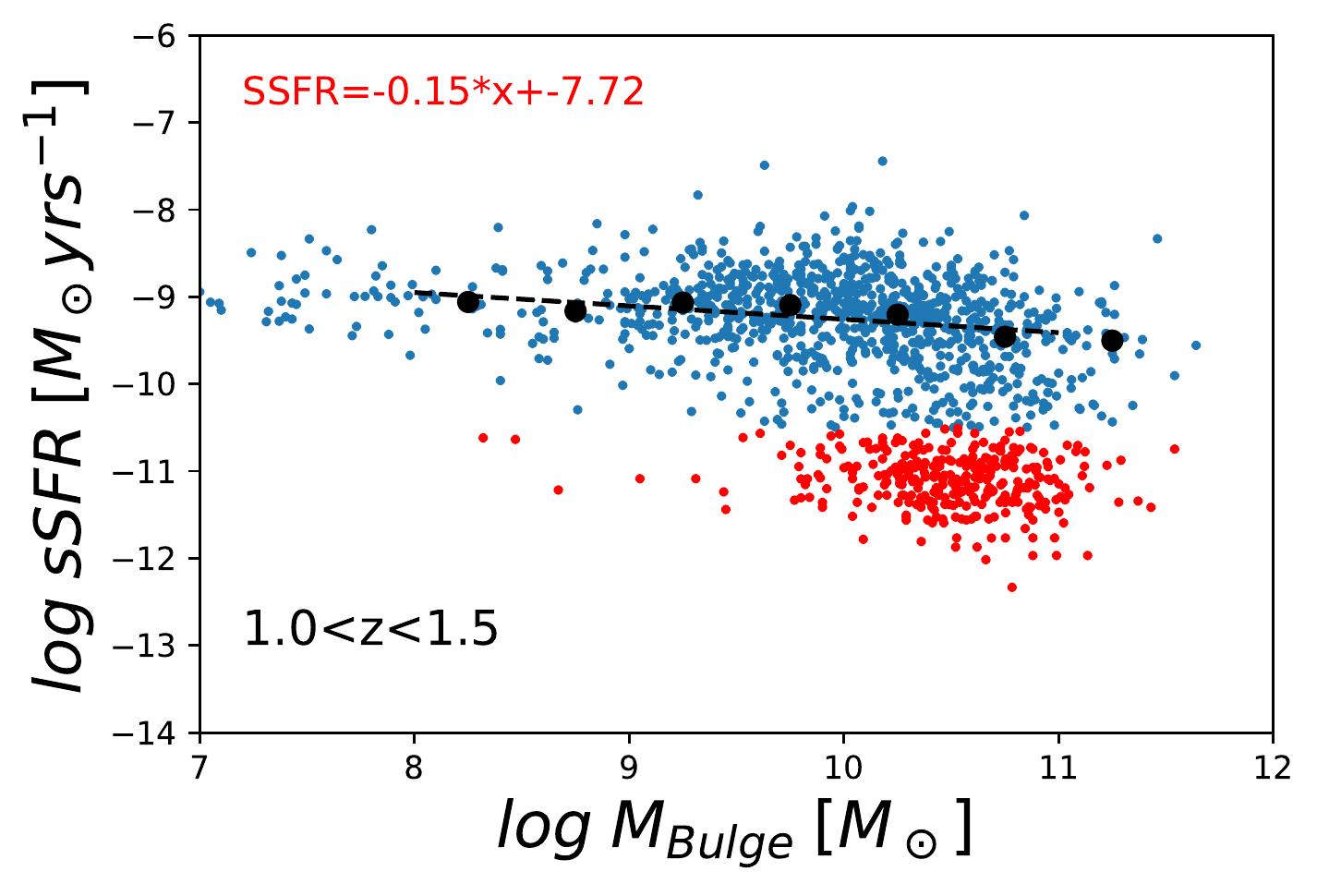}&
\includegraphics[width=0.45\textwidth]{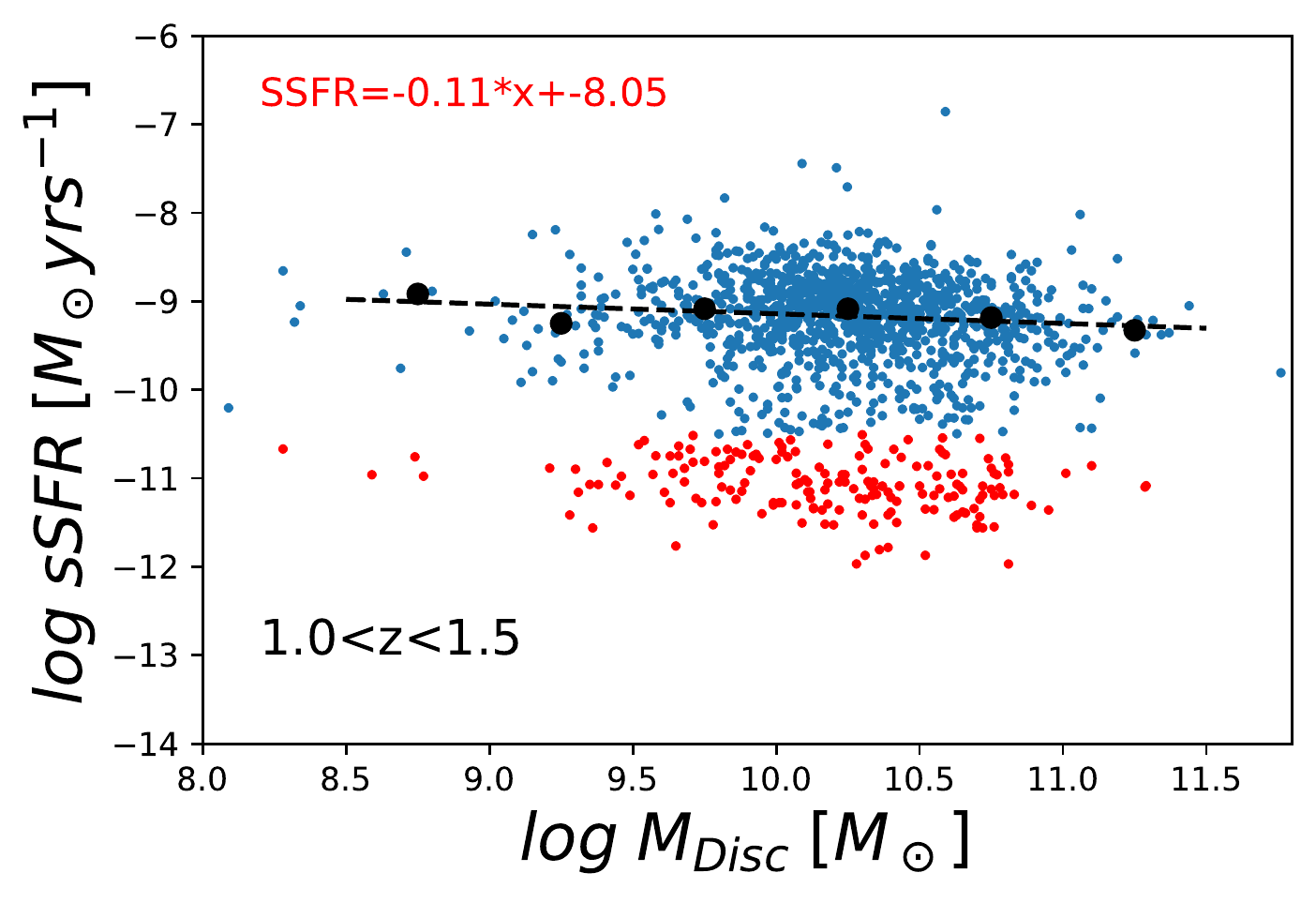}\\
\end{array}$
\caption{Star formation-morphology-quenching. Same as the previous Figure.}
\label{fig:quenching4}
\end{figure*}

\begin{figure*}
\centering
$\begin{array}{c c}
\includegraphics[width=0.45\textwidth]{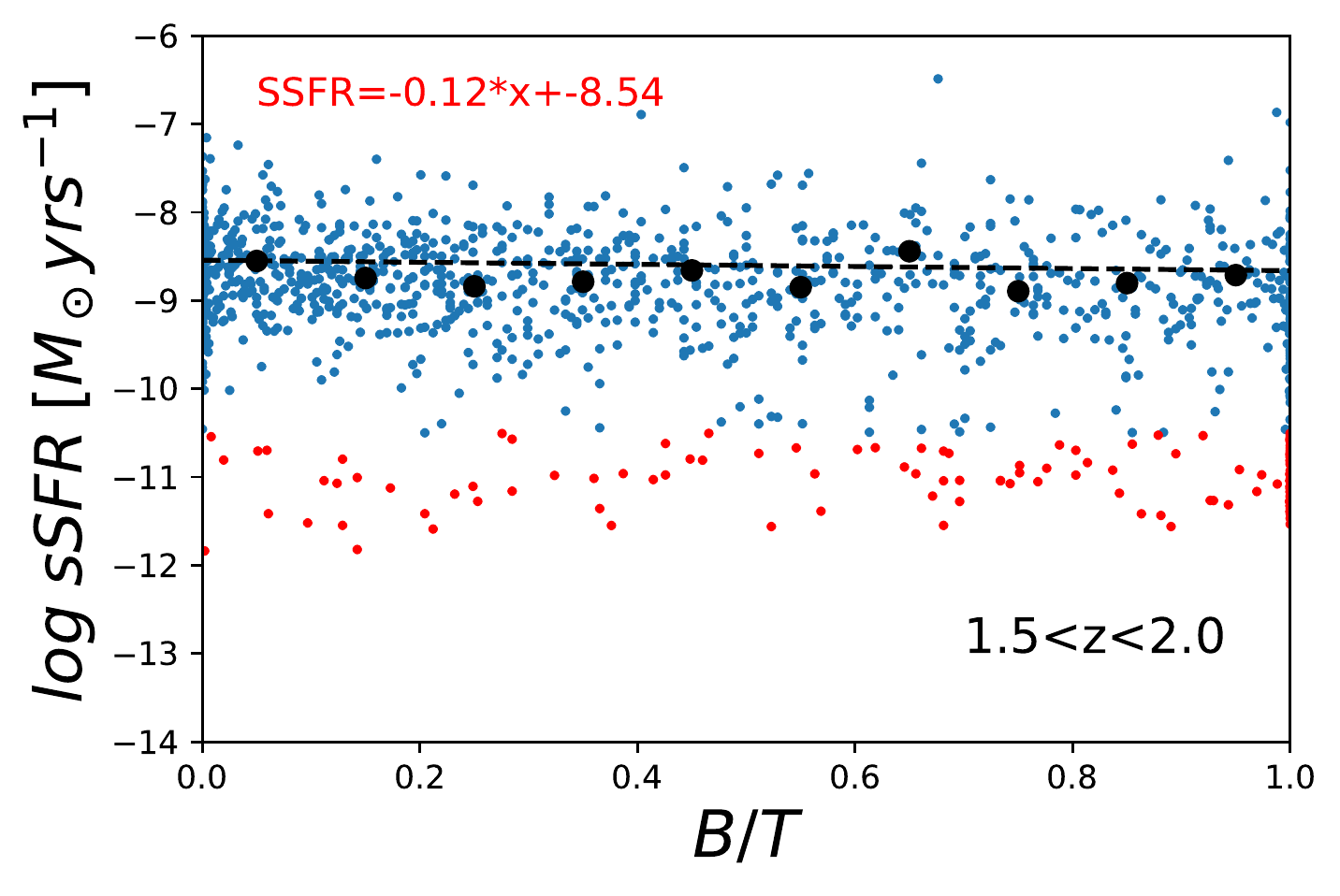}&
\includegraphics[width=0.45\textwidth]{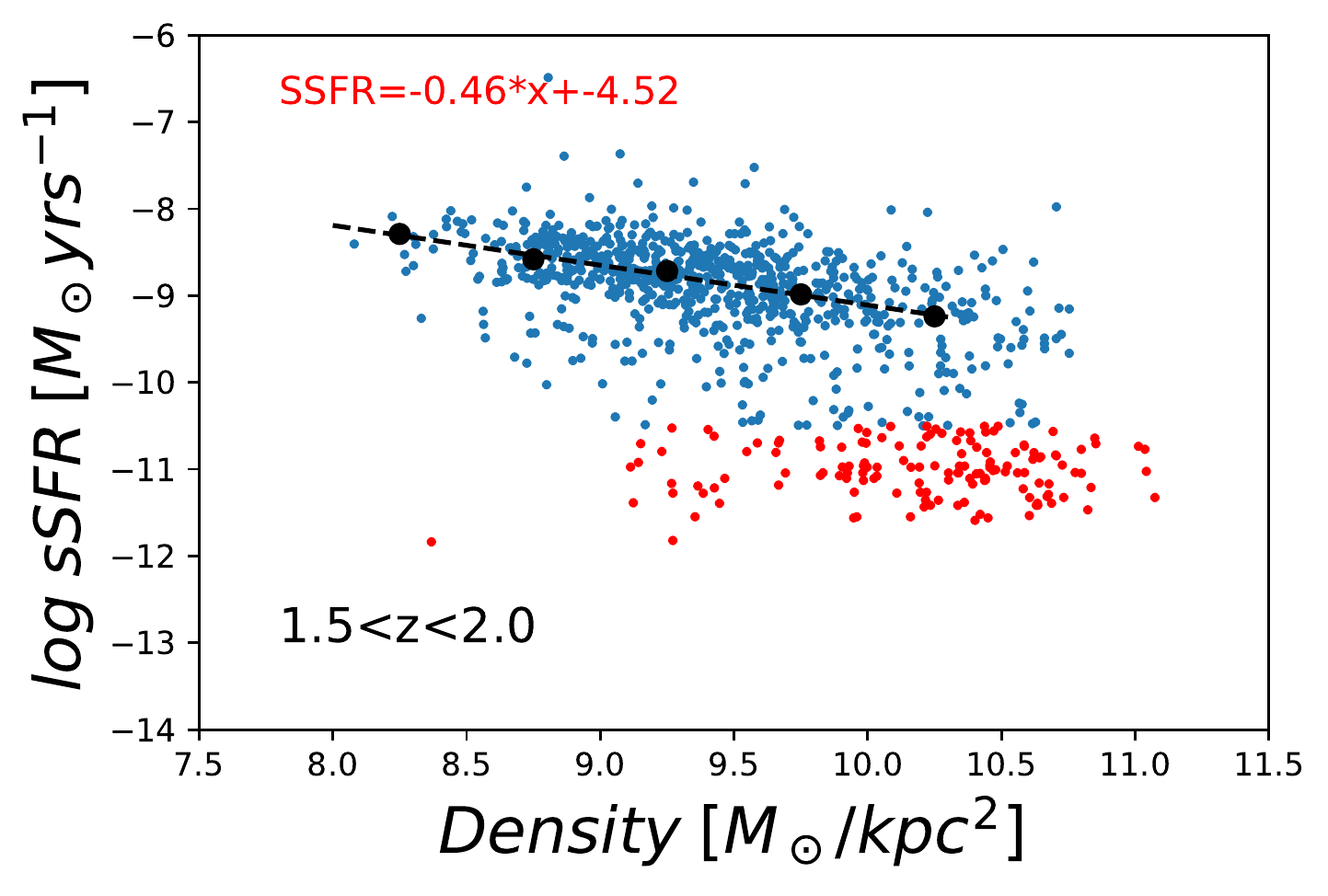}\\
\includegraphics[width=0.45\textwidth]{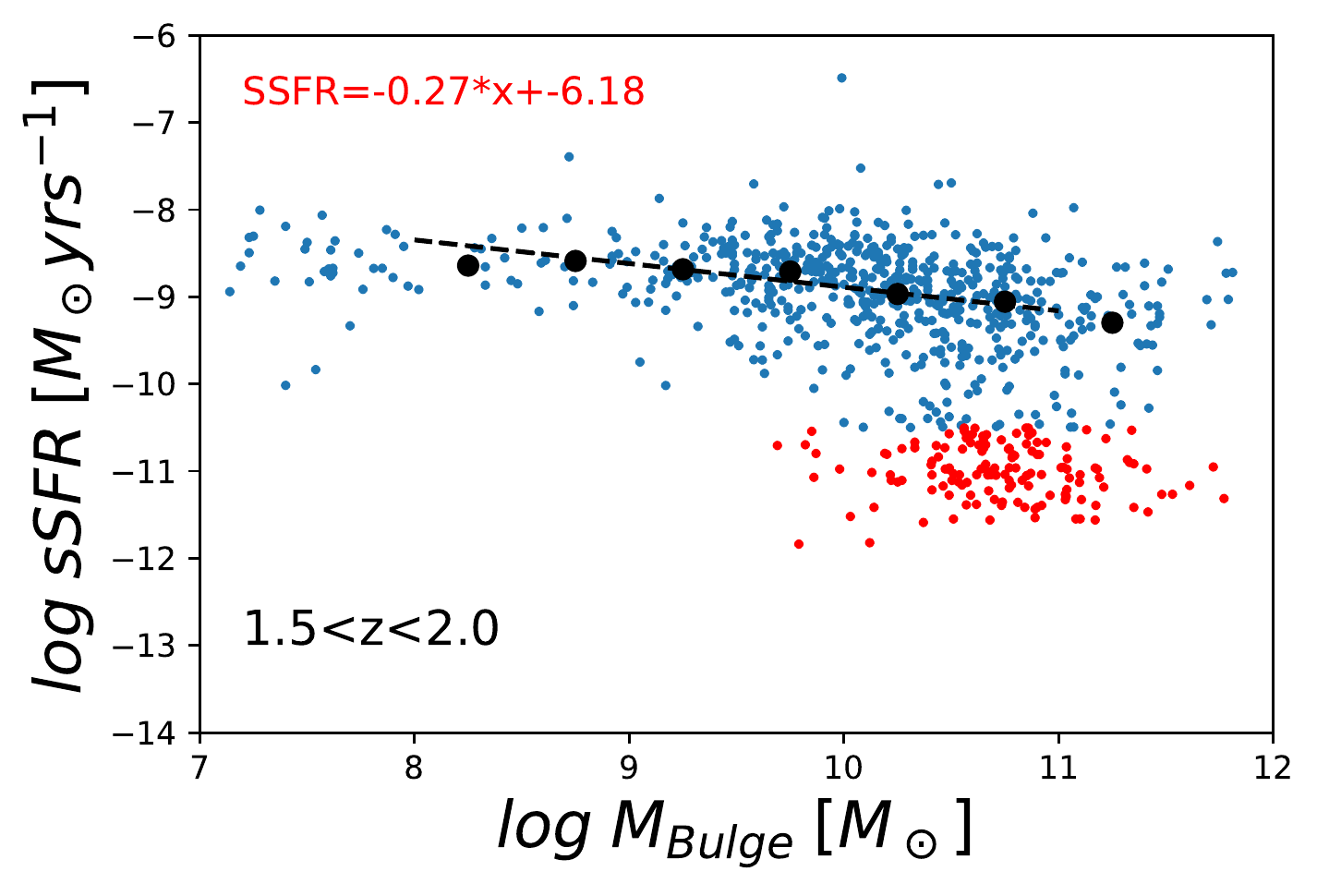}&
\includegraphics[width=0.45\textwidth]{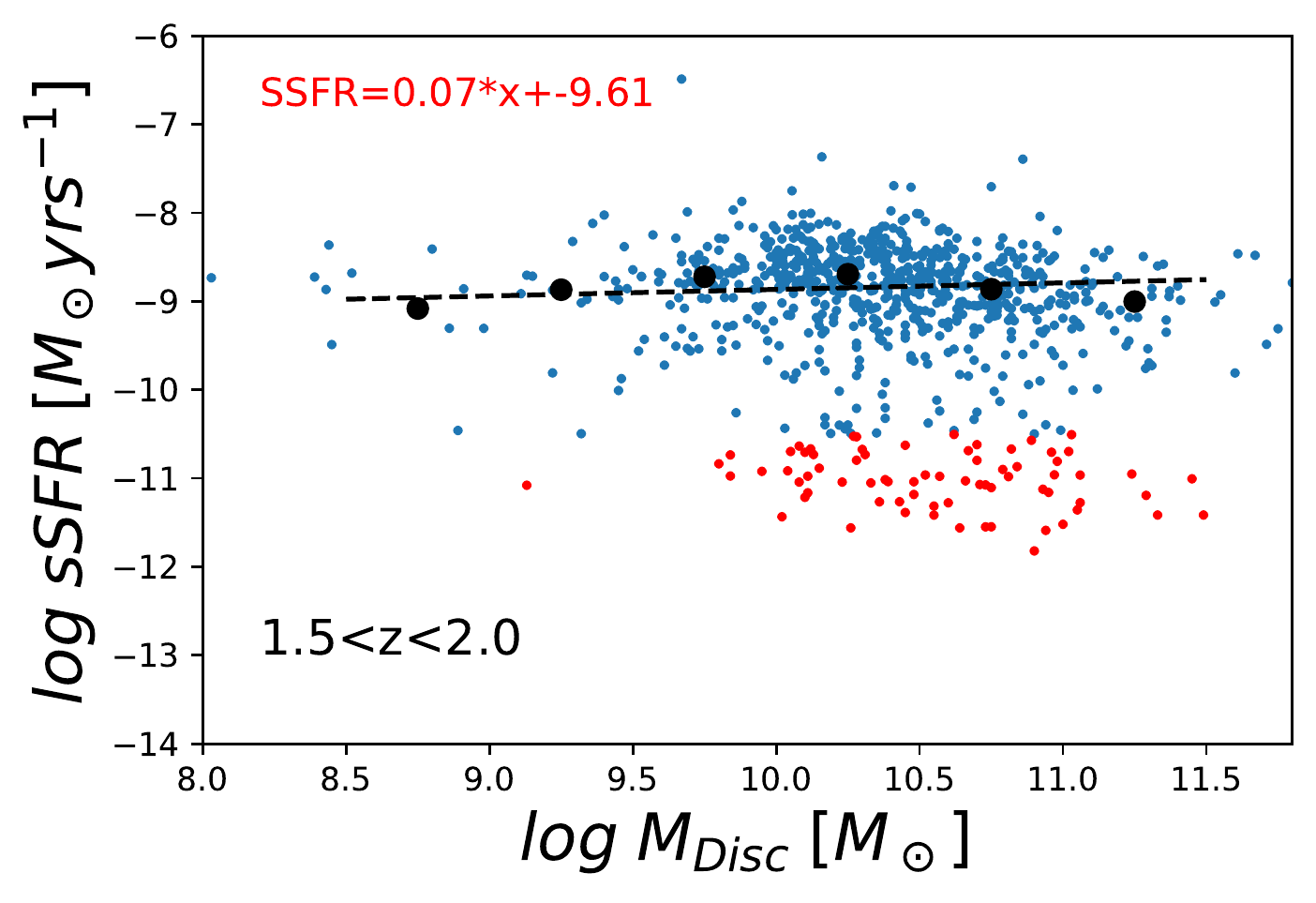}\\
\end{array}$
\caption{Star formation-morphology-quenching. Same as the previous Figure.}
\label{fig:quenching5}
\end{figure*}

In this section we aim to explore in details the correlations between the sSFR and the structural poperties of galaxies in order to understand which parameter is the one that strongly affect the star-formation activity of galaxies. For this reason, the analysis is focused on the star forming population only. The ensamble of Figures \ref{fig:quenching2},\ref{fig:quenching3},\ref{fig:quenching4},\ref{fig:quenching5} shows the distribution of the sSFR against the \BTM, Stellar mass density (D), $M_{Bulge}$ and $M_{Disc}$, at each redshift bin covered by the sample used in this work. Red numbers in each panels report the results of the linear best-fit that is done only on the star forming population. 

\section{Red discs and blue bulges}
\label{Ap:RedBlue}
\begin{figure*}
\centering
$\begin{array}{c c}
\includegraphics[width=0.5\textwidth]{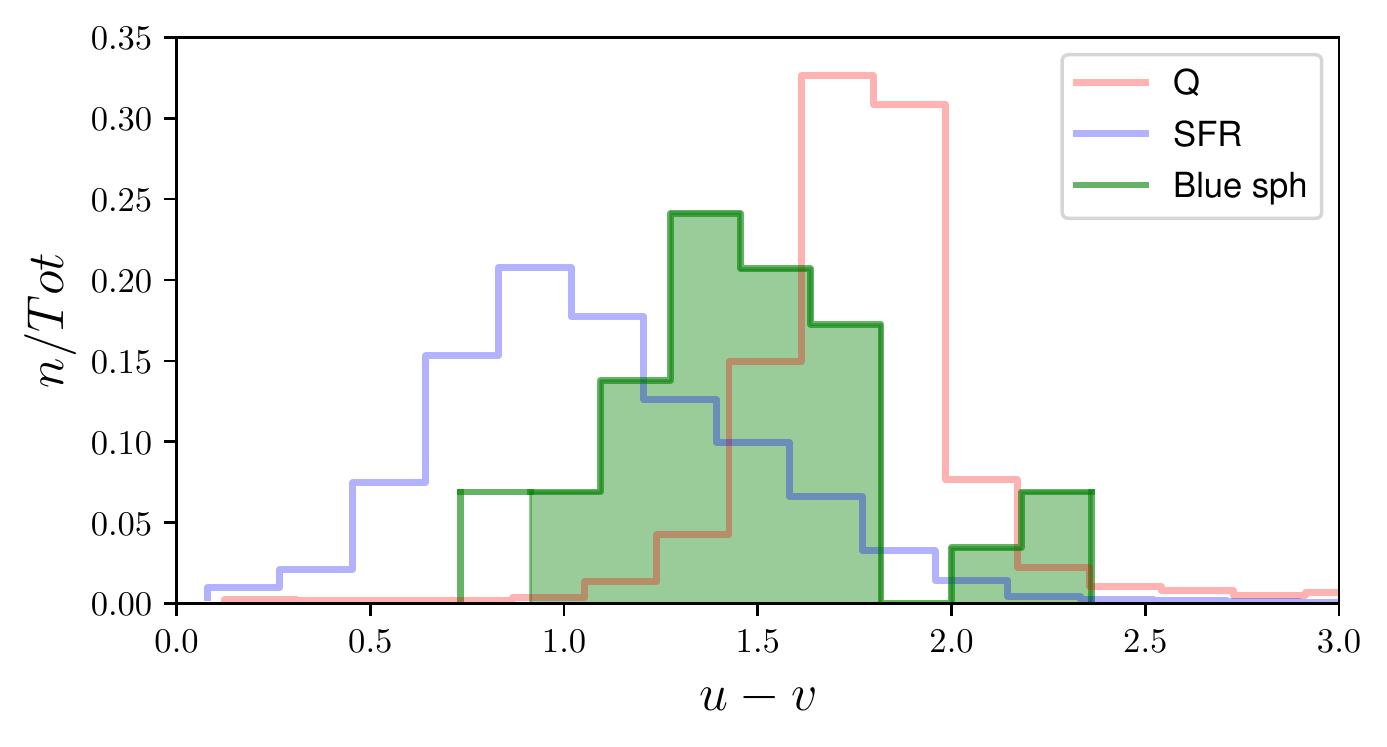}&
\includegraphics[width=0.495\textwidth]{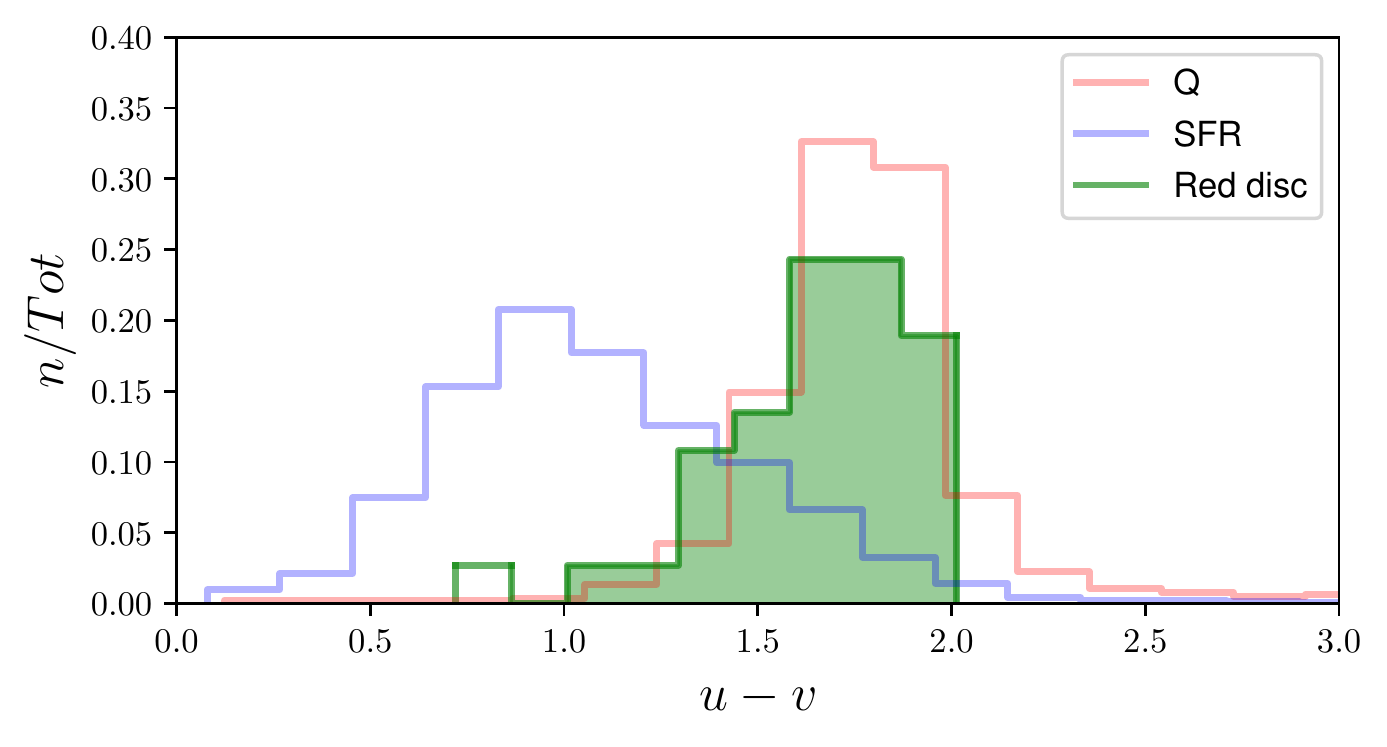}\\
\includegraphics[width=0.485\textwidth]{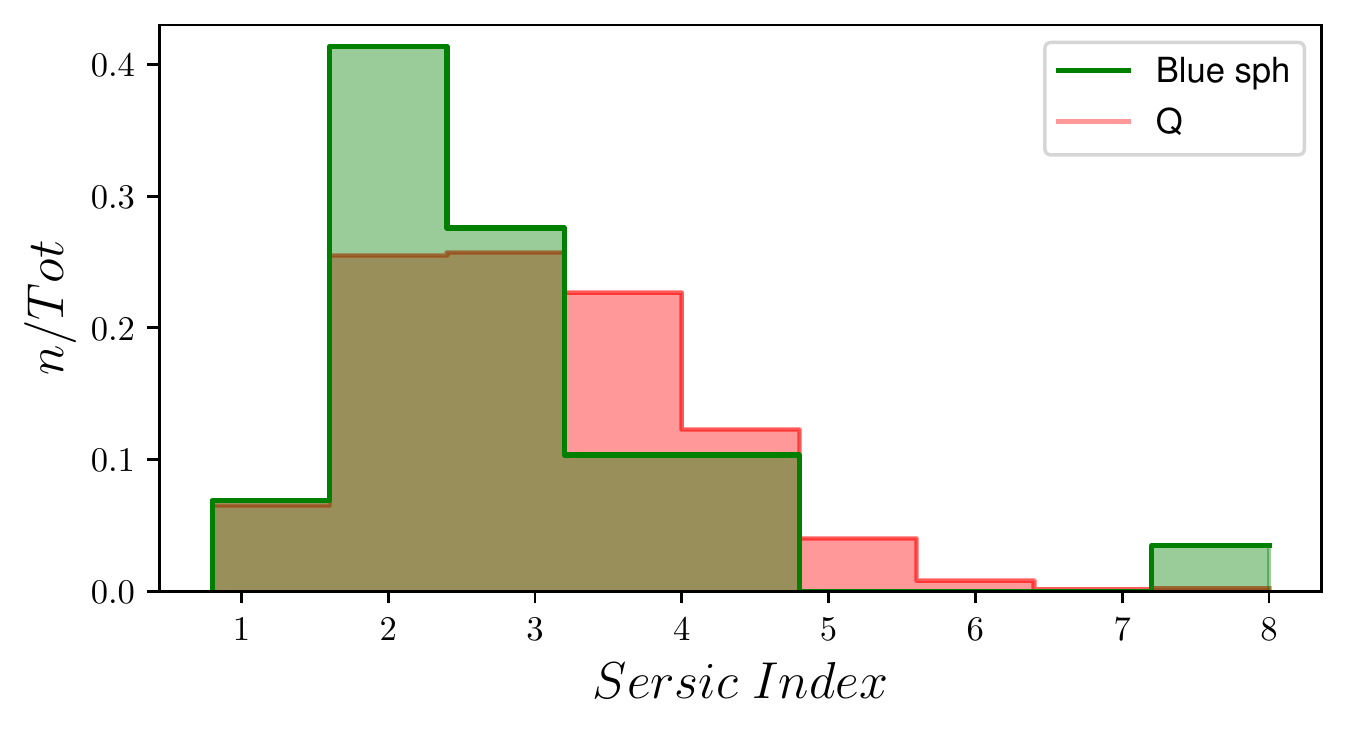}&
\includegraphics[width=0.5\textwidth]{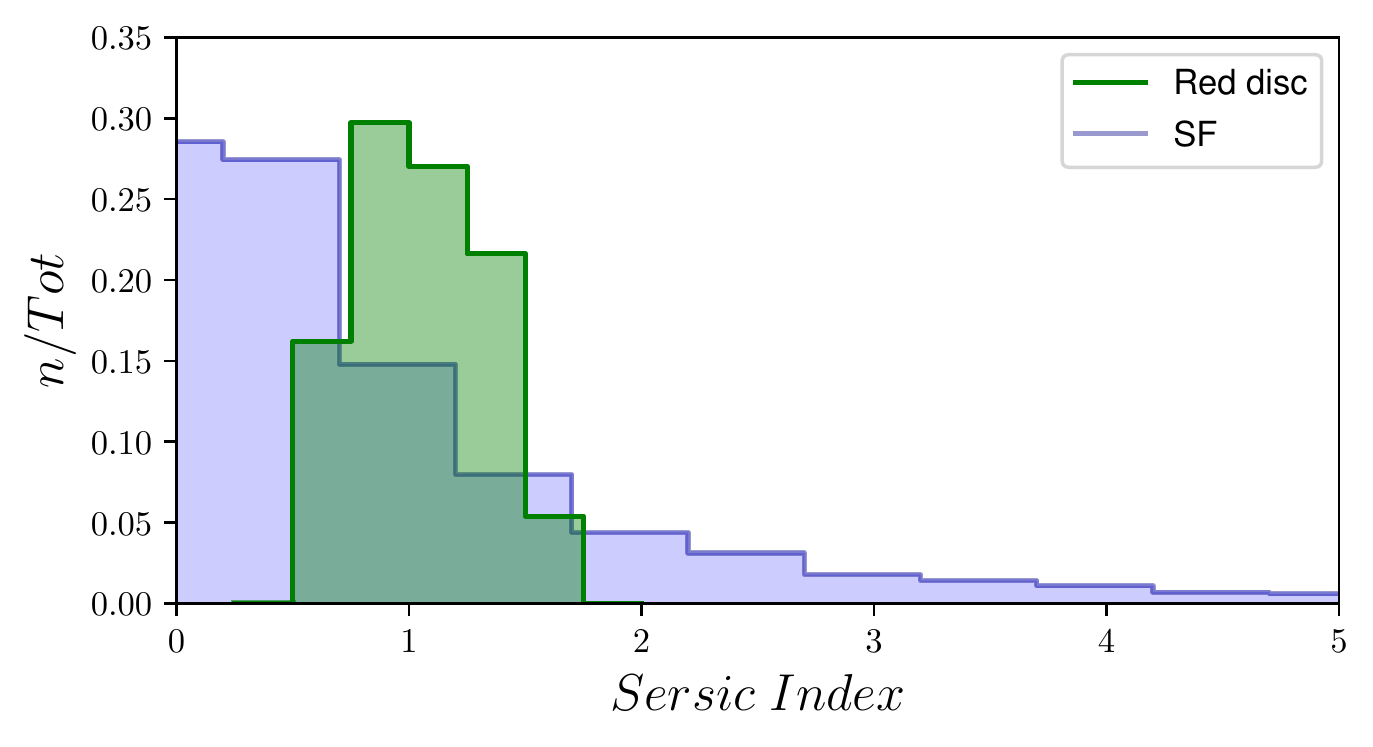}\\
\end{array}$
\caption{Properties of 'blue' bulges and 'red' discs are shown respectivevly on the left and right colum of panels. Top panels: distribution of the the U-V rest frame colors. Bottom panels: distribution of the \Sersic index from the single component fit. As a comparison colors and  \Sersic index of the SF(blue) and Q(red) populations are shown. In those latter cases properties of the surface brightness profile of the whole galaxy are shown.}
\label{fig:sph_disk}
\end{figure*}

In this section we analyze in details stellar and structural properties of the two sub-samples of galaxies that are composed by: - blue bulges or elliptical, - red discs. Left column panels of Figure \ref{fig:sph_disk} show distribution of U-V rest-frame colors and the sersic index of \emph{blue spheroids} and blue bulges. Same analysis for \emph{red discs} is reported on the right columns sequence of plots. It is interesting to notice that \emph{blue spheroids} have a color distribution that is in between the star-foming and the quiescent populations, while their structure(sersic index) is similar to the elliptical populations. Otherwise, the subsample of the red discs, as expected, follows the same sersic index of spirals and pure disks, being passive and showing red rest-frame colors as the quenced population.

\begin{figure}
$\begin{array}{c}
\includegraphics[width=0.41\textwidth]{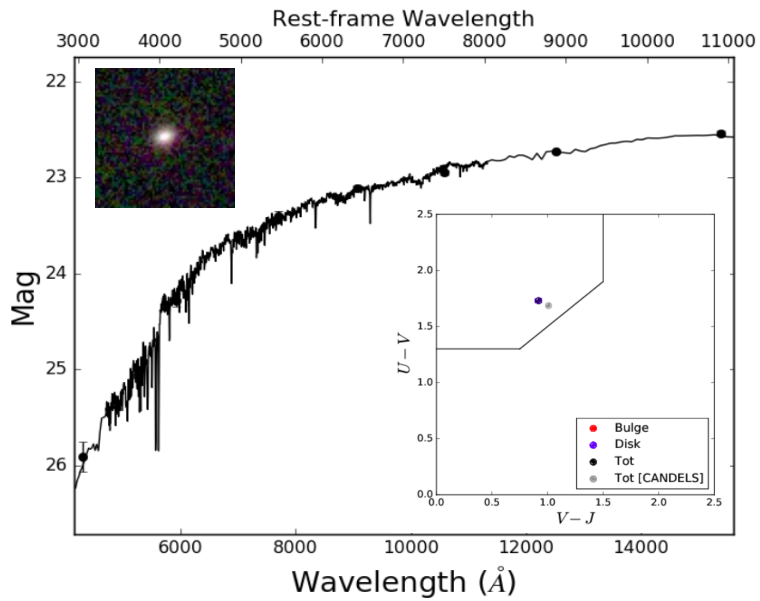}\\
\includegraphics[width=0.4\textwidth]{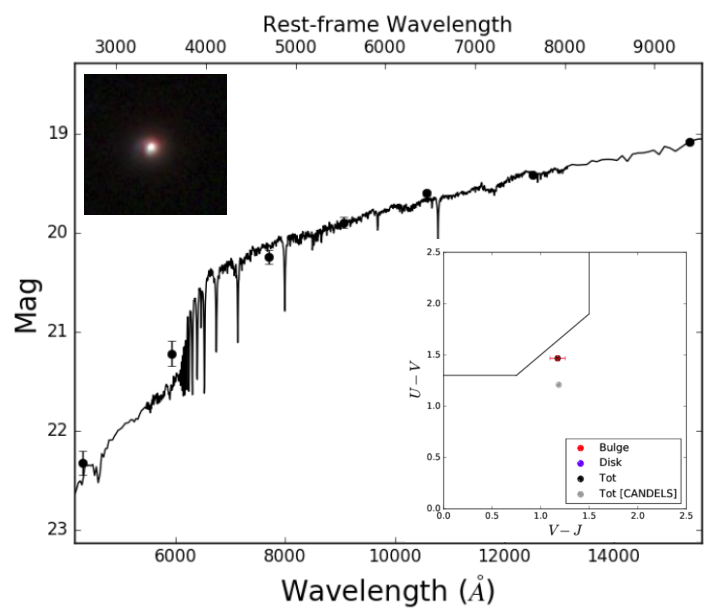}\\
\centering
\includegraphics[width=0.4\textwidth]{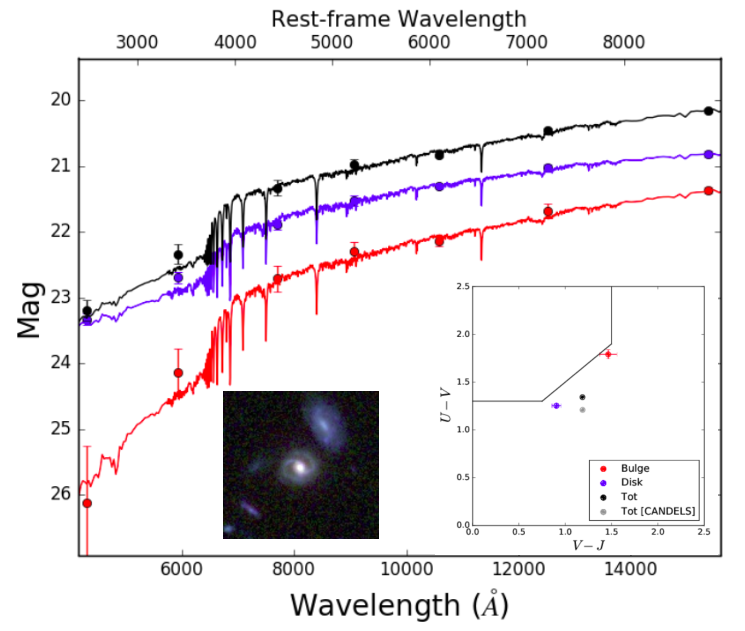}
\end{array}$
\caption{Top panel: red disky galaxy. No relevant bulge is detected by the machine learning classification and from the single model fit. The main plot shows the resulting  behaviour of the spectral energy distribution (theoretical behaviour retrieved with FAST). Black points are the seven bans from CANDELS. The right corner plot shows the position of the galaxy in the UVJ rest-frame colors plane. The ensamble of information confirms the passive nature of the galaxy.
Middle panel: blue spheroid. The SED shows the typical shape of a star forming galaxy, in agreement with the galaxy position in the star forming cloud in UVJ color plot. 
Bottom panel: galaxy hosting a blue bulge. Blue and red lines/points represent respectively disc and bulge photometry. The same color legend is applied in the UV,VJ rest-frame colors plot. }
\label{fig:example}
\end{figure}

\section{Robustness of the U,V,J rest-frame colors estimation}
\label{apx:UVJ}

\begin{figure}

$\begin{array}{c}
\includegraphics[width=0.42\textwidth]{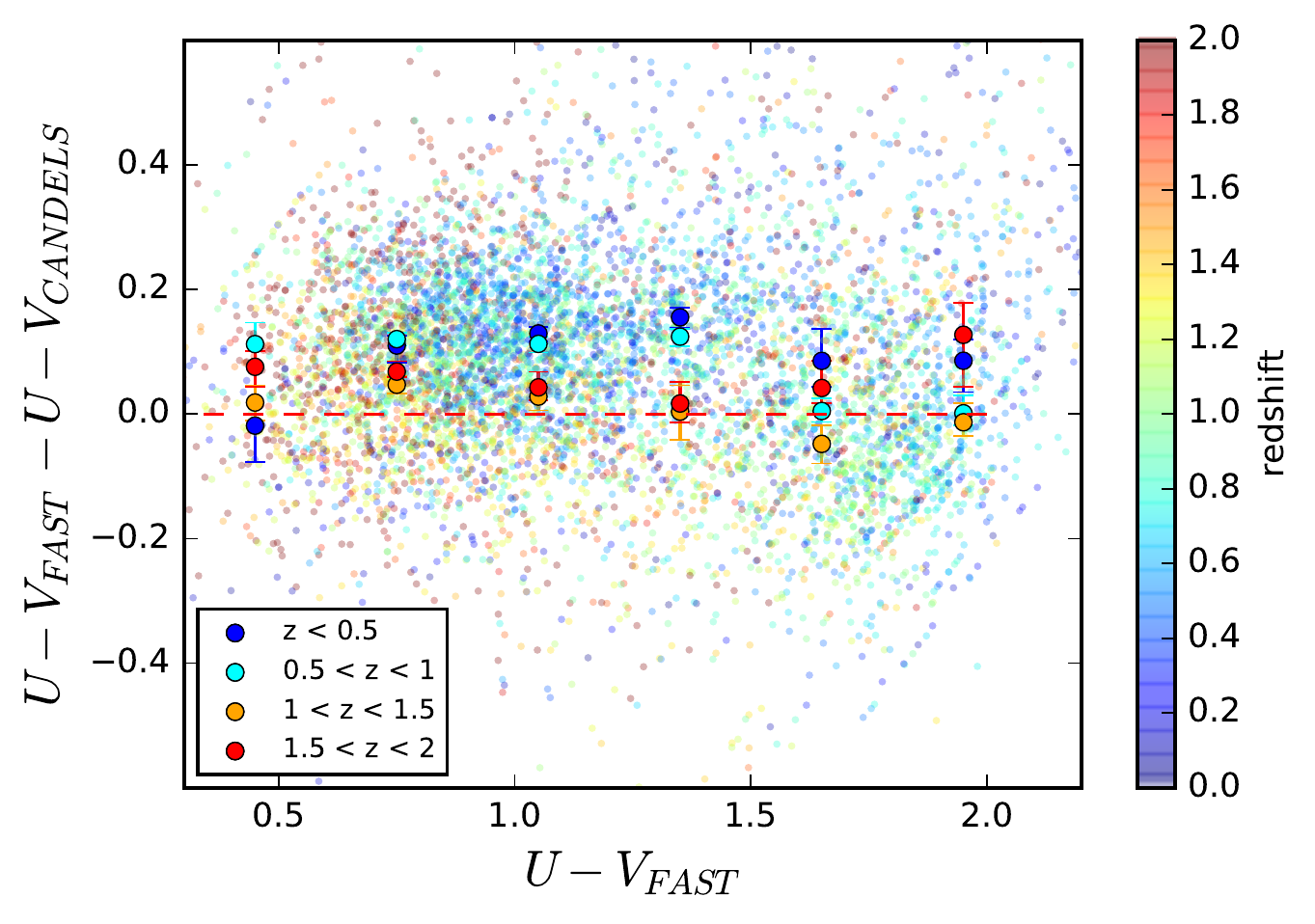}\\
\includegraphics[width=0.42\textwidth]{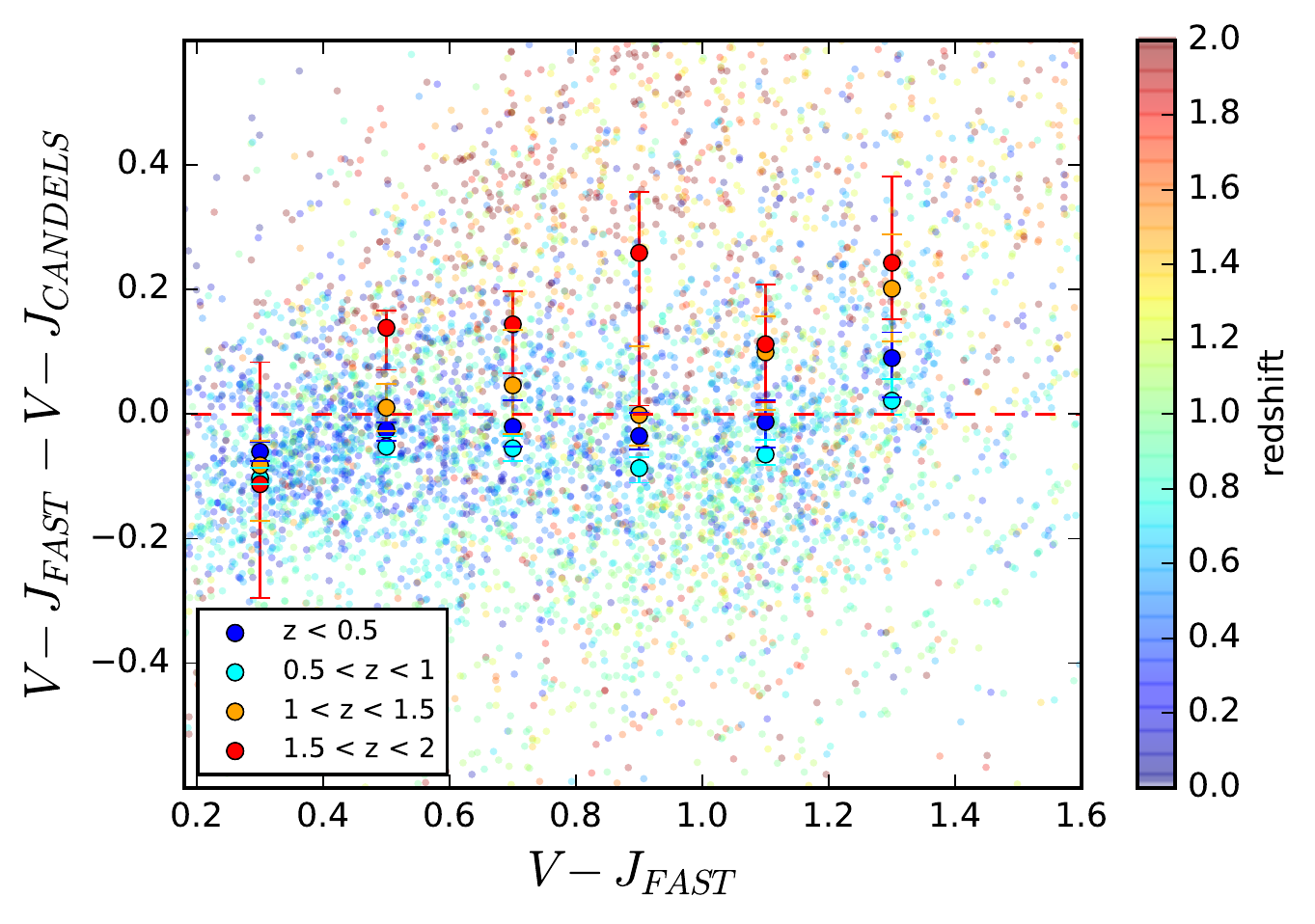}\\
\includegraphics[width=0.42\textwidth]{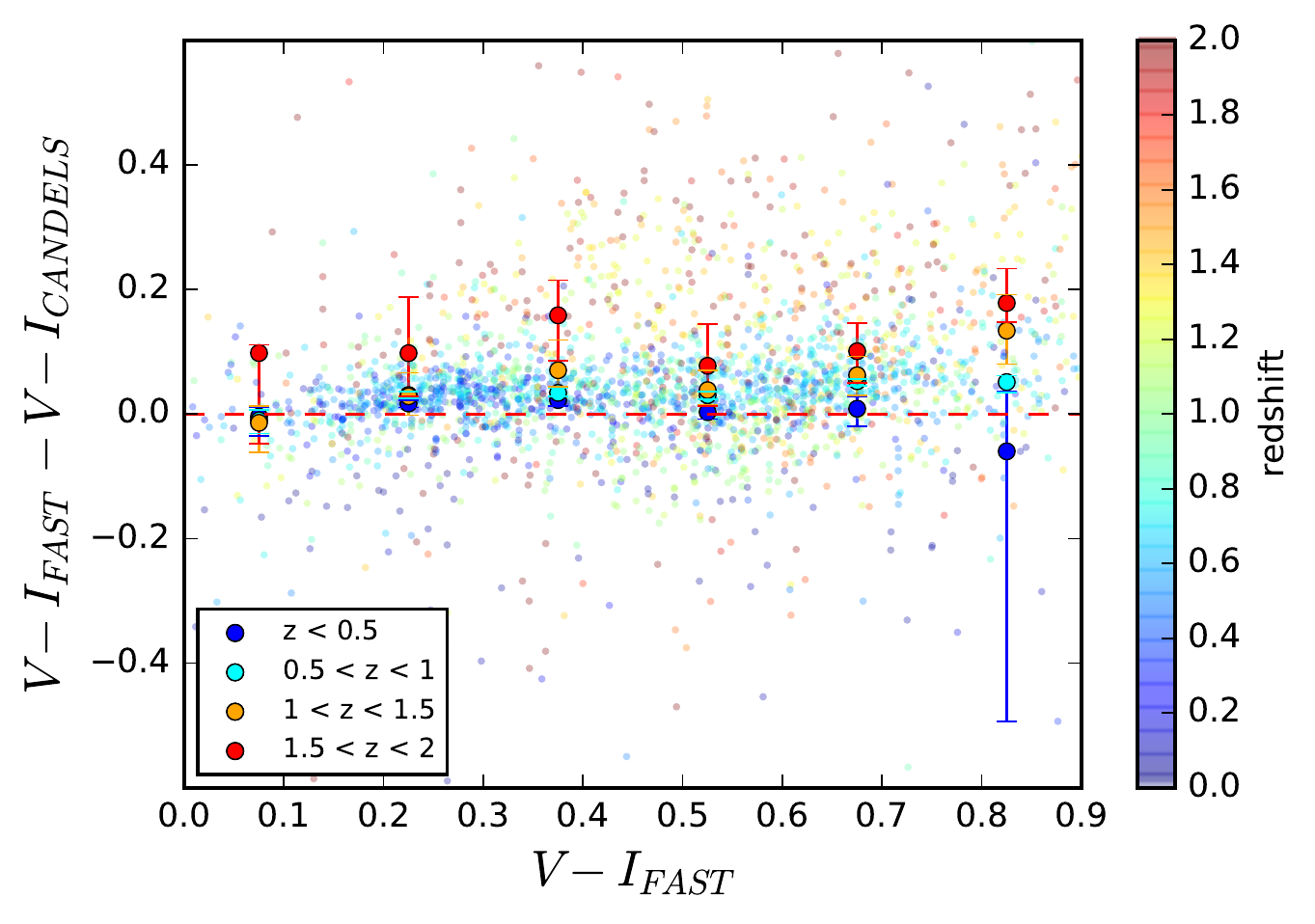}\\
\end{array}$
\caption[UVJ test]{U-V,V-J rest-frame colors comparison between CANDELS and DM18 catalogs. The mean bias is <0.1 dex and the scatter is (..) }
\label{fig:UVJ_test}
\end{figure}

To test the goodness of the colors estimation, we compare results with the rest-frame colors from the official CANDELS catalog, as it is shown in Figure \ref{fig:UVJ_test}. The label $CANDELS$ and $FAST$ are meant to identify respectively colors taken from the CANDELS catalog and the one derived from the present analysis. The median scatter is lower then <0.1 dex. However, the dispersion increases for $z>1$.
The J band rest-frame, in this redshift range, falls around $3600$ nm while the last observed filter is the H-band (1600 nm). Outside the wavelength coverage of the dataset, the theoretical SED can introduce errors cause since it is not well constrained by observations. To solve this issue, we used an alternative color, replacing the J band with a bluer one, the I band. \citealp{Wang2017}, show that the U-V, V-I plane has the same properties and presents similar behavior for SFR and extinction coefficient as the standard U-V,V-J. Consequently, it can be used, at high redshift, to disentangle star forming and quiescent galaxies. The last plot of Figure \ref{fig:UVJ_test} shows the comparison with the new color combination. The dispersion factor, in the latest redshift bin, is smaller and of the order of 0.1 dex.

\bsp	
\label{lastpage}
\end{document}